\documentclass[openright]{book}

\usepackage{amssymb}

\usepackage{amsmath}

\usepackage{graphicx}

\usepackage{listings}
\begin{document}

\frontmatter

\begin{titlepage}
\begin{center}
{{\Large{\textsc{Alma Mater Studiorum\\Universit\`a degli Studi di
Bologna}}}} \rule[0.1cm]{13cm}{0.1mm}
\rule[0.5cm]{13cm}{0.6mm}
\end{center}
\vspace{2mm}
\begin{center}
\Large{\bf Dottorato di Ricerca in Fisica}
\end{center}
\begin{center}
\Huge{\bf Entanglement Entropies in One-Dimensional Systems}\\
\end{center}
\vspace{1mm}
\begin{center}
{\bf Settore Concorsuale di Afferenza: 02/A2\\
Settore Scientifico Disciplinare: FIS/02\\
Coordinatore: Prof. Fabio Ortolani}
\end{center}
\vspace{2mm}
\begin{center}
\large{\bf Presentata da:}\\
\LARGE{\bf Luca Taddia}
\end{center}
\vspace{20mm}
\par
\noindent
\begin{minipage}[t]{0.5\textwidth}
{\large{\bf Relatrice:\\
Prof.ssa Elisa Ercolessi
}}
\end{minipage}
\hfill
\begin{minipage}[t]{0.5\textwidth}\raggedleft
{\large{\bf Co-relatori:\\
Prof. Germ\'an Sierra\\
\&\\
Dr. Marcello Dalmonte\\}}
\end{minipage}
\newline
\vspace{5mm}
\begin{center}
\rule[0.1cm]{13cm}{0.1mm}
\rule[0.5cm]{13cm}{0.6mm}
\end{center}
\begin{center}
{\large{\bf Esame Finale - Ciclo XXV\\
Anno 2013}}
\end{center}
\end{titlepage}

\titlepage

\tableofcontents

\mainmatter


\chapter{Introduction}

In the last fifty years, condensed-matter physics, i.e., the branch of physics studying the ways matter organizes, both from a macroscopical and a microscopical point of view, has become one of the most popular fields of research in physics. Actually, because of the broadness of the subject, it is difficult even to give a comprehensive list of the phenomena that condensed-matter physicists study: this list includes, among the others, solid-state physics \cite{AshcroftMermin1976}, superconductivity and superfluidity \cite{Annett2004}, Bose-Einstein condensation \cite{PitaevskiiStringari2003}, high-temperature superconductivity \cite{Schrieffer2007}, quantum Hall effect \cite{PrangeGirvin1987} and other exotic states of matter \cite{Csontos2010}. In the nineties of the last century, the field underwent an impressive boost, because of to the discovery of new quantum-optical technologies, allowing the realization of systems exhibiting such phases in a fully controllable way \cite{BlochDalibardZwerger2008}, and even of new systems, especially in low spatial dimensions \cite{HHMDRGDPN2010}.

From a microscopical point of view, all such system are described by Hamiltonians depending on a set of parameters. For different values of such parameters, the Hamiltonian predicts different behaviors; the value of the parameters for which there is a change in these behaviors are known as {\it phase transitions}, and the physics is there particularly interesting \cite{MorandiNapoliErcolessi2001}. The detection of phase transitions has always been one of the main goals of condensed matter physicists; as we will see in section \ref{QPT}, a lot of information about phase transitions at low finite temperature can be deduced by the corresponding  behavior at $T=0$, where quantum fluctuations dominate over the thermal ones \cite{Sachdev2011, Vojta2003}. Beside this general fact, people are interested in understanding what kind of physics those phase transitions display, since, even when occurring in the same system, they may have very different observable behaviors, as it happens for the spin-1/2 XY chain (see section \ref{XY}, and reference \cite{Franchini2011}).

Standard techniques for the detection of phase transitions involve, in the quantum case, the estimate of the energy of the ground and first excited states, and of correlation functions \cite{Giamarchi2003}. In the last ten years, a large amount of interest has been devoted to the use of quantum-informatic concepts \cite{NielsenChuang2000}, as concurrence \cite{Wootters1998}, fidelity \cite{ZanardiPaunkovic2006}, and {\it entanglement entropies}, to get information about the physics of a system, critical or not \cite{AFOV2008}. A special role in these studies has been played by R\'enyi entanglement entropies (see \cite{CalabreseCardyDoyon2009} and references therein), since they have been shown to contain information about the global correlations of the system. In particular, in one-dimensional conformal systems \cite{DiFrancescoMathieuSenechal1997}, they have been shown by Calabrese and Cardy, in a celebrated work, to display a universal behavior \cite{CalabreseCardy2004}, immediately linkable to the universality class of the problem. The power of these quantities has been enhanced by their easy accessibility from numerical techniques, in particular from density-matrix-renormalization-group computations \cite{White1992, White1993}, especially when the system has just one spatial dimension.

However, even when the critical system has one spatial dimension, the Calabrese-Cardy formula for entanglement entropies is not the end of the story, for many reasons. The first is that not all critical systems are conformal, and in such cases entanglement entropies have been shown to display very peculiar behaviors \cite{FIJK2007,EEFR2011}. Moreover, being conformal field theory just an effective description of the low-energy physics of the system, a plenty of corrections to the conformal behavior of R\'enyi entanglement entropies arises, both due to finite-size effects (if the system is finite) \cite{LSCA2006,CCEN2010} and to lattice effects (if the system is defined on a lattice, as it usually happens for numerical studies) \cite{BerganzaAlcarazSierra2012}.

The emergence of such corrections can be a bother or an advantage, depending on the ignorance or not of their analytical form, when trying to estimate physical quantities from numerical simulations. In the first case, the unknown corrections behave as a noise, typically worsening the quality of the fits \cite{Xavier2010}. The problem can be easily solved when a formula is given for such corrections: if this is the case, the estimation can still be done with good results and, if the correction contains some additional information about the physics of the systems, it can be used to understand something more about it. It is therefore very important, for these two reasons, to be able to analytically compute such corrections.

This thesis is divided in 3 main chapters. In chapter \ref{prel}, we briefly introduce a number of more or less standard concepts, that we will use in the remaining chapters: they spread from general physical concepts, such as quantum phase transitions (section \ref{QPT}) and Luttinger liquids (section \ref{Bosonization}), mathematical frameworks, such as lattice models (section \ref{lattice}) and conformal field theory (section \ref{CFT}), numerical methods, such as the density-matrix renormalization group algorithm (section \ref{DMRG}), to entanglement entropies (section \ref{Entanglement}). Appropriate references will be given in the corresponding sections.

In chapter \ref{Crossovers} we show how the corrections to the conformal behavior of entanglement entropies can lead, due to the knowledge of their analytical form, to accurate estimations of physical quantities. Based on the work of Calabrese and collaborators \cite{CCEN2010} we develop in section \ref{section method} a new method, based on density-matrix renormalization group simulations, for the detection of the Luttinger parameter of a Luttinger-liquid system. In section \ref{XXZcheck} we check the method against a known model, the critical spin-1/2 XXZ chain, whose Luttinger parameter is known analytically; in section \ref{Hubbardcheck} we test it on a different model, the deep attractive Hubbard model, where the Luttinger parameter is known from numerical exact studies: in any case, we find an excellent agreement of our predictions with the exact results, except for particular situations, that will be diffusely discussed. In section \ref{K_dipoles}, we use our method, and other numerical independent ones, to detect the Luttinger parameter in a critical chain of hard-core dipolar bosons and to estimate the position in the parameter space of a crossover, i.e., of a transition between different critical phases, between a superfluid phase and a charge-density-wave one. In section \ref{K_3-2} we examine the critical properties of the non-integrable spin-3/2 XXZ chain, i.e., central charge, sound velocity and Luttinger parameter, by means of different methods. In the last section, \ref{E3-2}, we study some properties of the entanglement entropies, both for ground and excited states, in the same model. All the work of this section was done in collaboration with M. Dalmonte and E. Ercolessi; the content of sections \ref{section method}, \ref{XXZcheck}, \ref{Hubbardcheck} and \ref{K_dipoles} was published in reference \cite{DalmonteErcolessiTaddia2011}, while the one of sections \ref{K_3-2} and \ref{E3-2} in reference \cite{DalmonteErcolessiTaddia2012}.

In chapter \ref{CFT+OBC} we adapt the conformal-field-theory approach developed by F.C. Alcaraz, M.I. Berganza and G. Sierra in \cite{AlcarazBerganzaSierra2011}, for the computation of the entanglement entropies of excited states, to the ground and first excited state of conformal systems with general conformal boundary conditions. The general formalism is given in sections \ref{replica} and \ref{GCBC}. In section \ref{IsingCFT} we perform the explicit computations in the $c=1/2$ minimal conformal field theory, and we test it against density-matrix renormalization group simulations, using the spin-1/2 critical Ising model; in section \ref{FFCFT} we do the same for the $c=1$ compactified bosonic conformal field theory, where the lattice realization is given by the spin-1/2 XX chain. In any case, we find an excellent agreement between the conformal predictions and the density-matrix renormalization group data, confirming that we are able to analytically compute the entanglement entropies for general conformal boundary conditions. The work was done in collaboration with G. Sierra, F.C. Alcaraz and J.C. Xavier and published in reference \cite{TXAS2013}.

In chapter \ref{conc} we briefly resume the content of the thesis and discuss some possible interesting developments.


\chapter{Preliminaries}\label{prel}

In this chapter, we briefly review what stands at the basis of the original work of this thesis. In section \ref{QPT} we introduce quantum phase transitions, and we motivate the study of critical phases at $T=0$. In section \ref{lattice} we give some examples of one-dimensional lattice models, that will be the subject of study of the remaining of the work. In section \ref{Bosonization} we introduce the concept of Luttinger liquid, a recurrent phase in one dimension, and we briefly describe the field-theoretical structure of such a system. In section \ref{CFT} we review the basic concepts of conformal field theory and of boundary conformal field theory. In section \ref{Entanglement} we introduce R\'enyi entropies and their connection with one-dimensional systems and conformal field theory. Finally, in section \ref{DMRG} we give a description of the numerical method we use to study lattice systems, the density-matrix renormalization group algorithm.

 
\section{Quantum Phase Transitions}\label{QPT}

A generic physical system, microscopically described by a quantum Hamiltonian, can exhibit different kinds of macroscopic behavior, depending on the values of a set of control parameters, like temperature, pressure, etc.. Those behaviors are called {\it phases} \cite{MorandiNapoliErcolessi2001}. If a system can display two or more of such phases, there must exist at least one value of the control parameters that represent a transition between them: we have in this case a {\it phase transition}; when such points organize along a line, we have a {\it critical line}. One can distinguish between different phases by defining, for each one, an {\it order parameter}, i.e., a physical quantity that is non-zero in the corresponding phase and null in the other ones. This concept is related to the one of {\it spontaneous symmetry breaking}: systems that exhibit a symmetry in the Hamiltonian can have a low-energy configuration that explicitly violates the symmetry itself, i.e., they can {\it order}; an order parameter is a mathematical object reflecting this order.

At high temperature the situation is simple, since the phase diagram can be described just by means of classical statistical mechanics \cite{MorandiNapoliErcolessi2001}. The physics of the system is, in this case, governed by {\it classical fluctuations} around the mean value, whose typical energy scale is of order $K_BT$, being $K_B$ the Boltzmann constant and $T$ the temperature. As $T$ lowers, the situation becomes more complicated, since different kinds of fluctuations arise, namely {\it quantum} ones \cite{Sachdev2011,Vojta2003}, which originate from the quantum nature of the system. Their energetic scale is of the order of $\hbar\,\omega$, being $\omega$ the typical low-excitation frequency of the spectrum.

Fluctuations can manifest themselves both in space and in time, on typical scales known as {\it correlation length} $\xi$ and {\it correlation time} $\tau$ \cite{Sachdev2011}. Approaching a critical point, they both tend to diverge as power laws, in a way that is characteristic of the critical point itself. In particular, it is customary to characterize the divergence in the following way: 
\begin{equation}\label{correlation_length_time}
 \xi\sim\left|t\right|^{-\nu},\;\tau\sim\left|t\right|^{-\nu z}
\end{equation}
being $\nu$ and $z$ the {\it correlation-length} and {\it dynamic critical exponents}, and $t$ a control parameter, i.e., the reduced temperature $(T-T_c)/T_c$ in the classical case ($T_c$ is the critical temperature corresponding to the critical point), and something else in the quantum case, e.g., pressure or magnetization.

\begin{figure}[t]
 \begin{minipage}{\textwidth}
  \centering
  \includegraphics[width=0.5\textwidth]{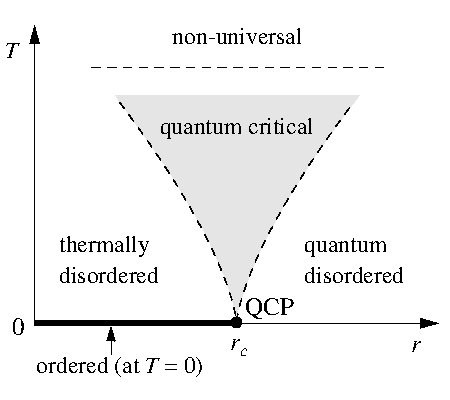}
 \end{minipage}
 \caption{Typical phase diagram of a 3D system. Reprint with permission from \cite{Vojta2003}.}\label{Vojta}
\end{figure}

A typical phase diagram one can encounter, at least in three spatial dimensions (3D), is the one of figure \ref{Vojta}: this is the case, e.g., for a 3D ferromagnet \cite{Sachdev2011}. A critical line is present, separating an {\it ordered} phase from a disordered one. The line ends on the $T=0$ axis, in what is called a {\it quantum critical point}: thermal fluctuations are there completely suppressed and the system is fully quantum. The quantum criticality of course propagates even at finite temperature, in the region where quantum and thermal fluctuations are of comparable importance: this region is called {\it quantum critical}. The boundaries of the quantum critical region are determined by the condition $K_BT\sim\hbar\,\omega$, that, according to (\ref{correlation_length_time}), reads as
\begin{equation}
 K_BT\sim\left|r-r_c\right|^{\nu z}
\end{equation}
following figure \ref{Vojta}'s notation. It is therefore clear that studying the behavior of quantum critical points is not just an academic task, but has a fundamental physical relevance even at the finite-temperature level.

In one spatial dimension (1D), the situation is in general different from the depicted one. The reason is the famous Mermin-Wagner-Hohenberg (MWH) theorem \cite{MerminWagner1966,Hohenberg1967}: in low spatial dimensions, a system with sufficiently short-ranged interactions cannot spontaneously break any continuous symmetry of its Hamiltonian, if not at zero temperature. Therefore, at finite temperature, any true order is forbidden, and an intermediate situation between order and disorder is only possible: it is called {\it quasi-long-range order}, and it is characterized by power-law vanishing spatial correlations \cite{Giamarchi2003}. However, there is something more than the MWH theorem, that constitutes a further peculiarity of 1D: in many cases, long-range-order is forbidden even at $T=0$, due to quantum fluctuations, that become especially important. This is the case for, e.g., Luttinger liquids (see section \ref{Bosonization}), superfluids, antiferromagnets, but {\it not} for ferromagnets (see section \ref{XXZ}). Therefore, since in many cases one cannot define order parameters even at $T=0$, one has to compare directly the decays of different correlation functions to see which one is the slowest decaying, and characterizes the {\it dominant order} of the phase; we will see several examples of this in the following. Just to establish a notation, transition between different dominant orders are called {\it crossovers}: in such situations, no energy gap between the ground and the first excited state opens or closes across the transition. To conclude, the physics of 1D systems at zero temperature is very peculiar, and leads to exotic properties of quantum matter, even at finite temperature.


\section{Lattice Models}\label{lattice}

Lattice models \cite{Takahashi1999}, i.e., models in which degREE's of freedom are not continuous but discrete, are of unvaluable importance in physics, and particularly in condensed-matter and solid-state physics. First of all, they can describe the behavior of real solids: an example is the celebrated Hubbard model (see section \ref{Hubbard}); second, they constitute an ultra-violet regularization of continuous models, both in condensed-matter and high-energy physics, that are often plagued by divergences \cite{Rothe2012}; third, in recent years, optical lattices and ultracold atoms and molecules have allowed physicists to experimentally realize versions of such models in a fully controllable way (see, e.g., \cite{KinoshitaWengerWeiss2004}): they have therefore increased a lot their experimental relevance. In the following, we will consider some of these models, basically the ones we will treat in the rest of the work.


\subsection{Free Fermions}\label{FF}

In this section, we will focus on the Hamiltonian \cite{LiebSchultzMattis1961}
\begin{equation}\label{H_ab}
 H_{(a,b)}=-\frac{1}{2}\left\{\sum_{j=1}^{L-1}\left[c_j^\dagger c_{j+1}+c_{j+1}^\dagger c_j\right]+a\,c_1^\dagger c_1+b\,c_L^\dagger c_L\right\}
\end{equation}
where $c_j,\ j=1,\cdots,L$ is a spinless fermionic annihilation operator and $a,b\in\mathbb{R}$; for each couple $(a,b)$ we have a different model, which we will denote by the couple itself. We are considering, instead of the easier periodic case, the current open Hamiltonian, useful for the considerations of chapter \ref{CFT+OBC}. Few of these models have a simple analytical solution, that we are going to explicit. Before beginning the analysis of the single cases, we note however that $H_{(a,b)}$ commutes with the fermion number operator $\hat{N}=\sum_{j=1}^Lc_j^\dagger c_j$, that is therefore fixed. We will typically work with ground states with exactly $N=L/2$ fermions, the so-called \emph{half-filled} case, for reasons that will be discussed in chapter \ref{CFT+OBC}.


\subsubsection{(0,0) Case}
We will give the solution in this case in some detail, to explain the procedure, that is common to all the discussed ones. Our goal is to write the Hamiltonian in the form
\begin{equation}
 H_{(0,0)}=\sum_m^L\epsilon_md_m^\dagger d_m
\end{equation}
being $d_m$ the diagonalizing fermionic annihilation operator; we call $c$ the column vector having the $c_j$'s as entries, and $d$ the analog for the $d_m$'s. We note that we have not yet specified the range spanned by the $m$'s: it will contain $L$ elements but, in principle, before performing the computation, we cannot know the exact range. What we look for is a square matrix $F$ such that
\begin{equation}\label{d}
 d^\dagger=F\,c^\dagger
\end{equation}
and preserving the fermionic nature of the operators, i.e., we suppose $F$ to be a unitary matrix.

The matrix $F$ is easily found by requiring $\left|m\right>\equiv d_m^\dagger\left|0\right>$ to be an eigenstate of $H_{(0,0)}$ of eigenvalue $\epsilon_m$, $\left|0\right>$ being its vacuum. It is easily found that the entries $F_{mj}$ must satisfy the system of equations
\begin{equation}\label{system}
 \begin{cases}
  -\frac{1}{2}\left(F_{m,j-1}+F_{m,j+1}\right)=\epsilon_mF_{mj}\\
  F_{m0}=F_{m,L+1}=0
 \end{cases}
\end{equation}
with $j\in\{1,\cdots,L\}$.

We solve it by the ansatz
\begin{equation}
 F_{mj}\equiv A\,e^{ip_mj}+B\,e^{-ip_mj}
\end{equation}
where the constants $A$ and $B$ and the momenta $p_m$ have to be fixed by imposing the conditions (\ref{system}). From the second line of (\ref{system}) we find $B=-A$ and
\begin{equation}
 p_m=\frac{\pi\,m}{L+1}\label{momenta}
\end{equation}
with $m\in\mathbb{Z}$, so that we have
\begin{equation}
 F_{mj}=A\,\sin\frac{\pi\,m\,j}{L+1}
\end{equation}
Instead, from the first of (\ref{system}) we find the spectrum:
\begin{equation}\label{spectrum}
 \epsilon_m=-\cos p_m
\end{equation}
We note that the first equation of (\ref{system}) is independent of $(a,b)$: therefore, the spectrum will be in any case of the form (\ref{spectrum}). Moreover, we are now able to choose the values of the $m$'s. Indeed, we have the lowest energy eigenvalue at $m=0\mod L+1$; however, we have to discard it, because it leads to a null $F_{mj}$. Therefore, we can choose the range of the $m$'s as a set of $L$ contiguous integers avoiding $0\mod L+1$: the simplest choice is $\{1,\cdots,L\}$. Finally, $A$ is found by requiring $F$ to be unitary (in this case, orthogonal), i.e.,
\begin{equation}
 \sum_{j=1}^LF^*_{mj}F_{nj}=\delta_{mn}
\end{equation}
translating into $A=\sqrt\frac{2}{L+1}$. We conclude
\begin{equation}
 d_m^\dagger=\sqrt\frac{2}{L+1}\sum_{j=1}^L\sin\frac{\pi\,m\,j}{L+1}c_j^\dagger
\end{equation}
It is straightforward to check, from their explicit form, that the $d_m$'s satisfy fermionic commutation relations, once that the $c_j$'s do. Moreover, we note that $F$ is real, and therefore othogonal rather than unitary.

\begin{figure}[t]
 \begin{minipage}{\textwidth}
  \includegraphics[width=0.5\textwidth]{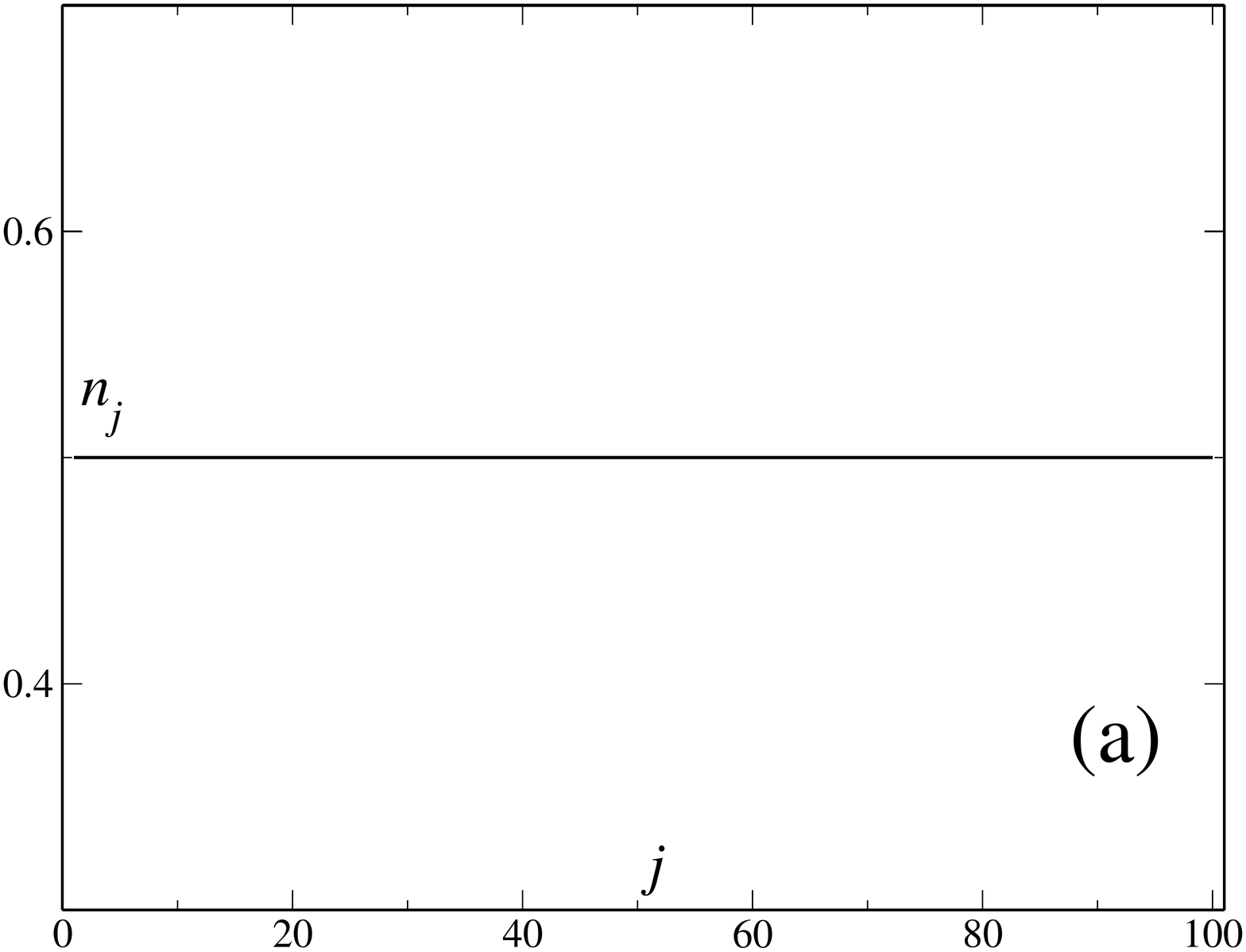}
  \includegraphics[width=0.5\textwidth]{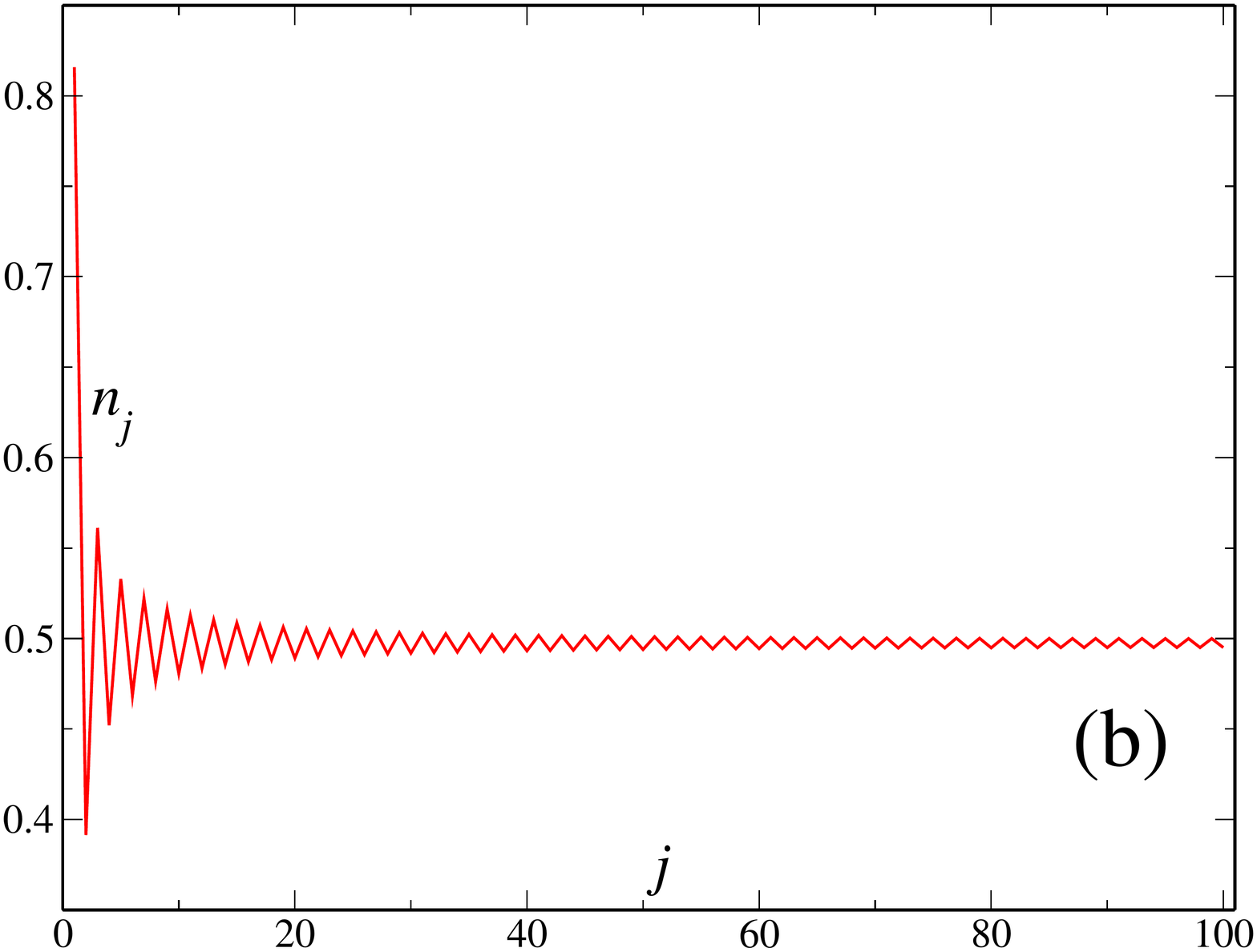}
  \includegraphics[width=0.5\textwidth]{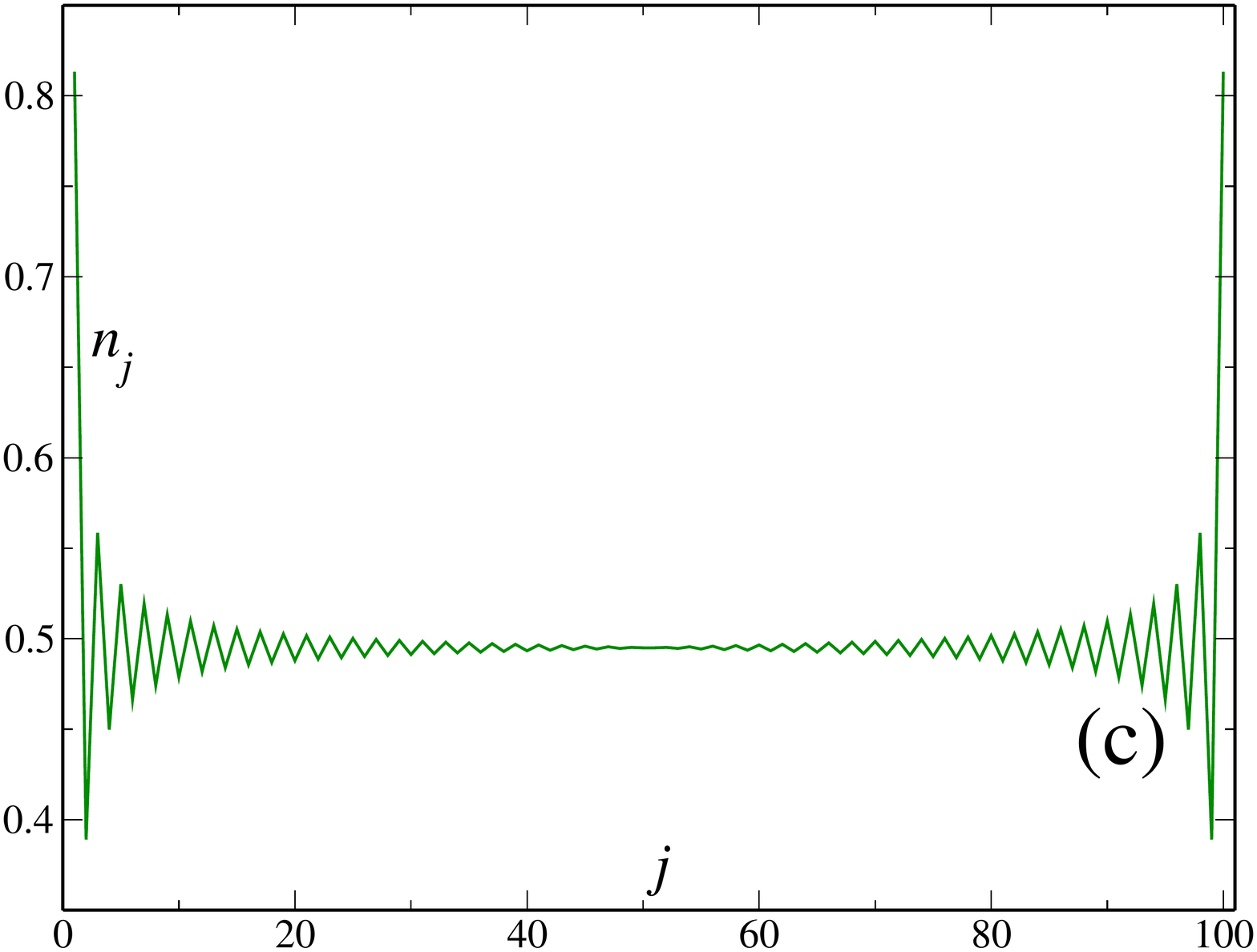}
  \includegraphics[width=0.5\textwidth]{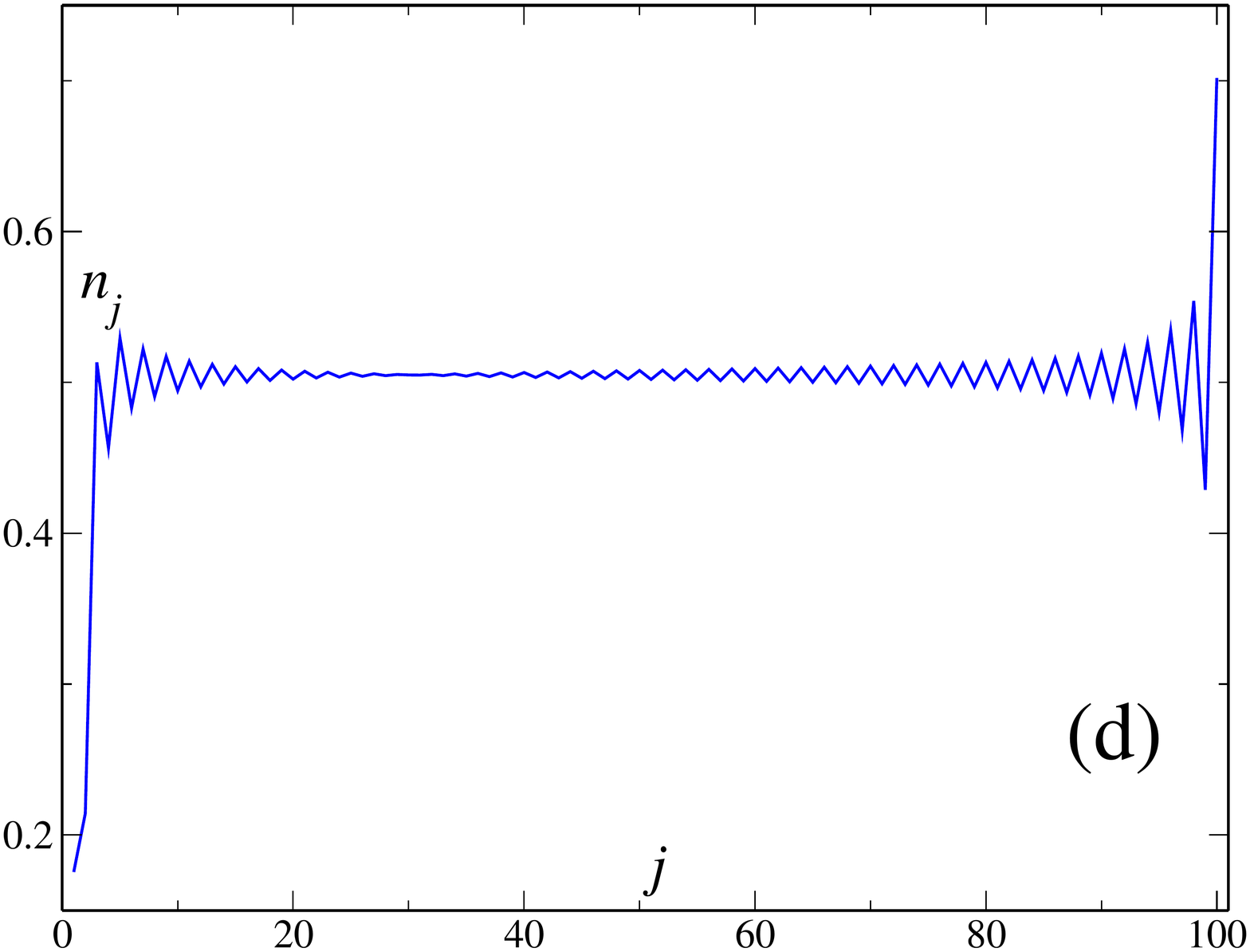}
 \end{minipage}
 \caption{Occupation profiles in the (0,0) (a), (1,0) (b), (1,1) (c) and (2, 1/2) (d) cases, with $L=100$.}\label{Occupations_FF}
\end{figure}

Let us now consider a state with fixed fermion number. Let $I$ be any subset of $\{1,\cdots,L\}$; we then define
\begin{equation}
 \left|I\right>\equiv\prod_{m\in I}d_m^\dagger\left|0\right>
\end{equation}
As already mentioned, we choose to stay in the half-filled sector, i.e., the ground state $\left|GS\right>$ shall correspond to $I=GS\equiv\{1,\cdots,L/2\}$. We can now compute any correlation function on the state $I$. For future use, we consider
\begin{equation}
 C^I_{jk}\equiv\left<I\right|c_j^\dagger c_k\left|I\right>
\end{equation}
We note that $C^I_{jj}$ is nothing but the occupation profile on the state $\left|I\right>$. To carry out the computation, we need to invert (\ref{d}). Being $F$ orthogonal, the inversion is trivial, and we get
\begin{equation}
 c_j^\dagger=\sqrt\frac{2}{L+1}\sum_{m=1}^L\sin\frac{\pi\,m\,j}{L+1}d_m^\dagger
\end{equation}
Making use of the fact that
\begin{equation}
 \left<I\right|d_m^\dagger d_n\left|I\right>=\delta_{mn}\begin{cases}
  1, & m\in I\\
  0, & m\notin I
 \end{cases}
\end{equation}
we immediately get
\begin{equation}
 C^I_{jk}=\frac{2}{L+1}\sum_{m\in I}\sin\frac{\pi\,m\,j}{L+1}\sin\frac{\pi\,m\,k}{L+1}
\end{equation}
For the ground state, we get explicitly
\begin{equation}
 C_{jk}^{GS}=\begin{cases}
  \frac{1}{2}, & j=k\\
  \frac{1}{2(L+1)}\left[\frac{\sin\frac{\pi(j-k)}{2}}{\sin\frac{\pi(j-k)}{2(L+1)}}-\frac{\sin\frac{\pi(j+k)}{2}}{\sin\frac{\pi(j+k)}{2(L+1)}}\right], & j\neq k
 \end{cases}
\end{equation}
We remark that the occupation profile is constant (see figure \ref{Occupations_FF}(a)).

We derive now the thermal partition function
\begin{equation}
 Z_{(0,0)}\equiv\mbox{Tr}\left[e^{-\beta H_{(0,0)}}\right]
\end{equation}
for the half-filled ground state of the (0,0) model, being $\beta$ the inverse temperature of the system. In order to do this, we position the Fermi momentum of the system at $k_F=\pi/2$; with this convention, the energy levels are indexed by an half-odd $n$ between $-(L-1)/2$ and $(L-1)/2$: we have $m\equiv(L+1)/2+n$. In terms of the new indexes, the energies are
\begin{equation}
 \tilde{\epsilon}_n=\sin\frac{\pi n}{L+1}
\end{equation}
We split the sum in the negative- and positive-$n$ part:
\begin{equation}
 \begin{split}
 H_{(0,0)} &=\sum_{n=1/2}^{(L-1)/2}\tilde{\epsilon}_n\tilde{d}_n^\dagger\tilde{d}_n+\sum_{n=-(L-1)/2}^{-1/2}\tilde{\epsilon}_n\tilde{d}_n^\dagger\tilde{d}_n=\\
 &=\sum_{n=1/2}^{(L-1)/2}\tilde{\epsilon}_n\left(\tilde{d}_n^\dagger\tilde{d}_n-\tilde{d}_{-n}^\dagger\tilde{d}_{-n}\right)
 \end{split}
\end{equation}
Defining the {\it hole} operators as
\begin{equation}
 \tilde{f}_n\equiv\tilde{d}_{-n}^\dagger
\end{equation}
one has
\begin{equation}\label{H_ph}
 H_{(0,0)}=\sum_{n=1/2}^{(L-1)/2}\tilde{\epsilon}_n\left(\tilde{d}_n^\dagger\tilde{d}_n+\tilde{f}_n^\dagger\tilde{f}_n\right)+A_L
\end{equation}
where $A_L\equiv-\sum_{n=1/2}^{(L-1)/2}\tilde{\epsilon}_n$. The partition function is therefore, being the Hamiltonian a two-species free Hamiltonian \cite{NegeleOrland1998},
\begin{equation}
 Z_{(0,0)}=e^{-\beta A_L}\prod_{n=1/2}^{(L-1)/2}\left(1+e^{-\beta\tilde\epsilon_n}\right)^2
\end{equation}

We are interested now in the {\it thermodynamic limit} $L\gg1$. The linearized spectrum is
\begin{equation}
 \tilde{\epsilon}_n\simeq\frac{\pi n}{L},\ n\in\left\{1/2,3/2,\cdots\right\}
\end{equation}
Let us start from the value of $A_L$. It is divergent, but we regularize it with the $\zeta$-function technique \cite{DiFrancescoMathieuSenechal1997}. We get
\begin{equation}
 A_L=-\sum_{m=1}^{L/2}\tilde{\epsilon}_{m-1/2}\simeq-\frac{\pi}{L}\left[\zeta(-1,-1/2)+1/2\right]=-\frac{\pi}{24L}
\end{equation}
Defining $q\equiv e^{-\frac{\beta\pi}{L}}$, we have
\begin{equation}
 e^{-\beta A_L}\simeq q^{-1/24}
\end{equation}
For the trace, we use the \emph{Jacobi triple-product} identity \cite{DiFrancescoMathieuSenechal1997}
\begin{equation}
 \prod_{n=1/2,3/2,\cdots}^{+\infty}\left(1+t\,q^n\right)\left(1+q^n/t\right)=\frac{\sum_{n\in\mathbb{Z}}q^{\frac{n^2}{2}}t^n}{\prod_{n=1}^{+\infty}(1-q^n)}
\end{equation}
so that
\begin{equation}
 Z_{(0,0)}=q^{-1/24}\frac{\sum_{n\in\mathbb{Z}}q^{\frac{n^2}{2}}}{\prod_{n=1}^{+\infty}(1-q^n)}=\frac{1}{\eta(q)}\sum_{n\in\mathbb{Z}}q^{\frac{1}{2}n^2}
\end{equation}
being $\eta(q)$ the {\it Dedekind-$\eta$ function} $\eta(q)\equiv q^{1/24}\prod_{n=1}^{+\infty}(1-q^n)$ \cite{DiFrancescoMathieuSenechal1997}.


\subsubsection{(1,0) Case}
The diagonalization procedure is exactly the same as in the previous case. Again, we come to (\ref{system}), but with different constraints:
\begin{equation}
 \begin{cases}
  -\frac{1}{2}\left(F_{m,j-1}+F_{m,j+1}\right)=\epsilon_mF_{mj}\\
  F_{m0}=F_{m1}\\
  F_{m,L+1}=0
 \end{cases}
\end{equation}
This time, we get
\begin{equation}
 p_m=\frac{\pi(m+1/2)}{L+1/2}
\end{equation}
with $m\in\mathbb{Z}$, and
\begin{equation}
 F_{mj}=\sqrt\frac{2}{L+1/2}\cos\frac{\pi(m+1/2)(j-1/2)}{L+1/2}
\end{equation}
and therefore $m$ can be chosen to be in $\{0,\cdots,L-1\}$. The correlator $C_{jk}^{GS}$ is easily shown to be
\begin{equation}
 C_{jk}^{GS}=\begin{cases}
  \frac{1}{2L+1}\left[L-\frac{(-1)^j}{2\sin\frac{\pi(2j-1)}{2(2L+1)}}\right], & j=k\\
  \frac{1}{2(2L+1)}\left[\frac{\cos\frac{\pi(j+k)}{2}}{\sin\frac{\pi(j+k-1)}{2(2L+1)}}-\frac{\sin\frac{\pi(j+k)}{2}}{\cos\frac{\pi(j+k-1)}{2(2L+1)}}+\right.\\
+\left.\frac{\sin\frac{\pi(j-k)}{2}}{\sin\frac{\pi(j-k)}{2(2L+1)}}-\frac{\cos\frac{\pi(j-k)}{2}}{\cos\frac{\pi(j-k)}{2(2L+1)}}\right], & j\neq k
 \end{cases}
\end{equation}
In this case, the occupation profile is not constant (see figure \ref{Occupations_FF}(b)): the first site is very likely filled, the second is likely empty, and so on, with an oscillating behavior, converging to the bulk value 1/2 leaving the boundary. We finally remark that the case (0,1) is easily obtained from the present one, just by making the replacement $F_{mj}\rightarrow F_{m,L+1-j}$.

Even in this case, we consider the partition function for the ground state of the system, $Z_{(1,0)}$. Since the momenta indices belong now to the set $\{0,\cdots,L-1\}$, we write now $m\equiv\frac{L-1}{2}+n$, with $n$ a half-odd between $-\frac{L-1}{2}$ and $\frac{L-1}{2}$. In this case, we come to a slight modification of equation (\ref{H_ph}), given by
\begin{equation}
 H_{(0,0)}=\sum_{n=1/2}^{(L-1)/2}\left(\tilde{\epsilon}_n\tilde{d}_n^\dagger\tilde{d}_n+\tilde{\delta}_n\tilde{f}_n^\dagger\tilde{f}_n\right)+A_L
\end{equation}
where $\tilde{\epsilon}_n\equiv\sin\frac{\pi(n-1/4)}{N+1/2}$, $\tilde{\delta}_n\equiv-\tilde{\epsilon}_{-n}$, and $A_L\equiv-\sum_{n=1/2}^{\frac{L-1}{2}}\tilde{\delta}_n$. The procedure for computing the thermal partition function is almost the same as in the previous case, but now, in the Jacoby triple-product identity we choose $t\equiv q^{-1/4}$. Then, by exactly the same steps as before, we get, in the thermodynamic limit,
\begin{equation}
 Z_{(1,0)}=\frac{1}{\eta(q)}\sum_{n\in\mathbb{Z}}q^{\frac{1}{2}\left(n-\frac{1}{4}\right)^2}
\end{equation}


\subsubsection{(1,1) Case}
The boundary constraints, in this case, are given by
\begin{equation}
 \begin{cases}
  F_{m0}=F_{m1}\\
  F_{m,L+1}=F_{mL}
 \end{cases}
\end{equation}
so that the momenta take the form
\begin{equation}\label{momenta(1,1)}
 p_m=\frac{\pi\,m}{L}
\end{equation}
with $m\in\mathbb{Z}$. We find, for $F_{mj}$,
\begin{equation}
 F_{mj}=A_m\cos\frac{\pi\, m(j-1/2)}{L}
\end{equation}
with $A_m$ to be determined by the orthogonality of $F$. First, we evince from this formula that we can choose $m\in\{0,\cdots,L-1\}$; moreover, differently from the previous cases, it comes out that $A_m$ is not the same for every $m$: we obtain
\begin{equation}
 A_m=\begin{cases}
  \frac{1}{\sqrt{L}}, & m=0\\
  \sqrt\frac{2}{L}, & m=1,\cdots,L-1
 \end{cases}
\end{equation}
$c$, as a function of $d$, takes the form
\begin{equation}
 c_j=\frac{1}{\sqrt{L}}d_0+\sqrt\frac{2}{L}\sum_{m=1}^{L-1}\cos\frac{\pi\,m(j-1/2)}{L}
\end{equation}
The first addend, independent of $j$, is an example of \emph{zero-mode} \cite{DiFrancescoMathieuSenechal1997}. $C_{jk}^{GS}$ is, in this case,
\begin{equation}
 C_{jk}^{GS}=\begin{cases}
  \frac{1}{2}-\frac{1}{2L}+\frac{1}{2L}\sin\frac{\pi(2j-1)}{2}\cot\frac{\pi(2j-1)}{2L}, & j=k\\
  \frac{1}{2L}\left[\sin\frac{\pi(j-k)}{2}\cot\frac{\pi(j-k)}{2L}-\cos\frac{\pi(j-k)}{2}+\right.\\
  +\left.\sin\frac{\pi(j+k-1)}{2}\cot\frac{\pi(j+k-1)}{2L}-\cos\frac{\pi(j+k-1)}{2}\right], & j\neq k
 \end{cases}
\end{equation}
The occupation profile is shown in figure \ref{Occupations_FF}(c), and the partition function is (the computation is analoguos to the one of the (1,0) case)
\begin{equation}
 Z_{(1,1)}=\frac{1}{\eta(q)}\sum_{n\in\mathbb{Z}}q^{\frac{1}{2}\left(n+\frac{1}{2}\right)^2}
\end{equation}


\subsubsection{$\boldsymbol{(a,1/a)}$ Case}
This case is more general than the previous, since in principle $a\in\mathbb{R}\setminus\{0\}$. The boundary constraints look now
\begin{equation}
 \begin{cases}
  F_{m0}=a\,F_{m1}\\
  F_{m,L+1}=\frac{1}{a}F_{mL}
 \end{cases}
\end{equation}
The momenta take the form (\ref{momenta(1,1)}), while $F$ can be found to be
\begin{equation}\label{F_(a,1/a)}
 F_{mj}=\sqrt\frac{2}{L\left(a^2-2a\cos\frac{\pi\,m}{L}+1\right)}\left[a\,\sin\frac{\pi\,m(j-1)}{L}-\sin\frac{\pi\,m\,j}{L}\right]
\end{equation}
and therefore, for $a\neq 1$, we can choose $m\in\{1,\cdots,L\}$. Moreover, it is easy to see that for $a=1$ we obtain the $F$ we had in the (1,1) case. Equation (\ref{F_(a,1/a)}) can be put in a more suggestive form by introducing the phase $\theta_m$ in the following way:
\begin{equation}
 \begin{cases}
  \cos\frac{\pi\,m\,\theta_m}{L}\equiv\frac{a\cos\frac{\pi\,m}{L}-1}{\sqrt{a^2-2a\cos\frac{\pi\,m}{L}+1}}\\
  \sin\frac{\pi\,m\,\theta_m}{L}\equiv\frac{a\sin\frac{\pi\,m}{L}}{\sqrt{a^2-2a\cos\frac{\pi\,m}{L}+1}}
 \end{cases}
\end{equation}
so that $F$ looks
\begin{equation}
 F_{mj}=\sqrt\frac{2}{L}\sin\frac{\pi\,m\left(j-\theta_m\right)}{L}
\end{equation}
The correlation function $C_{jk}^{GS}$ has not an explicit form; we just have the general formula
\begin{equation}\label{correlation function}
 C_{jk}^{GS}=\sum_{m=1}^{L/2}F_{mj}F_{mk}
\end{equation}
The occupation profile, not analytically computable, is shown in figure \ref{Occupations_FF}(d): we note that the effect of a stronger boundary term on the first site is to deplete the site itself. Finally, we remark that the $(1/a,a)$ case is obtained from the current one, as in the (1,0) case, by the replacement $F_{mj}\rightarrow F_{m,L+1-j}$.


\subsection{Quantum Spin Chains}

Spin chains are among the most studied models of 1D condensed-matter physics \cite{Takahashi1999}. The reasons are multiple: first of all, they can be viewed as effective theories for interesting many-body problems, for instance of the Hubbard model (see section \ref{Hubbard}); they are the simplest examples of models capturing many-body effects related to magnetism, and nevertheless they often possess very rich phase diagrams, similar to the ones of real systems; the simplest spin chains, in one dimension, are exactly solvable, and, for instance, can be useful to test predictions of effective approaches, such as quantum field theories (see chapters \ref{Crossovers} and \ref{CFT+OBC}), and numerical techniques, such as density-matrix renormalization group techniques (see section \ref{DMRG}); finally, in recent times some quantum spin chains have been proven to be experimentally feasible in quantum-optical setups (see, e.g., \cite{KCKIEFLDM2010}). In the following, we will review known results about some of the simplest quantum spin chains, that will be useful in the rest of the work.


\subsubsection{Spin-1/2 XY Chain}\label{XY}
\begin{figure}[t]
 \begin{minipage}{\textwidth}
  \centering
  \includegraphics[width=0.5\textwidth]{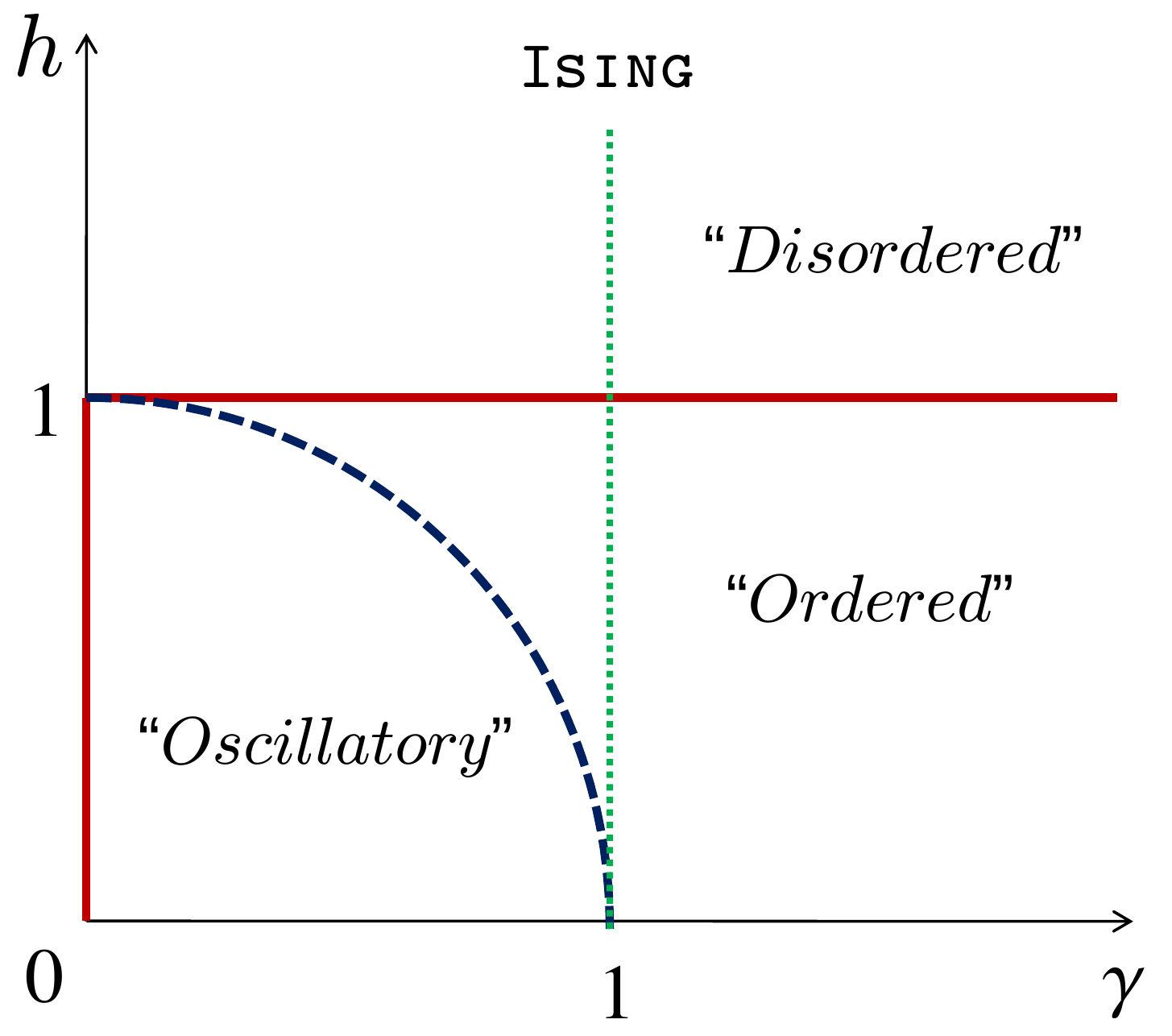}
 \end{minipage}
 \caption{Phase diagram of the spin-1/2 XY chain. Reprint with permission from \cite{Franchini2011}.}\label{Franchini}
\end{figure}

The spin-1/2 XY chain Hamiltonian is \cite{LiebSchultzMattis1961}
\begin{equation}\label{XY_ham}
 H_{\gamma h}=-\frac{1}{2}\sum_{j=1}^L\left[\frac{1+\gamma}{2}\sigma_j^x\sigma_{j+1}^x+\frac{1-\gamma}{2}\sigma_j^y\sigma_{j+1}^y+h\sigma_j^z\right]
\end{equation}
where $\gamma,\,h\in\mathbb{R}$ and $\vec{\sigma}_j$ contains the Pauli matrices as entries; periodic boundary conditions (PBC) are assumed. For some values of $(\gamma,h)$ the model reduces to some simpler ones: e.g., for $\gamma=0$, it becomes the \emph{XX model} \cite{Takahashi1999}, whose physics is the same as the one of free fermions of section \ref{FF}, while for $\gamma=1$ we have the famous \emph{Ising model} \cite{Mussardo2007}. The model is exactly solvable by the sequence Jordan-Wigner $\rightarrow$ Fourier $\rightarrow$ Bogolyubov transformations \cite{LiebSchultzMattis1961, Franchini2011}: $H_{\gamma h}$ is equivalent (up to a constant additive term) to a quadratic Hamiltonian of spinless fermions of the form
\begin{equation}\label{XY_fer}
 H_{\gamma h}^{\pm}=-\frac{1}{2}\sum_{j=1}^N\left[\left(c_{j+1}^\dagger c_j+c_j^\dagger c_{j+1}\right)+\gamma\left(c_{j+1}c_j+c_j^\dagger c_{j+1}^\dagger\right)-2hc_j^\dagger c_j\right]
\end{equation}
with antiperiodic/periodic boundary conditions depending on whether one is considering a state with an even/odd number of down spins (this quantity can be seen to be a good quantum number \cite{LiebSchultzMattis1961}). We just remark, for future use, that, if we would have chosen the spin chain to be open, the fermions should also have been open, with no quantum numbers. The single particle spectrum can be shown to be \cite{LiebSchultzMattis1961,Franchini2011}
\begin{equation}
 \epsilon_m=\sqrt{\left(h-\cos p_m\right)^2+\gamma^2\sin^2 p_m}
\end{equation}
with $p_m=\frac{2\pi(m+1/2)}{L},\,m\in\mathbb{Z}$ in the antiperiodic case. The model is critical, i.e., the single particle spectrum is gapless, if $(\gamma=0,h\leq1)$ (XX universality class) or $(\gamma,h=1)$ (Ising universality class) \cite{Franchini2011}: the two cases display very different physics, as we will show. In figure \ref{Franchini} we show the phase diagram of the model: we restrict it to the region $(\gamma\geq 0,h\geq 0)$, because of the invariance of (\ref{XY_ham}) under the canonical transformations
\begin{equation}\label{canonical}
 \begin{cases}
  \sigma_j^x\rightarrow -\sigma_j^x\\
  \sigma_j^y\rightarrow -\sigma_j^y\\
  \sigma_j^z\rightarrow\sigma_j^z\\
 \end{cases},\,\,\,\,\,\,\,\,\,\begin{cases}
  \sigma_j^x\rightarrow\sigma_j^y\\
  \sigma_j^y\rightarrow -\sigma_j^x\\
  \sigma_j^z\rightarrow\sigma_j^z\\
 \end{cases}
\end{equation}


\subsubsection{Spin-1/2 XXZ Chain}\label{XXZ}
We consider now the Hamiltonian
\begin{equation}\label{XXZ_OBC}
 H_\Delta^{OBC}=-\frac{1}{4}\sum_{j=1}^{L-1}\left(\sigma_j^x\sigma_{j+1}^x+\sigma_j^y\sigma_{j+1}^y-\Delta\sigma_j^z\sigma_{j+1}^z\right)
\end{equation}
where $\Delta\in\mathbb{R}$ is the \emph{anisotropy} coefficient: it differs from the (0,0) XX model for the third term. Open boundary conditions (OBC) are assumed, as it is obvious from (\ref{XXZ_OBC}). The physics of the above Hamiltonian is very different from the one of $H_{\gamma h}$: it is easily shown, via a Jordan-Wigner transformation, that it can be mapped to an {\it interacting} Hamiltonian of spinless fermions, thus much less trivial than the one of the previous section \cite{Giamarchi2003}; however, differently from Hamiltonian (\ref{XY_ham}), this one commutes with the $z$-component of the total spin $\sigma^z\equiv\sum_{j=1}^L\sigma_j^z$. The model is solvable by \emph{Bethe ansatz} \cite{Takahashi1999}, and can be shown to be critical for $|\Delta|\leq 1$, and in the same universality class of the free fermions of section \ref{FF} for $-1<\Delta\leq 1$ \cite{Giamarchi2003}. For $\Delta<-1$ the system orders ferromagnetically (here the ground state is a true factorized state), while for $\Delta>1$ it tends to order antiferromagnetically \cite{Takahashi1999}.


\subsubsection{Spin-3/2 XXZ Chain}\label{spin-3/2}
Generalizations of the Hamiltonian (\ref{XXZ_OBC}) have been widely considered in literature (see, e.g., \cite{Xavier2010}); in particular, we consider here the case in which $\vec{\sigma}$ is not a spin-1/2 operator but a spin-3/2 one:
\begin{equation}
 \vec{\sigma}=\left(\begin{array}{c}
 \left(\begin{array}{cccc}
  0 & \sqrt{3} & 0 & 0 \\
  \sqrt{3} & 0 & 2 & 0 \\
  0 & 2 & 0 & \sqrt{3} \\
  0 & 0 & \sqrt{3} & 0
 \end{array}\right)\\
 i\left(\begin{array}{cccc}
  0 & -\sqrt{3} & 0 & 0 \\
  \sqrt{3} & 0 & -2 & 0 \\
  0 & 2 & 0 & -\sqrt{3} \\
  0 & 0 & \sqrt{3} & 0
 \end{array}\right)\\
 \left(\begin{array}{cccc}
  3 & 0 & 0 & 0 \\
  0 & 1 & 0 & 0 \\
  0 & 0 & -1 & 0 \\
  0 & 0 & 0 & -3
 \end{array}\right)
 \end{array}\right)
\end{equation}
As in the spin-1/2 case, the Hamiltonian commutes with $\sigma^z$. Contrary to the spin-1/2 case, the spin-3/2 XXZ chain is not exactly solvable, and therefore one has to rely only on approximated or numerical techniques. In particular, at the isotropic point $\Delta=1$, the model has been conjectured by Haldane, in a wider contest \cite{Haldane1982, Haldane1983a, Haldane1983b}, and then analytically (see \cite{Auerbach1994} for field-theoretical arguments) and numerically confirmed \cite{HWHM1996} to be massless, and to belong to the same universality class of the spin-1/2 case. Leaving the isotropic point, the model is critical, as in the spin-1/2 case, for $|\Delta|\leq 1$ \cite{AlcarazMoreo1992}, and for $-1<\Delta\leq 1$ it belongs to the same universality class of  the spin-1/2 XXZ chain. We remark that the isotropic model finds experimental realizations in quasi-1D anti-ferromagnets of magnetic ions, as CsVCl$_3$ \cite{IKET1995} and AgCrP$_2$S$_6$ \cite{MutkaPayenMolini1993}.

The considered Hamiltonians are not the most general ones we can consider: one can add, for instance, non-nearest neighbour interactions, leading in some cases to frustrated ground states (see, e.g., \cite{MajumdarGhosh1969}), or non-linear terms, as in the spin-1 $\lambda-D$ chain \cite{BotetJullienKolb1983}. We will not consider such spin chains in this thesis.


\subsection{Hubbard Model}\label{Hubbard}

Instead of considering spinless fermions, one could work with {\it spinful} fermions, i.e., fermions of two different species, as real electrons. Systems involving such particles are of incredible importance in condensed-matter physics. The simplest interacting model involving spinful fermions is the celebrated \emph{Hubbard model} \cite{Hubbard1963} (see, e.g., \cite{Montorsi1992} for a review):
\begin{equation}\label{Hubbard_Hamiltonian}
 H_U=\sum_{j}\left[-\sum_{\sigma=\uparrow,\downarrow}\left(c^\dagger_{\sigma j}c_{\sigma,j+1}+c^\dagger_{\sigma,j+1}c_{\sigma j}\right)+Un_{\uparrow j}n_{\downarrow j}\right]
\end{equation}
where the index $\sigma=\,\uparrow,\ \downarrow$ distinguishes operators related to the different fermion species; operators with different polarization simply anticommute, and $n_\sigma$ is defined in analogy with the spinless case. This Hamiltonian is exactly solvable by Bethe ansatz \cite{EFGKK2005}, and displays several phases: e.g., in the attractive ($U<0$) half-filled ($n_\uparrow=n_\downarrow=1/4$) case it undergoes spin-charge separation into a gapped spin sector and a gapless charge one, as it can be argued from field-theoretical considerations \cite{Giamarchi2003}; we will come back to these considerations in section \ref{Hubbardcheck}.


\subsection{Dipolar Bosons}
Let us consider now the Hamiltonian
\begin{equation}\label{dipoles}
 H_D=-\sum_j\left(b_j^\dagger b_{j+1}+b_{j+1}^\dagger b_j\right)+D\sum_{j<k}\frac{n_jn_k}{|j-k|^3}
\end{equation}
where $b_i$ is a hard-core annihilation bosonic operator. The model is not exactly solvable; as far as we know, a complete phase diagram is not available. What is known is that in the repulsive regime ($D>0$) the model is gapless, at least starting from some $\bar{D}$ \cite{DalmonteErcolessiTaddia2011}. Moreover, as we will argue in deeper detail in section \ref{K_dipoles}, in the same gapless phase we have a transition between different dominant order, i.e., a transition between different dominant correlations, the {\it superfluid} (SF) and the {\it charge-density wave} (CDW) one.  Hamiltonian (\ref{dipoles}) is known to describe one-dimensional tubes of polar molecules or magnetic atoms with dipole moment aligned perpendicularly to the tube via a dc electric field \cite{LMSLP2009} loaded onto an optical lattice \cite{BlochDalibardZwerger2008}; in such systems, the interaction strength $D$ can be tuned by, e.g., tuning the depth of the optical potential.


\section{Bosonization and Luttinger Liquids}\label{Bosonization}

As it can argued from very general statements, due to the Mermin-Wagner theorem \cite{MerminWagner1966}, 1D physics is very peculiar. This fact can be understood by means of very simple arguments, like the following. Let us consider, e.g., non-interacting electrons on a line. If an electron tries to propagate along the line, sooner or later it has to meet another electron, and because of the Pauli principle, it induces on it some kind of motion, and so on. Therefore, in contrast with higher dimensional systems, it is easy to figure out that in 1D only {\it collective} excitations are possible. From a technical point of view, this statement is equivalent to say that the {\it Landau theory} of Fermi liquids \cite{NegeleOrland1998} breaks down at $d=1$, and no quasi-particle picture is available: the ground state is very different from a free-fermion ground state.

However, it is easy to understand what is happening. Let us consider, for simplicity, a (0,0) free fermionic system, whose Hamiltonian is given by equation (\ref{H_ab}) with $a=b=0$ (interactions, both repulsive or attractive, do not change very much the situation). As we argued, the spectrum of the theory is given by equation (\ref{spectrum}), where the momenta are the ones of equation (\ref{momenta}), with $m=1,\cdots,L$. At half-filling, for large $L$, the spectrum is, at first order, linear close to the Fermi momentum $p_{L/2}$: if we now write $\epsilon_m\equiv\epsilon_{L/2+n}$, we have
\begin{equation}
 \epsilon_{L/2+n}=\sin\frac{\pi(n-1/2)}{L+1}\simeq\frac{\pi(n-1/2)}{L}
\end{equation}
for large $L$ and small $n$, i.e., close to the Fermi surface. It is natural to expect that the small-$n$ physics across the Fermi level, i.e., the so-called {\it low energy physics} is the most important one \cite{Giamarchi2003}; we look therefore at {\it particle-hole} excitations, i.e., excitations involving the passage of an electron from an energy level under the Fermi level to one above. Their energy is of the form
\begin{equation}
 E(q)=\frac{\bar{n}+q-1/2}{L}-\frac{\bar{n}-1/2}{L}=\frac{\pi q}{L}
\end{equation}
where $\bar{n}<0$ is a negative integer labeling an energy level beneath the Fermi one: the dispersion relation of the couple is massless, and therefore it can be interpreted as a massless bosonic particle. It can be shown (see, e.g., \cite{Giamarchi2003}) that the full spectrum of the theory is reproducible just by means of this kind of excitations, and therefore, although the system was originally fermionic, we end up with {\it bosonic} physics! This fact is peculiar of 1D.

This qualitative picture can be made quantitative by means of the technique known as {\it bosonization} \cite{Haldane1981,Giamarchi2003}. Starting from a lattice model, a recipe is available to obtain a continuum {\it quantum field theory} (QFT) of bosonic nature, describing the low-energy physics of the model. For interacting spinless fermions, in many cases, one ends up with a Hamiltonian of the type
\begin{equation}\label{H_LL}
 H_{LL}=\frac{u}{2\pi}\int dx\left[K\left(\partial_x\theta(x)\right)^2+\frac{1}{K}\left(\partial_x\phi(x)\right)^2\right]
\end{equation}
where $\phi$ is the field describing the system and $\theta$ is linked to its conjugate field by the relation $\Pi\equiv\partial_x\theta/\pi$. It is easy, from these relations, to recognize in equation (\ref{H_LL}) a massless relativistic QFT \cite{DiFrancescoMathieuSenechal1997}, where $u$ plays the role of the light velocity, that, in this contest, is called {\it sound velocity}; $K$ is a coefficient, known as {\it Luttinger parameter}, related to the couplings of the microscopic Hamiltonian \cite{Giamarchi2003}. The correspondence between the original theory and the QFT is, in most cases, not exact, but well describes the large-scale properties of the system. Apart from the free ones, other terms can arise in the Hamiltonian from the bosonization procedure, depending on the form of the interaction: we will come back to this point in chapter \ref{Crossovers}.

Equation (\ref{H_LL}) is a very efficient tool for the computation of physical properties, encoded in correlation functions. In particular, it comes out that some of them strongly depend on the Luttinger parameter of the system. For instance, for free spinless fermions, the density-density correlation function is computed to be \cite{Giamarchi2003}
\begin{equation}
 \left<n(r)n(0)\right>=\frac{K}{2\pi^2}\frac{y_\alpha^2-x^2}{\left(x^2+y_\alpha^2\right)^2}+\frac{2}{(2\pi\alpha)^2}\cos(2k_Fx)\left(\frac{\alpha}{r}\right)^{2K}
\end{equation}
being $r\equiv\sqrt{x^2+(u\tau)^2}$, $\tau$ the imaginary time, $\alpha$ an ultraviolet cutoff regularizing the QFT and $y_\alpha\equiv u\tau+\alpha\,\mbox{sgn}\,\tau$. A correlation function of this form defines the phase of the system, known as {\it Luttinger liquid} \cite{Giamarchi2003}. Another typical correlation function is the superconducting one \cite{Giamarchi2003}, i.e.,
\begin{equation}
 \left<O_{SU}(r)O^\dagger_{SU}(0)\right>=\frac{1}{\left(\pi\alpha\right)^2}\left(\frac{\alpha}{r}\right)^{2/K}
\end{equation}
where $O_{SU}(r)=\psi^\dagger(r)\psi^\dagger(r+a)$, being $\psi^\dagger$ the slowly-oscillating part of the continuum analog of $c^\dagger$ and $a$ the lattice site, preventing $O$ from being null. It is interesting to stress that when the dominant decay exponent of the density-density correlation is larger, the one of the superconducting correlation is smaller, and {\it viceversa}: there is competition between the two phases. When the first is smaller, we say that the phase of the system is a {\it charge density wave}, while in the opposite case we have a superconducting phase \cite{Giamarchi2003}. This is to say that we can use the correlation functions exponent to build a sort of phase diagram, distinguishing between different dominant orders. We have not true quantum phase transitions separating them, since these correlations clearly indicate the presence of quasi-long-range order, but we can say that we have transition between dominant orders, or {\it crossovers} (see section \ref{QPT}).

A QFT having Hamiltonian (\ref{H_LL}) is an example of {\it conformal field theory} \cite{DiFrancescoMathieuSenechal1997}, as we will explain in detail in section \ref{CFT}. These theories are particularly important in 1D, because their symmetries allow their exact solution: one can therefore exploit theit full power to compute any correlation function. Being more specific, Luttinger liquids are very special conformal field theories, i.e., $c=1$ free compactified scalar bosonic massless theories \cite{DiFrancescoMathieuSenechal1997}, and a lot is known about them. We will come back to these arguments in the following chapters.

Finally, we stress the interplay between analytical and numerical techniques in 1D systems. As we have seen, the physics is ruled by $u$ and $K$, that distinguish between different regions of the phase diagram. However, in most cases they are not exactly computable, and one has to rely on exact solutions, when available, or on perturbative techniques, that are, in 1D, often plagued by divergences \cite{Giamarchi2003}. Therefore, in many cases, the only way one has to extract $u$ and $K$ is by numerical studies: we will come back in detail to this point in chapter \ref{Crossovers}.


\section{Conformal Field Theory}\label{CFT}

In this section we give some basic notions of conformal field theory \cite{BelavinPolyakovZamolodchikov1984, DiFrancescoMathieuSenechal1997} we will need in the remaining of the work. The material contained in this section is taken, unless explicitly stated, from the book by Di Francesco, Mathieu and S\'en\'echal \cite{DiFrancescoMathieuSenechal1997}.


\subsection{Basic Concepts}

A {\it conformal field theory} (CFT) is a QFT endowed of an additional symmetry, the one under conformal transformations, whose effect is to modify the background metrics just with a local scale factor; the conformal group always contains the Poincar\'e group as a subgroup. In arbitrary spatial dimension $d\geq 2$ the conformal group has dimension $\frac{(d+3)(d+2)}{2}$; however, for $d=1$, its dimension becomes {\it infinite}, and this fact allows for the complete solvability of the theory. More specifically, if one writes the coordinates of the background, that we take now to be the full plane $\mathbb{R}^2$, as complex coordinates $z,\ \bar{z}$, the conformal group is exactly the group of {\it holomorphic} (and {\it antiholomorphic}) functions, suggesting the power of the complex formalism in CFT.

The conformal group, in 1D, acts on the fields $\Phi$ of the theory in the following way. Let $\epsilon$ be a infinitesimal conformal coordinate change: then, the consequent variation of the field is
\begin{equation}
 \delta_\epsilon\Phi(w)=-[Q_\epsilon,\Phi(w)]
\end{equation}
being $Q_\epsilon$ the conformal charge associated to the conformal symmetry, given by
\begin{equation}
 Q_\epsilon\equiv\frac{1}{2\pi i}\oint dz\ \epsilon(z)T(z)
\end{equation}
and being $T(z)$ the holomorphic part of the {\it stress-energy tensor} of the theory (for every holomorphic quantity there is, on the complex plane, the antiholomorphic corresponding). Expanding $T(z)$ in Laurent series, i.e., defining
\begin{equation}
 T(z)\equiv\sum_{n\in\mathbb{Z}}z^{-n-2}L_n,\ \epsilon(z)=\sum_{n\in\mathbb{Z}}z_{n+1}\epsilon_n
\end{equation}
one has
\begin{equation}
 Q_\epsilon=\sum_{m\in\mathbb{Z}}\epsilon_nL_n
\end{equation}
Therefore, the $L_n$ are the generators of the local conformal transformations, composing the celebrated {\it Virasoro algebra}
\begin{equation}\label{virasoro}
 \begin{split}
  [L_n,L_m] &=(n-m)L_{n+m}+\frac{c}{12}n(n^2-1)\delta_{n+m,0}\\
  [L_n,\bar{L}_m] &=0\\
  [\bar{L}_n,\bar{L}_m] &=(n-m)\bar{L}_{n+m}+\frac{c}{12}n(n^2-1)\delta_{n+m,0}
 \end{split}
\end{equation}
The second of equations (\ref{virasoro}) indicates the decoupling of the two subalgebras. As it should be clear, the stress-energy tensor plays a key role in CFT. A consequence of its importance is its appearance in the so called {\it Ward identity}, encoding the consequences of conformal invariance on correlation functions:
\begin{equation}\label{Ward}
 \delta_{\epsilon,\bar{\epsilon}}\left<X\right>=-\frac{1}{2\pi i}\oint_Cdz\ \epsilon(z)\left<T(z)X\right>+\frac{1}{2\pi i}\oint_Cd\bar{z}\ \bar{\epsilon}(\bar{z})\left<\bar{T}(\bar{z})X\right>
\end{equation}
where $X$ is a string of operators, depending on some coordinates $w_j,\ \bar{w}_j$, and $C$ a path circling all such coordinates.

Different CFTs are usually distinguished by some real numbers, namely, the {\it central charge} $c$ and the {\it conformal dimensions of primary operators}. While giving a precise definition of $c$ is a quite technical task (it is related to the number of degREE's of freedom of the theory), primary operators are very simply defined: they are the operators that create from the vacuum the highest-weight states of representations of the conformal group, called {\it Verma modules}. If $w\rightarrow z(w)$ is a conformal transformation, a holomorphic primary field $\phi_h$ is defined by the behavior
\begin{equation}\label{primary}
 \phi_h'(z)=\left(\frac{dz}{dw}\right)^{-h}\phi_h(w)
\end{equation}
where $h$ is the (holomorphic) {\it conformal dimension} of the field, while the highest-weight state of each Verma module is given by
\begin{equation}
 \left|h\right>\equiv\lim_{z\rightarrow 0}\phi_h(z)\left|0\right>
\end{equation}
in the framework of {\it radial quantization} \cite{DiFrancescoMathieuSenechal1997} ($z=0$ corresponds to the point at $\tau=-\infty$). For most CFTs, once one knows the central charge and the primary operators of the theory, one has all the needed information to solve it. These states can be shown to be the lowest-excited states of the theory. In condensed-matter theory, CFTs containing just a finite number of Verma modules are especially important, and they are called {\it minimal} CFTs. The simplest minimal CFT, that will be widely considered in chapter \ref{CFT+OBC}, has central charge $c=1/2$, and the primary operators, of conformal dimensions 0 (the associated primary field is the {\it identity}, that is present in every CFT), 1/2 and 1/16. In the next section, a very special role will be played by a specific concept, the one of {\it fusion rules}, i.e., the way primary operators behave when inserted in correlation functions. For the $c=1/2$ minimal CFT, we have
\begin{equation}
 \begin{split}
  \sigma\times\sigma &= \mathbb{I}+\epsilon\\
  \sigma\times\epsilon &= \sigma\\
  \epsilon\times\epsilon &= \mathbb{I}
 \end{split}
\end{equation}
and the meaning is that when two operators in the left-hand side are both inserted in the same correlation function they can be substituted with what is on the right-hand side. The missing fusion rules, involving the identity, are trivial. It is customary to define the {\it fusion coefficients} $\mathcal{N}^k_{ij}$ as
\begin{equation}
 \phi_i\times\phi_j=\sum_k\mathcal{N}^k_{ij}\phi_k
\end{equation}

In the remaining of the work, we will use a nice relation about the energy of the ground state, $E_0$, and of an excited state of conformal dimension $h$, $E_h$, at {\it finite size} $L$:
\begin{equation}\label{E_h}
 E_h-E_0=\frac{\pi u}{L}h
\end{equation}
being $u$ the sound velocity in the system. In particular, it constitutes a way to estimate the conformal dimension of an excited state from numerical computations. Another useful formula gives information about the ground-state energy at finite size \cite{BloteCardyNightingale1986, Affleck1986}:
\begin{equation}\label{E_0}
 E_0=e_\infty L+e_s-\frac{\pi cu}{6\delta L}+O(L^{-2})
\end{equation}
where $\delta=1/4$ for periodic/open boundary conditions, and being $e_\infty$ and $e_s$ respectively the bulk and boundary energy contributions (see even eq. (\ref{epsilon_CFT})). Therefore, looking at the $O(1/L)$ term, one can get information about the product $cu$.


\subsection{Boundary Conformal Field Theory}\label{BCFT}

When the manifold a CFT is defined on has a boundary, the operator content of the theory is usually different from the one on the complex plane. Let us consider the simplest open situation, the {\it upper-half plane} $\mathbb{H}\equiv\{z\in\mathbb{C}|\ \mbox{Im}z\geq 0\}$. The imposition of the conformal invariance on the boundary strongly constrains the theory, i.e., reduces of one half the number of independent Virasoro generators. This fact has several consequences on the physics. One is that correlation functions of fields on $\mathbb{H}$ can be computed by the {\it mirror-image} trick, i.e., for, e.g., a two-points function,
\begin{equation}
 \left<\phi_{h_1,\bar{h}_1}(z_1,\bar{z}_1)\phi_{h_2,\bar{h}_2}(z_2,\bar{z}_2)\right>_{\mbox{uhp}}=\left<\phi_{h_1}(z_1)\bar{\phi}_{\bar{h}_1}(z_1^*)\phi_{h_2}(z_2)\bar{\phi}_{\bar{h}_2}(z_2^*)\right>_{\mathbb{C}}
\end{equation}
where $\bar{h}_i$ are the antiholomorphic conformal dimensions of the fields and $\bar{\phi}_{\bar{h}}$ is the antiholomorphic part of the field $\phi_{h,\bar{h}}$. A second important consequence is the arising of severe constraints on the operator content of the theory. In fact, if a system is defined on a strip having on its edges some conformal boundary conditions, that we will call $\alpha$ and $\beta$, its thermal partition function will take the form
\begin{equation}
 Z_{\alpha\beta}(q)=\sum_hn^h_{\alpha\beta}\chi_h(q)
\end{equation}
where $q\equiv e^{-\pi\beta/L}$, and $\chi_h$ is the conformal character relative to the Verma module associated with the primary operator of conformal dimension $h$; the non-negative integers $n^h_{\alpha\beta}$ are therefore the number of times the Verma module associated to $h$ appears in the spectrum of the CFT (this statement is highly non-trivial; see \cite{DiFrancescoMathieuSenechal1997} for further details). It was shown by Cardy \cite{Cardy1989} that such integers are simply given by
\begin{equation}\label{boundary-fusion}
 n^h_{\alpha\beta}=\mathcal{N}^h_{\alpha\beta}
\end{equation}
where $\mathcal{N}^h_{\alpha\beta}$ are the fusion coefficients introduced in the previous section. Actually, equation (\ref{boundary-fusion}) does not hold for any conformal boundary condition one considers, but just for some special boundary conditions $\tilde{h}$, in isomorphic correspondence with the primary operators of the theory. In the CFT picture, one can think to exchange the roles of space and time: for boundary conditions, it corresponds to evolve a {\it boundary state}, on one edge of the system, to the other. For the $c=1/2$ minimal CFT, these boundary states are given by
\begin{equation}\label{BS_Ising}
 \begin{split}
  &\left|\tilde{0}\right>=\frac{1}{\sqrt{2}}\left|0\right>+\frac{1}{\sqrt{2}}\left|\epsilon\right>+\frac{1}{\sqrt[4]{2}}\left|\sigma\right>\\
  &\left|\tilde{\frac{1}{2}}\right>=\frac{1}{\sqrt{2}}\left|0\right>+\frac{1}{\sqrt{2}}\left|\epsilon\right>-\frac{1}{\sqrt[4]{2}}\left|\sigma\right>\\
  &\left|\tilde{\frac{1}{16}}\right>=\left|0\right>-\left|\epsilon\right>
 \end{split}
\end{equation}
where $\left|0\right>$, $\left|\epsilon\right>$ and $\left|\sigma\right>$ are the so-called {\it Ishibashi states} \cite{DiFrancescoMathieuSenechal1997}, i.e., special states belonging to the Verma modules of $\mathbb{I}$, $\epsilon$ and $\sigma$. The physical meaning of the states (\ref{BS_Ising}) will be discussed in section \ref{IsingCFT}.

In the contest of entropy computations, a quantity is particularly interesting: the Affleck-Ludwig {\it boundary entropy} \cite{AffleckLudwig1991}. It was argued that the thermodynamic entropy of an open system described by a CFT should contain a zero-temperature contribution $\ln g$, where $g\equiv g_1g_2$ factorizes in the product of two terms, each one associated to a boundary. Once one knows the boundary state $\left|j\right>$ associated to the boundary $j$, the relative $g_j$ is simply given by \cite{Saleur1998}
\begin{equation}\label{ALg}
 g_j=\left<j|0\right>
\end{equation}
Therefore, in the minimal $c=1/2$ CFT, the boundary entropies should be simply $g_{\tilde{0}}=g_{\tilde{\frac{1}{2}}}=\frac{1}{\sqrt{2}}$, $g_{\tilde{\frac{1}{16}}}=1$.


\section{Entanglement Entropies}\label{Entanglement}

In recent years, a lot of attention has been devoted to the concept of {\it entanglement} \cite{EinsteinPodolskyRosen1935}, one of the most peculiar aspects of the quantum theory. Its importance has been pointed out in many different fields, such as black-hole physics \cite{HolzheyLarsenWilczek1994}, quantum information and computation \cite{NielsenChuang2000} and so on. However, we will focus here on its applications in condensed-matter physics and QFT; in particular, we will restrict to the concept of entanglement entropy, for reasons that will become evident in the following.


\subsection{Definitions}

Let us consider a 1D system of length $L$, and divide it in two intervals, namely $A$ and $B$ (this is not the most general configuration, but it is the one we will use in the remaining of the thesis). Of course, from the density matrix $\rho$ of the whole system, one can define the reduced density matrix of $A$ or $B$, i.e., $\rho_{A/B}\equiv\mbox{Tr}_{B/A}[\rho]$ \cite{NielsenChuang2000}. One can now define the quantity:
\begin{equation}\label{VNE}
 S(A)\equiv-\mbox{Tr}_A\left[\rho_A\ln\rho_A\right]
\end{equation}
This quantity is commonly known as (\emph{bipartite}) \emph{Von Neumann Entanglement Entropy} (VNEE) \cite{NielsenChuang2000}, being just the von Neumann entropy relative to the interval $A$. A remarkable property of this quantity is that it is not extensive: in particular, we have
\begin{equation}\label{Area_law_precursor}
 S(A)=S(B)
\end{equation}
and therefore it just depends on the bipartition of the system. This suggests that the VNEE has to be related to the boundary separating $A$ and $B$: this fact, that holds even for $d>1$, takes the name of \emph{area law} \cite{EisertCramerPlenio2010}, and states that, under certain assumptions, the VNEE of a bipartition is proportional to the area of the boundary separating the intervals. However, as we will see, violations of the area law are common in 1D physics, especially if the theory describing the system is critical \cite{CalabreseCardy2004}.

The VNEE can be generalized in several ways. The most commonly used is the \emph{R\'enyi Entanglement Entropy} (REE) \cite{Renyi2007}, defined as:
\begin{equation}\label{RE}
 S_n(A)\equiv\frac{1}{1-n}\ln\mbox{Tr}_A\left[\rho_A^n\right]
\end{equation}
with $n>0$. The above quantities have been proved \cite{CalabreseLefevre2008} to characterize the full entanglement spectrum \cite{LiHaldane2008} of 1D density matrices; moreover, they spring up quite naturally, as we will see in detail, from CFT computations (see, e.g., \cite{CalabreseCardy2004} and section \ref{replica}). Even these quantities satisfy, as it can be easily seen, the property (\ref{Area_law_precursor}), and obey therefore, at least in certain cases, the area law, as VNEE does.


\subsection{Entanglement Entropies and Criticality}\label{EE_crit}

In this section, we will focus on the behavior of REE's in a quantum system that is undergoing a second-order QPT \cite{MorandiNapoliErcolessi2001}: in such cases the system is described by a CFT \cite{DiFrancescoMathieuSenechal1997}; if the system is 1D, then the CFT is in (1+1)D, and we know that the infinite-dimensional Virasoro algebra allows us to completely solve the theory (see section \ref{CFT}). Therefore, one may expect that analytical computations can be done even for REE's. This is what actually happens: indeed, it has been shown by Holzhey, Larsen and Wilczek \cite{HolzheyLarsenWilczek1994} and then by Calabrese and Cardy \cite{CalabreseCardy2004,CalabreseCardy2009} that, for a conformal system of size $L$, the leading behavior of REE's takes the form (see also section \ref{replica})
\begin{equation}\label{CC}
 S_n(l,L)=S_n^{CC}(l,L)\equiv c_n^\eta+\frac{c}{6\eta}\left(1+\frac{1}{n}\right)\ln\left[\frac{\eta L}{\pi}\sin\frac{\pi l}{L}\right]
\end{equation}
where $l$ is the size of the subsystem $A$ (now containing, if present, one of the boundaries of the system), $\eta=1,\,2$ for PBC/OBC and $c$ is the central charge of the underlying CFT. At least three things are remarkable in this formula: first, the functional form (\ref{CC}), a logarithm, is very typical, and immediately allows one to understand, just looking at it, for instance from numerical data, if the system is in a critical phase or not\footnote{If the (infinite) system is in a massive phase, but close to a conformal point of central charge $c$, it has been shown by Calabrese and Cardy that its VNEE saturates in $l$ to the value \cite{CalabreseCardy2004}
\begin{equation}
 S_1=\frac{c}{6}\ln\xi\nonumber
\end{equation}
being $\xi$ the correlation length in the system. This is actually an efficient way of estimating $\xi$ from numerical data.}; second, the coefficient of the logarithm indicates what is the universality class the system is belonging to (moreover, as we will see in the following, formula (\ref{CC}) is one of the best methods available to get the central charge of the CFT from numerical simulations); third, formula (\ref{CC}) represents a clear violation of the area law, since it explicitly depends on the subsystem size $l$. Formula (\ref{CC}) has been proven to be correct for several critical systems, both analytically (see, e.g., \cite{JinKorepin2004}) and numerically (see, e.g., \cite{VLRK2003}). Of course, being CFT in lattice models just an approximation, corrections to equation (\ref{CC}) can be in principle present: they will be discussed in section \ref{corrections} and in the following chapters. Finally, we spend a word on the constants $c_n^\eta$. At least in the $n=1$ case, the two constant have proven, by Zhou and collaborators, to be correlated one to the other: in fact, we have \cite{ZBFS2006}
\begin{equation}\label{ALBE}
 c_1^{\eta=2}-\frac{c_1^{\eta=1}}{2}=\ln g
\end{equation}
being $\ln g$ the so-called \emph{Affleck-Ludwig boundary entropy} \cite{AffleckLudwig1991}, that depends only on the conformal boundary condition one has, as pointed out in section \ref{BCFT}. However, the situation is, in this case, slightly different from the one of thermodynamic entropy. In fact, thermodynamic entropy is a property of the whole system, that possesses two {\it true} boundaries, while entanglement entropies are properties of a bipartition: now, what in the previous case was the system, is just a part, possessing just one true boundary, since the second is the one separating the two subsystems of the bipartition.


\subsection{Non-Universal Corrections to Entanglement Entropies}\label{corrections}

As one can easily imagine, formula (\ref{CC}) is not the end of the story. In fact, compared to the lattice model, that is the object under study, CFT, coming from the bosonization procedure or other similar arguments, is just an approximation to most easily understand the physics of the model, and therefore the quantities that are computed using CFT do not perfectly reproduce experimental or numerical data.

\begin{figure}[t]
 \begin{minipage}{\textwidth}
  \includegraphics[width=0.5\textwidth]{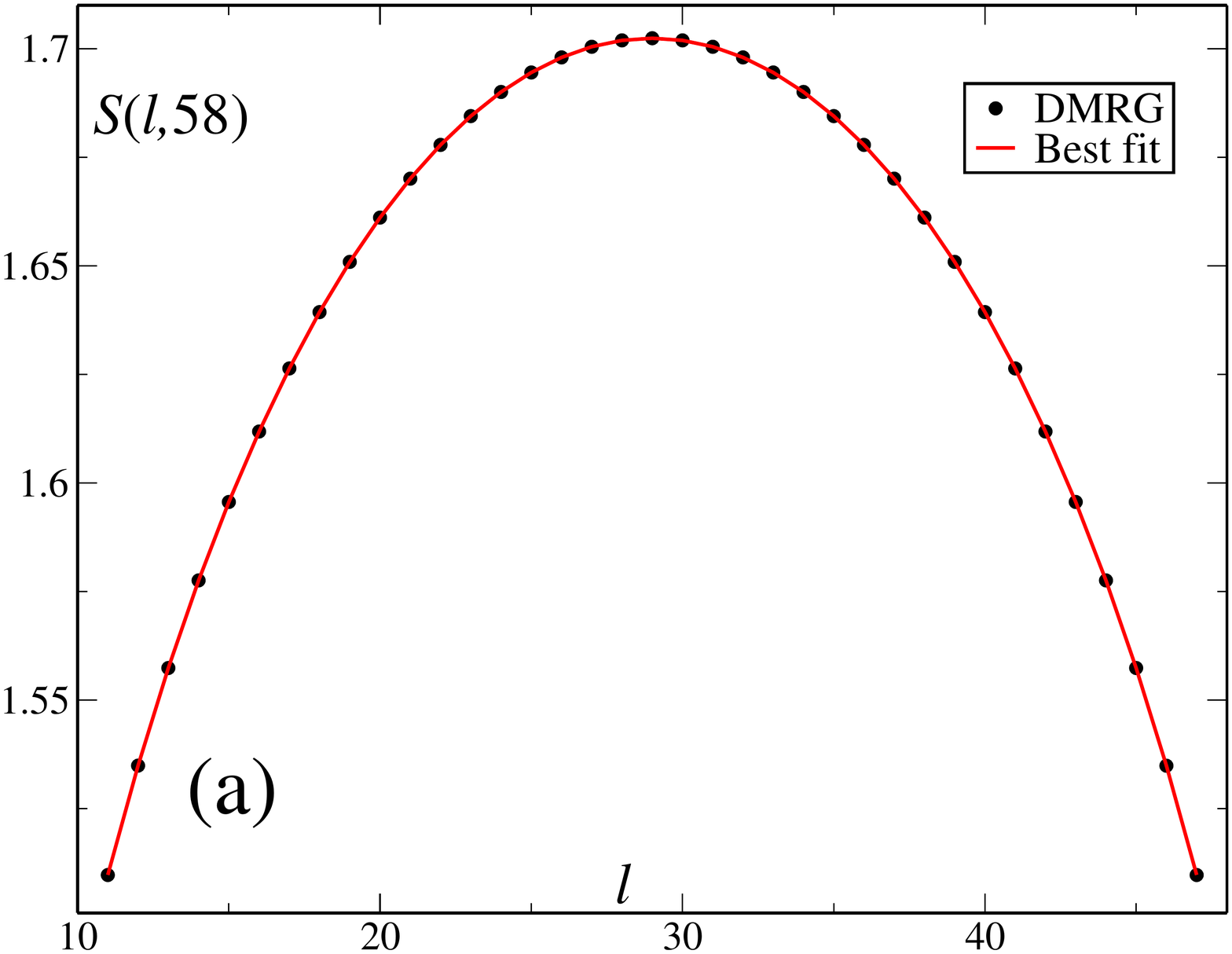}
  \includegraphics[width=0.5\textwidth]{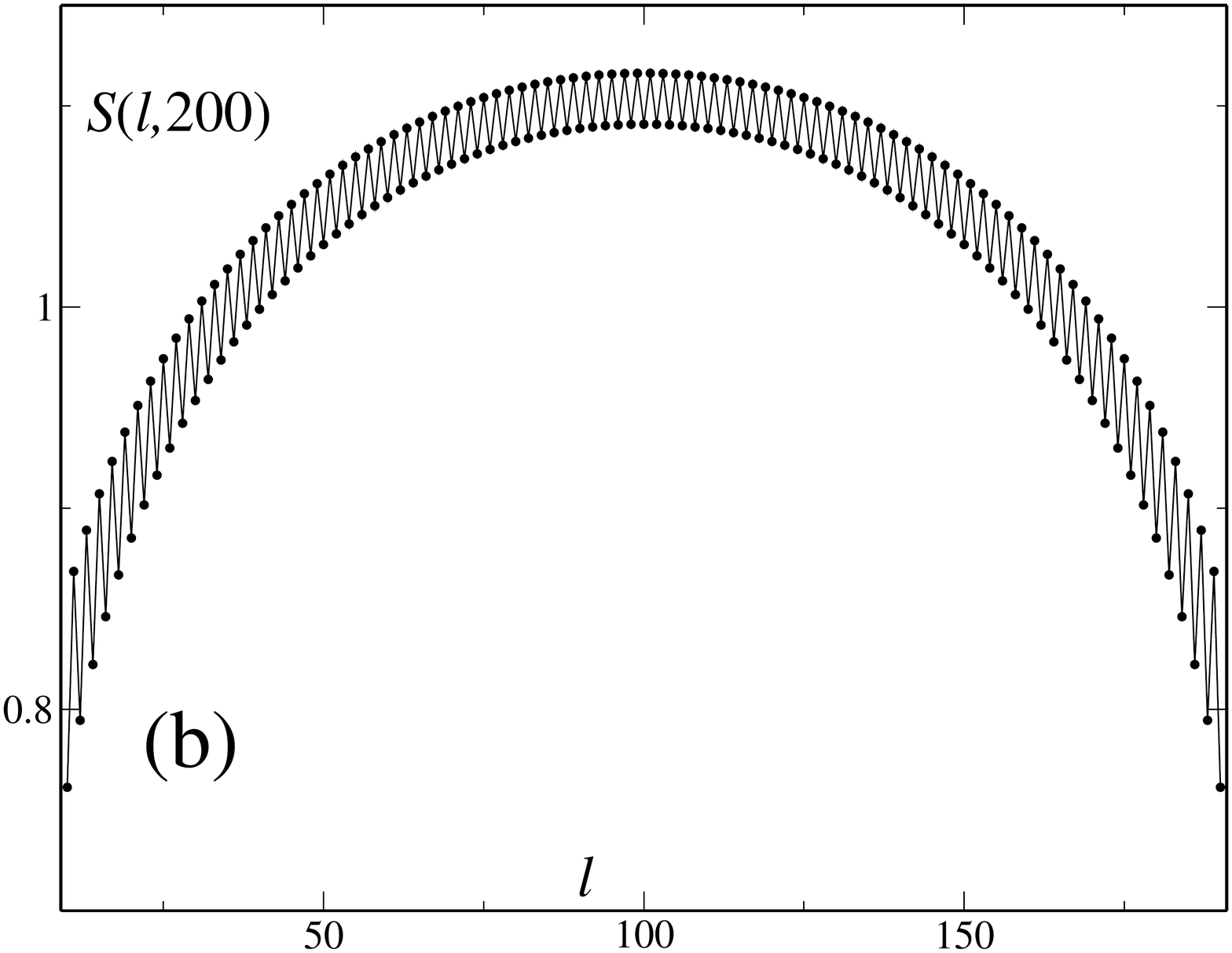}
 \end{minipage}
 \caption{DMRG-computed VNEE in the spin-1/2 XXZ chain with $\Delta=0.5$. (a) PBC: $L=58$, up to 1024 states per block and 4 sweeps. The best fit is done with formula (\ref{CC}), and gives $c_1^1=0.730$, $c=1.001$; (b) OBC: $L=200$, up to 512 states per block and 4 sweeps.}\label{VNE_XXZ1/2}
\end{figure}

This fact is particularly true for entanglement entropies, as it was first argued by Laflorencie and collaborators \cite{LSCA2006}. Let us first consider a critical spin-1/2 XXZ chain with periodic boundary conditions (PBC), i.e., the Hamiltonian
\begin{equation}
 H_\Delta^{PBC}=-\frac{1}{4}\sum_{j=1}^{L}\left(\sigma_j^x\sigma_{j+1}^x+\sigma_j^y\sigma_{j+1}^y-\Delta\sigma_j^z\sigma_{j+1}^z\right)
\end{equation}
with $-1<\Delta\leq 1$. Its VNEE can be computed via DMRG simulations: we report a typical result in figure \ref{VNE_XXZ1/2}(a). As it is explained in the caption, in this case formula (\ref{CC}) perfectly fits the numerical data, and therefore it seems there is no need of further considerations. However, the situation changes if one consider OBC, i.e., Hamiltonian (\ref{XXZ_OBC}): strong oscillating corrections are present, and make the extraction of the central charge from numerical data more complicated. Therefore, from a purely practical point of view, it would be nice to have an analytical expression for these corrections. Studying numerically (see appendix \ref{peschel_method} for details) the (0,0) free lattice fermions of section \ref{FF}, it was found in reference \cite{LSCA2006} that the VNEE follows the behavior $S(l,L)=S_{CC}(l,L)+(-1)^lS_A(l,L)$, with
\begin{equation}\label{XX_SA}
 S_A(l,L)\propto\frac{1}{\frac{L}{\pi}\sin\frac{\pi l}{L}}
\end{equation}
Moreover, studying the spin-1/2 XXZ chain with DMRG, it was found that, in its critical region, formula (\ref{XX_SA}) can be generalized to
\begin{equation}
 S_A(l,L)\propto\frac{1}{\left[\frac{L}{\pi}\sin\frac{\pi l}{L}\right]^K}
\end{equation}
being $K$ the Luttinger parameter of the model ($K=1$ for the (0,0) lattice fermions).

\begin{figure}[t]
 \begin{minipage}{\textwidth}
  \includegraphics[width=0.5\textwidth]{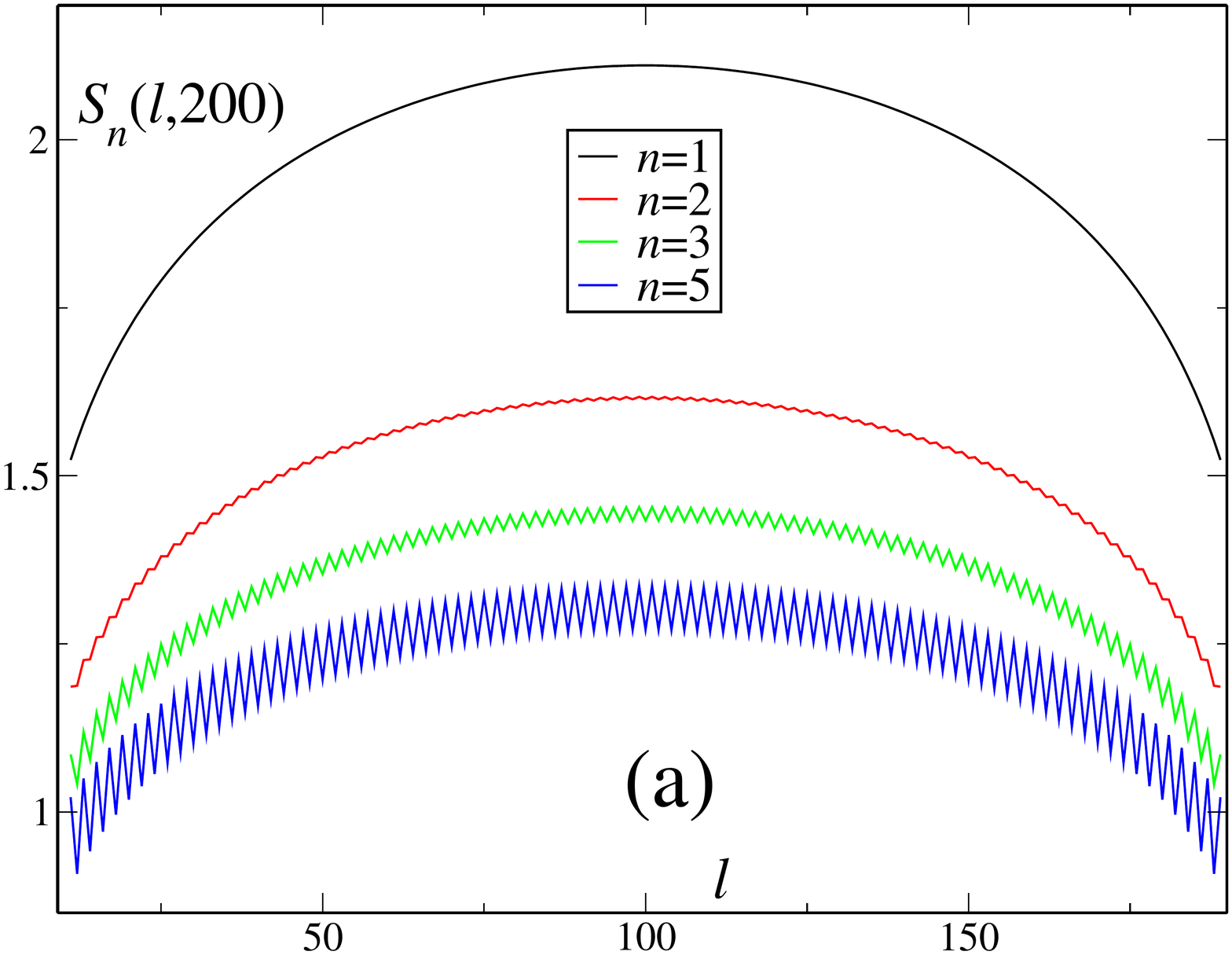}
  \includegraphics[width=0.5\textwidth]{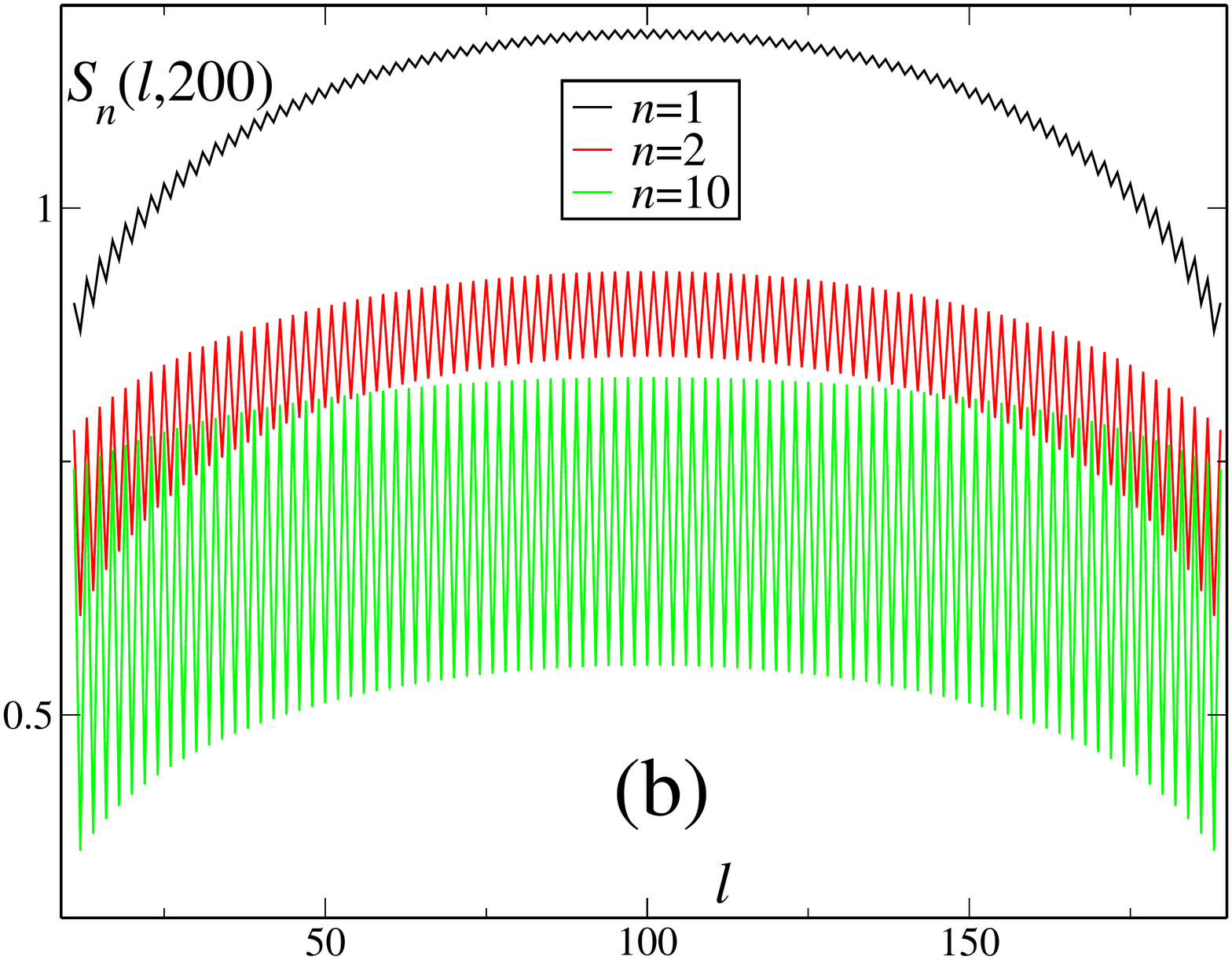}
 \end{minipage}
 \caption{REE's for the spin-1/2 XX chain, with $L=200$ and PBC (a), OBC (b). The computation was done with the method of appendix \ref{peschel_method}.}\label{REs_XX1/2}
\end{figure}

The landscape was extended to REE's by Calabrese and collaborators \cite{CCEN2010}. Both PBC and OBC were considered, for the spin-1/2 XX and XXZ chains: we report some data in figure \ref{REs_XX1/2}. Apart from the different behavior of the corrections with PBC and OBC (for OBC, the corrections are seen to decrease much slower than at PBC), the most significant feature of these REE's is that the oscillating corrections are present even for PBC, when $n>1$. In reference \cite{CalabreseEssler2010}, Calabrese and Essler computed this correction, for large $l$, in the infinite periodic spin-1/2 XX chain, showing that its coefficient should actually vanish for $n=1$. A combination of these analytical and numerical results for the spin-1/2 XXZ chain led the authors of \cite{CCEN2010} to conjecture the REE's to take, in a finite system, the form
\begin{equation}\label{S_LL}
 S_n(l,L)=S_n^{CC}(l,L)+S_n^{CCEN}(l,L)
\end{equation}
where
\begin{equation}\label{CCEN}
 S_n^{CCEN}(l,L)=\frac{f_n^\eta\left(\frac{l}{L}\right)\cos\left(2k_Fl+\omega\right)}{\left|\frac{2\eta L}{\pi}\sin k_F\sin\frac{\pi l}{L}\right|^\frac{2K}{\eta n}}
\end{equation}
Several explanations are to be given. First of all, the conjecture just holds for the Luttinger liquid universality class, as the exponent of the denominator of the expression, containing the Luttinger parameter, shows. The function $f_n(x)$ is thought to be a scaling function of its argument, and it is model-dependent. As previously mentioned, in the infinite spin-1/2 XX chain with PBC, it has been shown \cite{CalabreseEssler2010} that it takes, when $l$ is large, the form
\begin{equation}
 f_n^1=\frac{2}{1-n}\frac{\Gamma^2\left(\frac{1+1/n}{2}\right)}{\Gamma^2\left(\frac{1-1/n}{2}\right)}
\end{equation}
being $\Gamma(x)$ the Euler-$\Gamma$ function. This quantity can be seen to be null for $n=1$, as it is seen numerically in figure \ref{REs_XX1/2}(a), and different from zero for $n\neq 1$ (even when $n<1$). A similar calculation has been performed even in the OBC case by Fagotti and Calabrese \cite{FagottiCalabrese2011}. It was found that, in this case, one has a slightly but substantially different form:
\begin{equation}
 f_n^2=\frac{2}{1-n}\frac{\Gamma\left(\frac{1+1/n}{2}\right)}{\Gamma\left(\frac{1-1/n}{2}\right)}
\end{equation}
so that $f_1^2=-1\neq 0$, as expected from numerical simulations (see figure \ref{REs_XX1/2}(b)), and $f_n\neq 0$, at least for $n>1$.

This kind of corrections has been derived analytically, at least partially, in the frameworks of conformal perturbation theory \cite{CardyCalabrese2010} and \emph{replica trick} \cite{HolzheyLarsenWilczek1994,CalabreseCardy2004} by Cardy and Calabrese, even if a direct bosonization derivation is not yet available (very recently, Swingle, McMinis and Tubman \cite{SwingleMcMinisTubman2012} gave a proof of formula (\ref{CCEN}) that seems to be easily extendable to the general case). In particular, it was shown for infinite systems that the presence of relevant or irrelevant perturbations to the bulk CFT Hamiltonian can lead to \emph{unusual}, i.e., $n$-dependent corrections to the $n$-th REE: beside non-unusual corrections, proportional to $l^{2(2-\Delta)}$ ($l^{2-\Delta}$ for open systems), where $\Delta$ is the scaling dimension of the perturbation, two other types of corrections are present, proportional to $l^{-2\Delta/n}$ ($l^{-\Delta/n}$ for open systems) and $l^{2-\Delta-\Delta/n}$. If the perturbation is marginal, the situation is completely different, and leads to \emph{logarithmic} corrections proportional to $(\ln l)^{-2}$. We note, referring to the spin-1/2 XXZ chain, that the second of the above predicted exponents is compatible with (\ref{CCEN}), with $\Delta=K$, that is, the most relevant conformal dimension in the bosonic continuum theory \cite{Giamarchi2003}. It is therefore clear that having a good control of the kind of corrections that can arise can give decisive hints on the operator content of the effective CFT.  We finally remark that the corrections arising from \emph{boundary} perturbations have been computed, in a similar way, by Eriksson and Johannesson \cite{ErikssonJohannesson2011}, and are shown to be non-unusual for each perturbation one adds.

A different kind of study was performed by Xavier and Alcaraz in \cite{XavierAlcaraz2012}. The authors noted, by means of DMRG simulations, that oscillating finite-size corrections arise if there is at least one $U(1)$ symmetry, and are absent for models with just discrete ones; moreover, the operator they originate from should be the {\it energy} operator, as checked estimating $\Delta$ in different models.

It is worth mentioning that the parity oscillations considered so far are not a peculiarity of critical regions. Actually, Giampaolo and collaborators showed in reference \cite{GMDADSI2012} for the spin-1/2 XY chain that such corrections can emerge even in massive phases, and their presence or absence is related to the  closeness to a {\it factorizable} (i.e., product) ground state. This behavior is very interesting and needs further investigation.

Until now, we considered only non-universal corrections to the scaling of REE's, i.e., corrections that do not depend only on the universality class of the model: in fact, e.g., in the spin-1/2 XXZ chain, the Luttinger parameter $K$, the exponent of such corrections, explicitly depends on the anisotropy coefficient $\Delta$. Different types of corrections, this time \emph{universal}, can arise if one considers, instead of the ground state, an \emph{excited} state of a conformal system, as shown by Alcaraz, Berganza and Sierra \cite{AlcarazBerganzaSierra2011,BerganzaAlcarazSierra2012}: this case will be discussed in detail in chapter \ref{CFT+OBC}.


\section{The Density-Matrix Renormalization Group algorithm}\label{DMRG}

Even in 1D systems, where exact solutions of some complicated models are available \cite{Takahashi1999}, it is not possible to extract information about the general interacting case without making use of some approximations. Such approximations can be performed analytically (see section \ref{Bosonization}) or numerically. Among the latter, in the last twenty years, since its first introduction by White, the density-matrix renormalization group (DMRG) algorithm \cite{White1992,White1993} (see \cite{Schollwock2005} for a review) has revealed to be one of the cheapest and most reliable numerical 1D techniques. In this section we give a brief overview of the method, that will be widely used in the following. The DMRG code we will use in the remaining of the work, that is very versatile and can be applied to a great variety of lattice models, was developed in Bologna by F. Ortolani and collaborators, and was used by the members of the group in a number of studies (see, e.g., \cite{FanoOrtolaniZiosi1998} and, in the last two years, the works \cite{DDDBO2011,DalmonteZollerPupillo2011,RoscildeDegliEspostiBoschiDalmonte2012,DEMOV2012}).


\subsection{The Algorithm}

The DMRG algorithm is a particular kind of renormalization group algorithm, suited for the treatment of lattice models. Traditional renormalization schemes \cite{Wilson1975} operate iteratively in the following way: a system of size $L$ is divided into small subsystems, and then they are taken in couples; the Hamiltonian of each couple is then diagonalized. However, most of the times this diagonalization cannot be carried out exactly, and one has to choose some "information", that he thinks not to be fundamental, to be discarded. In the standard Wilson approach, the prescription is to keep just the $M$ eigenstates corresponding to the lowest lying eigenvalues of the couple Hamiltonian, where $M$ is chosen by the user. Starting from these eigenvalues and eigenstates, a new "renormalized" couple Hamiltonian is built, and each one is now treated as a single system, whose Hamiltonian is a $M\times M$ diagonal matrix. The procedure is iterated until all couples are merged to the whole system. Despite the success of Wilson's approach to the Kondo problem \cite{Wilson1975}, the method resulted inefficient to describe other systems, mainly for two reasons: first, for the assumption that, in general, the most significant states for the physics of a system are the ones associated with the lowest-lying energy eigenvalues; second, because each couple is diagonalized separately from the others, and therefore just eigenstates vanishing on the boundary of the couple blocks are kept.

These problems were solved with the introduction of DMRG \cite{White1992,White1993,Schollwock2005}: it was shown that the optimal decimation prescription, i.e., the optimal (the meaning of this word can be intended in many ways; see \cite{Schollwock2005}) way to choose the states to keep, is given not in term of the Hamiltonian but of the reduced density matrix of the considered couple (or {\it block}). Moreover, the boundary-conditions problem was avoided by the choice of diagonalizing, rather than a single-block Hamiltonian, the Hamiltonian of a {\it superblock} containing the block as a subsystem, finding a desired eigenstate, called {\it target state}. Another key feature of the algorithm is the renormalization procedure: it is not done by merging couples anymore, but by adding one site to the edge of each block in a particular way, a number of times sufficient to renormalize the whole system. A complete renormalization procedure is called {\it sweep}, and the precision of the obtained data typically betters with the number of performed sweeps. The renormalization scheme was determined empirically to optimize the goodness of the results, and is different for PBC and OBC. In particular, fixing the number of kept states per block, the obtained results are better for OBC than for PBC; however, a renormalization scheme has been found by Verstraete, Porras and Cirac, in order to have the same precision with the same number of kept states both with OBC and with PBC \cite{VerstraetePorrasCirac2004}.

DMRG gives, as results, the properties of some selected states, typically the ground state or the low excited states of a quantum lattice Hamiltonian \cite{White1992,White1993}. Several symmetries can be implemented in the algorithm, so that it is able to diagonalize the Hamiltonian in some specific sector. Among the quantities that can be computed, we include the total energy, correlation functions, and R\'enyi entropies (see below). The typical error on these quantities can be computed, and it is, at least, of the order of the sum of the discarded weights of the reduced density matrix, known as {\it truncation error}.

The algorithm has revealed incredibly powerful already at the beginning of its history, when it was used to determine the Haldane gap in the spin-1 Heisenberg chain \cite{WhiteHuse1993}. A far from being exhaustive list of the successes of DMRG is given in \cite{Schollwock2005}.

Finally, we remark that several ways to implement a DMRG algorithm to study time-dependent problems have been found in recent years, giving a huge boost to this new research field (see \cite{SchollwockWhite2006} for a review).


\subsection{DMRG and Entanglement Entropies}\label{Ent_DMRG}

Consider now equations (\ref{VNE}) and (\ref{RE}). It is straightforwardly seen that, in terms of the eigenvalues $\{w\}$ of the reduced density matrix of the subsystem $A$, they look respectively
\begin{equation}
 S(A)=-\sum_ww\ln w
\end{equation}
and
\begin{equation}\label{REE_DMRG}
 S_n(A)=\frac{1}{1-n}\ln\sum_ww^n
\end{equation}
Since $\{w\}$ is the natural outcome of the DMRG algorithm, these quantities are very cheaply computed using this method.

Apart from this practical motivation, we mention that entanglement entropies have turned out to be important in DMRG theory: we do not discuss this subject here, and we refer to \cite{Schollwock2005} for further insights.


\chapter{Detection of Crossovers by Entanglement Entropies}\label{Crossovers}

In this chapter, we see how the previously defined REE (see section \ref{Entanglement}) allows for an accurate estimation of the Luttinger parameter of a Luttinger liquid (see section \ref{Bosonization}). What reported in this chapter was developed by myself, M. Dalmonte and E. Ercolessi in \cite{DalmonteErcolessiTaddia2011,DalmonteErcolessiTaddia2012}. In section \ref{section method}, we describe the numerical method we developed; in section \ref{XXZcheck} and \ref{Hubbardcheck} we test it against two integrable models, the spin-1/2 XXZ chain and the deep-attractive Hubbard model; in section \ref{K_dipoles} we use it to make predictions on a chain of hard-core dipolar bosons; in section \ref{K_3-2} we analyze the critical properties of the spin-3/2 XXZ chain; in section \ref{E3-2} we address two interesting questions about the REE's in that model.


\section{The Method}\label{section method}

The idea of this chapter is very simple, but its practical realization is, as we will see, non-trivial: we want to estimate the Luttinger parameter of a Luttinger liquid system from DMRG simulations using REE's, i.e., by means of equation (\ref{S_LL}). This task is of great practical significance, since accurate estimations of $K$ from correlation functions are in many cases hard to perform \cite{CreffieldHauslerMacDonald2001}. The use of REE's, that are the quantities DMRG can furnish with best precision (see section \ref{Ent_DMRG}), will, as we shall see, provide a reliable practical tool for the Luttinger-parameter detection.

The first idea that, of course, occurs, is to use directly (\ref{S_LL}) to fit numerical data, and to get $K$ in this simple way. Unfortunately, many problems arise. First, we note that the oscillations are, in some cases, small in amplitude when compared to the leading contributions. Second, there is a conceptual difficulty: the functional form of $f_n^\eta(x)$ is in principle unknown, even if there are numerical clues that it is actually a constant \cite{LSCA2006,CCEN2010,XavierAlcaraz2011a}. Third, the number of fit parameters, i.e., four ($c_n^\eta$, $c(L)$, $f_n^\eta$, $K(L)$), is quite large, and one has therefore to consider large systems. Fourth, as usual, one has to take a finite-size scaling (FSS) (see, e.g., \cite{DiFrancescoMathieuSenechal1997}) of the finite-size data, and if the precision of the estimations is not so good, the procedure may give bad results. This technique was employed by Xavier and Alcaraz in \cite{XavierAlcaraz2011a}, manifesting all its intrinsic limitations.

The method we are going to introduce is tailored to solve these problems (or, at least, some of them). Let us consider the $n$-th REE difference
\begin{equation}\label{method}
 dS_n(L)\equiv S_n\left(\frac{L}{2},L\right)-S_n\left(\frac{L}{2}-\frac{\pi/2}{k_F},L\right)
\end{equation}
An explanation is in order. The first term is just the half-chain $n$-th REE: if $L$ is even, the oscillating term of the REE has there a maximum, since there $\cos(2k_Fl)=1$ (we will assume, unless stated, $\omega=0$). The second term is just the $n$-th REE computed at a subsystem size $l$ at which $\cos(2k_Fl)=-1$, and therefore $dS_n(L)$, being the difference of the two, must always be almost two times larger than the oscillation amplitude: our method allows, in a way, to "enhance" such an amplitude.

For large $L$, formula (\ref{method}) can be simplified, in order to give
\begin{equation}\label{Useful_formula}
 dS_n(L)=\frac{\pi^4c\left(1+\frac{1}{n}\right)}{48\eta k_F^2}\frac{1}{L^2}+\frac{\cos(k_FL)}{L^{\frac{2K}{\eta n}}}\left[a+\frac{b}{L}+o\left(\frac{1}{L}\right)\right]
\end{equation}
where $c$ and $K$ are the thermodynamic limit values of the central charge and the Luttinger parameter respectively (a scaling form of $c(L)$ and $K(L)$ has been supposed, and it simply results in subleading terms, contained in the $o$ in the square bracket), and $a$, $b$ are constants related to the values of $f_n(x)$ and its first derivative at $x=1/2$ ($f_n(x)$, being a scaling function of its argument, has been expanded in a neighbour of 1/2). If we neglect the terms that are $o(L^{-1})$ in the square bracket, we are left with just three unknown parameters, since $c$ is known to be one for Luttinger liquids. Therefore, what we expect is to use the quantity (\ref{method}), very easily obtained from DMRG simulations, as a function of $L$ to extrapolate the thermodynamic value of the Luttinger parameter $K$. In the next sections, we will see how to apply the technique to physical models and show the corresponding results.


\section{First Check: the Spin-1/2 XXZ Chain}\label{XXZcheck}

We start by looking at the critical spin-1/2 XXZ chain (see section \ref{XXZ})
\begin{equation}
 H_\Delta=-\sum_{j}\left[S_j^xS_{j+1}^x+S_j^yS_{j+1}^y-\Delta S_j^zS_{j+1}^z\right]
\end{equation}
at PBC/OBC and total size $L$: the QFT description of the model is known to be a Luttinger liquid \cite{Giamarchi2003}. The Luttinger parameter of the corresponding Luttinger liquid is known by Bethe ansatz (BA), and, being it a bulk property, is the same in both cases:
\begin{equation}\label{K_BA}
 K=\frac{\pi}{2\cos^{-1}(-\Delta)}
\end{equation}
Moreover, we are interested in the value of $k_F$: in our case it is $\pi/2$, as a consequence of the Luttinger theorem \cite{Giamarchi2003}.

\begin{figure}[t]
 \centering
 \includegraphics[width=0.5\textwidth]{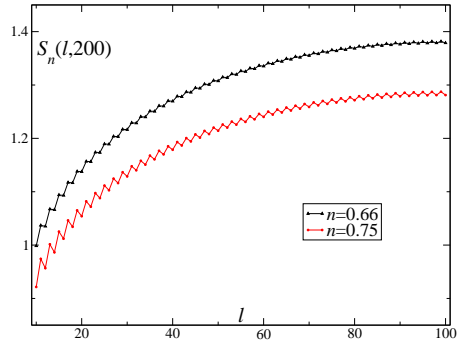}
 \caption{REE's with $n<1$ for the spin-1/2 XXZ chain with OBC, $\Delta=0.5$ and $L=200$.\label{REs_n<1_XXZ1/2}}
\end{figure}

The REE's have been shown to scale as (\ref{S_LL}) \cite{LSCA2006,CCEN2010} (see also figure \ref{VNE_XXZ1/2}); however what is new is that, for OBC, oscillations are present even when $n<1$ (see figure \ref{REs_n<1_XXZ1/2}: we plot just the region $l<L/2$, since the Hamiltonian is symmetric under reflection $j\rightarrow L-j$). Even for $n\geq 1$, we observe that the REE's have the predicted shape. However, we observe that the precision of the REE's computed with DMRG lowers as $n<1$ lowers, and therefore we just plot $S_{0.66}$ and $S_{0.75}$. The reason is that, as it can be seen from equation (\ref{REE_DMRG}), lowering $n$ below 1, the discarded weights in the DMRG truncation procedure become progressively more important, and therefore, in order to achieve a better precision for the corresponding REE's, one should keep a larger number of block per states.

\begin{figure}[t]
 \begin{minipage}{\textwidth}
  \includegraphics[width=0.5\textwidth]{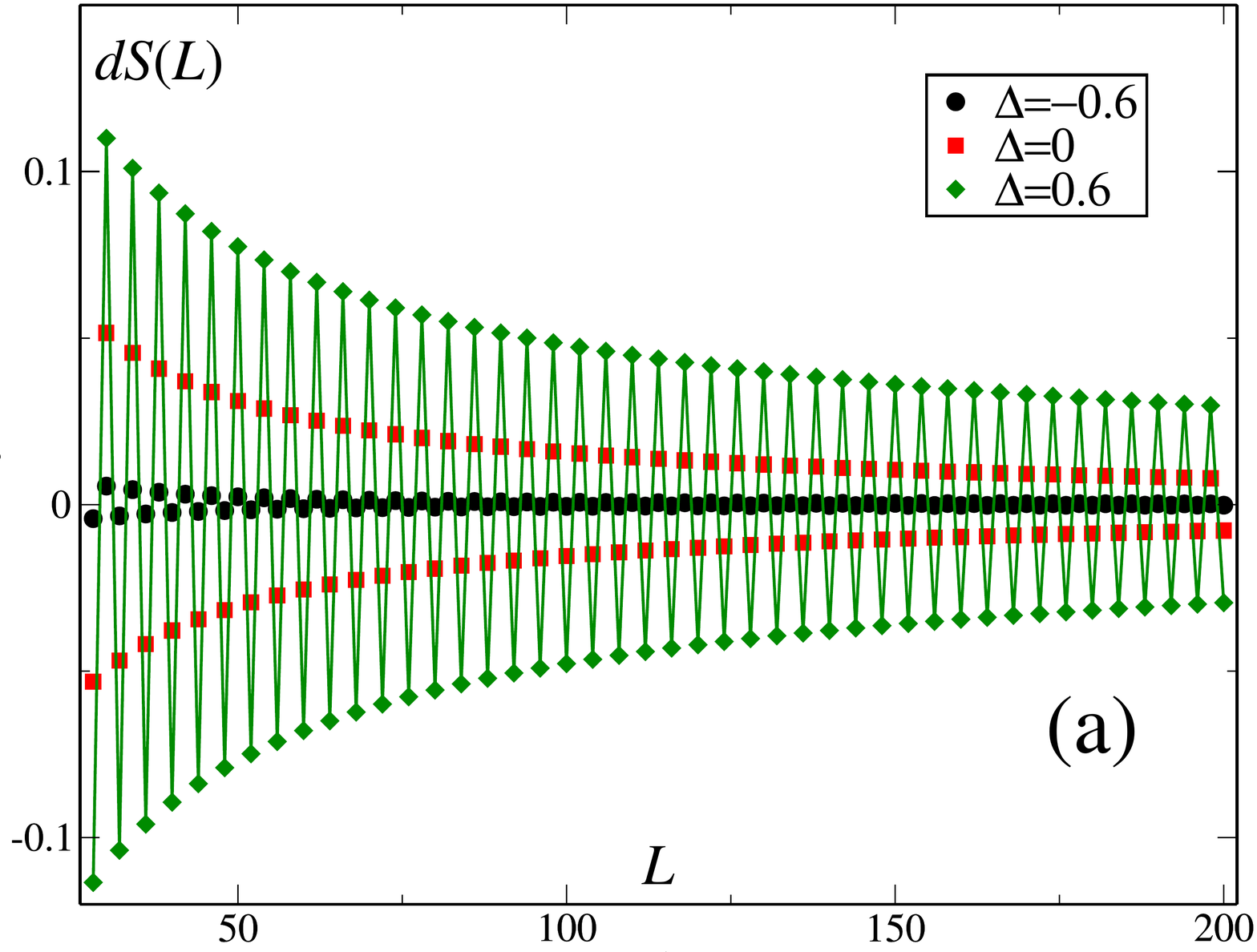}
  \includegraphics[width=0.5\textwidth]{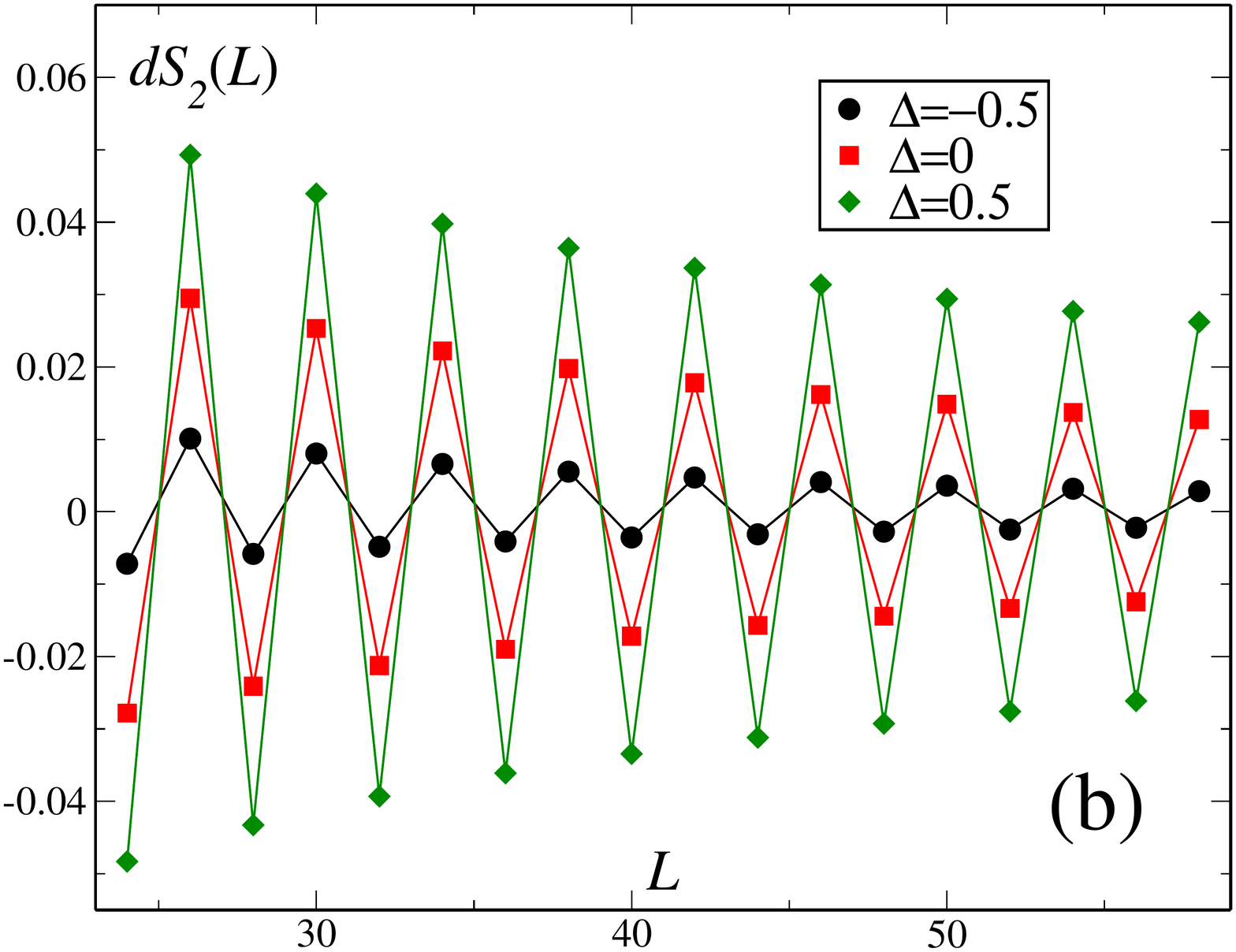}
 \end{minipage}
 \begin{minipage}{\textwidth}
  \centering
  \includegraphics[width=0.5\textwidth]{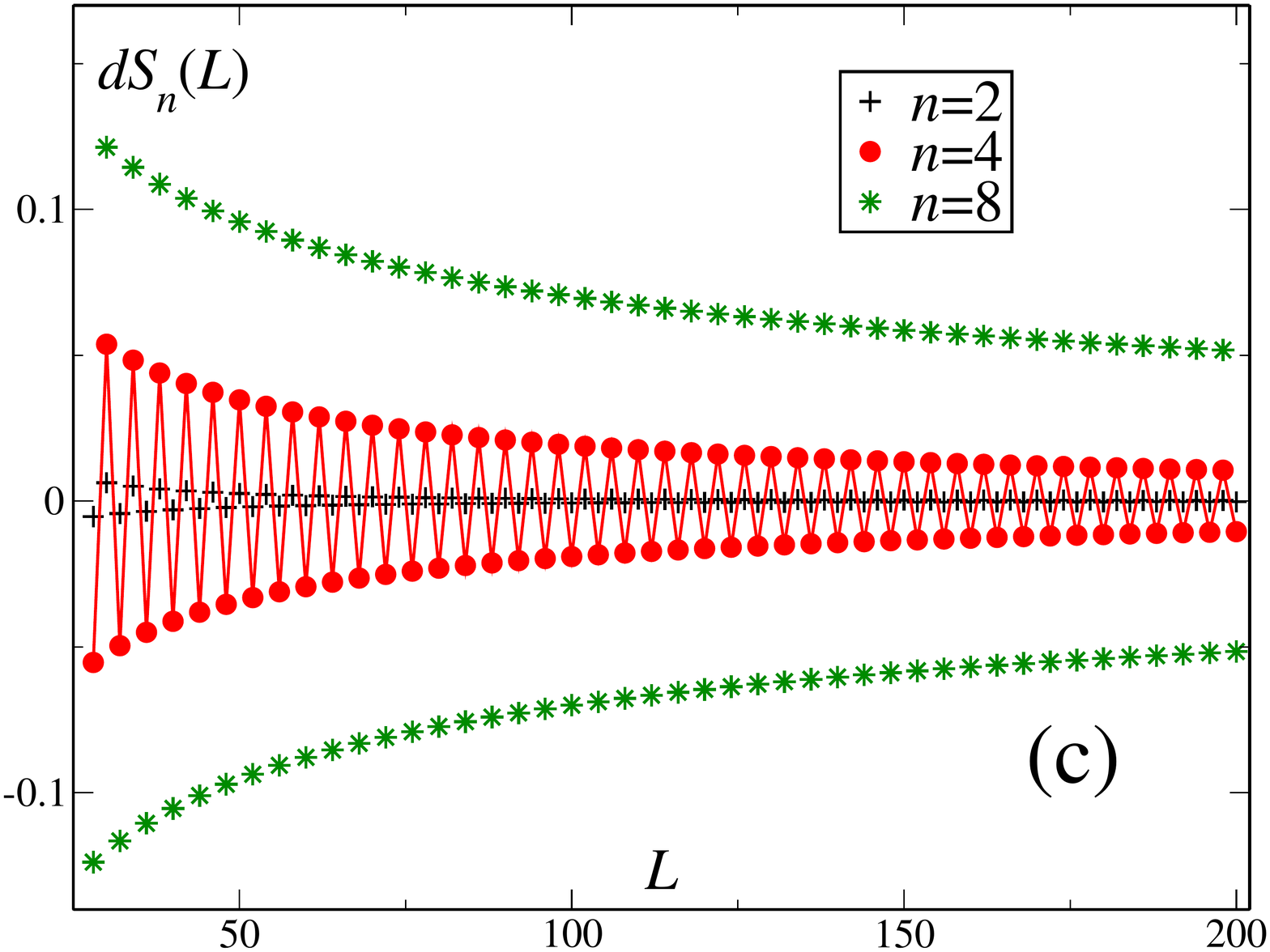}
 \end{minipage}
 \caption{$dS_n(L)$ in the spin-1/2 XXZ chain for different values of $\Delta$ at OBC, $n=1$, even $L\in[28,200]$ (a) and PBC, $n=2$, even $L\in[24,58]$ (b). (c): $dS_n(L)$ at $\Delta=-0.9$, for different values of $n$. Lines are best fits of the data with formula (\ref{Useful_formula}).}\label{dS_n_XXZ1/2}
\end{figure}

Since the behavior is the expected one, we can apply our method. Our DMRG data are produced at even $L\leq 200/60$, keeping up to 512/1024 states per block and applying 4 sweeps, in order to have a truncation error of $10^{-9}/10^{-8}$ for OBC/PBC; REE's with $0.5\leq n\leq 100$ are computed. In figure \ref{dS_n_XXZ1/2} we plot $dS_n$ as a function of $L$, keeping respectively $n$ and $\Delta$ fixed. Looking at these pictures, we note two interesting features of $dS_n$: first, at fixed $n$, its amplitude increases as $\Delta$ increases, reflecting the antiferromagnetic nature of the oscillations \cite{LSCA2006}; second, at fixed $\Delta$, the amplitude increases with $n$.

A key task in the procedure is to establish which value of $n$ gives the best estimate of $K$. In order to do it, we define the quantity $\delta\equiv K_{num}-K_{ana}$ (being $K_{num}$ and $K_{ana}$ the values of $K$ we get from DMRG and formula (\ref{K_BA})) and, keeping $\Delta$ fixed, we consider $\left|\delta\right|$. The results, for OBC and $\Delta=0.5$, are shown in figure \ref{kappa_figure}(a), and the shape is similar for almost all $\Delta$ (see below). We note that our method provides a value of $K$ that is very close to the predicted one in a wide range of $n$, namely for $n\in[1,3]$, and therefore is quite robust to the choice of $n$. Actually, we get the best value of $K$ for $n=0.66$, but since there the precision of $S_{0.66}$ is not so good, we think the result is a coincidence. Therefore, in the following, we will (almost) always use $n\in[1,3]$, taking the $n$ that gives the best fit accuracy.

Another interesting point, especially from the point of view of the computational cost, is to decide how many data points we need to have an accurate estimate. To do this, we consider $L\in[52,52+L_{fit}]$ at $\Delta=\pm0.5$, varying $L_{fit}$, i.e., the number of kept points in the fit procedure. The result is shown in figure \ref{kappa_figure}(b). It is clear from the picture that, at least in a region that is quite distant from the border of the critical phase, the method is robust to the choice of $L_{fit}$, and the results are, at OBC, good at 1\% of precision already at $L_{fit}\lesssim30$, while, for lower $L_{fit}$, strong finite-size effects, combined with errors due to the fitting procedure, affect the results. Therefore one could, in principle, apply our method using a quite narrow range of system sizes.
\begin{figure}[t]
 \begin{minipage}{\textwidth}
  \includegraphics[width=0.5\textwidth]{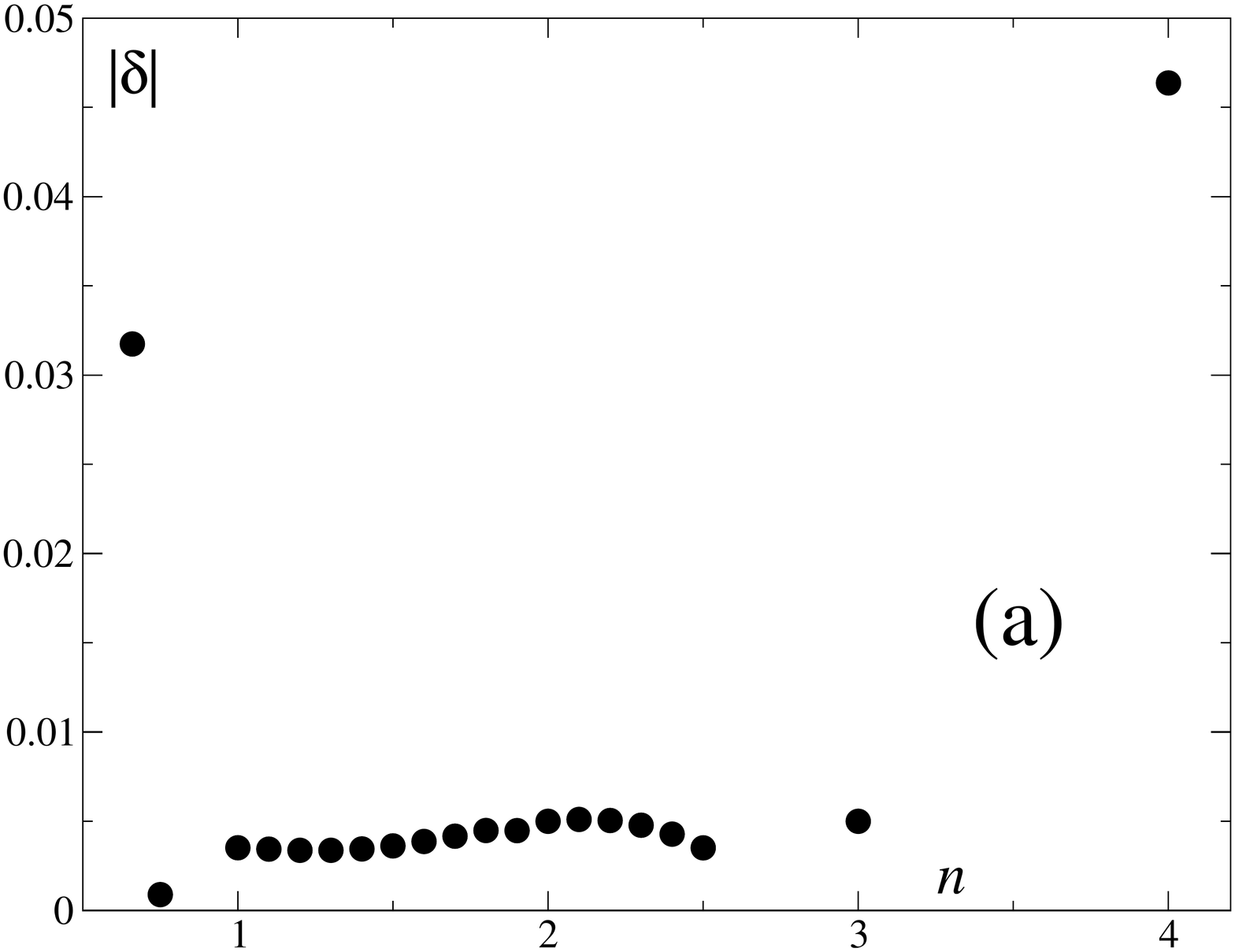}
  \includegraphics[width=0.5\textwidth]{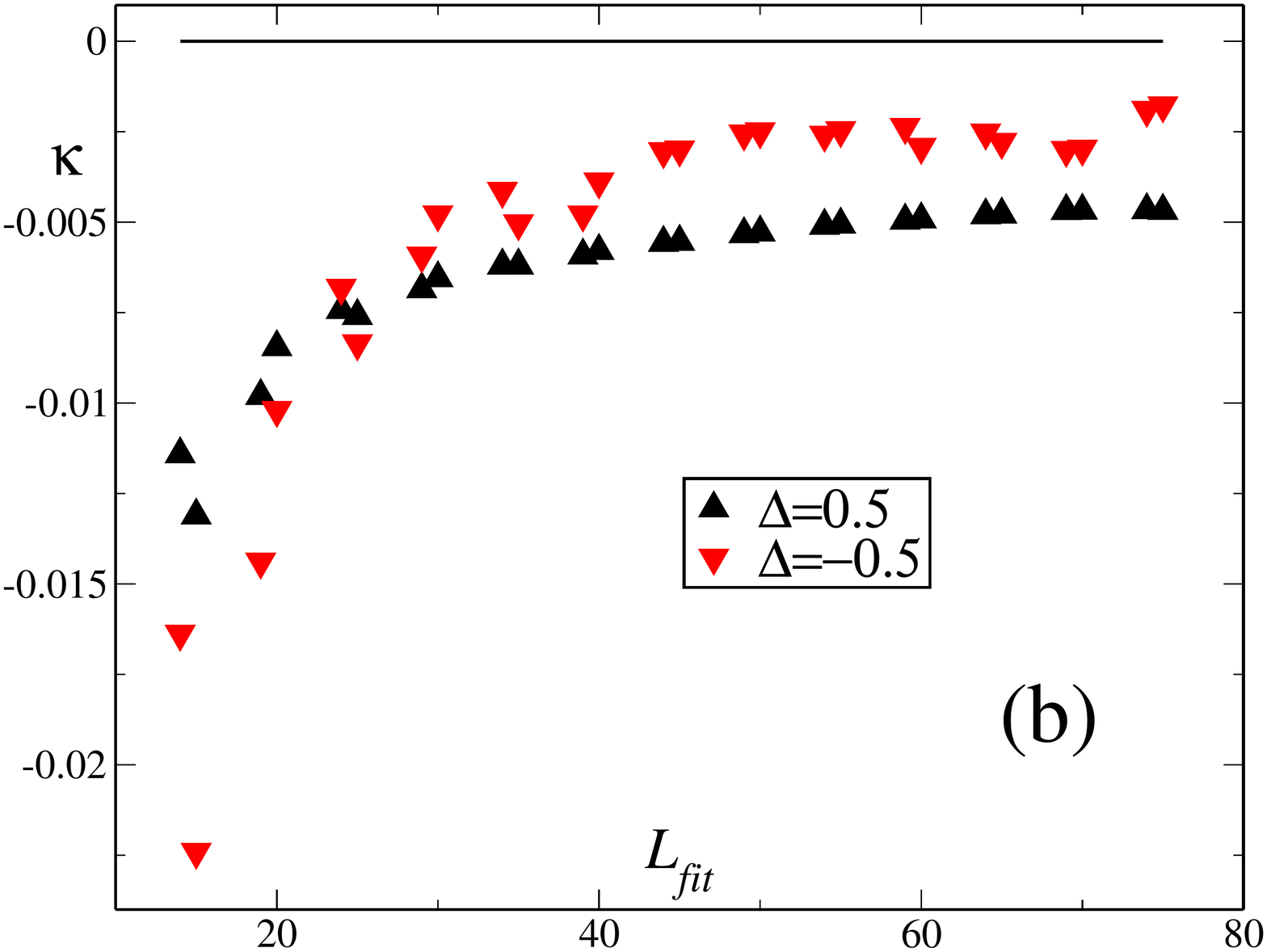}
  \caption{Deviation of the measured value of $K$ from the predicted one, as a function of $n$, at OBC, $\Delta=0.5$, $L\in[28,200]$ (a), and as a function of $L_{fit}$ (see text), at OBC and $\Delta=\pm0.5$ (b), in the spin-1/2 XXZ chain ($\kappa\equiv\delta/K_{ana}$).\label{kappa_figure}}
 \end{minipage}
\end{figure}

\begin{figure}[t]
 \includegraphics[width=\textwidth]{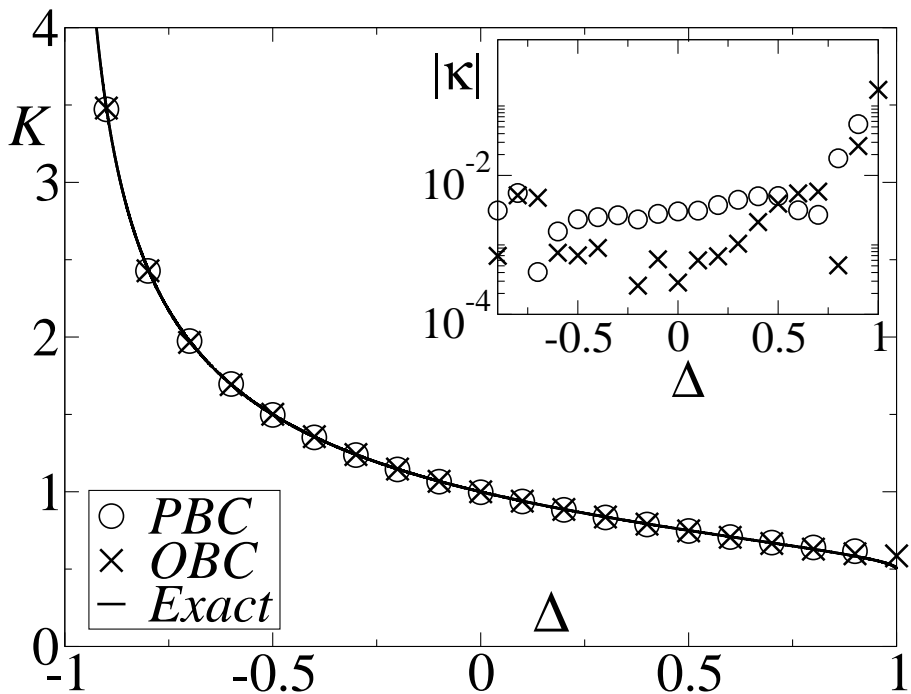}
 \caption{Estimation of $K(\Delta)$ for the spin-1/2 XXZ chain, and comparison with formula (\ref{K_BA}). Inset: absolute relative deviation $\left|\kappa\right|\equiv\left|K_{num}/K_{ana}-1\right|$ as a function of $\Delta$.\label{K-Delta}}
\end{figure}

We plot in figure \ref{K-Delta} the main result of this section, i.e., the behavior of $K(\Delta)$ that we get from the numerics. The $\Delta$-step we consider is 0.1, starting from $\Delta=-0.9$ to $\Delta=1$. The fits of the numerical data with formula (\ref{Useful_formula}) are in general very good, and the resulting value of $K$ is in most cases in excellent agreement with the analytical prediction, both at OBC and at PBC. We used, to extract $K$, $dS_1$ for $\Delta\geq-0.6$, $dS_4$ for $\Delta=-0.7,-0.8$ and $dS_8$ for $\Delta=-0.9$: the reason is, as previously stated, that the oscillations are so small for $\Delta$ close to -1 that we have to consider larger $n$ in order to make them appreciable. The uncertainty on $K$ is computed by fitting the data twice, once by using $L\in[62,200]$ and then $L\in[64,200]$, and taking (half of) the difference of the two results: the uncertainty, estimated this way, is always of the order of $10^{-3}$ or lower. Exceptions to these behaviors arise close to $\Delta=1$, due to the fact that $\Delta=1$ is a {\it Berezinzkij-Kosterlitz-Thouless} (BKT) transition point\cite{ItzyksonDrouffe1989}, where irrelevant perturbations to the Hamiltonian (\ref{H_LL}) become marginal and induce {\it logarithmic} corrections to observable quantities \cite{HWHM1996}, that we are not able to treat in an efficient way.

Therefore, our first check gives us the hope that our method can work in non-trivial cases, at least away from BKT transitions. In the next section we will test it on a different, more complicated model.


\section{Second Check: the 1D Attractive Hubbard Model}\label{Hubbardcheck}

We consider now the 1D attractive Hubbard model, i.e., Hamiltonian (\ref{Hubbard_Hamiltonian}) with $U<0$ (see section \ref{Hubbard}). As we already said, this model displays, in the deep attractive regime and at half filling, a spin-charge separation in a gapped spin and a gapless charge sector \cite{Giamarchi2003}, and only this last one is expected to be a Luttinger liquid. Therefore, REE's will contain a contribution from both. However, we expect we can still apply our method, since the contribution from the gapped sector will result in a constant (with respect to $l$; see section \ref{EE_crit}) that will cancel in the computation of $dS_n$. A check that the method works in this case is very important, since it demonstrates that it can be applied to multi-species systems, as generalized Hubbard models \cite{BCDDEBEO2010}, integer spin chains \cite{BotetJullienKolb1983}, spin ladders \cite{BDRS1993} and mixtures \cite{DEMOV2012}.

\begin{figure}[t]
 \begin{minipage}{\textwidth}
  \includegraphics[width=0.5\textwidth]{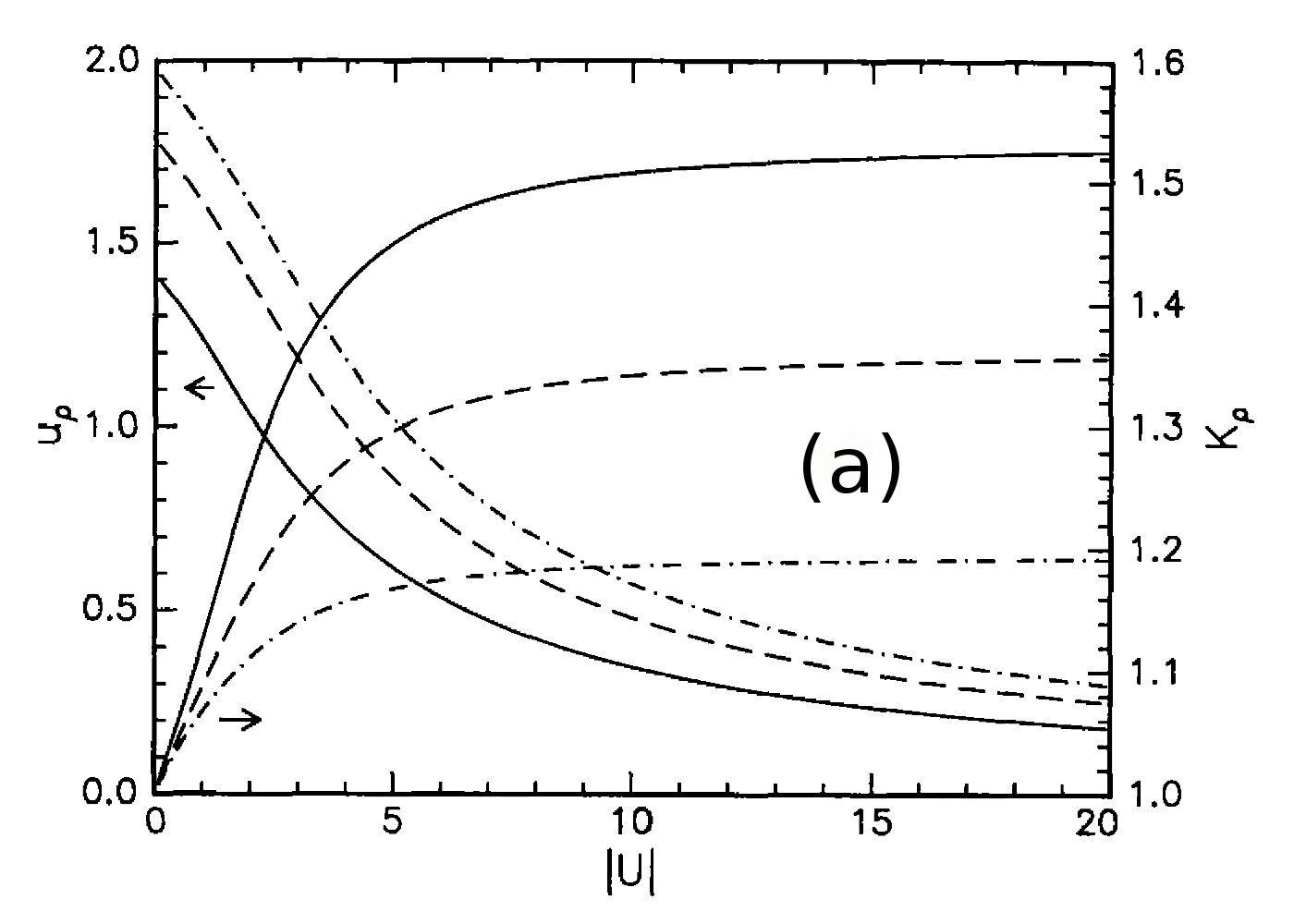}
  \includegraphics[width=0.5\textwidth]{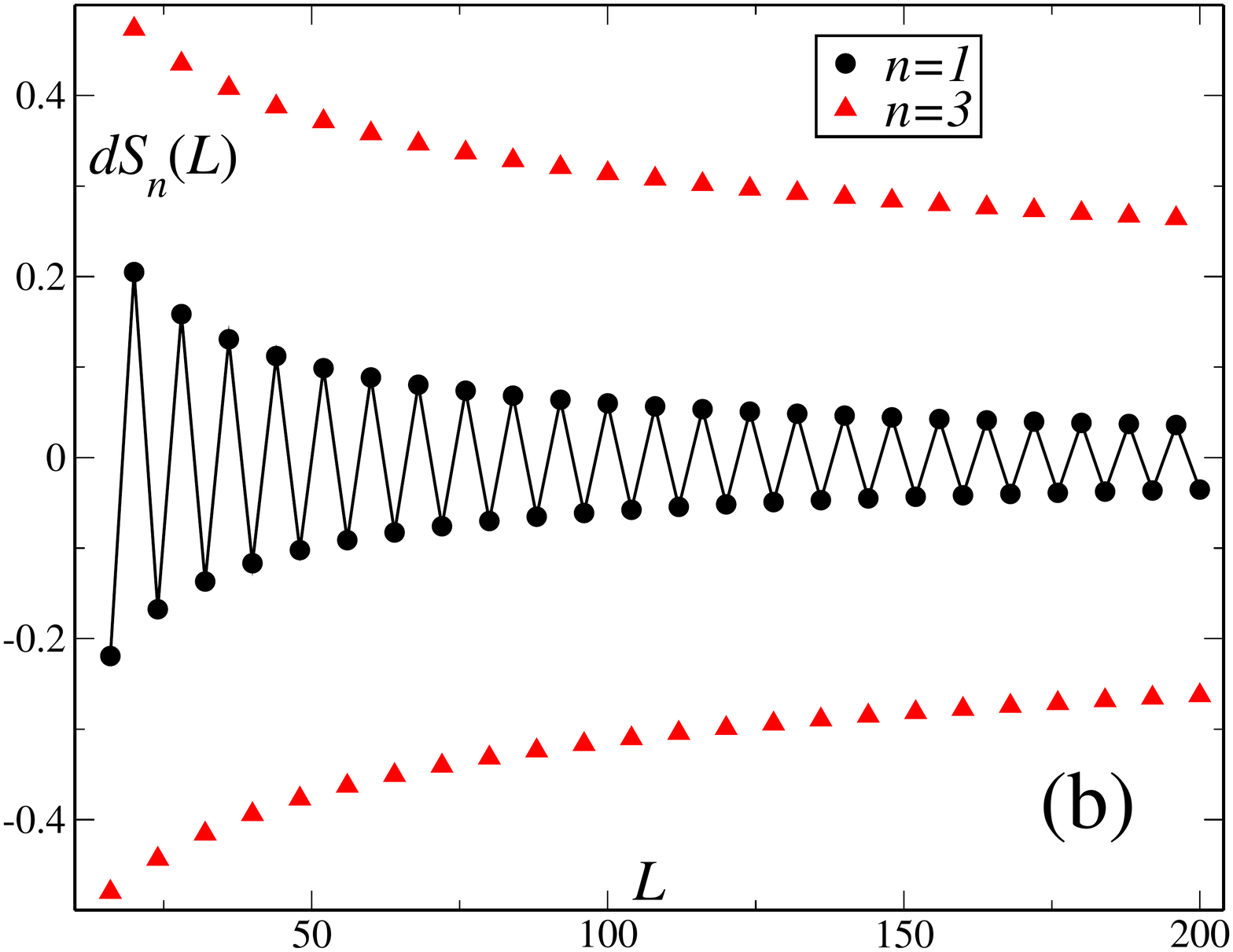}
  \caption{(a): charge sound velocity and Luttinger parameter as a function of $U<0$ in the 1D Hubbard model (reprint with permission from \cite{GiamarchiShastry1995}). (b): $dS_n$ in the 1D Hubbard model, for $U=-21$ and OBC. The solid line is a best fit with formula (\ref{Useful_formula}).\label{Hubbard-K}}
 \end{minipage}
\end{figure}

The check is possible because the charge Luttinger parameter $K_\rho$ has been computed via numerical BA as a function of $U<0$ by Giamarchi and Shastry \cite{GiamarchiShastry1995}: we report the result in figure \ref{Hubbard-K}(a). Our computations are carried out by DMRG, with even system sizes up to $L=200/72$, 512/1156 states per block and 5/4 sweeps at OBC/PBC, for $U=-15,-18,-21$, with truncation errors up to $10^{-9}/10^{-8}$; typical shapes of $dS_n$ are reported in figure \ref{Hubbard-K}(b). Since in this case $k_F=\pi/4$, we considered just system sizes that are multiples of 4. From the shown curves, we get $K$, that we compare with $K_{ana}=K_\rho/2$ \cite{Giamarchi2003}. What we get from DMRG for $n=1$ is in good agreement (up to 1\%) with the result of reference \cite{GiamarchiShastry1995}, demonstrating that our method can give good results even for multi-species systems. What plays a key role here is the fact that the gapped sector just gives a constant contribution to the REE's, resulting in a null contribution to $dS_n$.


\section{First Application: 1D Hard-Core Dipolar Bosons}\label{K_dipoles}

As a first application of our method, we consider a system of hard-core dipolar bosons, i.e., equation (\ref{dipoles}), with both OBC and PBC. It is physically reasonable that, increasing $D>0$, one can induce a crossover between two dominant orders, i.e., between the SF dominant order \cite{Giamarchi2003}, characterized by the correlation function
\begin{equation}
 B(i,x)\equiv\left<b_i^\dagger b_{i+x}\right>
\end{equation}
that can be computed, by bosonization (see section \ref{Bosonization}), to scale as
\begin{equation}
 B(i,x)\simeq x^{-\frac{1}{2K}}
\end{equation}
and the CDW dominant order \cite{Giamarchi2003}, characterized by the correlation
\begin{equation}\label{density-density-connected}
 D(i,x)=\left<n_in_{i+x}\right>-\left<n_i\right>\left<n_{i+x}\right>\simeq\frac{K}{2\pi x^2}+\cos(2k_Fx)x^{-2K}
\end{equation}
Comparing the two formulas, it is clear that this crossover should happen for $K=1/2$. Our task is therefore to estimate the $D$ for whick $K=1/2$.

\begin{figure}[t]
 \begin{minipage}{\textwidth}
  \includegraphics[width=0.5\textwidth]{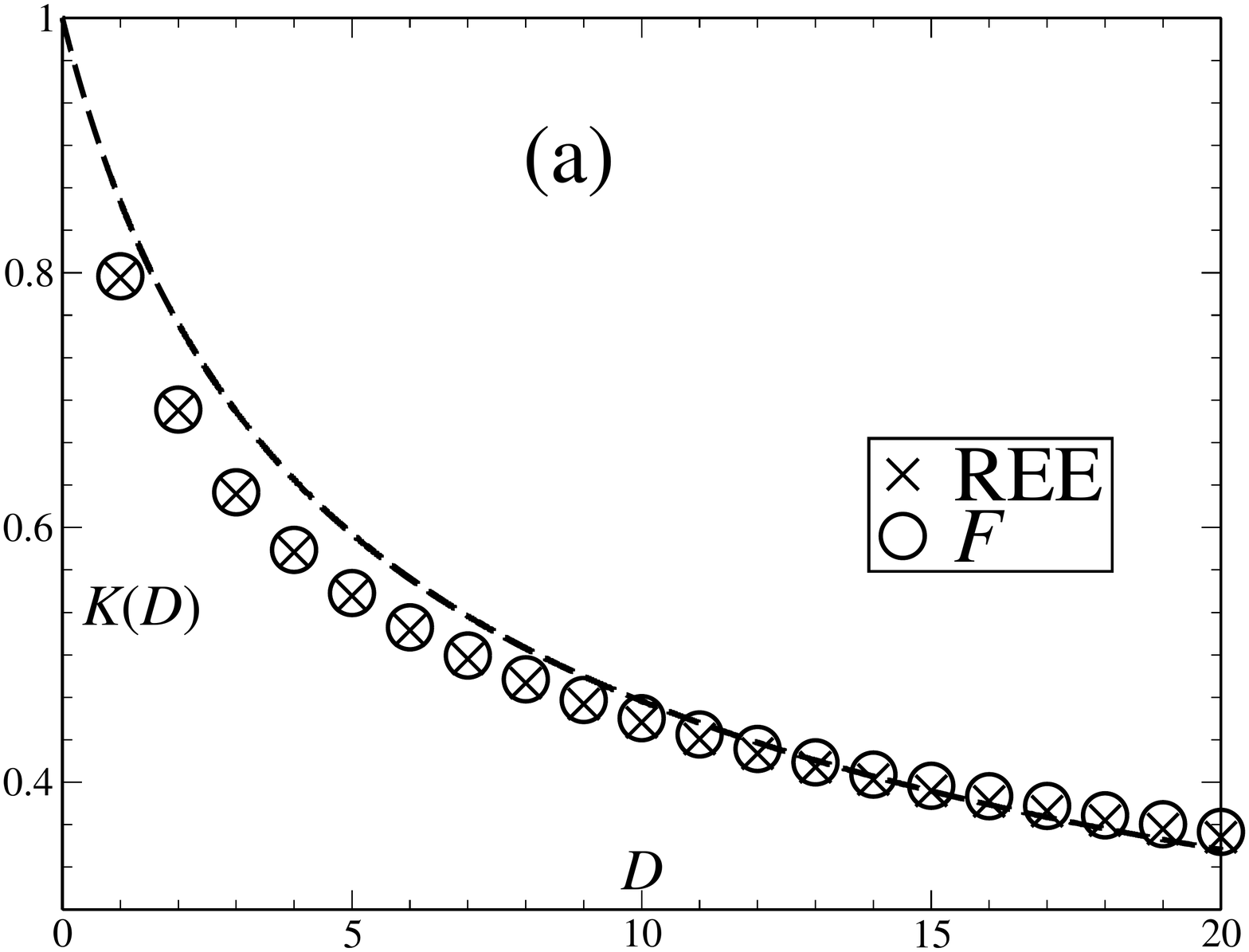}
  \includegraphics[width=0.5\textwidth]{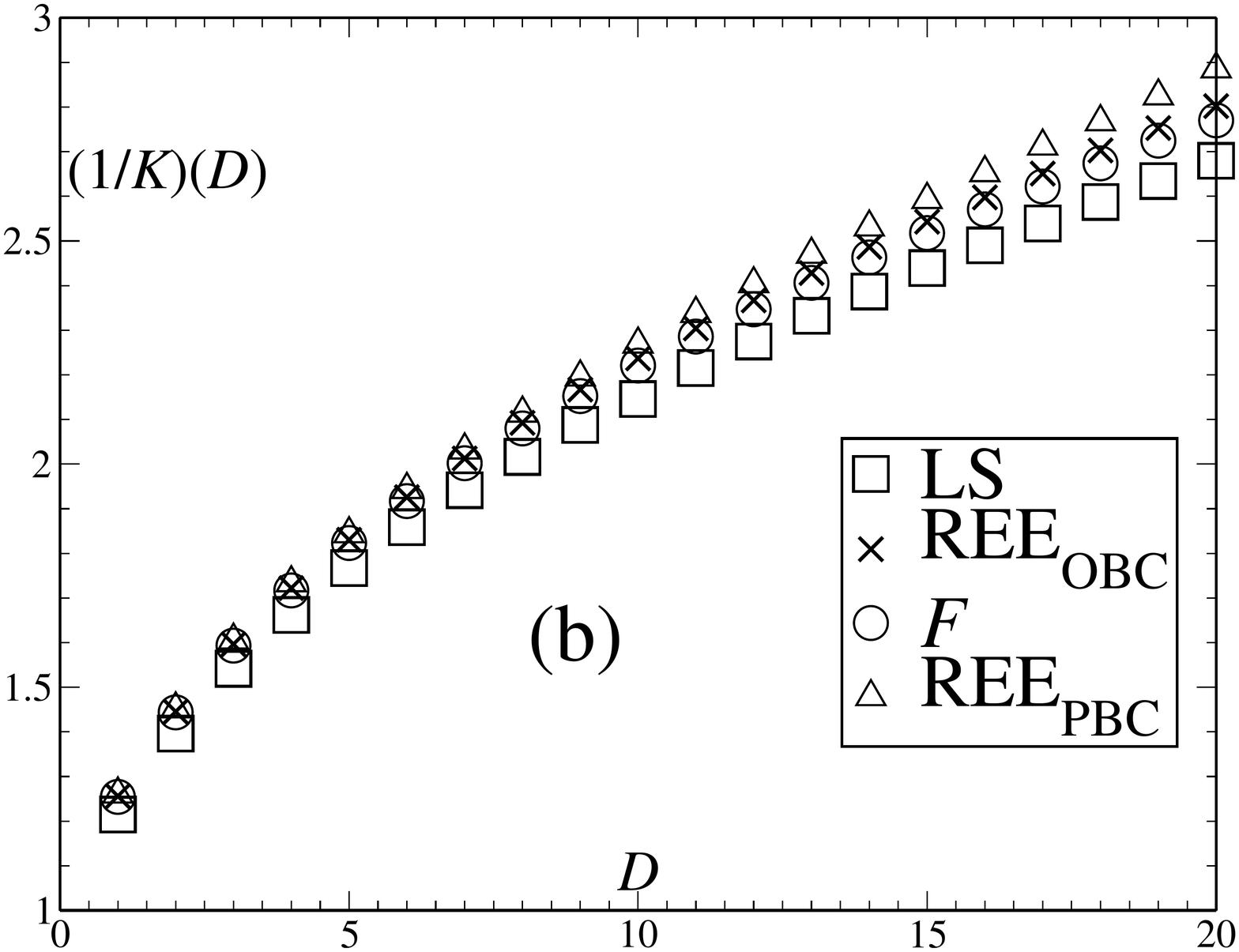}
  \caption{Estimates of the Luttinger parameter of 1D dipolar hard-core bosons as a function of the dipolar interaction strength $D$. The dashed line in panel (a) is an analytical result in the continuum by Dalmonte, Pupillo and Zoller \cite{DalmontePupilloZoller2010}, and REE's at OBC are used.\label{DB_K-D}}
 \end{minipage}
\end{figure}

This time, the model is not integrable, and is known to display, in the repulsive region, Luttinger-liquid physics \cite{DalmontePupilloZoller2010}; we choose quarter filling, i.e., $k_F=\pi/4$. Therefore we need other methods to compare the results we get from ours. In particular, we will use the so-called {\it level spectroscopy} (see appendix \ref{level spectroscopy}) and the {\it bipartite-fluctuations} method (see appendix \ref{bipartite fluctuations}), both with DMRG data. For computational-cost reasons, the interaction is truncated to the fifth nearest neighbour. Overall results will be shown in figure \ref{DB_K-D}.

\begin{figure}[t]
 \begin{minipage}{\textwidth}
  \centering
  \includegraphics[width=0.5\textwidth]{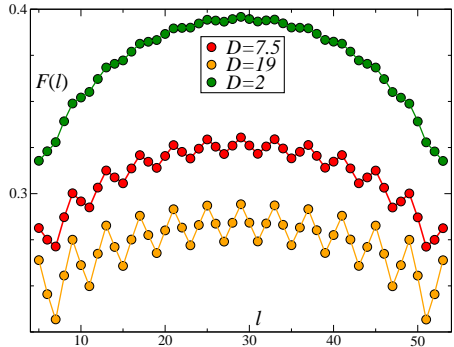}
 \end{minipage}
 \caption{Bipartite density fluctuations for 1D hard-core dipolar bosons and three different values of $D$ (PBC, $L=58$); solid lines are best fits (see text).}\label{DB_density fluctuations}
\end{figure}

Let us start describing the bipartite-fluctuations procedure we used. The conserved quantity in this case is the total particle number, and therefore we considered {\it density} fluctuations. They take the form, relatively to a subsystem $A$ of size $l$,
\begin{equation}
 F(l)=\sum_{i<j}^lD(i,j)
\end{equation}
where $D(i,j)$ is the connected correlation defined in equation (\ref{density-density-connected}) (see equation (\ref{bipfluc})). These quantities are easily computable by DMRG; we show their typical shapes in figure \ref{DB_density fluctuations}, for $L=58$ and PBC. However, direct use of formula (\ref{LeHur formula}) to get $K$ is not convenient, because of the strong oscillations that $F$ displays, in particular at large $D$ (see figure \ref{DB_density fluctuations}). However, as suggested by Song, Rachel and Le Hur \cite{SongRachelLeHur2010}, such oscillations should be analogous in form to the ones of formula (\ref{CCEN}), and therefore we use it to fit the data. Explicitly, we have, for PBC,
\begin{equation}
 F(l,L)=a_0+\frac{K(L)}{\pi^2}\ln\left[\frac{L}{\pi}\sin\frac{\pi l}{L}\right]+\frac{a_1\cos(2k_Fl+\omega)}{\left(\frac{L}{\pi}\sin\frac{\pi l}{L}\right)^{2K(L)}}
\end{equation}
being $a_0$, $a_1$, $\omega$ and $K(L)$ the four fit parameters we have in this case; in particular, the insertion of $\omega$ is necessary for the goodness of the fitting procedure, and it fixes to a value very close to $\pi/2$, indicating that the oscillating factor is of $\sin$-type, rather than $\cos$-type. In this case there is no need of FSS, since the values at $L=58$ are already very close to the ones we get from the other methods.

Next, we describe the level-spectroscopy procedure we employed. First of all, we used PBC, in order to avoid the effects of the boundaries; moreover, being the typical values of $e_{GS}$ very slowing depending on $L$, and therefore needing great precision, we performed our calculations with exact diagonalization (see, e.g., \cite{DRCMNAO1994}). Therefore, we just studied system sizes from $L=8$ up to 24: in figure \ref{LS_DB}(a) we plot the typical shapes of $e_{GS}(L)$. Getting $u$ from these fits (assuming $c=1$; see appendix \ref{level spectroscopy}) and the compressibility $\kappa$ from (\ref{kappa}) (the data was obtained by multi-target DMRG \cite{DegliEspostiBoschiOrtolani2004}), one can estimate $K$ by (\ref{K=upk}) (see figure \ref{LS_DB}(b)). The results systematically overestimate the ones obtained by other methods, and we impute this fact to some kind of finite-size effect, linked to the small system sizes we consider.
\begin{figure}[t]
 \begin{minipage}{\textwidth}
  \includegraphics[width=0.5\textwidth]{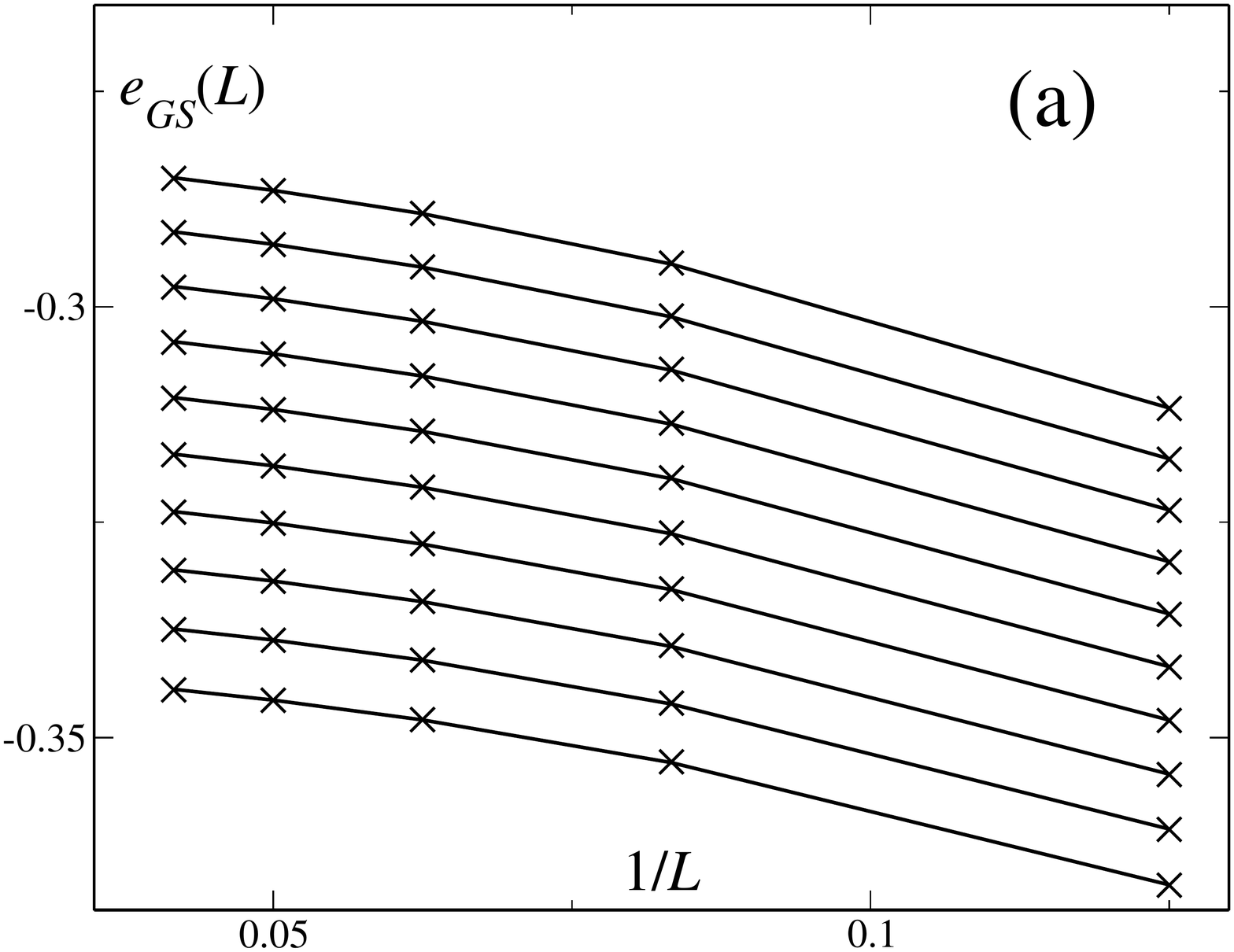}
  \includegraphics[width=0.5\textwidth]{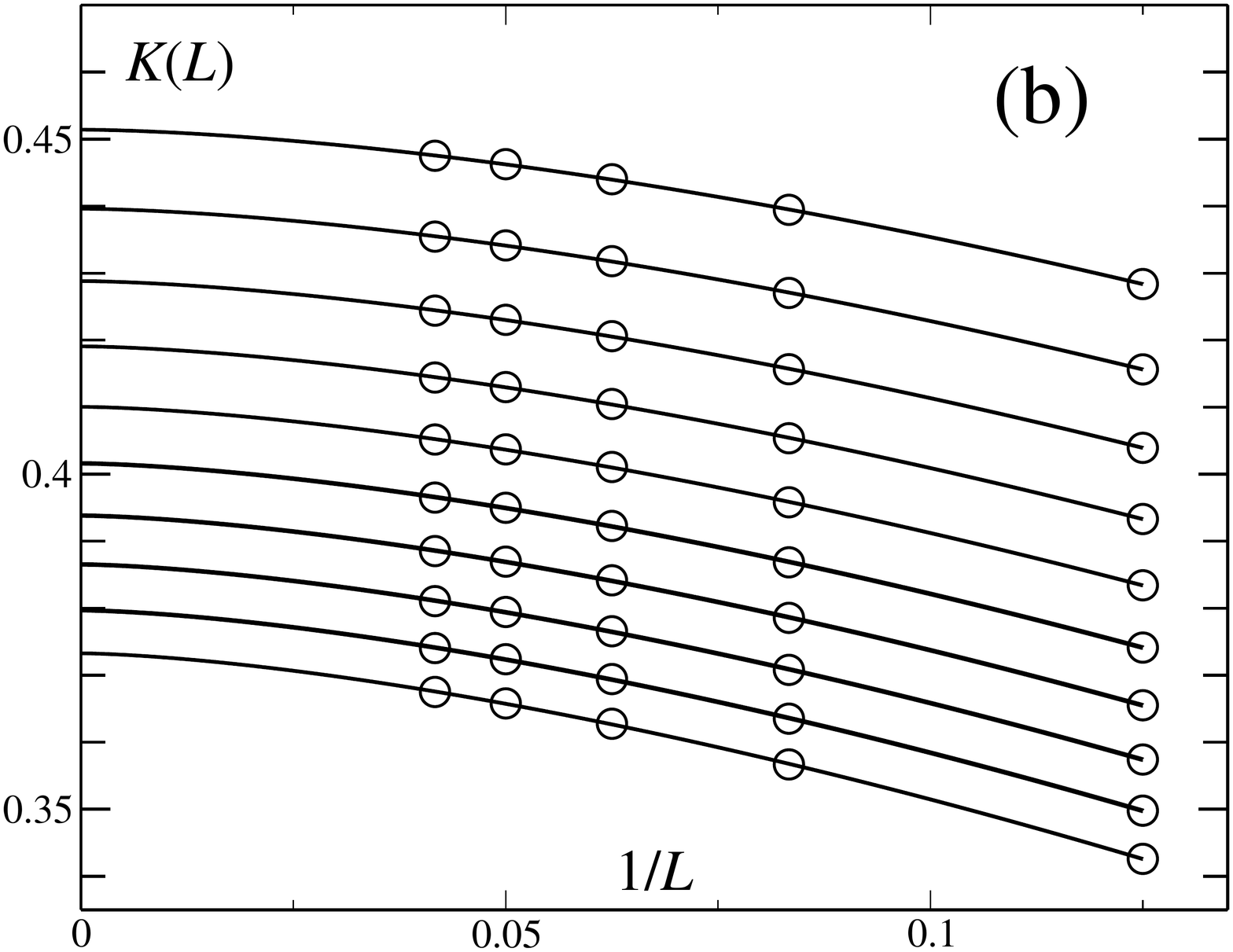}
  \caption{Level spectroscopy for 1D dipolar bosons, up to $L=24$. (a): ground-state energy density as a function of $L$, for values of $D$ from 20 to 11 (top to bottom). (b): the finite-size Luttinger parameter of the system as a function of $L$, for values of $D$ from 11 to 20 (top to bottom). Solid lines are best fits.\label{LS_DB}}
 \end{minipage}
\end{figure}

\begin{figure}[t]
 \begin{minipage}{\textwidth}
  \centering
  \includegraphics[width=0.5\textwidth]{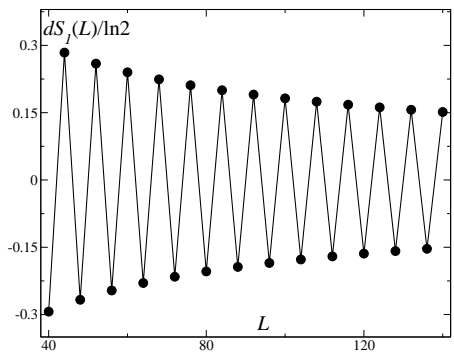}
 \end{minipage}
 \caption{$dS_1$ for 1D hard-core dipolar bosons as a function of $L$ for $D=5$ at OBC; the solid line is a best fit.}\label{dS_n_DB}
\end{figure}

Finally, we use the method we developed previously, both at OBC and PBC, with system sizes up to 140/60, 512/1024 states per block and 4 sweeps in order to have a typical truncation error of $10^{-9}/10^{-8}$ at the final size. A typical shape of $dS_n(L)$ is plotted in figure \ref{dS_n_DB}; we considered just system sizes that are multiples of 4, because of the quarter filling.

\begin{figure}[t]
 \includegraphics[width=\textwidth]{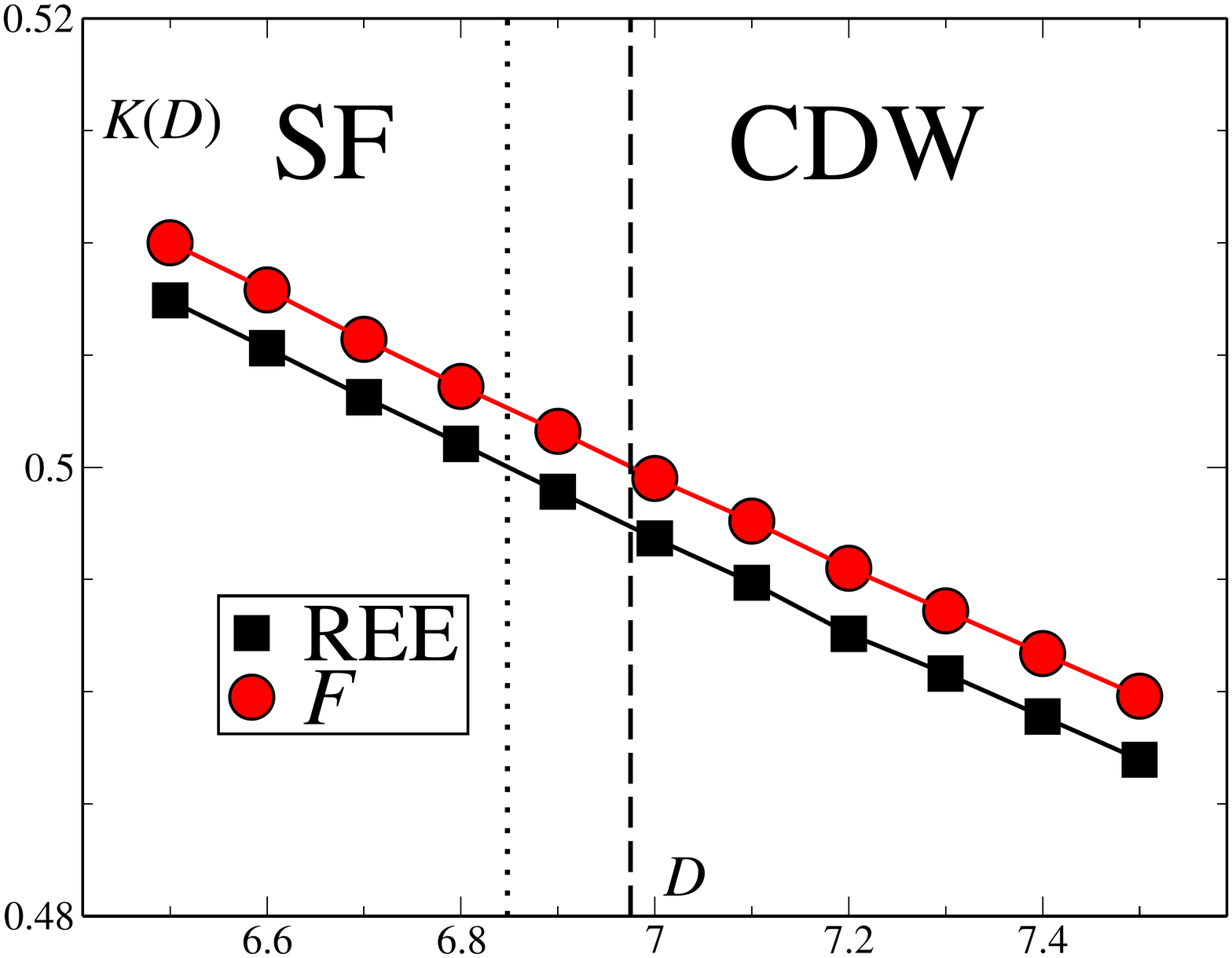}
 \caption{Estimation of the $D$ corresponding to the SF-CDW crossover for 1D dipolar hard-core bosons. The used REE data is taken at OBC. \label{D-crossover}}
\end{figure}

We are now in order to give a final estimation of $K(D)$, comparing the results we obtained with the three different methods, as displayed in figure \ref{DB_K-D}. As one can see, the results are generally in good agreement each other, fact that strengthens our hope that our method can be in general accurate. Coming to the first aim of our study, we estimate the SF-CDW crossover, that should occur at the $D$ for which $K=1/2$. In figure \ref{D-crossover} we show that this $D$ should be around 6.9-7.0; our method and the bipartite-fluctuations one give very close results.


\section{Second Application: the Spin-3/2 XXZ Chain}\label{K_3-2}

As a second application, we consider the spin-3/2 XXZ chain (see section \ref{spin-3/2}). For $-1<\Delta\leq 1$ the model is in the Luttinger liquid universality class but, being it non-integrable, the sound velocity and the Luttinger parameter are only numerically accessible. In this section we perform a systematic investigation of the dependence of $u$ and $K$ on the anisotropy coefficient $\Delta\in]-1,1]$.

We start with the estimation of the central charge of the theory, that in the thermodynamic limit is known to be exactly one; however, assuming it to be one would lead to a systematic error in the evaluation of the sound velocity. We estimate $c$ by computing, for PBC, the VNEE by DMRG, with even system sizes in the interval $[28,60]$, up to 1024 states per block and 4 sweeps, and maximum truncation error of $10^{-6}$. In particular, $S(l,L)$ does not display any oscillation, and we can safely consider the quantity
\begin{equation}
 S(L/2,L)=c^1_1+\frac{c(L)}{3}\ln\frac{L}{\pi}
\end{equation}
where we conjecture $c(L)$ to be of the form $c(L)=c+a_0L^{a1}$, being $c$, $a_0$ and $a_1$, together with $c_1^1$, free fit parameters. The thermodynamic limit of $c(L)$, is therefore $c_0$, and the best fit of our data indicate that $c=1.00$ up to a 2\% error in the whole critical region, error that lowers in the negative-$\Delta$ region. We used, together with this one, different methods to estimate $c$ via FSS, i.e., the ones by Nishimoto \cite{Nishimoto2011} and Xavier \cite{Xavier2010}, but we obtained comparable results. We thus conclude that our numerical data are of good quality, and we can therefore go on with our analysis.

\begin{figure}[t]
 \begin{minipage}{\textwidth}
  \centering
  \includegraphics[width=0.5\textwidth]{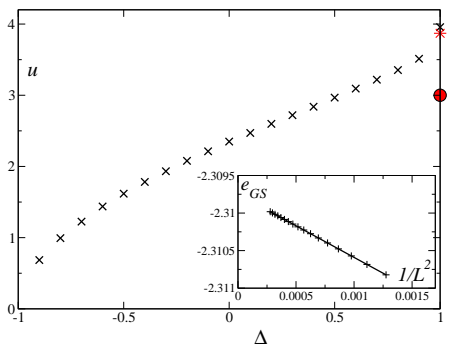}
 \end{minipage}
 \caption{Estimation of the sound velocity as a function of the anisotropy coefficient in the spin-3/2 XXZ chain; the red circle is a result from spin-wave theory, $u_{SW}(\Delta=1)=3$, and the red star is a previous DMRG estimate \cite{HWHM1996}. Inset: typical FSS of $e_{GS}$, at $\Delta=-0.5$.\label{sound_velocity}}
\end{figure}

The computed central charges can be used to extract the sound velocities from relation (\ref{epsilon_CFT}). To do it, we compute with DMRG the ground-state energy density at PBC for different even sizes $L\in[28,60]$. Then, by using relation (\ref{epsilon_CFT}) (the large amount of sizes we consider allows us to fit $\epsilon_{GS}$ as $a_0+a_1/L^2+a_2/L^{a_3}$) and the previously estimated values of the central charge, we can get $u$ as a function of $\Delta\in]-1,1]$ with a spacing of 0.1: the result is shown in figure \ref{sound_velocity}. As we can see from the figure, our result is in good agreement with a past DMRG study by Hallberg and collaborators \cite{HWHM1996}, and in bad agreement with the spin-wave result by Affleck \cite{Affleck1989}, in which quantum fluctuations are treated approximately. The inclusion of logarithmic corrections to (\ref{epsilon_CFT}) close to the Heisenberg point does not lead to appreciable differences. We finally remark that the dependence of $u$ on $\Delta$ is monotonically increasing, and therefore reaches its maximum at the BKT transition, in analogy with the spin-1/2 case \cite{GogolinNersesyanTsvelik1998}.

We are now in the position to extract the Luttinger parameter from the level-spectroscopy procedure (see appendix \ref{level spectroscopy}). The procedure is very easily applied for $|\Delta|\leq 0.9$ and PBC, giving good results; however, at $\Delta=1$, formula (\ref{kappa}) has to be adjusted by the insertion of logarithmic corrections \cite{AffleckBonner1990,HWHM1996}. This way, we obtain a value of $K$ of about $0.499\pm0.005$, in excellent agreement with previous numerical analysis \cite{HWHM1996}. Since the value of $K$ is the same as in the spin-1/2 case, this result constitutes a further proof of the Haldane conjecture \cite{Haldane1982, Haldane1983a, Haldane1983b}, stating that all the half-odd Heisenberg chains belong to the same universality class, the one of the SU(2) Wess-Zumino-Novikov-Witten theory (see, e.g., \cite{DiFrancescoMathieuSenechal1997}).

\begin{figure}[t]
 \begin{minipage}{\textwidth}
  \includegraphics[width=0.5\textwidth]{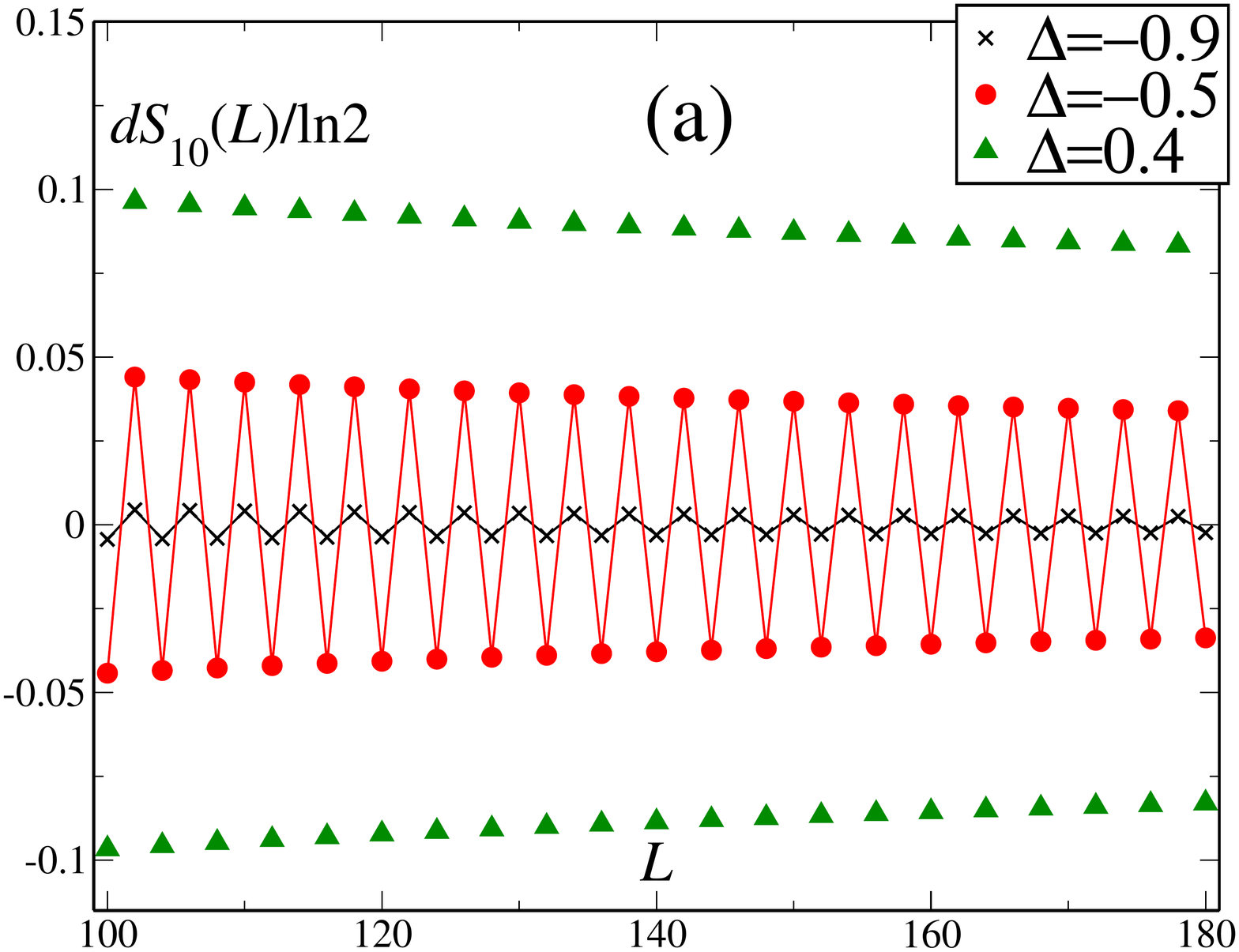}
  \includegraphics[width=0.5\textwidth]{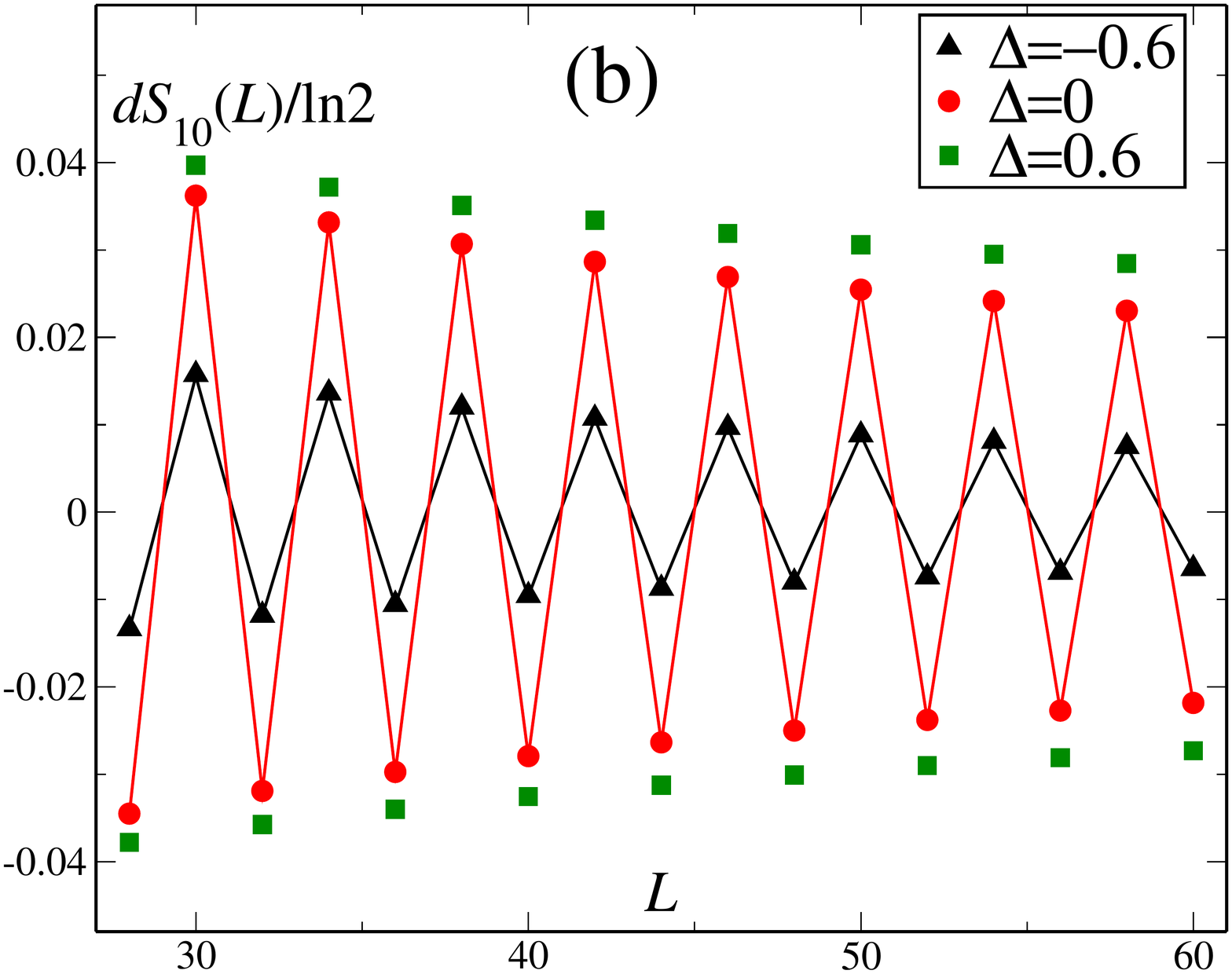}
  \caption{$dS_{10}(L)$ for the spin-3/2 XXZ chain at OBC (a) and PBC (b), at different values of $\Delta$. Solid lines are best fits with equation (\ref{Useful_formula}).\label{XXZspin3-2_dS_n}}
 \end{minipage}
\end{figure}

The second method we use is the REE-based one developed by us. In this case, besides PBC, we use DMRG data at OBC, with even system sizes in the range $[100,180]$, using up to 1024 states per block and 4 sweeps, in order to achieve truncation errors of $10^{-8}$ or lower; typical $dS_n(L)$ shapes are shown in figure \ref{XXZspin3-2_dS_n}. In this case, $k_F=\pi/2$, and the REE's-oscillation size is very small, as already noted by Xavier and Alcaraz \cite{XavierAlcaraz2011a}; therefore, we have to consider large $n$'s in order to fit the data: we usually consider $n=10$, both at OBC and PBC (however, for $\Delta>0$, where the oscillations are bigger, we can obtain comparable results even using smaller $n$). Some problems arise close to the isotropic point $\Delta=1$, where different kind of corrections are present, due to the fact that the Umklapp operator of the QFT description becomes there marginal and has to be taken into account: so, logarithmic corrections are important and cannot be neglected, causing a 5\%-15\% of deviation from the other methods estimations. Finally, since the oscillations are very small for large negative $\Delta$, our method is, in that regime, difficult to use; however, for OBC, where the oscillations exponent is half of the one at PBC, one can still obtain results in good agreement with the other methods.

Finally, we perform an analysis based on the bipartite-fluctuations method (see appendix \ref{bipartite fluctuations}). Here, since the Hamiltonian commutes with the total magnetization $\sigma^z=\sum_{j=1}^L\sigma_j^z$, we can use its bipartite fluctuations to implement the method. As usual, we have to perform DMRG simulations with PBC, taking even $L\in[28,60]$. Here, we choose to perform directly a FSS analysis, by considering $F(L/2,L)$ as a function of $L$: typical fits are shown in figure \ref{XXZspin3-2_F(L-2,L)}. As usual, logarithmic corrections arise close to $\Delta=1$, preventing accurate estimations.
\begin{figure}[t]
 \begin{minipage}{\textwidth}
  \centering
  \includegraphics[width=0.5\textwidth]{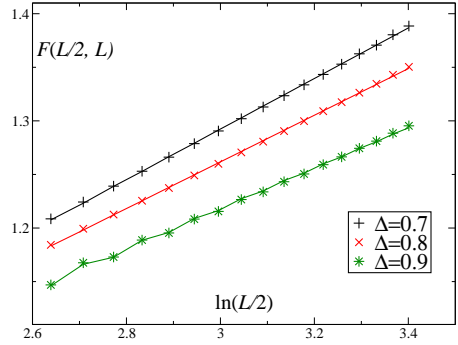}
 \end{minipage}
 \caption{Bipartite fluctuations of the magnetization at half-system size for the spin-3/2 XXZ chain, as a function of $L$; solid lines are best fits with equation (\ref{LeHur formula}) + oscillating corrections (see section \ref{K_dipoles}).\label{XXZspin3-2_F(L-2,L)}}
\end{figure}

\begin{figure}[t]
 \includegraphics[width=\textwidth]{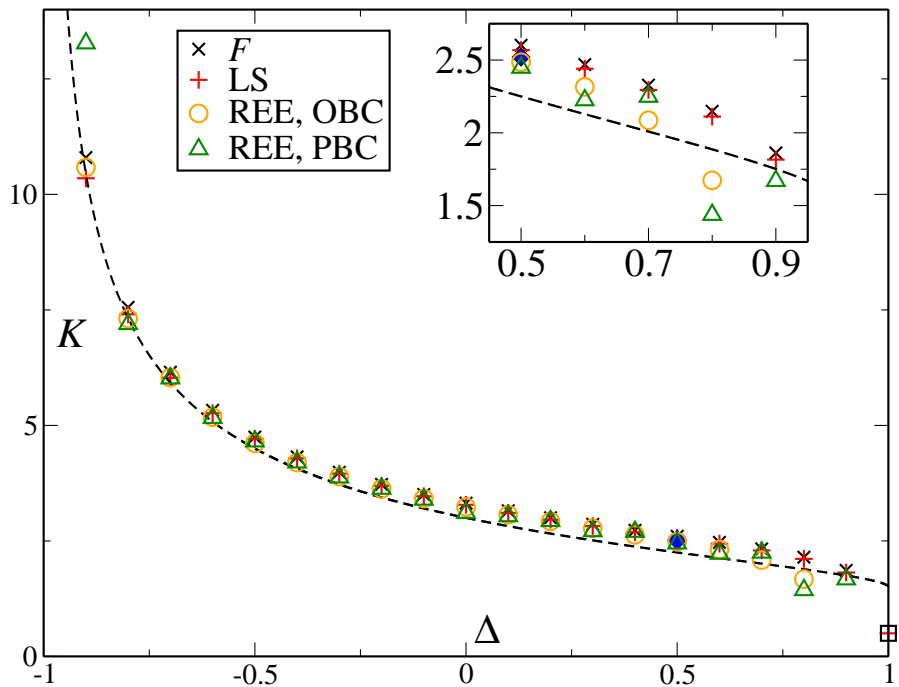}
 \caption{Estimate of the Luttinger parameter for the critical spin-3/2 XXZ chain, with different methods. The dashed line is the conjectured relation (\ref{AMC}); the blue diamond (at $\Delta=0.5$) and the black square (at $\Delta=1$) denote the level spectroscopy and exact results by Xavier \cite{Xavier2010} and Schulz \cite{Schulz1986} respectively. Inset: magnification of the region $\Delta\geq 0.5$. In both panels, the numerical errors are smaller than the typical point size, except for $\Delta=-0.9$ using the REE method with PBC: here the error is of the order of 2.\label{XXZspin3-2_K-comparison}}
\end{figure}

Overall results for the estimation of $K$ are shown in figure \ref{XXZspin3-2_K-comparison}. As it is easily seen, the results from the different methods are generally in good agreement with each other, except close to the boundaries of the critical region: the reasons were illustrated before, and are of various nature. The dashed line in the figure is a conjectured formula, originally formulated by Alcaraz and Moreo \cite{AlcarazMoreo1992} for the ferromagnetic critical region, stating that
\begin{equation}\label{AMC}
 K(\Delta)_S=2SK(\Delta)_{1/2}
\end{equation}
being $K(\Delta)_S$ the Luttinger parameter at anisotropy $\Delta$ in the spin-$S$ XXZ chain. As one can see, the conjecture is in a good (at least qualitative) agreement with the numerical data even in the $\Delta>0$ region.

To conclude, our method seems to be appliable in a wide variety of situations, and its limits are mainly linked to the presence of BKT and ferromagnetic transitions.


\section{Entanglement Entropies in the Spin-3/2 XXZ Chain}\label{E3-2}

In this section we switch to a partially disconnected problem, i.e., to the entanglement properties of the spin-3/2 XXZ chain.

\begin{figure}[t]
 \begin{minipage}{\textwidth}
  \includegraphics[width=0.5\textwidth]{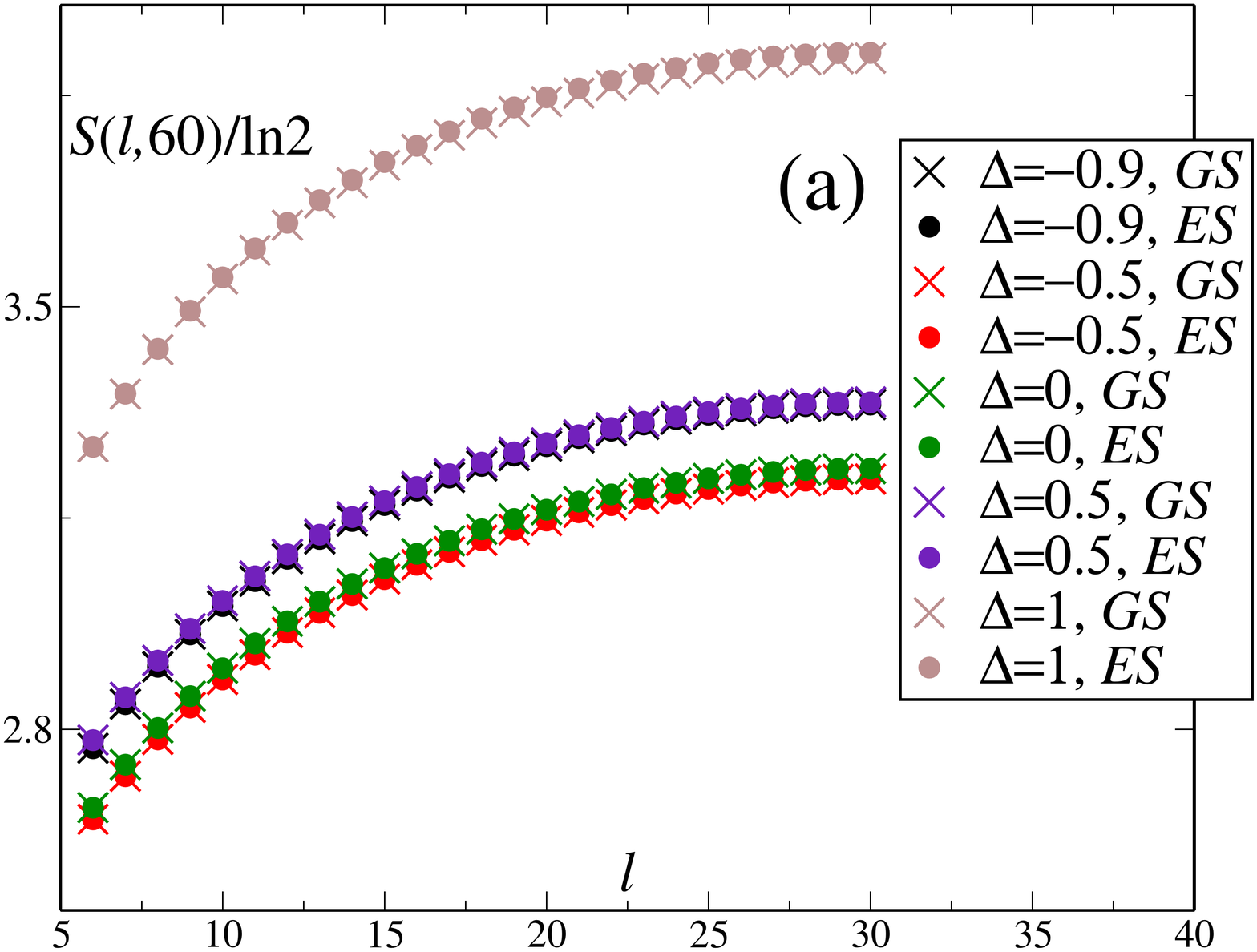}
  \includegraphics[width=0.5\textwidth]{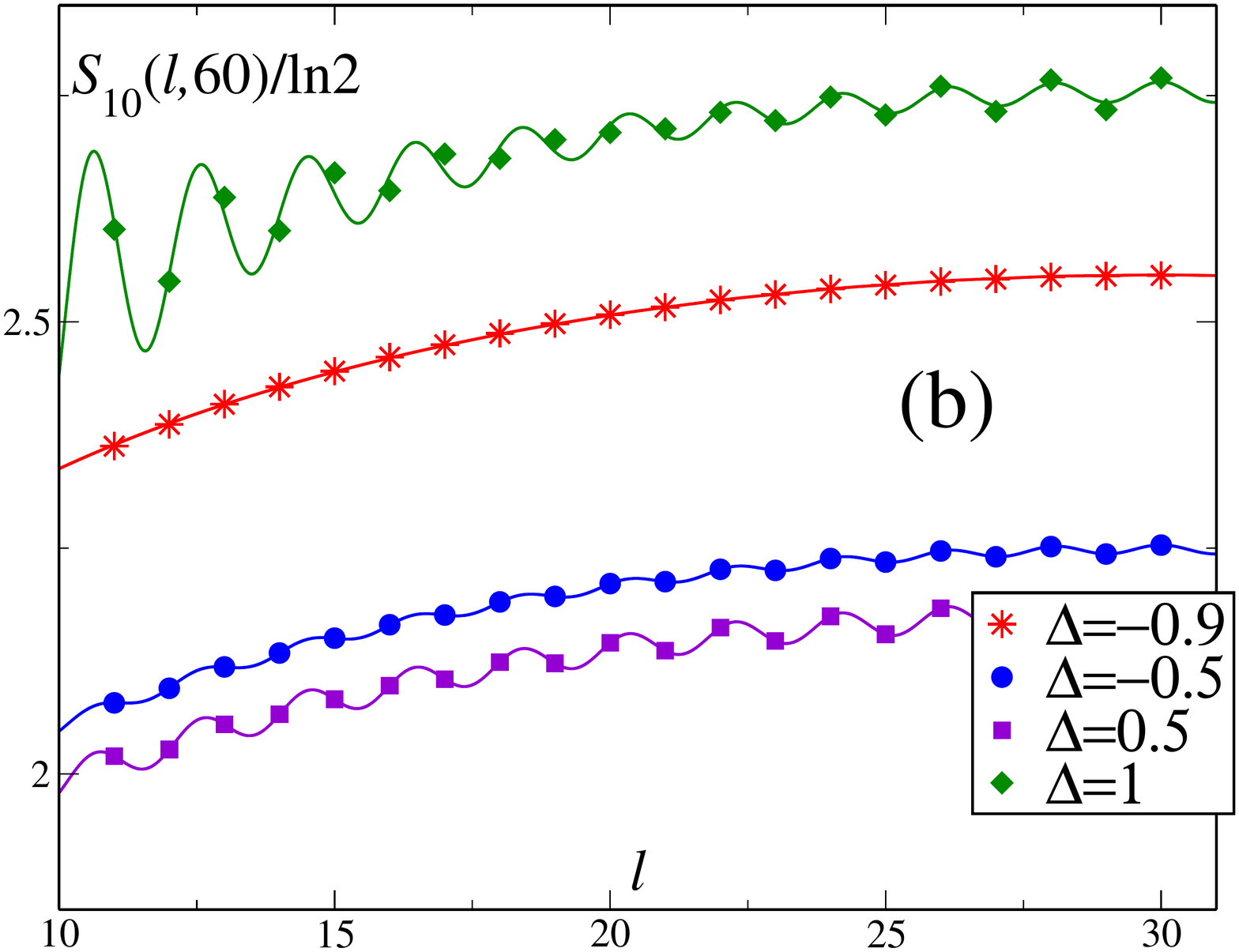}
  \caption{REE's for the spin-3/2 XXZ chain. (a): VNEE of ground and vertex-excited states at different values of $\Delta$; (b): $n=10$ REE for the vertex-excited state, at different values of $\Delta$ (solid lines are best fits with formula (\ref{CCEN}); see text). \label{exc-3-2}}
 \end{minipage}
\end{figure}

In the first part of the section, we numerically verify, by means of DMRG simulations, the prediction (\ref{F}) for the REE's of excited states in a periodic system described by a CFT (see chapter \ref{CFT+OBC} and \cite{AlcarazBerganzaSierra2011,BerganzaAlcarazSierra2012}), in a non-trivial model, i.e., the critical spin-3/2 XXZ chain, that is known to belong to the Luttinger liquid universality class. As we know, the model is not integrable, and therefore constitues a nice test for the cited formula, that up to now has been tested just on integrable models \cite{AlcarazBerganzaSierra2011,BerganzaAlcarazSierra2012,EloyXavier2012}. We considered the excited state that, in the $c=1$ CFT picture, is created by a {\it vertex operator} \cite{DiFrancescoMathieuSenechal1997}; if the ground state is defined as the ground state in the $\sigma^z=0$ sector, the excited states under exam is the ground state in the $\sigma^z=1$ sector (we recall that the Hamiltonian commutes with $\sigma^z$); the CFT prediction for this kind of state is, at PBC, that no universal correction should appear. Our simulations were performed by DMRG in the entire critical region $-1<\Delta\leq 1$, for $L=60$ and PBC. In figure \ref{exc-3-2}(a) we consider the VNEE of the ground and excited state, as a function of $l$, for different values of the anisotropy: in any case, we find an amazing superposition for the curves of ground and excited states, meaning that no universal corrections arise. Moreover, we checked this behavior with exact diagonalization (see, e.g., \cite{DRCMNAO1994}) for small system sizes up to $L=12$.

We note that, in the $n=1$ case, oscillations are absent, as it should be at PBC. For them to arise, we have to consider different $n$'s. In figure \ref{exc-3-2}(b), we plot the $n=10$ REE of the excited state as a function of $l$: it is evident that the oscillations have now a different shape with respect to the usual ones we have for the ground state of the $\sigma^z=0$ sector. Actually, formula (\ref{CCEN}) remarkably holds even in this case, as it has been checked for different values of $n$ and $\Delta$: the main difference with the ground state behavior is the Fermi momentum, that, being now $k_F=31\pi/60$, makes the oscillations shape looking uncommensurate. As usual, the agreement is worst close to the Heisenberg point $\Delta=1$, due to the fact that logarithmic corrections arise, as emphasized all along the present chapter.

\begin{figure}[t]
 \begin{minipage}{\textwidth}
  \centering
  \includegraphics[width=0.5\textwidth]{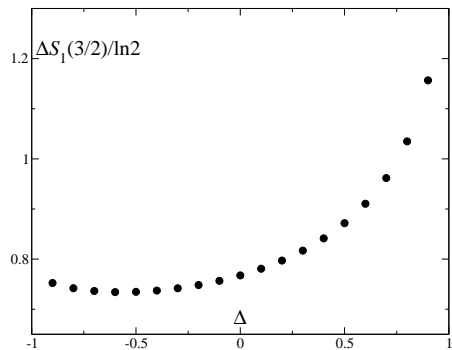}
 \end{minipage}
 \caption{$\Delta$-dependence of $\Delta S(3/2)$; the values are computed by taking the mean value of the quantity in the range $l\in[11,30]$, to eliminate its weak $l$-dependence.\label{Xavier-conj}}
\end{figure}

In the second part of the section, we analize a conjecture, formulated by Xavier in reference \cite{Xavier2010}. By means of DMRG simulations, he found that, both under OBC and PBC, the following relation links the VNEE's of the ground states of half-odd spin-$S$ XXZ chains:
\begin{equation}\label{xc}
 \Delta S(S)\equiv S(l,L)_S-S(l,L)_{S-1}=\frac{1}{2S-1}+\epsilon_S
\end{equation}
where $S\geq 5/2$ and $\epsilon_S\rightarrow 0$ in the thermodynamic limit: the above defined quantity is therefore universal. We verify that for $S=3/2$ equation (\ref{xc}) does not hold: we plot the $\Delta$-dependence of $\Delta S(3/2)$ in figure (\ref{Xavier-conj}). Our result numerically agREE's with what found in reference \cite{Xavier2010} for $\Delta=1/2$; moreover, we show that $\Delta S(3/2)$ develops an unexpected dependence on $\Delta$.


\chapter{Entanglement Entropies in Open Conformal Systems}\label{CFT+OBC}

In this chapter, we are going to study the effect of general conformal boundary conditions (see section \ref{BCFT}) on the entanglement entropies of a system described by a conformal field theory. Our aim is to fill a hole in the literature, since just one conformal boundary condition, the simplest one, was taken into account, by Calabrese and Cardy \cite{CalabreseCardy2004} (numerical studies are available too; see \cite{ZBFS2006,AffleckLaflorencieSorensen2009}). In section \ref{replica} we will derive the Calabrese-Cardy (\ref{CC}), with $\eta=2$, in a different way with respect to reference \cite{CalabreseCardy2004}; in section \ref{GCBC} we will derive formula (\ref{chiralF}) for the corrections to (\ref{CC}), when general conformal boundary conditions are chosen; in sections (\ref{IsingCFT}) and (\ref{FFCFT}) we will compute the explicit shape of such correction and verify it numerically in two non-trivial models, i.e., the $c=1/2$ minimal conformal field theory and the $c=1$ compactified free massless boson. The work was done by myself, J.C. Xavier, F.C. Alcaraz and G. Sierra \cite{TXAS2013}.


\section{The Free-Free Case}\label{replica}

In this section, we adapt the computation that was first performed by Calabrese and Cardy \cite{CalabreseCardy2004} (see also \cite{CalabreseCardy2009}) to the case in which the system has two {\it free} spatial edges (FF boundary conditions). The result for the REE's of the ground state is already known to be (\ref{CC}) with $\eta=2$; however, the technique we use is slightly different from the original one, allowing a straightforward extension to general conformal boundary conditions.

The goal is to compute the $n$-th REE (\ref{RE}) of the ground state in a 1D system of size $L$ with FF boundary conditions at $T=0$, described by a CFT; moreover, we suppose $A$ to share its left boundary with the system. It was shown by Holzhey, Larsen and Wilczek that, for the ground state, \cite{HolzheyLarsenWilczek1994}
\begin{equation}\label{Trrho^n}
 \mbox{Tr}_A[\hat{\rho}_A^n]=\frac{Z_n(A)}{Z^n}
\end{equation}
being $Z$ the partition function, and $Z_n(A)$ the partition function of a system composed of $n$ copies of the original one on a $n$-sheeted Riemann surface $\mathcal{R}_n$, obtained by identifying the subsystem $A$ of the $j$-th copy with the one of the $(j+1)$-th (the $n$-th A is "sewn" to the first): the extremes of the interval $A$ become therefore, in this picture, {\it branch points}.

Let us call $w\equiv \tau+ix$, with $\tau\in\mathbb{R}$ and $x\in[0,L]$, the complex coordinate on each sheet of $\mathcal{R}_n$, and $0,\ il$ the spatial extremes of $A$; let us choose $T\equiv 0$, i.e., $\beta=+\infty$. By the conformal transformation 
\begin{equation}\label{conformal transformation}
 z(w)\equiv\left[\frac{\sinh\frac{\pi(w-il)}{2L}}{\sinh\frac{\pi(w+il)}{2L}}\right]^{1/n}
\end{equation}
$\mathcal{R}_n$ is mapped to the unit disc $|z|\leq 1$, that we will call $\mathbb{D}$. This transformation affects $Z_n(A)$, being it a 0-points correlation function, according to the conformal Ward identity (\ref{Ward}) (actually, according to its finite version). Therefore, being the effect of $z(w)$ the same on each sheet, (\ref{Trrho^n}) must be proportional to the contour integration of $\left<T(w)\right>_{\mathcal{R}_n}$ (+ its antiholomorphic counterpart) elevated to the $n$-th power.

The advantage of performing the transformation (\ref{conformal transformation}) is that it ensures rotational and translational invariance. The transformation of $T(w)$ is \cite{DiFrancescoMathieuSenechal1997}
\begin{equation}\label{T-transformation}
 T(w)=\left(z'\right)^2T(z)+\frac{c}{12}\{z,w\}
\end{equation}
where $z'\equiv\frac{dz}{dw}$ and $\{z,w\}$ is the {\it Schwartzian derivative}
\begin{equation}
 \{z,w\}\equiv\left(\frac{z''}{z'}\right)'-\frac{1}{2}\left(\frac{z''}{z'}\right)^2
\end{equation}
Since we are now on the unit disc, we have that $\left<T(z)\right>_{\mathbb{D}}=0$, and
\begin{equation}\label{<T(w)>}
 \left<T(w)\right>_{\mathcal{R}_n}=\frac{c}{12}\{z,w\}=-\frac{\pi^2c}{24L^2}+\frac{\pi^2(1-\frac{1}{n^2})c}{96L^2}\frac{\sinh^2\frac{\pi il}{L}}{\sinh^2\frac{\pi(w+il)}{2L}\sinh^2\frac{\pi(w-il)}{2L}}
\end{equation}

Let us consider now the correlation function on the original space time $\mathbb{R}\times i[0,L]$, that we call {\it strip},
\begin{equation}\label{TPhistrip}
 \left<T(w)\Phi(w',\bar{w'})\right>_{\mbox{strip}}
\end{equation}
where $\Phi$ is a primary field of conformal dimensions $h,\bar{h}$. By the transformation
\begin{equation}\label{conformal transformation 2}
 w=\frac{L}{\pi}\ln z
\end{equation}
the strip is mapped to the upper-half plane \cite{DiFrancescoMathieuSenechal1997}, a geometry that is more suitable to calculations. The transfomation (\ref{conformal transformation 2}) changes $T(w)$, according to (\ref{T-transformation}), into
\begin{equation}
 T(w)=-\frac{\pi^2c}{24L^2}+\frac{\pi^2}{L^2}\ z\ T(z)
\end{equation}
and the primary field, according to (\ref{primary}), into
\begin{equation}
 \Phi(w,\bar{w})=\left(\frac{L}{\pi}z\right)^{-h}\left(\frac{L}{\pi}\bar{z}\right)^{-\bar{h}}\Phi(z,\bar{z})
\end{equation}
so that (\ref{TPhistrip}) takes the form
\begin{equation}\label{TPhi}
 \begin{split}
  \left<T(w)\Phi(w',\bar{w'})\right>_{strip} & =-\frac{\pi^2c}{24L^2}\left(\frac{L}{\pi}\right)^{-h-\bar{h}}z'\bar{z'}\left<\Phi(z',\bar{z'})\right>_{\mbox{uhp}}+\\
  & +\frac{\pi^2}{L^2}z\left(\frac{L}{\pi}\right)^{-h-\bar{h}}z'\bar{z'}\left<T(z)\Phi(z',\bar{z'})\right>_{\mbox{uhp}}
 \end{split}
\end{equation}
where $\left<\cdot\right>_{\mbox{uhp}}$ denotes the expectation value on the upper-half plane.

On the upper-half plane, the holomorphic and the antiholomorphic sectors are not independent anymore, and one can use the so called {\it mirror-image} trick (see section \ref{BCFT}). $\left<\Phi(z',\bar{z'})\right>_{\mbox{uhp}}$ looks now \cite{DiFrancescoMathieuSenechal1997}
\begin{equation}
 \left<\Phi(z',\bar{z'})\right>_{\mbox{uhp}}=\left<\Phi_{h}(z')\Phi_{h}(z'^*)\right>_{\mathbb{C}}=\frac{1}{(z-z'^*)^{2h}}
\end{equation}
where $\Phi_h$ is a {\it chiral} field of conformal dimension $h$; $\left<T(z)\Phi(z',\bar{z'})\right>_{\mbox{uhp}}$ is instead \cite{DiFrancescoMathieuSenechal1997}
\begin{equation}
 \begin{split}
  \left<T(z)\Phi(z',\bar{z'})\right>_{\mbox{uhp}} & =\left<T(z)\Phi_{h}(z')\Phi_{h}(z'^*)\right>_{\mbox{uhp}}=\\
  & =\frac{h}{(z-z')^2(z-z'^*)^2(z'-z'^*)^{2h-2}}
 \end{split}
\end{equation}
and we have therefore, after some calculations,
\begin{equation}
 \left<T(w)\Phi(il)\right>_{\mbox{strip}}=\frac{\left(\frac{2L}{\pi}\right)^{-2h}}{\left(\sinh\frac{\pi il}{L}\right)^{2h}}\left[-\frac{\pi^2c}{24L^2}+\frac{\pi^2h}{4L^2}\frac{\sinh^2\frac{\pi il}{L}}{\sinh^2\frac{\pi(w-il)}{L}\sinh^2\frac{\pi(w+il)}{L}}\right]
\end{equation}
Since \cite{DiFrancescoMathieuSenechal1997}
\begin{equation}
 \left<\Phi(il)\right>_{\mbox{strip}}=\frac{\left(\frac{2L}{\pi}\right)^{-2h}}{\left(\sinh\frac{\pi il}{L}\right)^{2h}}
\end{equation}
we have demonstrated that
\begin{equation}
 \left<T(w)\right>_{\mathcal{R}_n}=\frac{\left<T(w)\Phi(il)\right>_{\mbox{strip}}}{\left<\Phi(il)\right>_{\mbox{strip}}}
\end{equation}
if and only if
\begin{equation}
 h=\frac{c\left(1-\frac{1}{n^2}\right)}{24}
\end{equation}

It is thus clear that $\mbox{Tr}_A[\hat{\rho}_A^n]$ behaves, under conformal transformations, as a {\it one-point correlation function} on the strip, and one has therefore \cite{DiFrancescoMathieuSenechal1997}, since the conformal Ward identity completely determines the conformal properties of a correlation function,
\begin{equation}
 \mbox{Tr}_A[\hat{\rho}_A^n]=\tilde{c}_n\left(\frac{2L}{\pi}\sin\frac{\pi l}{L}\right)^{2n\frac{c\left(1-/n^2\right)}{24}}
\end{equation}
being $\tilde{c}_n$ a constant that cannot be exactly evaluated this way. It is now immediate to check that the $n$-th REE (\ref{RE}) looks like (\ref{CC}) with $\eta=2$, that is what we wanted to show.


\section{The General Case}\label{GCBC}

Let us now suppose that our open system does not satisfy $FF$, but more general boundary conditions preserving its conformal invariance. For a general CFT the complete set of such boundary conditions is not known, except for specific CFTs, like minimal models \cite{Cardy1989} and the $c=1$ free boson \cite{Saleur1998}. These are exactly the cases we will examinate in sections \ref{IsingCFT} and \ref{FFCFT}.

Let us consider the situation of section \ref{replica}, with generic conformal boundary conditions on the lower and on the upper edges, that we call $a$ and $b$ respectively. Performing the replica trick, this system will be mapped to a $n$-sheeted Riemann surface consisting of the usual seam of $n$ copies of the system; then, by the mapping (\ref{conformal transformation}), the strip is mapped to the unit disc $\mathbb{D}$. Let us call $z_{n,k}^\pm$ the images of the points $w=\pm\infty$ of the $k$-th sheet: they are given by
\begin{equation}\label{z_disc}
 z_{n,k}^\pm=\exp\left[i\frac{\pi}{n}(\mp x+2k)\right]
\end{equation}
where $x\equiv l/L\in[0,1]$: in the unit disc picture, the boundary conditions change exactly at these points. As we saw in section \ref{BCFT}, the effect of the open geometry is to constraint and reduce the operator content of the theory, that becomes, in general, {\it chiral}, and the exact operator content itself is related to the {\it fusion rules} of the CFT. It may happen or not that the theory contains the Verma module of the identity: in this case, the correction to the REE's of the ground state should vanish (in many cases, however, as we shall see, it is actually a constant). However, if the Verma module of the identity is not present, the operator creating the ground state for the vacuum is a different one, and it behaves as an {\it excited state} on the vacuum.

The right way to compute REE's for excited states created by primary operators was first introduced by Alcaraz, Berganza and Sierra in \cite{AlcarazBerganzaSierra2011} and adapted by the same authors to the formalism we are using in \cite{BerganzaAlcarazSierra2012}. Skipping the computations, what one has is that $\mbox{Tr}_A\left[\hat{\rho}_{\Upsilon,A}^n\right]$ for a low-lying excited state (generated by a {\it primary} field) takes the general form
\begin{equation}
 \mbox{Tr}_A\left[\hat{\rho}_{\Upsilon,A}^n\right]=\frac{Z_n(A)}{Z^n}F_\Upsilon^{(n)}(A)
\end{equation}
where $\Upsilon(\tau,x)$ is the primary operator creating the excited state from the vacuum and $F_\Upsilon^{(n)}(A)$ is the correlation-functions ratio
\begin{equation}\label{F}
 F_\Upsilon^{(n)}(A)\equiv\frac{\left<\prod_{k=0}^{n-1}\Upsilon_k(-\infty,0)\Upsilon_k^\dagger(+\infty,0)\right>_{\mathcal{R}_n}}{\left<\Upsilon_0(-\infty,0)\Upsilon_0^\dagger(+\infty,0)\right>_{\mathcal{R}_1}^n}
\end{equation}
The $n$-th REE is therefore
\begin{equation}
 S_n(l,L)=S_n^{CC}(l,L)+\frac{1}{1-n}\ln F_\Upsilon^{(n)}(A)
\end{equation}
and the computation of the correction to the Calabrese-Cardy behavior reduces to a correlation-function computation.

The main consequence of the open geometry on a CFT is that it makes it {\it chiral} (see section \ref{BCFT}): all the operators we consider contain therefore just their holomorphic part. The explicit computation of (\ref{F}) is carried out by going to the unit disc, by using transformation (\ref{conformal transformation}). The $\Upsilon$'s are primary fields, and transform according to (\ref{primary}). We have, after performing all the derivatives,
\begin{equation}
 F_\Upsilon^{(n)}(A)=n^{-2nh}\left(\frac{\prod_{k=0}^{n-1}z_{n,k}^-z_{n,k}^+}{\left(z_{1,0}^-z_{1,0}^+\right)^n}\right)^h\frac{\left<\prod_{k=0}^{n-1}\Upsilon(z_{n,k}^-)\Upsilon^\dagger(z_{n,k}^+)\right>_{\mathbb{D}}}{\left<\Upsilon(z_{1,0}^-)\Upsilon^\dagger(z_{1,0}^+)\right>^n_{\mathbb{D}}}
\end{equation}
The prefactor of the correlation functions looks explicitly
\begin{equation}
 \left(\frac{\prod_{k=0}^{n-1}z_{n,k}^-z_{n,k}^+}{\left(z_{1,0}^-z_{1,0}^+\right)^n}\right)^h=e^{i2\pi(n-1)h}
\end{equation}
and the correlators on the unit disc are simply correlators on the complex plane, because of the chirality of the theory, that has already been taken into account: there is here no need of mirror-tricks. In the end, one is left with
\begin{equation}\label{chiralF}
 F_\Upsilon^{(n)}(A)=n^{-2nh}e^{i2\pi(n-1)h}\frac{\left<\prod_{k=0}^{n-1}\Upsilon(z_{n,k}^-)\Upsilon^\dagger(z_{n,k}^+)\right>_{\mathbb{C}}}{\left<\Upsilon(z_{1,0}^-)\Upsilon^\dagger(z_{1,0}^+)\right>^n_{\mathbb{C}}}
\end{equation}
which is the form we will use for computations in the next sections.


\section{First Case: the $c=1/2$ Minimal CFT}\label{IsingCFT}
The common task of the next two sections is the numerical test of the predictions of CFT, on the base of formula (\ref{chiralF}). Let us start by considering the simplest minimal CFT, i.e., the one with $c=1/2$. The boundary conditions are, in minimal theories, completely classified, as we saw in section \ref{BCFT}. In particular, for the $c=1/2$ minimal CFT, the fusion coefficients are given by
\begin{equation}\label{fusion}
 \begin{split}
  \mathcal{N}^h_{00} &=\delta^h_0\\
  \mathcal{N}^h_{0\frac{1}{16}} &=\delta^h_{\frac{1}{16}}\\
  \mathcal{N}^h_{0\frac{1}{2}} &=\delta^h_{\frac{1}{2}}\\
  \mathcal{N}^h_{\frac{1}{16}\frac{1}{16}} &=\delta^h_0+\delta^h_{\frac{1}{2}}\\
  \mathcal{N}^h_{\frac{1}{16}\frac{1}{2}} &=\delta^h_{\frac{1}{16}}\\
  \mathcal{N}^h_{\frac{1}{2}\frac{1}{2}} &=\delta^h_0\\
 \end{split}
\end{equation}
These rules immediately tell us that for the ground state with $\tilde{0}\tilde{0}$, $\tilde{\frac{1}{16}}\tilde{\frac{1}{16}}$ and $\tilde{\frac{1}{2}}\tilde{\frac{1}{2}}$ boundary conditions there should not be any correction to the REE's (at least, up to a constant, as we will see). Non-trivial corrections should appear in the other cases.

The first step is to individuate which 1D quantum lattice model corresponds to the CFT. It is known that the simplest of these lattice realizations is the 1D critical {\it Ising model} \cite{Mussardo2007}, i.e., the Hamiltonian (\ref{XY_ham}) with $h=\gamma=1$. The next step (and, indeed, the most complicated one) is to individuate the correspondence between the CFT boundary conditions and the lattice ones: it was found \cite{Cardy1986,SaleurBauer1989,Cardy1989,ZBFS2006} that the $\tilde{0}$ ($\tilde{\frac{1}{2}}$) CFT boundary condition corresponds to fix the $x$-component of $\vec{\sigma}$ to $+1$ ($-1$), while the $\tilde{\frac{1}{16}}$ boundary condition amounts to let the edge free. In particular, as we see in appendix \ref{Fixed_DMRG}, this kind of boundary condition can be implemented with DMRG in an exact way (actually, the REE's for $\tilde{\frac{1}{16}}\tilde{\frac{1}{16}}$ boundary condition, being the fermionic Hamiltonian (\ref{XY_fer}) quadratic, can be computed with the exact method of appendix \ref{peschel_method}). From now on, for sake of clarity, we will replace $\tilde{0}$, $\tilde{\frac{1}{2}}$ and $\tilde{\frac{1}{16}}$ by $+$, $-$ and $F$ respectively. In table \ref{table:ising}, we resume all the cases we are going to consider.
\begin{table}
 \caption{The open Ising model.}
 \centering
 \begin{tabular}{c c c c}
  \hline\hline
  $a$ & $b$ & $h$ & Figure \\
  \hline
  $+$ & $+$ & 0 & \ref{Ising++}\\
  $F$ & $F$ & 1/2 & \ref{Ising_exc_FF}\\
  $+$ & $F$ & 1/16 & \ref{Ising_+F_RE2}\\
  $+$ & $-$ & 1/2 & \ref{Ising+-}\\
  \hline
 \end{tabular}
 \label{table:ising}
\end{table}

Let us begin by considering the $FF$ case: we numerically compute, with the method of appendix \ref{peschel_method}, the VNEE for system sizes multiple of 20 in the interval $[60,180]$, and we fit each curve with formula (\ref{CC}): for each $L$, we obtain a value of $c_1^{\eta=2}(L)$ and $c(L)$. Taking the FSS of these estimations we have, in the end, $c_1^{\eta=2}=0.241043$, $c=0.499137$: this value of the central charge is very close to the theoretical one, i.e., $0.5$. In order to further check our data, we compute, using the same method, the VNEE with PBC, with a system size $L=1000$, and we extrapolate the value of $c_1^{\eta=1}=0.484163$: the difference $c_1^{\eta=1}-c_1^{\eta=2}/2$ is of the order of $10^{-2}$, confirming therefore relation (\ref{ALBE}) with $g=1$ , as expected: no boundary entropy is present in the $FF$ case. The $FF$ values of REE's will be taken as references for the other boundary conditions.

\begin{figure}[t]
 \begin{minipage}{\textwidth}
  \includegraphics[width=0.5\textwidth]{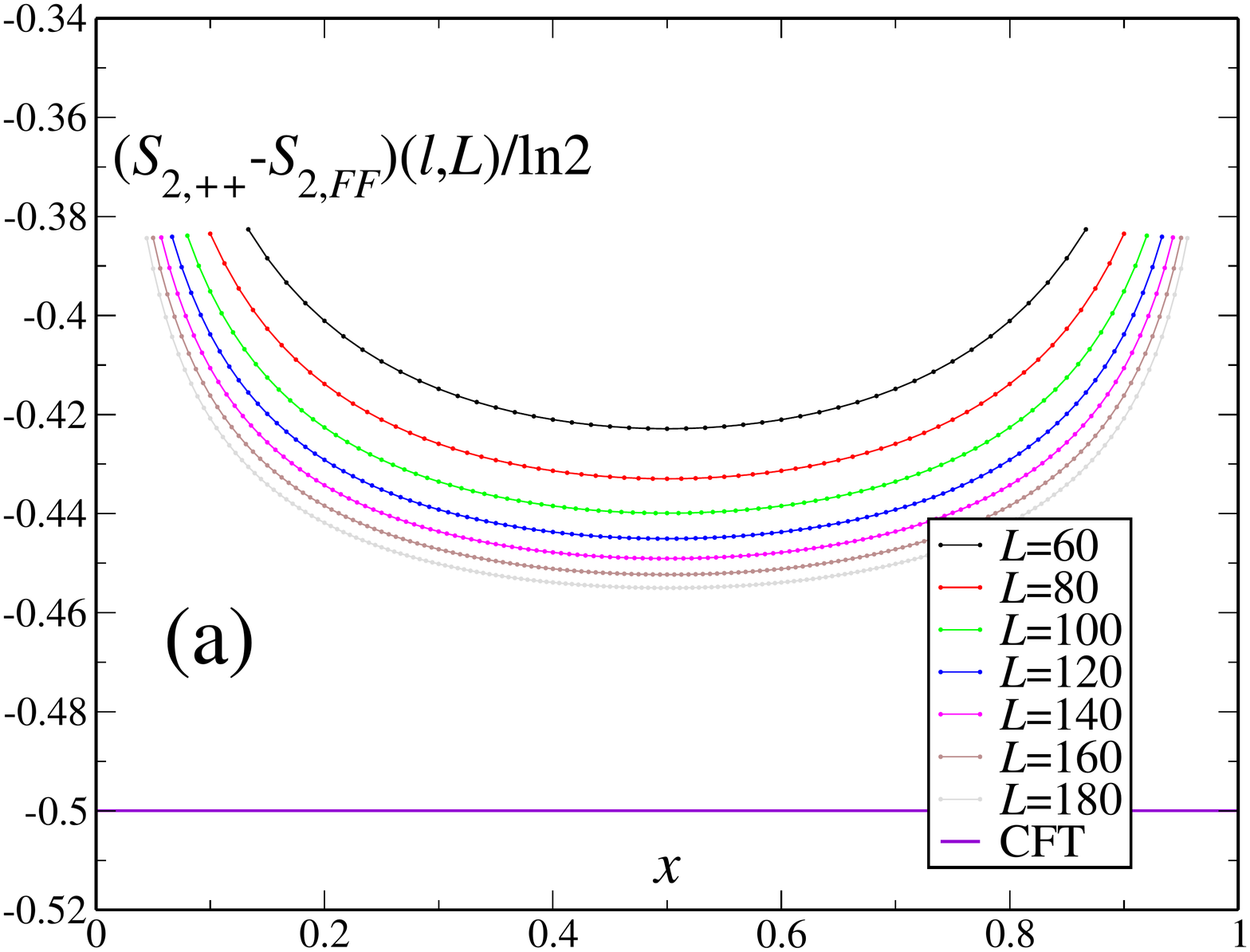}
  \includegraphics[width=0.5\textwidth]{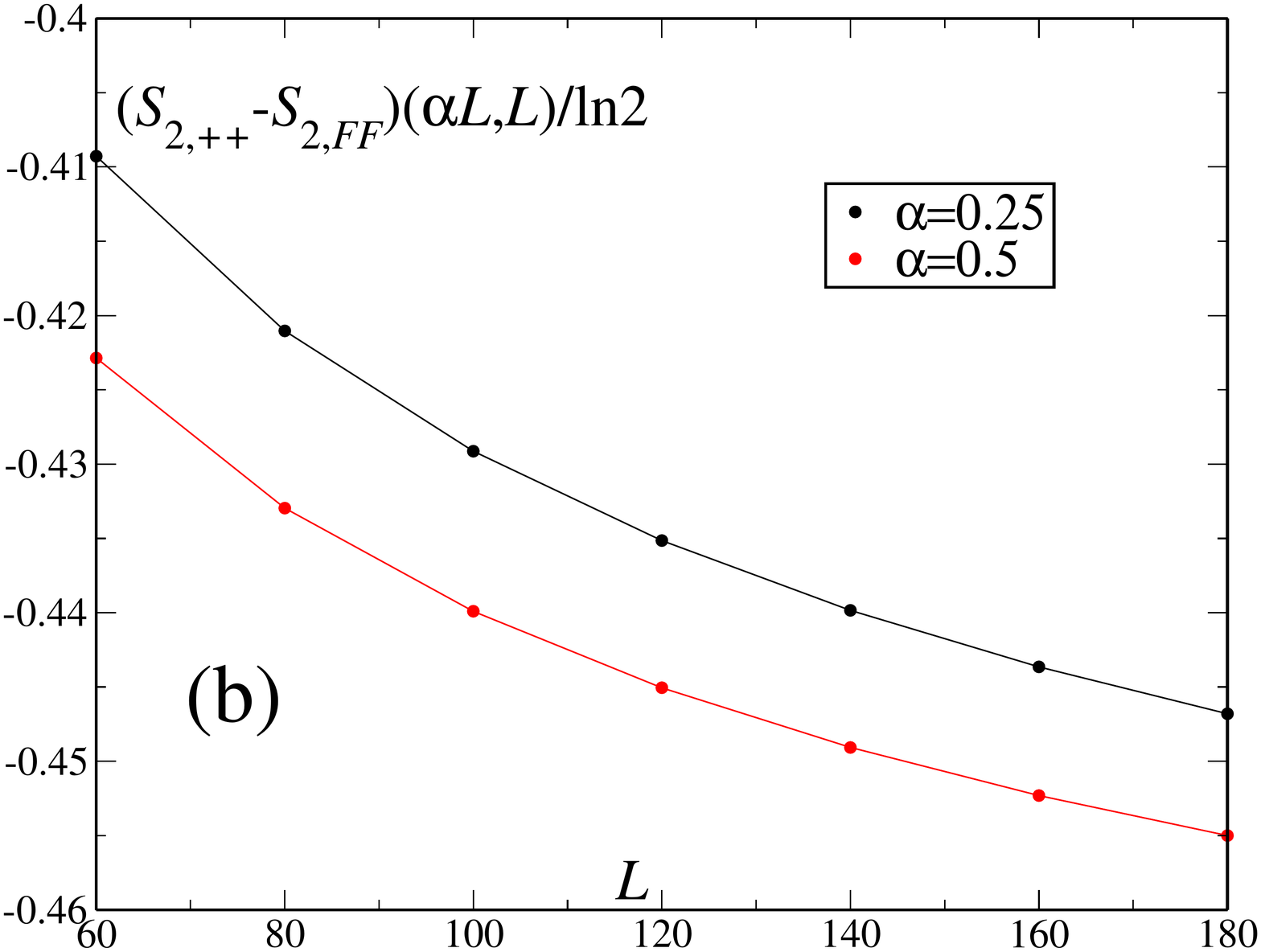}
  \includegraphics[width=0.5\textwidth]{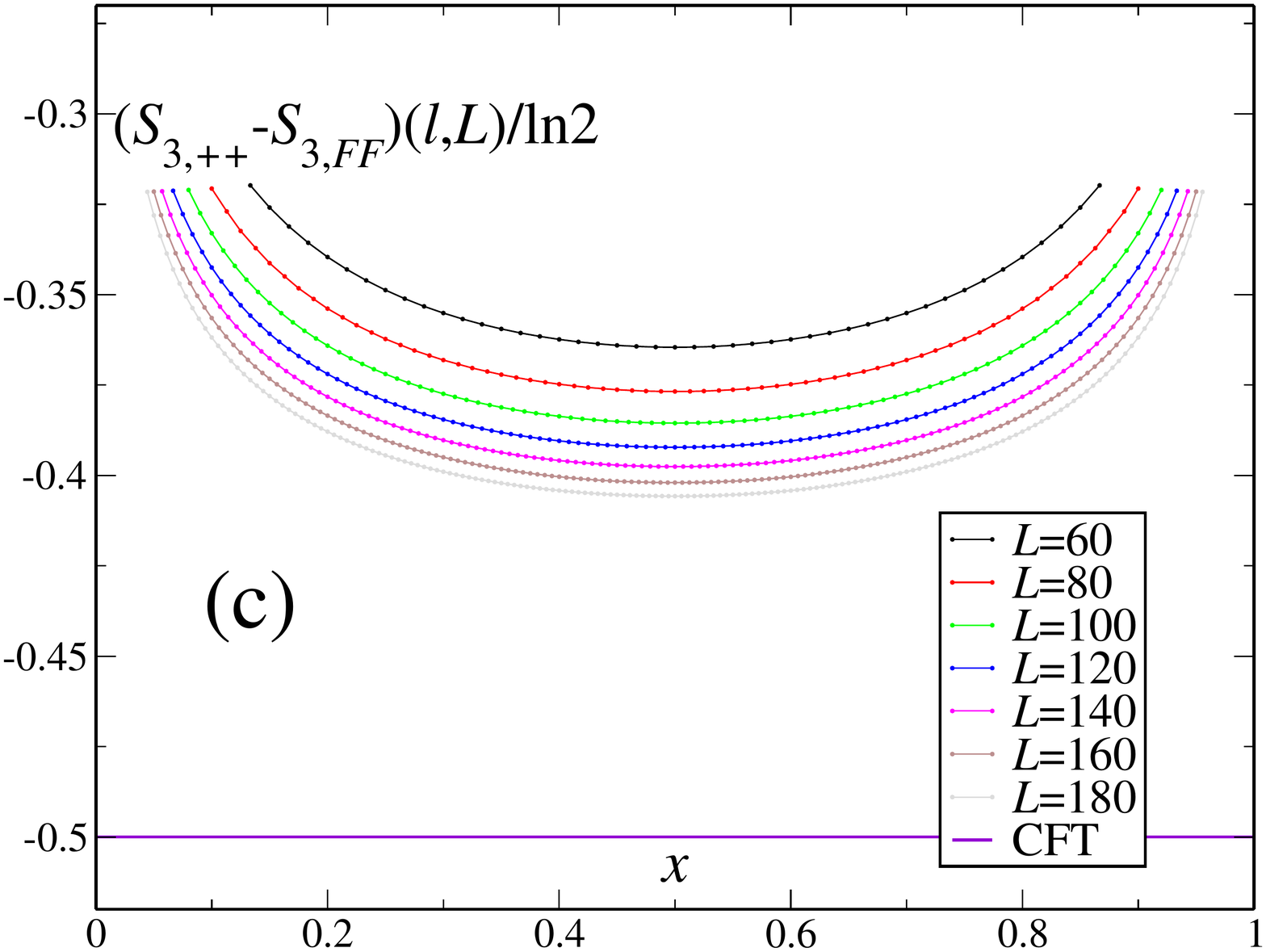}
  \includegraphics[width=0.5\textwidth]{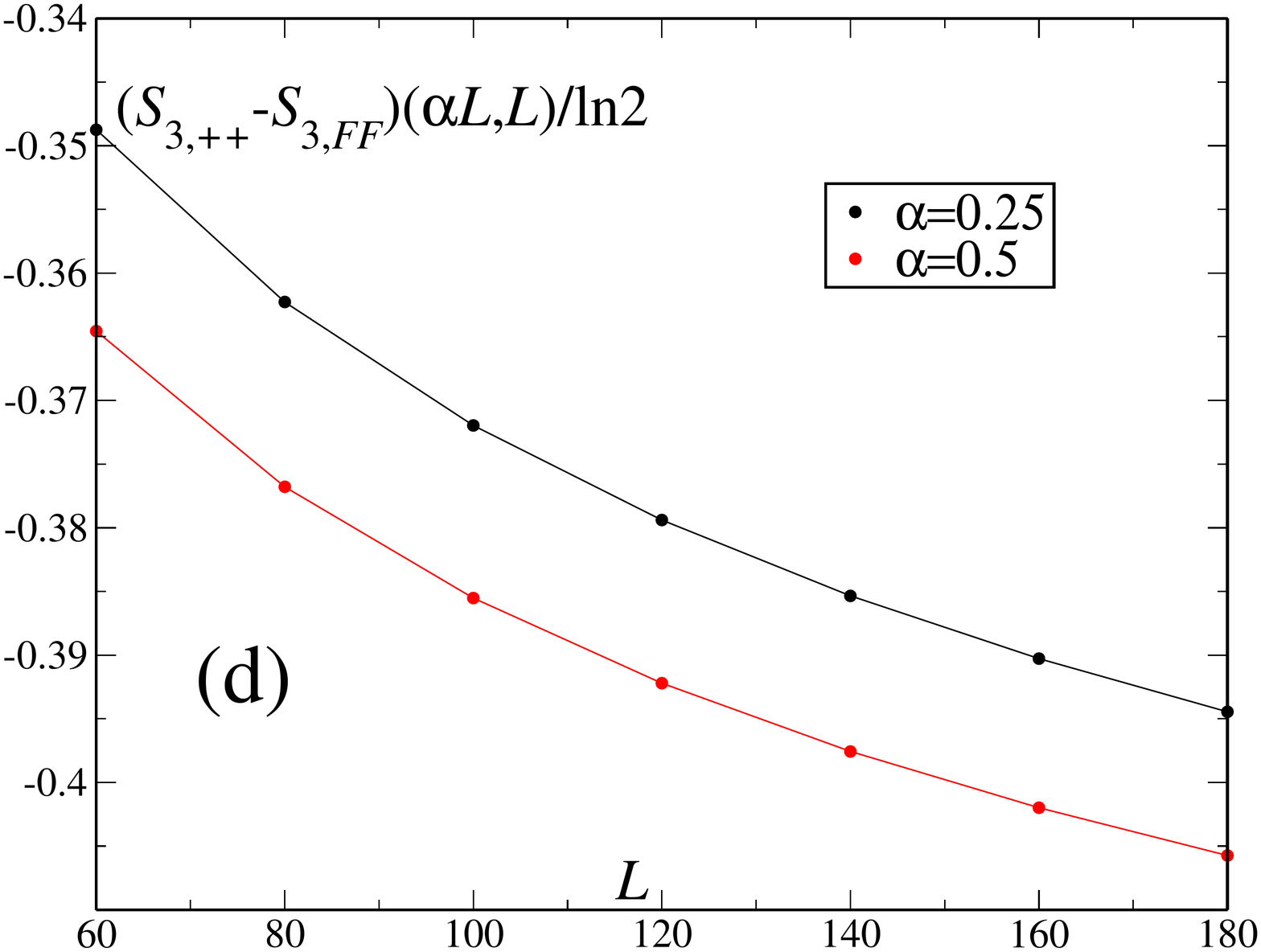}
 \end{minipage}
 \caption{Comparison between the DMRG computations and the CFT predictions for $n=2,\ 3$ REE's for the ground state of the critical Ising model with $++$ boundary conditions. (a): finite-size numerically-computed corrections to the $FF$ $S_2$, as a function of $x\equiv l/L$, for $x\in[8/L,1-8/L]$; (b): FSS of the previous difference for specific values of $x$, using the 5-parameters formula $y=a_0+\frac{a_1}{x^{a_2}}+\frac{a_3}{x^{a_4}}$ (solid lines); (c), (d): same analysis for $n=3$. In any case, the agreement with the constant value $-\frac{1}{2}\ln 2$ is good up to differences smaller than 1\%.}\label{Ising++}
\end{figure}

Let us examinate now the cases $++$ and $--$, where no non-trivial corrections should appear for the ground state. We just consider the $++$ case, by means of DMRG: the $--$ can be seen to be analogous to the $++$ one, by applying the first of transformations (\ref{canonical}) on the lattice model. We performed DMRG simulations by implementing the boundary conditions exactly, following appendix \ref{Fixed_DMRG}; we considered system sizes in the range $L=60-180$, and used 3 sweeps and up to 800 states per block, in order to achieve a typical truncation error of $10^{-12}$ or less (these will be the typical system sizes we will consider, and the typical truncations we will have in the section; we will not report this technical information anymore). Results are shown in figure \ref{Ising++}(a) and (c) for $n=2$ and $3$: the corrections seem to flatten, by increasing $L$, to a finite value. In figures \ref{Ising++}(b) and (d) we show by FSS (see caption) that this value is, in both cases, $-\frac{1}{2}\ln 2$, i.e., the value of the boundary entropy of one edge (see section \ref{BCFT}): the only arising correction is this constant one.

\begin{figure}[t]
 \begin{minipage}{\textwidth}
  \includegraphics[width=0.5\textwidth]{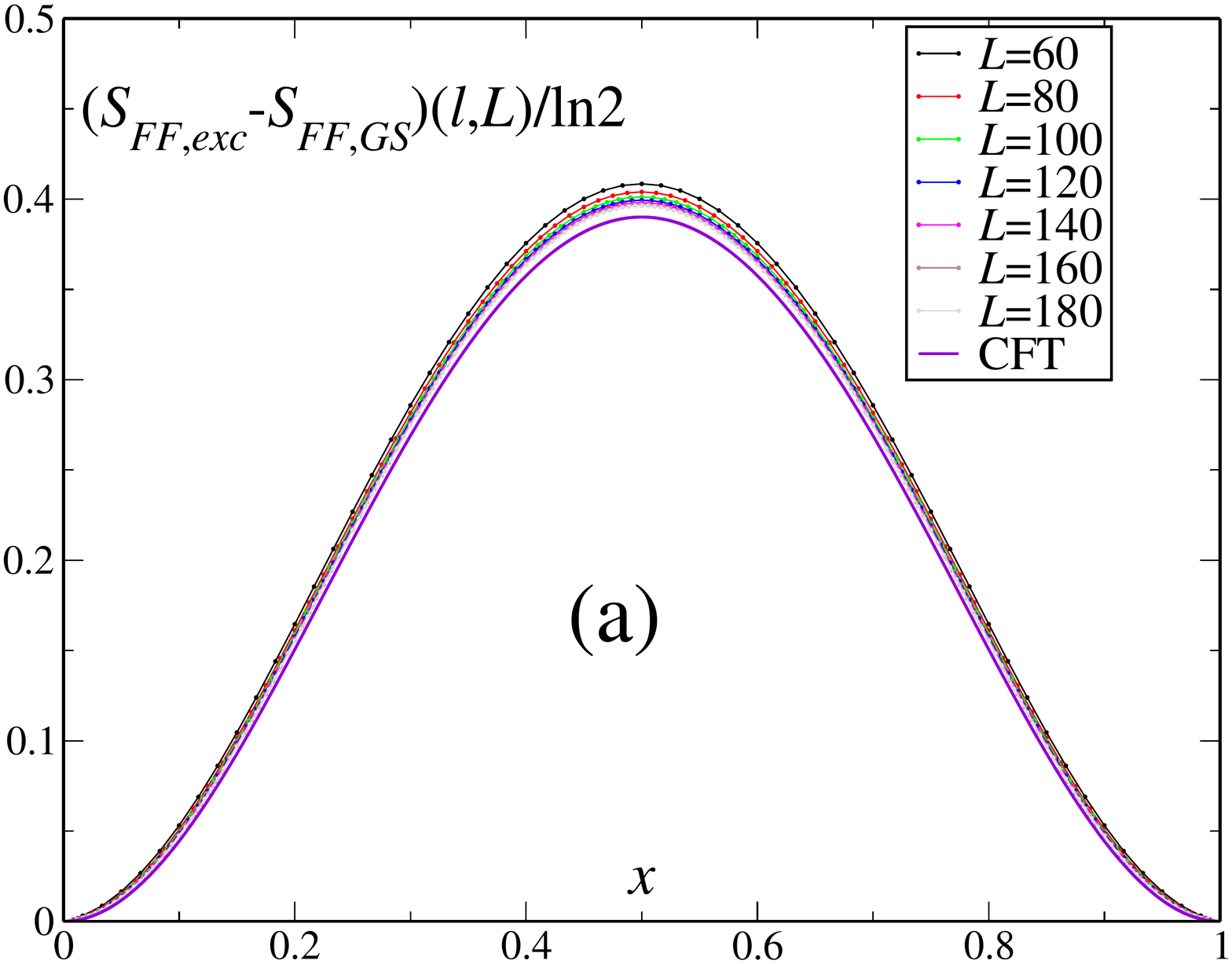}
  \includegraphics[width=0.5\textwidth]{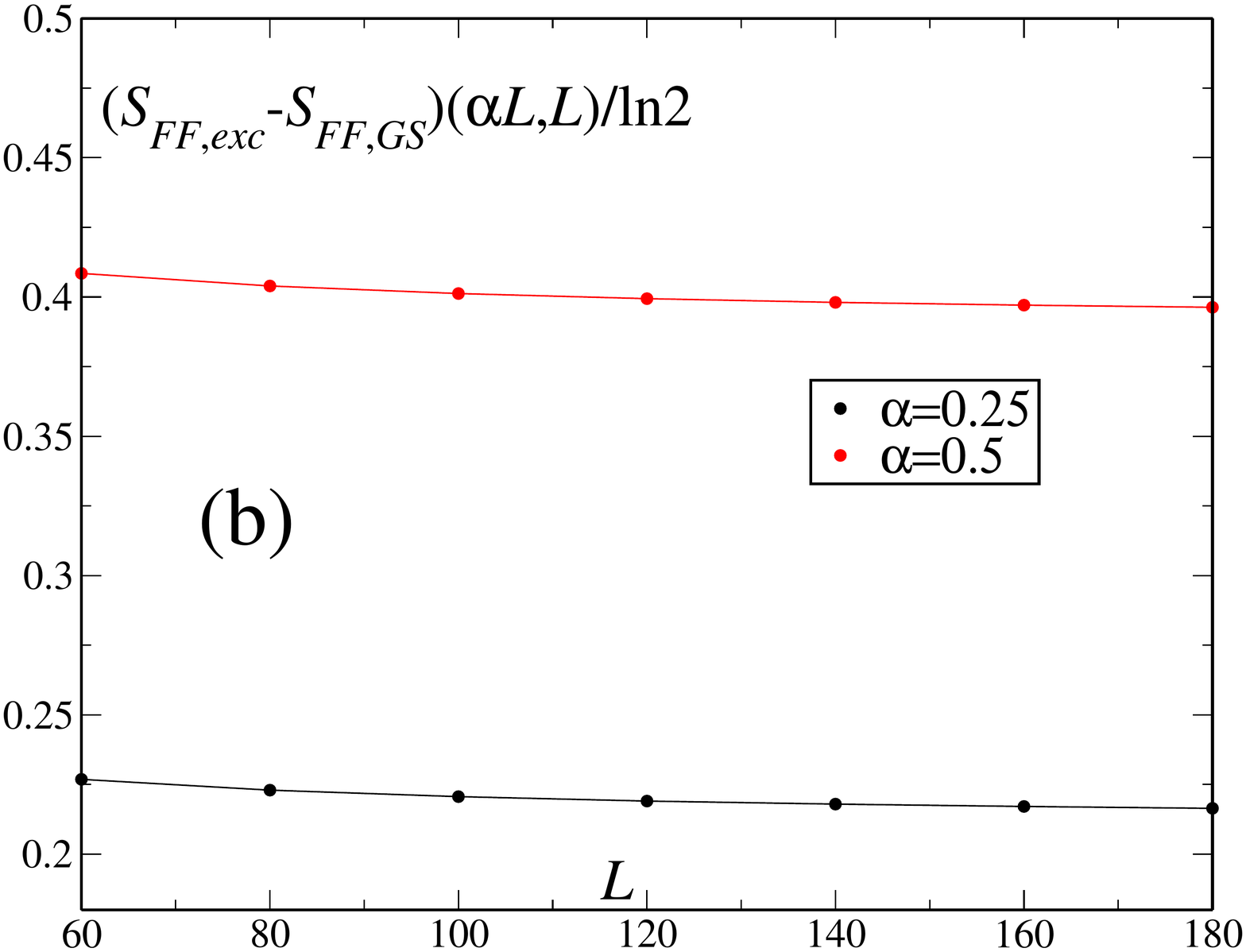}
  \includegraphics[width=0.5\textwidth]{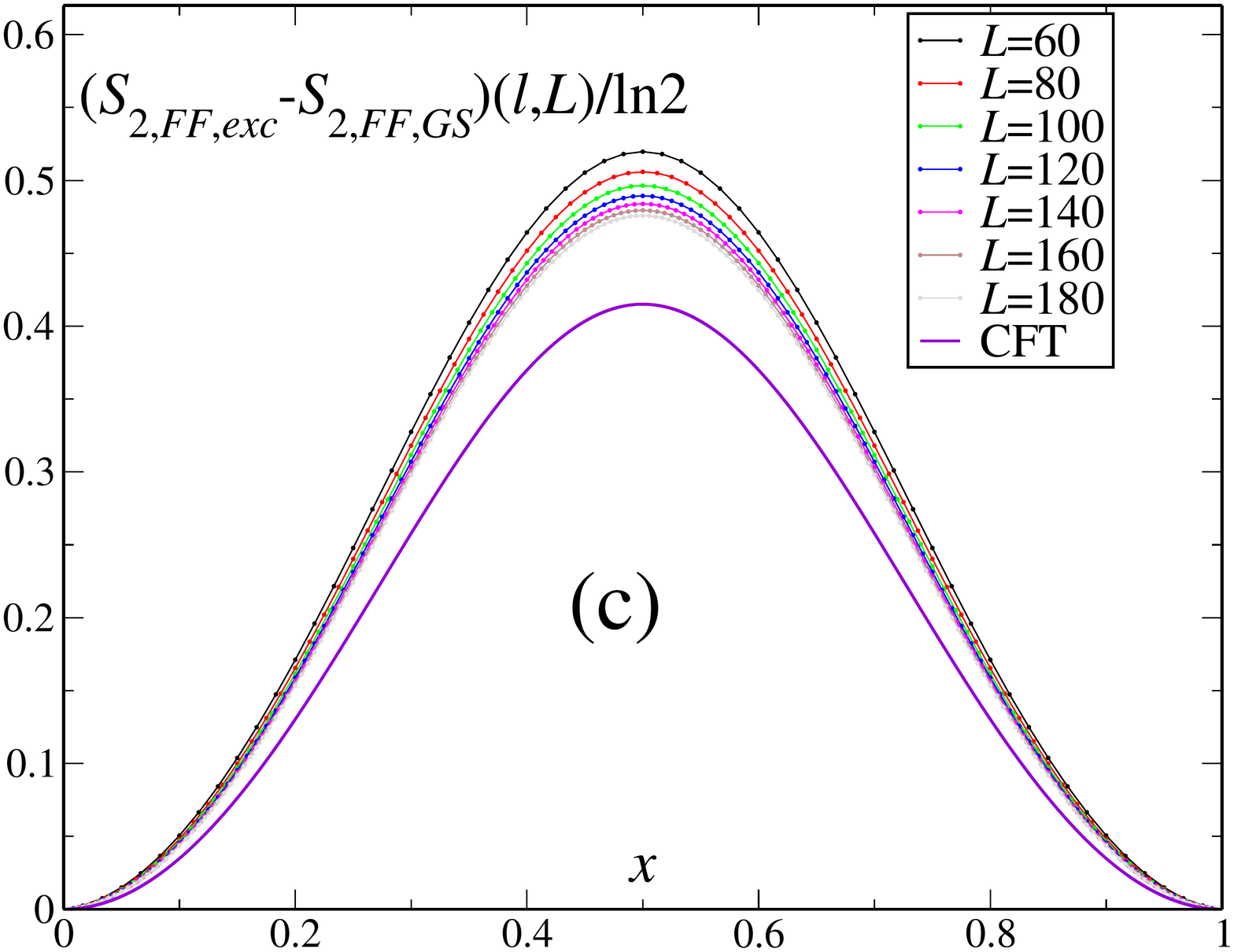}
  \includegraphics[width=0.5\textwidth]{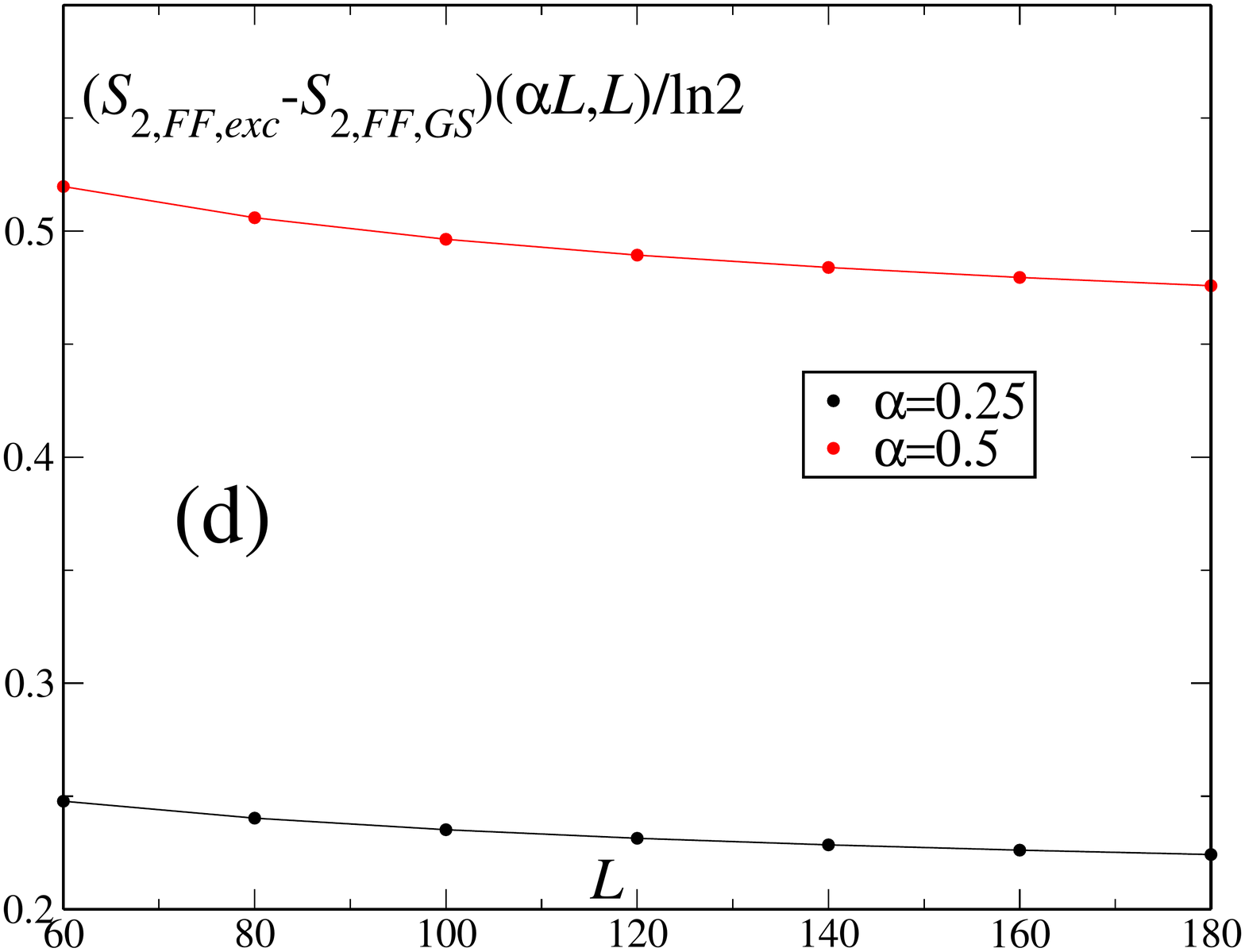}
 \end{minipage}
 \caption{Comparison between the numerical computation and the CFT prediction for the VNEE and the $n=2$ REE of the first excited state in the critical Ising model with $FF$ boundary conditions. (a): finite-size numerically-computed corrections to the $FF$ VNEE of the first excited state; (b): FSS of the previous difference for specific values of $l$, using the 5-parameters formula $y=a_0+\frac{a_1}{x^{a_2}}+\frac{a_3}{x^{a_4}}$ (solid lines); (c), (d): same analysis for $n=2$.}\label{Ising_exc_FF}
\end{figure}

In the $FF$ case, there is something more than what we just said: in fact, according to the fourth fusion rule of (\ref{fusion}), the spectrum includes a second conformal tower, the one of the $\epsilon$ operator, whose chiral counterpart is the Majorana operator $\chi$ \cite{DiFrancescoMathieuSenechal1997}: the first-excited-state REE's should have non-trivial corrections, that are of the form $F^{(n)}_\chi(x)$, according to (\ref{chiralF}). The conformal block of Majorana operators is known to be a Pfaffian (see appendix \ref{CB}), and therefore is, in principle, very easily computable. Moreover, it has been argued by Berganza, Alcaraz and Sierra \cite{BerganzaAlcarazSierra2012} that
\begin{equation}\label{Fs}
 F^{(n)}_\chi(x)=\sqrt{F_{i\partial\phi}^{(n)}(x)}
\end{equation}
where $\phi$ is the bosonic field of a $c=1$ CFT defined on the complex plane (see section \ref{FFCFT}). Remarkably, Essler, L\"auchli and Calabrese analytically continued $F^{(n)}_{i\partial\phi}$ to non-integer $n$ \cite{EsslerLauchliCalabrese2012}:
\begin{equation}\label{ECL}
 F^{(n)}_{i\partial\phi}(x)=\left\{\left[\frac{2\sin(\pi x)}{n}\right]^n\frac{\Gamma\left(\frac{1+n+n\csc(\pi x)}{2}\right)}{\Gamma\left(\frac{1-n+n\csc(\pi x)}{2}\right)}\right\}^2
\end{equation}
where $\Gamma(x)$ is the Euler $\Gamma$ function. We start by considering $n=1$, i.e., the VNEE. Since for $n=1$ both numerator and denominator in equation (\ref{RE}) vanish, we have to take the limit of the whole expression, and not to consider just $F^{(1)}_\chi$. What one has in the end is, as found by Essler, L\"auchli and Calabrese \cite{EsslerLauchliCalabrese2012}
\begin{eqnarray}\label{chi_1}
 \lim_{n\rightarrow 1}\frac{1}{1-n}\ln F^{(n)}_\chi(x)=\ln|2\sin(\pi x)|+\psi\left(\frac{1}{2\sin(\pi x)}\right)+\sin(\pi x)
\end{eqnarray}
being $\psi(x)$ the {\it digamma} Euler function $\psi(x)\equiv\frac{d}{dx}\ln\Gamma(x)$, i.e., the logarithmic derivative of the Euler $\Gamma$ function. For $n=2$ there is no problem, and we simply have, as shown by Berganza, Alcaraz and Sierra \cite{BerganzaAlcarazSierra2012},
\begin{equation}\label{chi_2}
 F^{(2)}_\chi(x)=\frac{7+\cos(2\pi x)}{8}
\end{equation}
To confirm these predictions, we compute the VNEE and the $n=2$ REE for the first excited state with the method of appendix \ref{peschel_method}, and we compare them with the CFT predictions (\ref{chi_1}) and (\ref{chi_2}), as shown in figure \ref{Ising_exc_FF}. As always, the finite-size agreement is not perfect, but taking the thermodynamic limit, that is equivalent to a continuum limit, we find in both cases an agreement well below 1\%.

We switch now to the $+F$ case (we do not consider the $-F$ case, that is equivalent to the $+F$ one for symmetry reasons): in this case, the spectrum just contains the conformal tower of $\sigma$, and therefore the ground-state REE's will receive corrections of the type $F^{(n)}_\sigma(x)$. The general conformal block of $\sigma$ fields has been computed for any even number of operators by Ardonne and Sierra \cite{ArdonneSierra2010}; we resume the result in appendix \ref{CB}. We start by considering the $n=1$ case, i.e., the denominator of equation (\ref{chiralF}). It is simply given by (see appendix \ref{CB})
\begin{equation}
 S_0^2(v_1,v_2)=v_{12}^{-1/8}
\end{equation}
so that, for $v_1=z_{1,0}^-$ and $v_2=z_{1,0}^+$,
\begin{equation}
 S_0^2(z_{1,0}^-,z_{1,0}^+)=2^{-1/8}e^{i\pi/16}\left[\sin(\pi x)\right]^{-1/8}
\end{equation}
Let us switch to the $n=2$ case: it is easy to see that formula (\ref{S_p^2n}), for $n=2$, explicitly looks
\begin{equation}
 S^4_p(v_1,v_2,v_3,v_4)=\frac{1}{\sqrt{2}}\prod_{a<b}v_{ab}^{-1/8}\sqrt{\sqrt{v_{13}v_{24}}+(-1)^p\sqrt{v_{14}v_{23}}}
\end{equation}
where $p=0,1$ labels the two possible conformal blocks. Placing the fields at the right coordinates (\ref{z_disc}), it reduces to
\begin{equation}\label{F^2_sigma}
 S^4_p(z_{2,0}^-,z_{2,0}^+,z_{2,1}^-,z_{2,1}^+)=[\sin(\pi x)]^{-1/4}\begin{cases}
  \cos\frac{\pi x}{4}, & p=0\\
  \sin\frac{\pi x}{4}, & p=1
 \end{cases}
\end{equation}
so that the two corresponding values of $F_{\sigma}^{(n)}$ are given by
\begin{equation}\label{F_sigma^2}
 F_{\sigma,p}^{(2)}=\begin{cases}
  \cos\frac{\pi x}{4}, & p=0\\
  \sin\frac{\pi x}{4}, & p=1
 \end{cases}
\end{equation}
Since our system is ordered at the left boundary, we expect the correction to be, in our case, the one for $p=0$. The $n=2$ REE is therefore
\begin{equation}\label{F_sigma^2}
 S_{2,+F}(l,L)-S_2^{CC}(l,L)=-\ln\cos\frac{\pi x}{4}
\end{equation}
In the $n=3$ case the computation proceeds in an analogous way. There, one deals with 6-points correlation functions of $\sigma$ fields, and therefore he would expect to have 4 independent ones. However, it can be seen, with a very boring computation, by means of equation (\ref{S_p^2n}), that we do not show here, that three of them are equal, and therefore one is left with just two independent conformal blocks. Analogously with the 4-point case, one has, in the end,
\begin{equation}\label{F_sigma^3}
 F_{\sigma,p}^{(3)}(x)=\begin{cases}
  \cos\frac{\pi x}{3}, & p=0\\
  \frac{1}{\sqrt{3}}\sin\frac{\pi x}{3}, & p=1
 \end{cases}
\end{equation}
and chooses, for the $+F$ case, the correction with $p=0$, so that
\begin{equation}\label{F_sigma^3}
 S_{3,+F}(l,L)-S_3^{CC}(l,L)=-\frac{1}{2}\ln\cos\frac{\pi x}{3}
\end{equation}

\begin{figure}[t]
 \begin{minipage}{\textwidth}
  \includegraphics[width=0.5\textwidth]{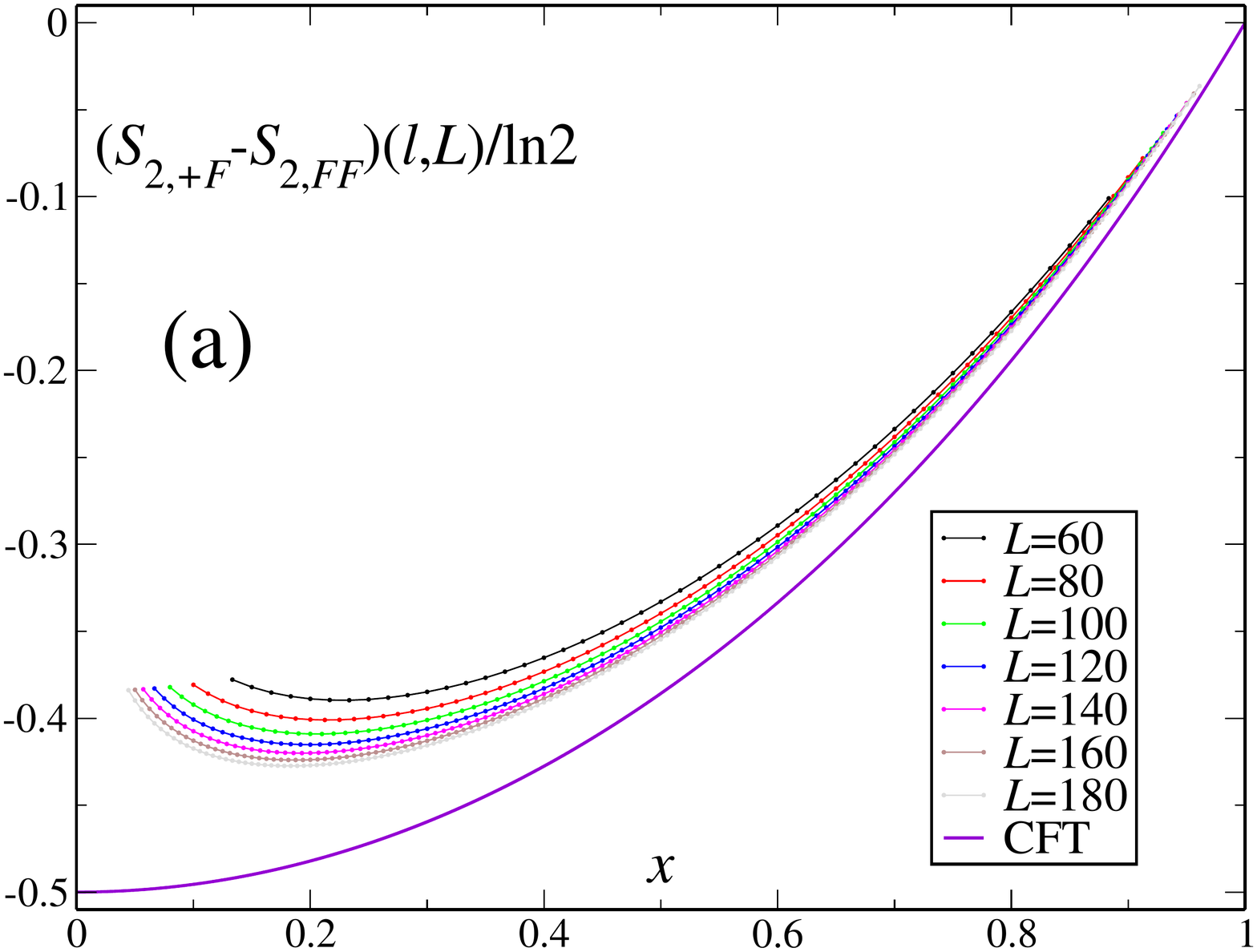}
  \includegraphics[width=0.5\textwidth]{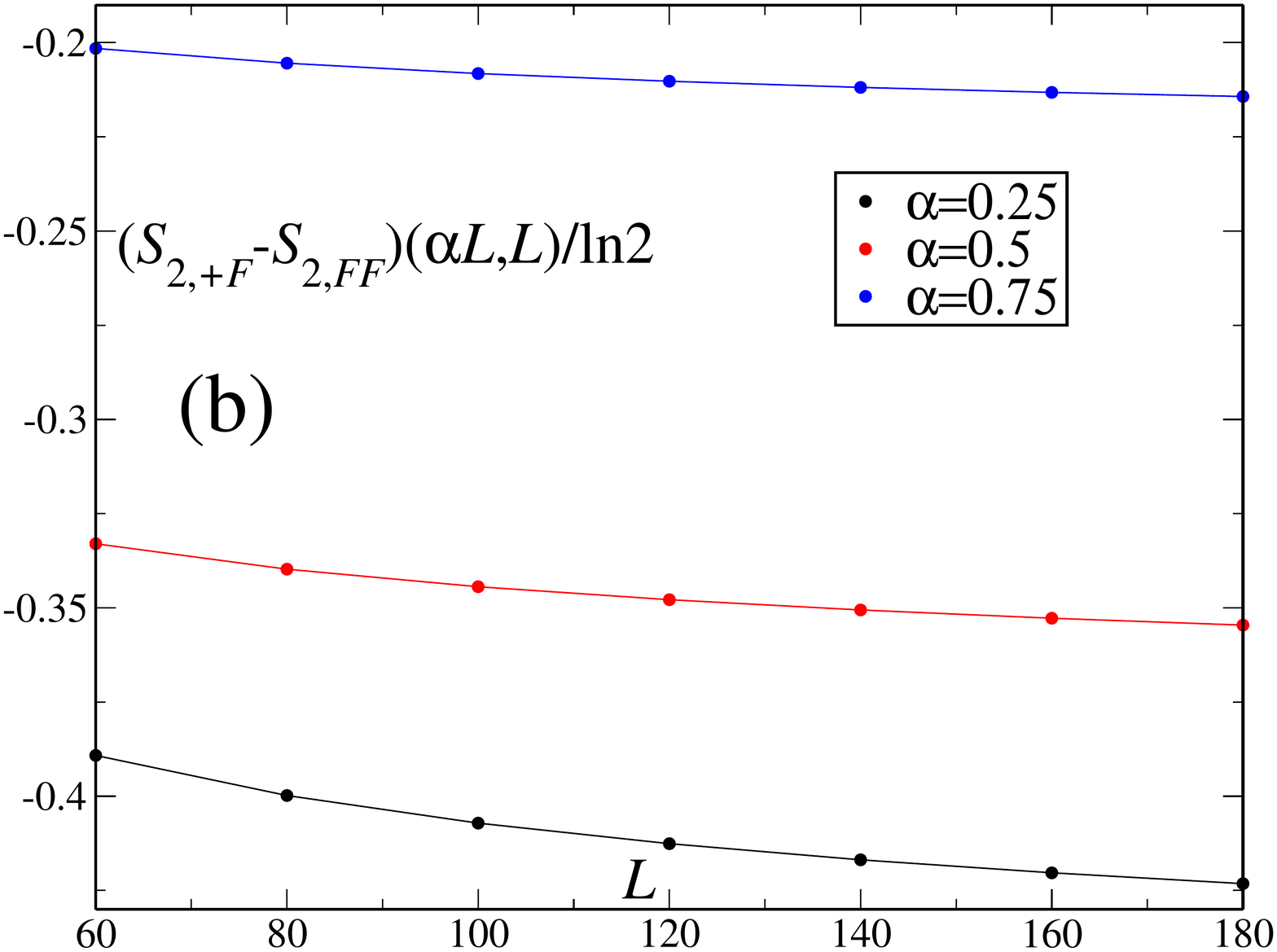}
  \includegraphics[width=0.5\textwidth]{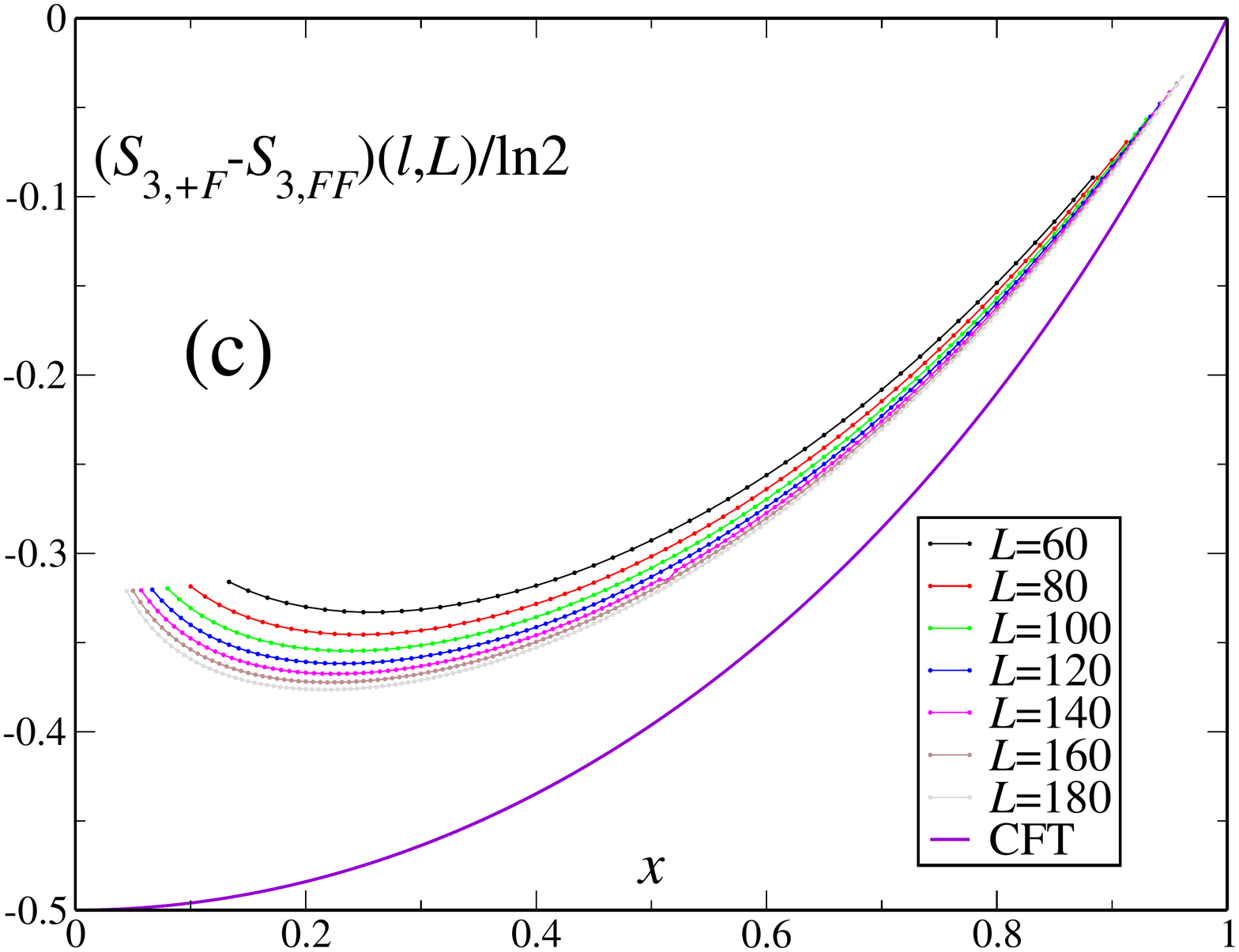}
  \includegraphics[width=0.5\textwidth]{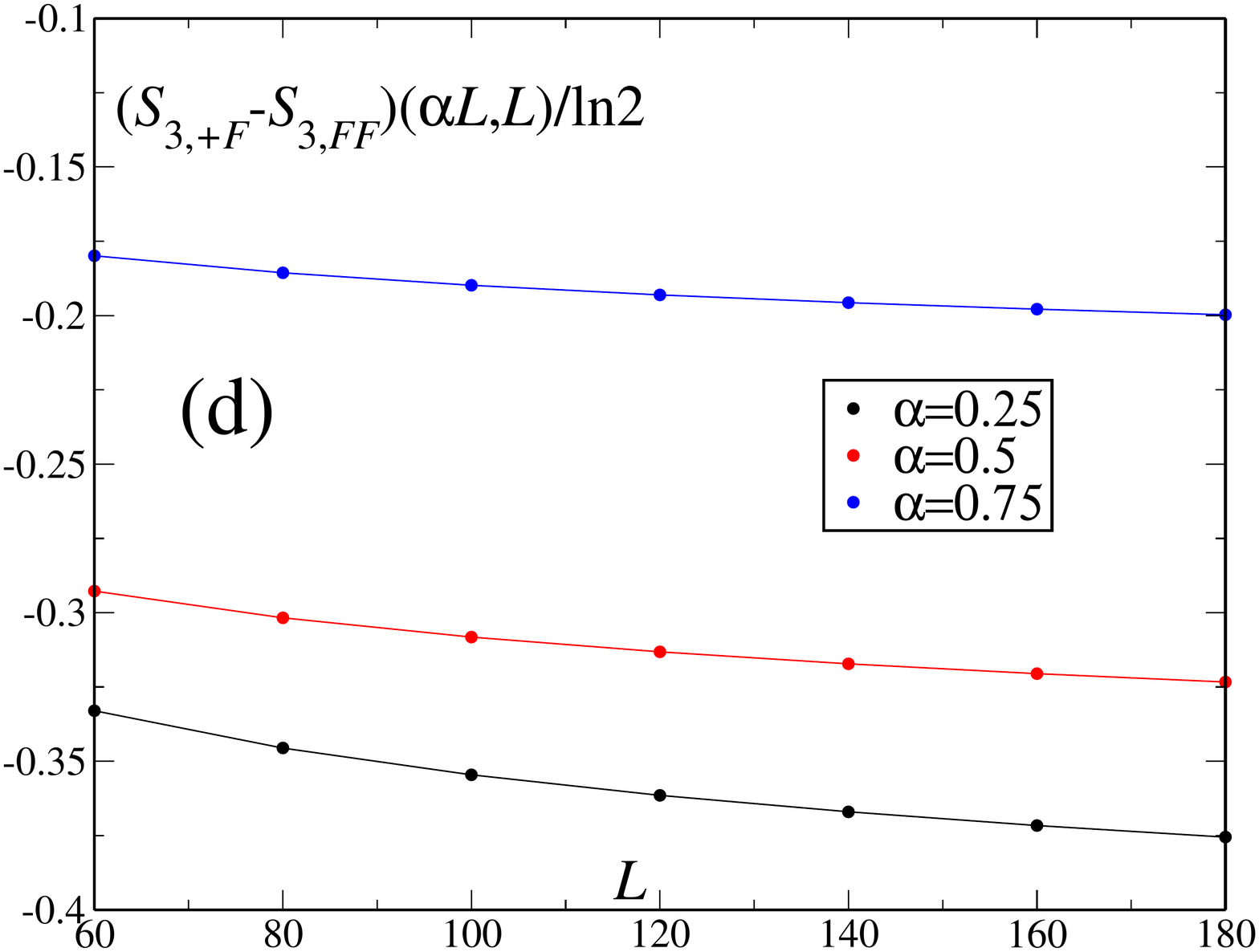}
  \caption{Comparison between the DMRG computations and the CFT predictions for the $n=2$ and $n=3$ REE's in the critical Ising model with $+F$ boundary conditions. (a): finite-size numerically-computed corrections to the $FF$ $S_2$; (b): FSS of the previous difference for specific values of $l$, using the 5-parameters formula $y=a_0+\frac{a_1}{x^{a_2}}+\frac{a_3}{x^{a_4}}$ (solid lines); (c), (d): same analysis for $n=3$.\label{Ising_+F_RE2}}
 \end{minipage}
\end{figure}

The $n=2,\,3$ REE's in the $+F$ have been computed by means of DMRG, implementing the boundary condition in an exact way, according to the recipe of appendix \ref{Fixed_DMRG}; the results are shown in figure \ref{Ising_+F_RE2}. As usual, the convergence of the numerical data to the CFT prediction is very slow, and one has to take a FSS to verify it. What we obtain is in excellent agreement with the CFT computation, with a precision always greater than 1\%. However, we had to add to (\ref{F_sigma^2}) and (\ref{F_sigma^3}) a constant value, i.e., the boundary entropy $-\frac{1}{2}\ln2$ relative to the $+F$ case, that our computations are not able to predict.

\begin{figure}[t]
 \begin{minipage}{\textwidth}
  \includegraphics[width=0.5\textwidth]{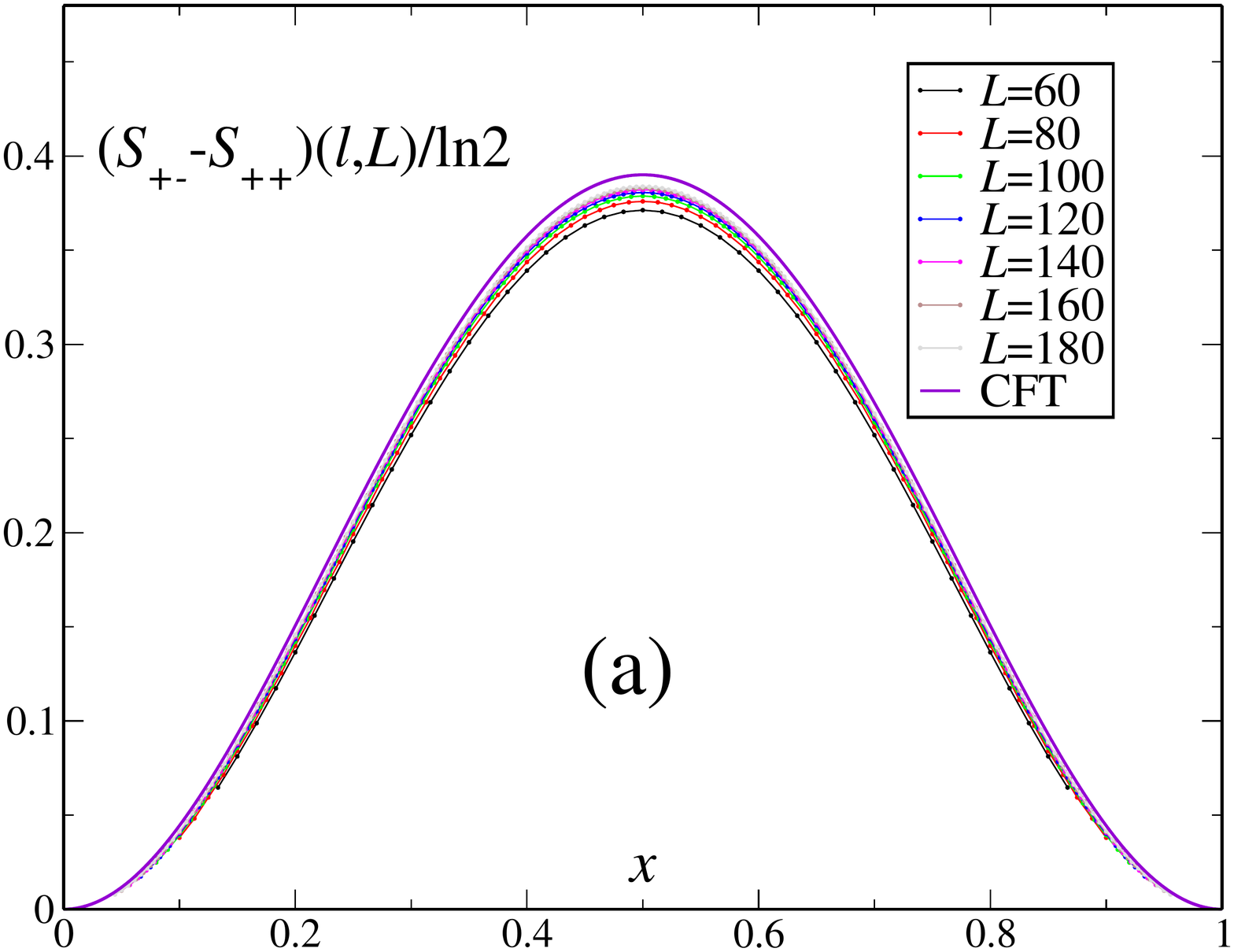}
  \includegraphics[width=0.5\textwidth]{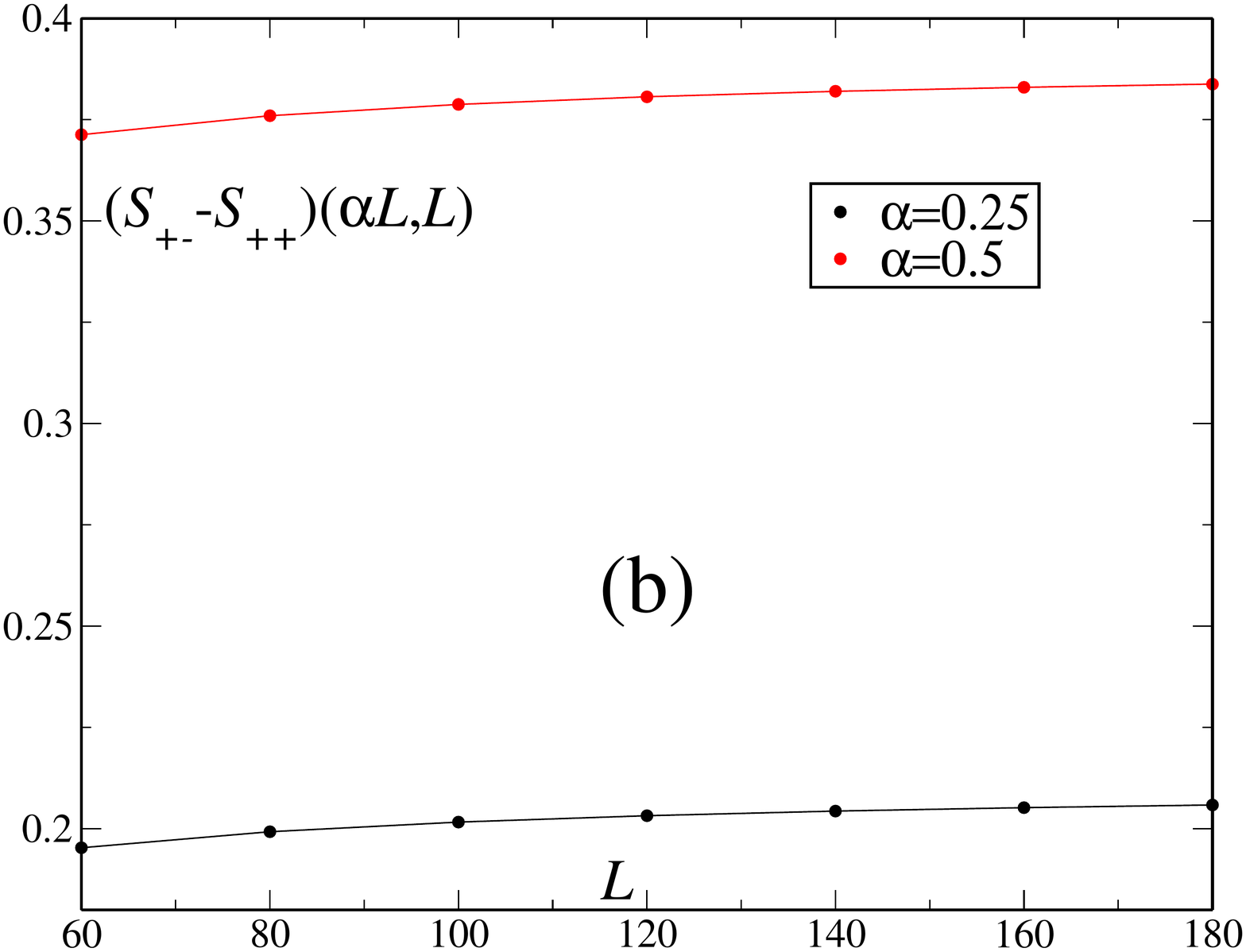}
  \includegraphics[width=0.5\textwidth]{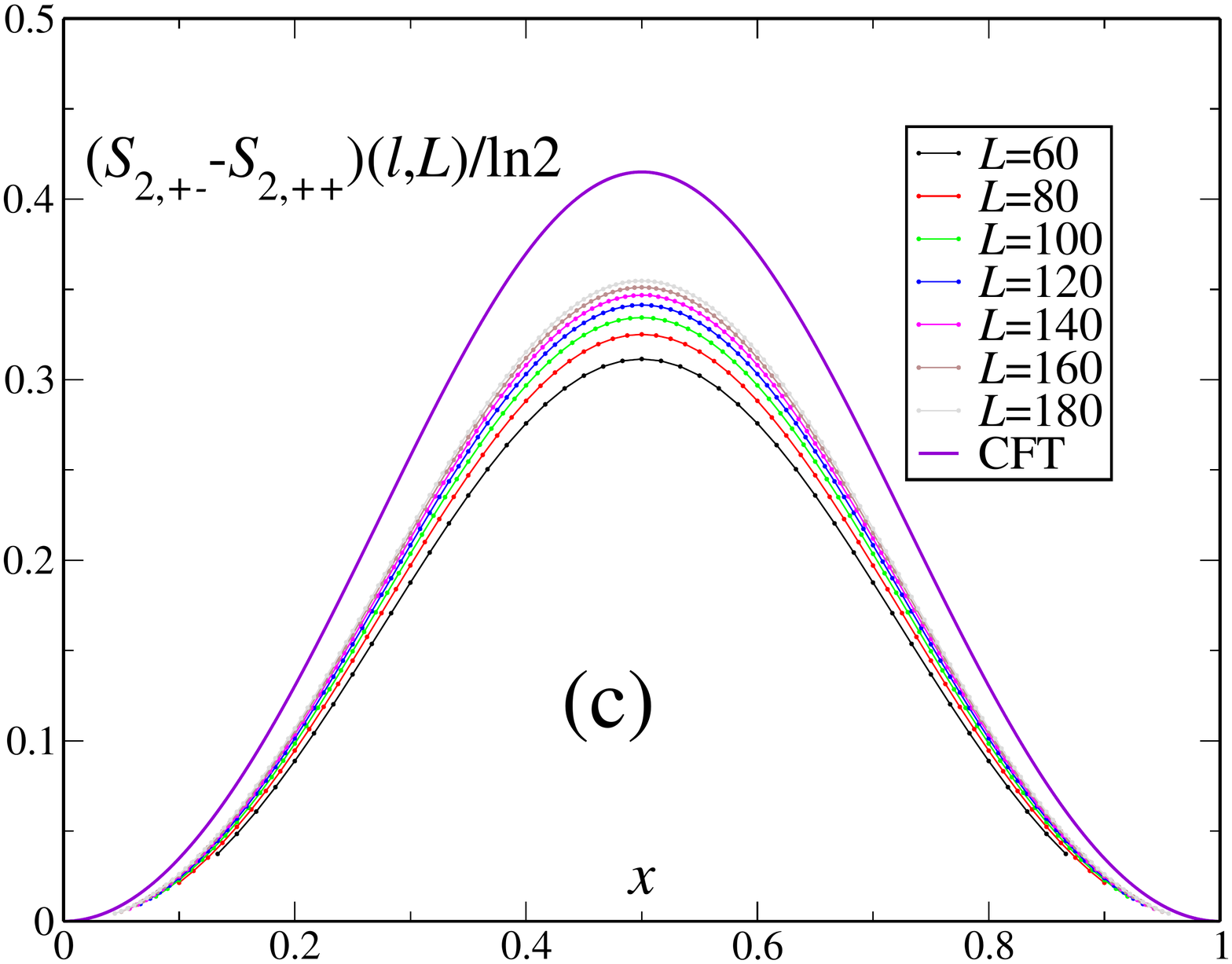}
  \includegraphics[width=0.5\textwidth]{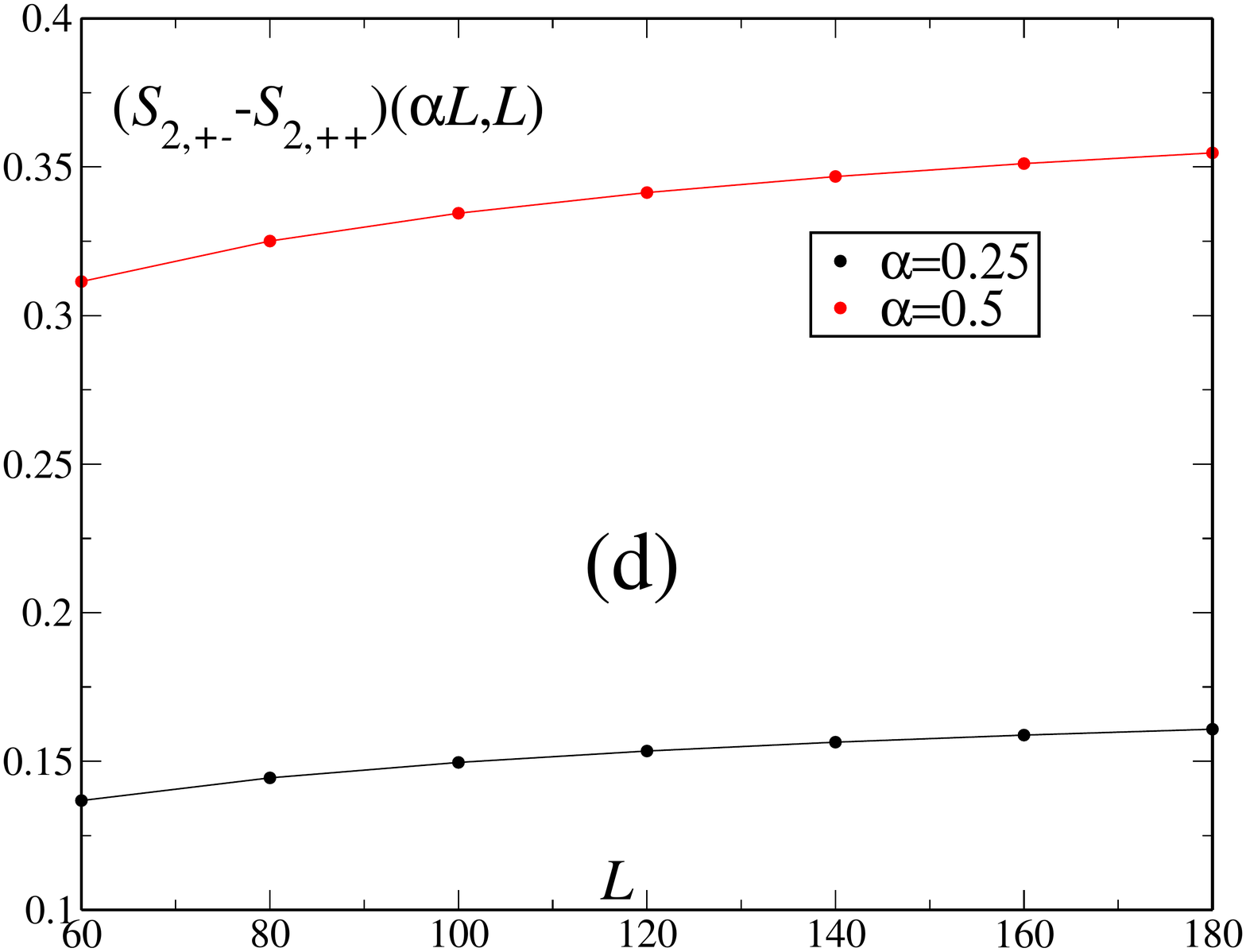}
 \end{minipage}
 \caption{Comparison between the numerical computations and the CFT predictions for the VNEE and the $n=2$ REE of the ground state in the critical Ising model with $+-$ boundary conditions. (a): finite-size numerically-computed corrections to the $+-$ VNEE of the ground state; (b): FSS of the previous difference for specific values of $l$, using the 5-parameters formula $y=a_0+\frac{a_1}{x^{a_2}}+\frac{a_3}{x^{a_4}}$ (solid lines); (c), (d): same analysis for $n=2$.}\label{Ising+-}
\end{figure}

The only case we still have to check is the $+-$ one: the operator content of the CFT is the conformal tower of the $\epsilon$ operator. Therefore, the correction to the REE's of the ground state shall be the same of the first excited state for $FF$ boundary conditions. We plot our DMRG results for VNEE and $n=2$ REE in figure \ref{Ising+-}(a) and (c), and their FSS's in (b) and (d): the agreement with the CFT prediction is good with a precision of $10^{-3}$ or, in most cases, less. We remark that, to make the correction more clean, we subtracted this time from the REE's the corresponding REE's of the ground state at $++$ boundary conditions: in fact, they both contain the same boundary-entropy contributions, that therefore cancels and leaves just the non-constant correction.

To resume, we were able to compute the corrections to the scaling (\ref{CC}), with exception of the constant boundary-entropy contributions, in all the situations allowed by our formalism, i.e., when the states originate from primary operators.


\section{Second Case: the $c=1$ Compactified Free Boson}\label{FFCFT}

The second case we are going to examine is the $c=1$ compactified massless free boson \cite{DiFrancescoMathieuSenechal1997}, i.e., Hamiltonian (\ref{H_LL}). Since the theory is not minimal, the situation is, {\it a priori}, more complicated than for the Ising model. However, it has been shown by Saleur \cite{Saleur1998} that there are just two possible conformal boundary conditions, known as {\it Dirichlet} ($D$) and {\it Neumann} ($N$). For Dirichlet boundary conditions the value of the scalar field is fixed to a constant value $\Phi_0$ at the edge, while for Neumann boundary conditions it is the dual of the scalar field that is fixed to a value $\tilde{\Phi}_0$. The corresponding $g$'s, defined in equation (\ref{ALg}), have been computed by Saleur \cite{Saleur1998} to be
\begin{equation}
 g_D=\frac{1}{\sqrt{2r\sqrt{\pi}}},\;\; g_N=\sqrt{r\sqrt{\pi}}
\end{equation}
where $r$ is the compactification radius of the boson (equation (120) in reference \cite{Saleur1998} is actually wrong; the one we are giving is the correct value). For $r=\frac{1}{2\sqrt{\pi}}$, that is exactly our case (see below), one has $g_D=1$ and $g_N=1/\sqrt{2}$, so that the corresponding boundary entropies are 0 and $-\frac{1}{2}\ln 2$.

First of all, let us consider the simplest lattice realization of the CFT: it is, as shown by Bilstein and Wehefritz, the modified XX Hamiltonian \cite{BilsteinWehefritz1999,Bilstein2000}
\begin{equation}\label{bilstein}
 H_B=-\sum_{j=1}^{L-1}\left(\sigma_j^x\sigma_{j+1}^x+\sigma_j^x\sigma_{j+1}^x\right)-\frac{1}{2}(\alpha_-\sigma_1^-+\alpha_+\sigma_1^++\alpha_z\sigma_1^z+\beta_-\sigma_L^-+\beta_+\sigma_L^++\beta_z\sigma_L^z)
\end{equation}
The sum term in the Hamiltonian is simply the open spin-1/2 XX chain (see section \ref{XY}); the boundary terms make it, in general, not Hermitian. With PBC and without boundary terms, the system is well known to be, at {\it half filling}, described by a $c=1$ compactified massless free boson, with compactification radius $r=\frac{1}{2\sqrt{\pi}}$ (see, e.g., \cite{Giamarchi2003}; in the book by Di Francesco, Mathieu and S\'en\'echal \cite{DiFrancescoMathieuSenechal1997}, the notation is slightly different, and the compactification radius is $R\equiv 2r\sqrt{\pi}$); the bulk theory shall be the same even in this case. An accurate exact analysis of the open model (\ref{bilstein}), performed by Bilstein \cite{Bilstein2000}, has pointed out that, varying the values of the boundary parameters, the partition function of the system is exactly the same as in a $c=1$ compactified free bosonic theory with $r=\frac{1}{2\sqrt{\pi}}$ and OBC \cite{Saleur1998}. Explicitly, when the boundary condition is $DD$, one has \cite{Saleur1998}
\begin{equation}\label{Z_DD}
 Z_{DD}(q,\Delta)=\frac{1}{\eta(q)}\sum_{n\in\mathbb{Z}}q^{\frac{1}{2}\left(\Delta+n\right)^2}
\end{equation}
where $\eta(q)$ is the Dedekind $\eta$ function and $\Delta\equiv\Phi_0-\Phi_0'$ is the difference of the field values at the boundaries; we remark that is what we obtained explicitly for the cases of section \ref{FF}, that belong therefore to the $DD$ class of boundary conditions. For $NN$ boundary conditions, one has
\begin{equation}\label{Z_NN}
 Z_{NN}(q,\Delta)=\frac{1}{\eta(q)}\sum_{n\in\mathbb{Z}}q^{2\left(n+\tilde{\Delta}/2\right)^2}
\end{equation}
where now $\tilde{\Delta}\equiv\tilde{\Phi}_0-\tilde{\Phi}_0'$ is the difference of the dual-field values at the boundaries. The partition function for mixed $ND$ boundary conditions is given by
\begin{equation}\label{Z_DN}
 Z_{ND}(q)=\frac{1}{2\eta(q)}\sum_{n\in\mathbb{Z}}q^{\frac{1}{4}\left(n-\frac{1}{2}\right)^2}
\end{equation}
where the values of the fields at the boundaries are absent \cite{Saleur1998}. In all cases, $q\equiv e^{-\pi\beta/L}$, being $\beta$ the inverse temperature, plays the role of a modular parameter \cite{DiFrancescoMathieuSenechal1997}, even if there is no modularity at all. We just use a part of the Bilstein's results \cite{Bilstein2000}: the simplest realization of $DD$ boundary conditions is given, as one can expect, by the choice
\begin{equation}\label{realizationDD}
 DD:\ \alpha_-=\alpha_+=\alpha_z=\beta_-=\beta_+=\beta_z=0
\end{equation}
that is the $(0,0)$ case considered in section \ref{FF}. $NN$ and $ND$ boundary conditions are, instead, simply realized by the non-trivial choices
\begin{eqnarray}\label{realizationNNND}
 NN :\ & & \alpha_z=\beta_z=0,\ \alpha_-=\alpha_+=\beta_-=\beta_+=2\\
 ND :\ & & \alpha_z=\beta_z=\beta_-=\beta_+=0,\ \alpha_-=\alpha_+=2
\end{eqnarray}
We remark this is just a realization, and we could have done different choices by following Bilstein \cite{Bilstein2000}.

As in the previous case, our task is to determine the operator content of the different theories, looking directly at the partition functions. In order to do it, let us briefly describe what happens for a $c=1$ compactified boson defined on a torus. It is known \cite{DijkgraafVerlindeVerlinde1988,DiFrancescoMathieuSenechal1997} that in that case, the partition function looks
\begin{equation}
 Z(q,\bar{q})=\frac{1}{\left|\eta(q)\right|^2}\sum_{n,m\in\mathbb{Z}}q^{p^2_{mn}}\bar{q}^{\bar{p}^2_{mn}}
\end{equation}
being $q,\ \bar{q}$ complex numbers related to the the modular parameter of the torus, and 
\begin{equation}
 p_{mn}\equiv\frac{n}{2\sqrt{\pi}r}+\sqrt{\pi}mr,\;\; \bar{p}_{mn}\equiv\frac{n}{2\sqrt{\pi}r}-\sqrt{\pi}mr
\end{equation}
Therefore, the conformal dimensions of the primary operators of the theory are $h_{mn}=\frac{1}{2}p_{mn}^2$, $\bar{h}_{mn}=\frac{1}{2}\bar{p}^2_{mn}$. In our case, the compactification radius is $r=\frac{1}{2\sqrt{\pi}}$, so that the $p$'s become
\begin{equation}
 p_{mn}=n+\frac{m}{2},\;\; \bar{p}_{mn}=n-\frac{m}{2}
\end{equation}
This particular choice of the radius enhances the symmetry of the theory, so that, instead of a countable infinity of primary fields, one is left with just four; this fact makes the theory a so called {\it rational} CFT \cite{DiFrancescoMathieuSenechal1997}. To see this, let us parameterize the $p$'s as
\begin{equation}
 p=2r+\frac{\lambda}{2},\;\; \bar{p}=2s-\frac{\lambda}{2},\;\;r,s\in\mathbb{Z},\;\;\lambda=0,1,2,3
\end{equation}
It is easy to see that this parameterization is exactly equivalent to the previous one. The partition function becomes
\begin{equation}
 Z(q,\bar{q})=\frac{1}{\left|\eta(q)\right|^2}\sum_{\lambda=0}^3\left|K_\lambda(q)\right|^2
\end{equation}
with
\begin{equation}\label{Kharacters}
 K_\lambda(q)\equiv\frac{1}{\eta(q)}\sum_{n\in\mathbb{Z}}q^{\frac{1}{8}(4n+\lambda)^2}
\end{equation}
Explicitly, the characters (\ref{Kharacters}) look
\begin{equation}
 \begin{split}
  K_0(q) &=\frac{1}{\eta(q)}\sum_{n\in\mathbb{Z}}q^{\frac{1}{2}\left(2n\right)^2}\\
  K_1(q) &=\frac{1}{\eta(q)}\sum_{n\in\mathbb{Z}}q^{\frac{1}{2}\left(2n+\frac{1}{2}\right)^2}\\
  K_2(q) &=\frac{1}{\eta(q)}\sum_{n\in\mathbb{Z}}q^{\frac{1}{2}\left(2n+1\right)^2}\\
  K_3(q) &=\frac{1}{\eta(q)}\sum_{n\in\mathbb{Z}}q^{\frac{1}{2}\left(2n-\frac{1}{2}\right)^2}
 \end{split}
\end{equation}
These are the characters corresponding to the four Verma modules of the rational CFT. The chiral algebra of the theory is given by
\begin{equation}
 \mathcal{A}=\{\mathbb{I},T(z),J(z)=i\partial\phi(z),W^\pm(z)=e^{\pm 2i\phi(z)}\}
\end{equation}
which, in addition to the stress tensor $T(z)$ and the U(1) current $J(z)$, contains the two charge fields $W^{\pm}(z)$, of conformal dimension $h_W=2$. The weight of the primary fields are $h_\lambda=0,\ 1/8,\ 1/2,\ 1/8$ for $\lambda=0,1,2,3$, so that $K_0$ is the character of the conformal tower of the identity. Indeed, the expansion of this character around $q=0$, given by
\begin{equation}
 q^{1/24}K_0(q)=1+q+4q^2+O(q^3)
\end{equation}
shows that that the linear term corresponds to the current $J(z)$, while the quadratic one corresponds to the four fields $T(z)$, $\partial^2\phi(z)$ and $W^{\pm}(z)$. The fusion coefficients of the theory are simply given by \cite{DiFrancescoMathieuSenechal1997}
\begin{equation}
 \mathcal{N}_{\lambda\mu}^\nu=\delta_{\lambda+\mu}^\nu
\end{equation}
where the indices are defined modulo 4.

\begin{figure}[t]
 \begin{minipage}{\textwidth}
  \includegraphics[width=0.5\textwidth]{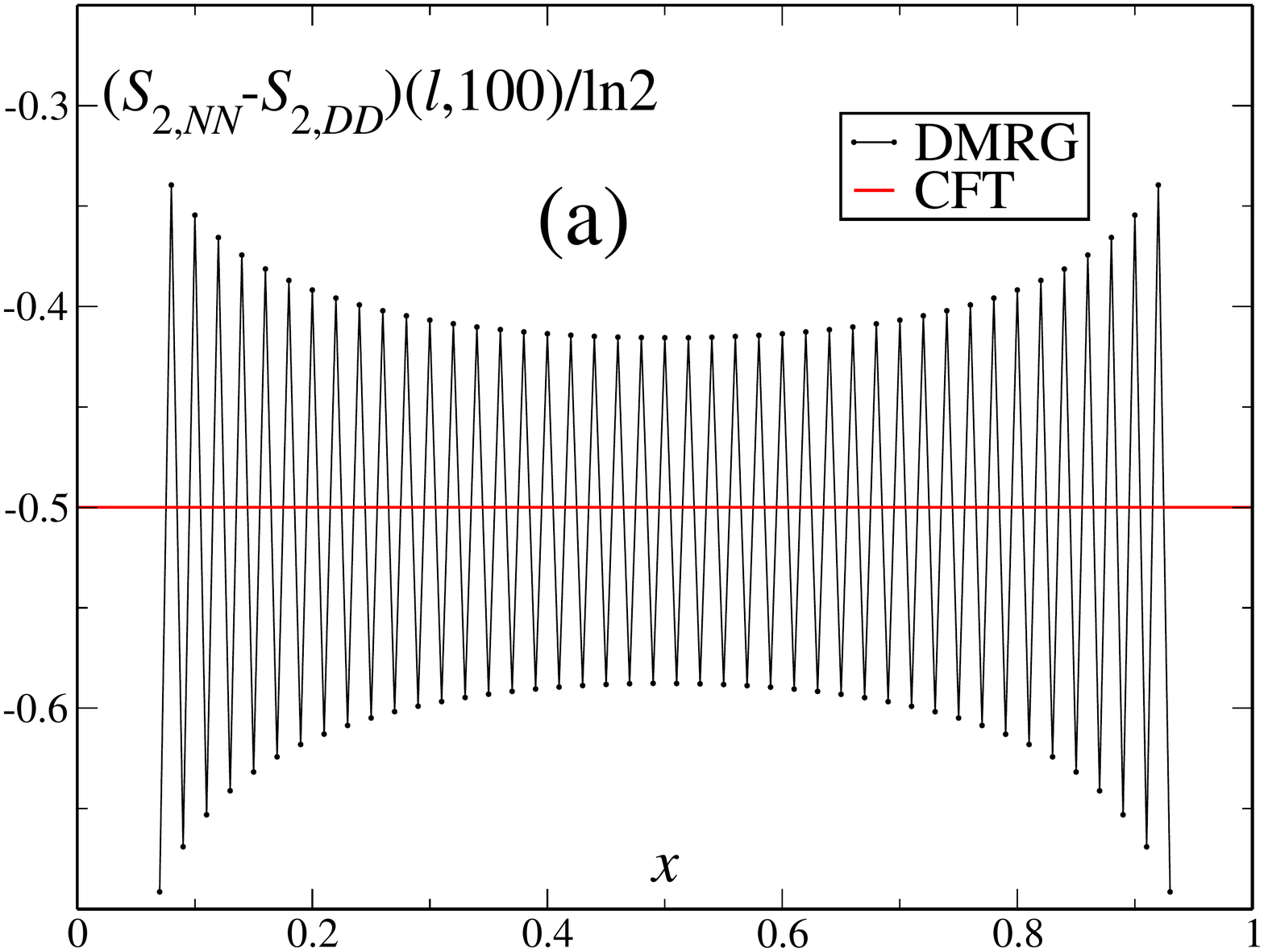}
  \includegraphics[width=0.5\textwidth]{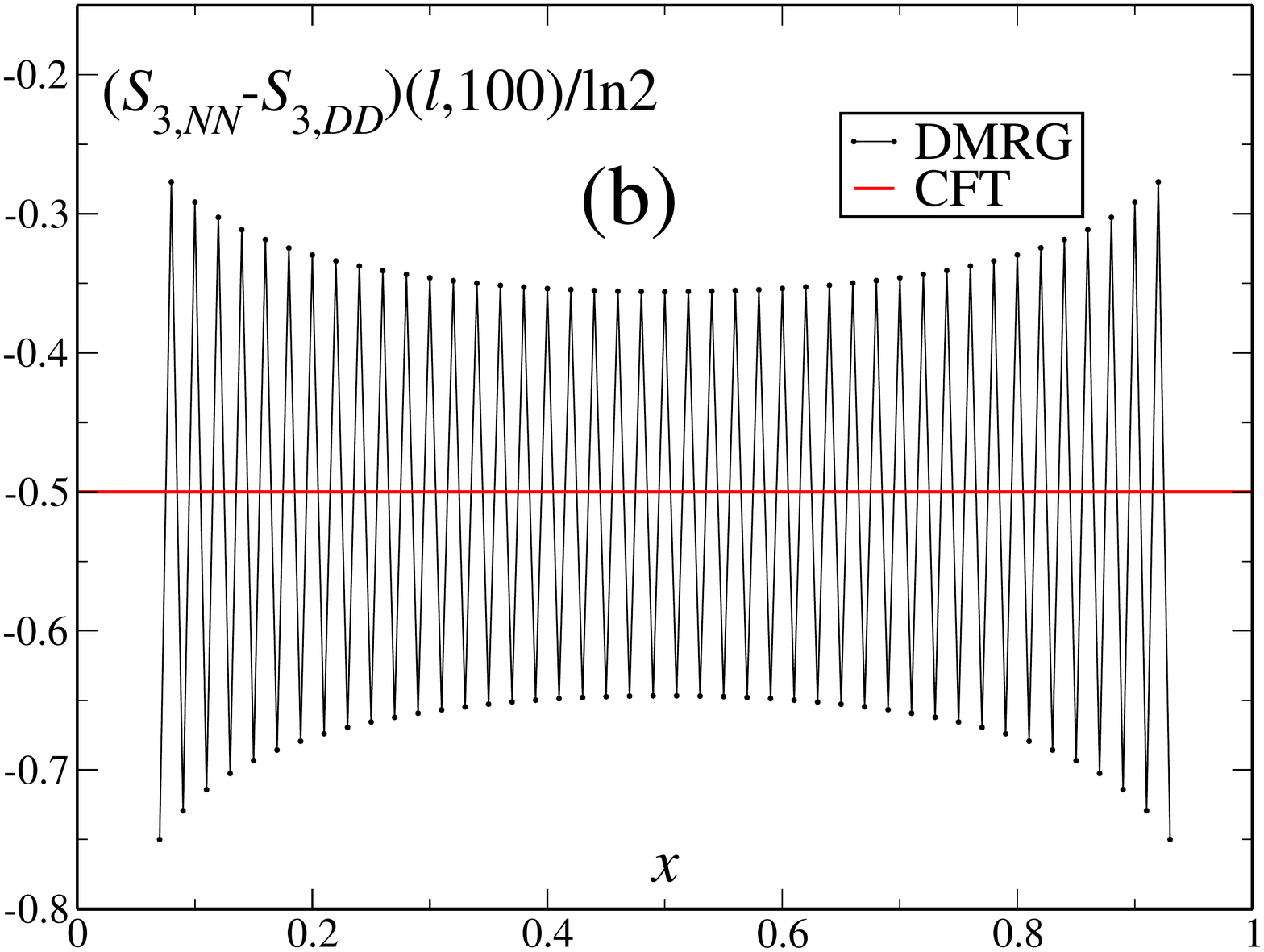}
  \caption{Numerical REE's for the ground state of Hamiltonian (\ref{bilstein}), with the choice (\ref{realizationNNND}), in the $NN$ case: in both cases the corrections oscillate around the constant value $-\frac{1}{2}\ln2$.\label{NN-DD}}
 \end{minipage}
\end{figure}

Let us come back to the open case: all the cases we will consider are listed in table \ref{table:bilstein}.
\begin{table}
 \caption{The open spin-1/2 XX chain.}
 \centering
 \begin{tabular}{c c c c}
  \hline\hline
  $a$ & $b$ & $h$ & Figure \\
  \hline
  $N$ & $N$ & 0 & \ref{NN-DD}\\
  $D$ & $D$ & 1/2 & \ref{REE_+1f}\\
  $D$ & $D$ & 1 & \ref{DD-ph}\\
  $N$ & $N$ & 1 & \ref{NN-exc}\\
  $N$ & $D$ & 1/16 & \ref{DN-DD}\\
  $N$ & $D$ & 9/16 & \ref{DN_exc-DD}\\
  \hline
 \end{tabular}
 \label{table:bilstein}
\end{table}
We start by considering the $DD$ and the $NN$ case. By direct comparison, it is easy to argue that the partition functions (\ref{Z_DD}) and (\ref{Z_NN}), with $\Delta=\tilde{\Delta}=0$, can be re-written as
\begin{equation}
 \begin{split}
  Z_{DD}(q) &=K_0(q)+K_2(q)\\
  Z_{NN}(q) &=K_0(q)
 \end{split}
\end{equation}
so that the $NN$ case is actually the simplest, containing just the identity sector. However, for both $DD$ and $NN$ boundary conditions, the ground state REE's should receive no corrections to the scaling (\ref{CC}), up to, as in the minimal-$c=1/2$ case, the constant boundary entropy, that our formalism seems not to be able to predict. The absence of universal non-constant corrections has been observed in the $DD$ case in several works (see, e.g., chapter \ref{Crossovers}); we are going to verify it numerically in the $NN$ case. We performed DMRG simulations at the system size $L=100$, using 3 sweeps and up to 1100 states per block (these are the typical number of sweeps and states per block we will use in the following), in order to have a truncation error of $10^{-10}$ or less; the results are shown in figure \ref{NN-DD}. The only corrections arising are an oscillating one, typical of $c=1$ systems (see, e.g., section \ref{corrections} and chapter \ref{Crossovers}) and a {\it constant} one, of the value of $-\frac{1}{2}\ln 2$ (it is exactly the boundary entropy associated to the $N$ boundary condition), as it happened in the $c=1/2$ case. The presence of the oscillating correction does not worry us so much, since it has been shown that it cannot be predicted just by means of CFT (see section \ref{corrections}): our formalism cannot therefore provide its specific form.

\begin{figure}[t]
 \begin{minipage}{\textwidth}
  \includegraphics[width=0.5\textwidth]{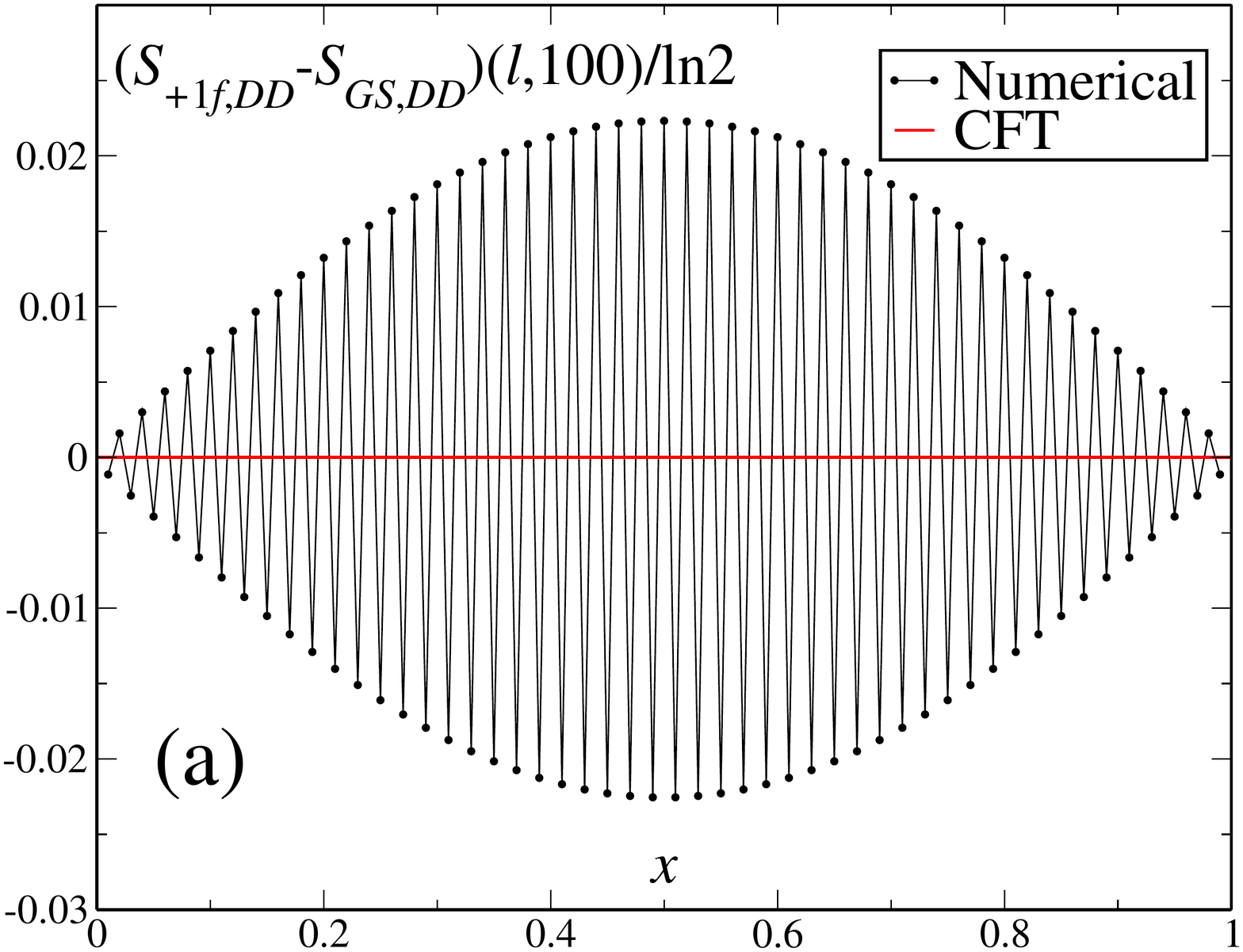}
  \includegraphics[width=0.5\textwidth]{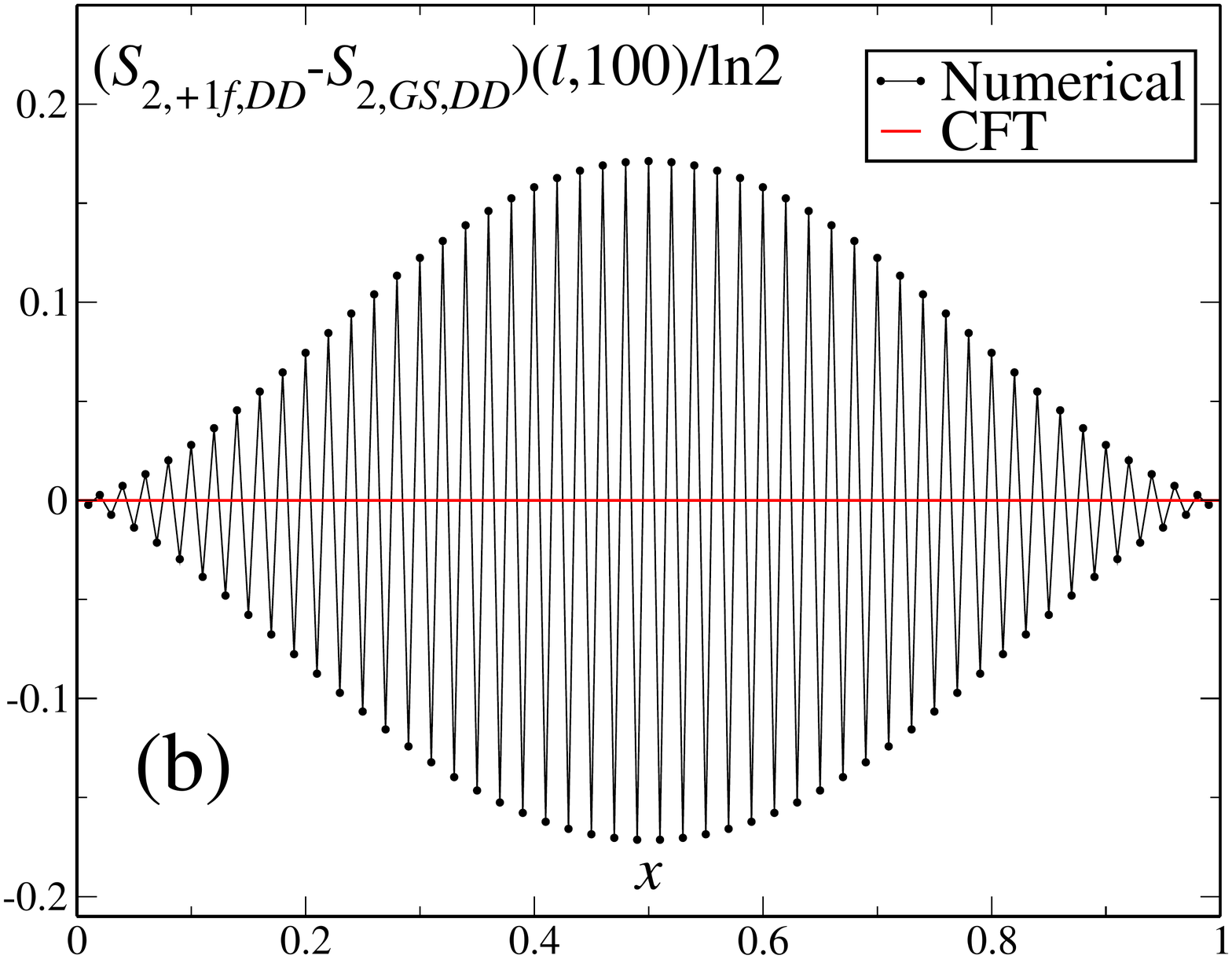}
  \caption{Numerical VNEE (a) and $n=2$ REE (b) of the ground state of the filling-$(1/2+1/L)$ sector of Hamiltonian (\ref{bilstein}) in the $DD$ case (\ref{realizationDD}), compared to the vanishing CFT prediction.\label{REE_+1f}}
 \end{minipage}
\end{figure}

In the $DD$ case, beside $K_0$, the partition function also contains the character $K_2$, whose primary field is a vertex operator of conformal weigth 1/2 \cite{DiFrancescoMathieuSenechal1997}. We proceed in analogy with the PBC case \cite{AlcarazBerganzaSierra2011,BerganzaAlcarazSierra2012}, and we consider the ground state in the filling-$1/2+1/L$ sector, i.e., the one with $L/2+1$ fermions. Its energy is given by (the single-particle spectrum is the one of the (0,0) Hamiltonian of section \ref{FF})
\begin{equation}
 E_{+1f}-E_{GS}=\sin\frac{\pi}{2(L+1)}
\end{equation}
that, expanded in powers of $1/L$, reads
\begin{equation}
 E_{+1f}-E_{GS}=\frac{\pi}{2L}+O(1/L^2)
\end{equation}
that identifies the conformal dimension of the operator creating the excitation as $1/2$ (see equation (\ref{E_h})), the same of the PBC case: this state is therefore the one we were looking for. The relative operator was predicted, by Alcaraz, Berganza and Sierra \cite{AlcarazBerganzaSierra2011}, not to bring any correction to the scaling of REE's. In figure \ref{REE_+1f} we show that this is the case, up to oscillating factors (the numerical data were produced by the method of appendix \ref{peschel_method}). We add that the considered state is degenerate: the ground state in the filling-$(1/2-1/L)$ sector has the same energy, and therefore is created by a vertex operator of conformal dimension 1/2, too. The REE's of this state are numerically seen to be exactly the ones of the ground state in the filling-$(1/2+1/L)$ sector, and therefore we do not show them here.

Moreover, both in the $DD$ and in the $NN$ case, there is a third excitation that is created by a primary operator. This fact can be seen by expanding $Z_{DD}(q)$ and $Z_{NN}(q)$ around $q=0$:
\begin{equation}\label{Z_DD,NN}
 \begin{split}
  q^{1/24}Z_{DD}(q) &=1+2q^{1/2}+q+O(q^2)\\
  q^{1/24}Z_{NN}(q) &=1+q+O(q^2)
 \end{split}
\end{equation}
We remark that, in the $DD$ case, the $q^{1/2}$ term corresponds to the states we just considered, i.e., the ones with one fermion more or less than the half filling, and has therefore a factor two on top. In both cases, the next term is $q$ with multiplicity one, telling us that there should be a state in the spectrum originating from an operator of conformal dimension 1, that is, in the CFT picture, $i\partial\phi$ \cite{DiFrancescoMathieuSenechal1997}. In the $DD$ case, the corresponding state is known to be the {\it particle-hole} excitation \cite{AlcarazBerganzaSierra2011,BerganzaAlcarazSierra2012}, i.e., the state created from the ground state by moving the closest-to-the-Fermi-level fermion one level up: let us verify it in our case. Its energy is given by
\begin{equation}
 E_{ph}-E_{GS}=2\sin\frac{\pi}{2(L+1)}
\end{equation}
Expanding it in powers of $1/L$, we have
\begin{equation}
 E_{ph}-E_{GS}=\frac{\pi}{L}+O(1/L^2)
\end{equation}
that, compared to equation (\ref{E_h}), tells us that the conformal dimension of the particle-hole exciting operator is 1, identifying it as $i\partial\phi$. In figure \ref{DD-ph} we show the VNEE and the $n=2$ REE, that have been computed by the method of appendix \ref{peschel_method}: the corrections should take the forms (\ref{chi_1}) and (\ref{chi_2}), multiplied by a factor 2, because what we are considering is now directly $F^{(n)}_{i\partial\phi}$, and not its square root. The numerical scaling of the corrections fully confirms the CFT prediction.
\begin{figure}[t]
 \begin{minipage}{\textwidth}
  \includegraphics[width=0.5\textwidth]{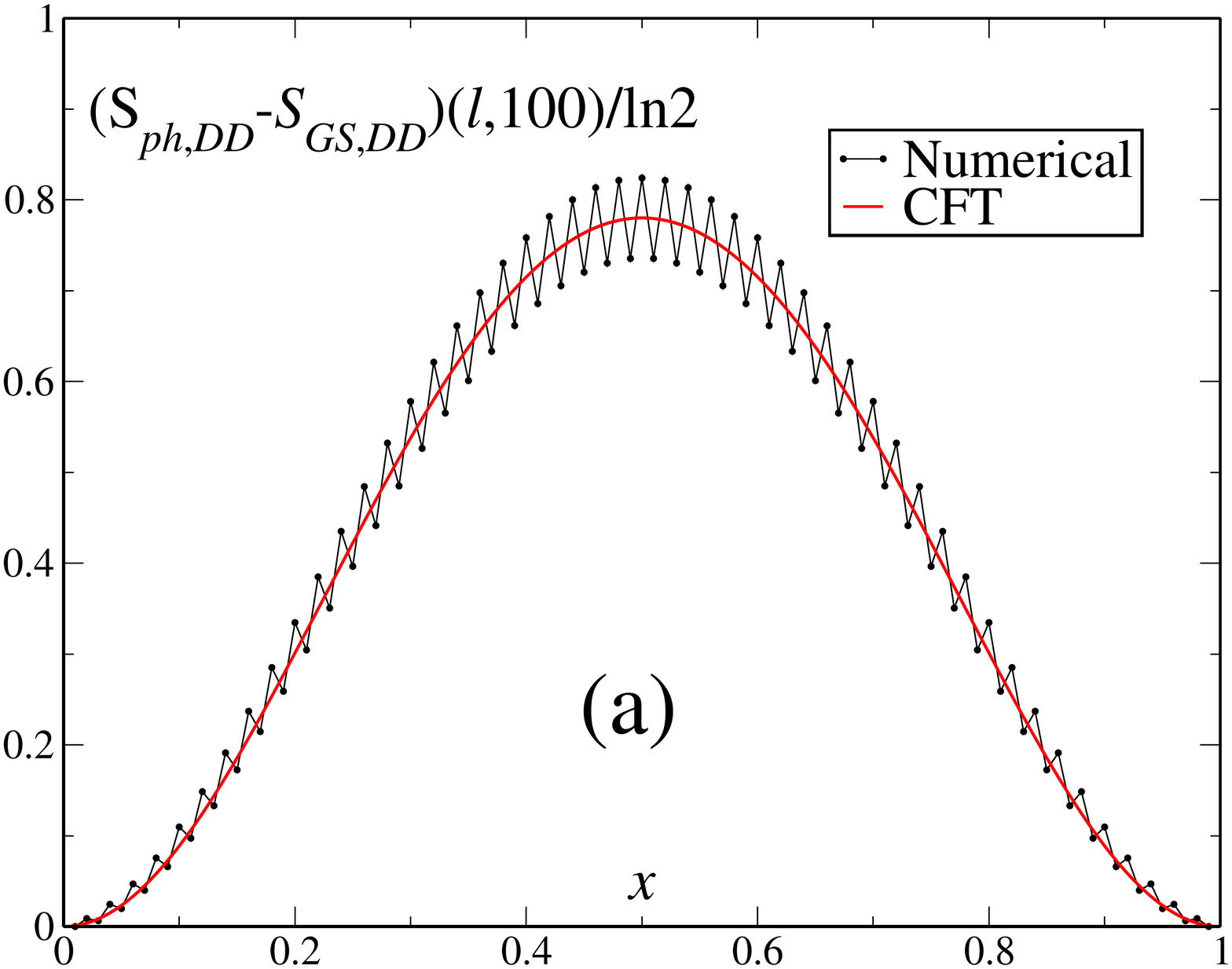}
  \includegraphics[width=0.5\textwidth]{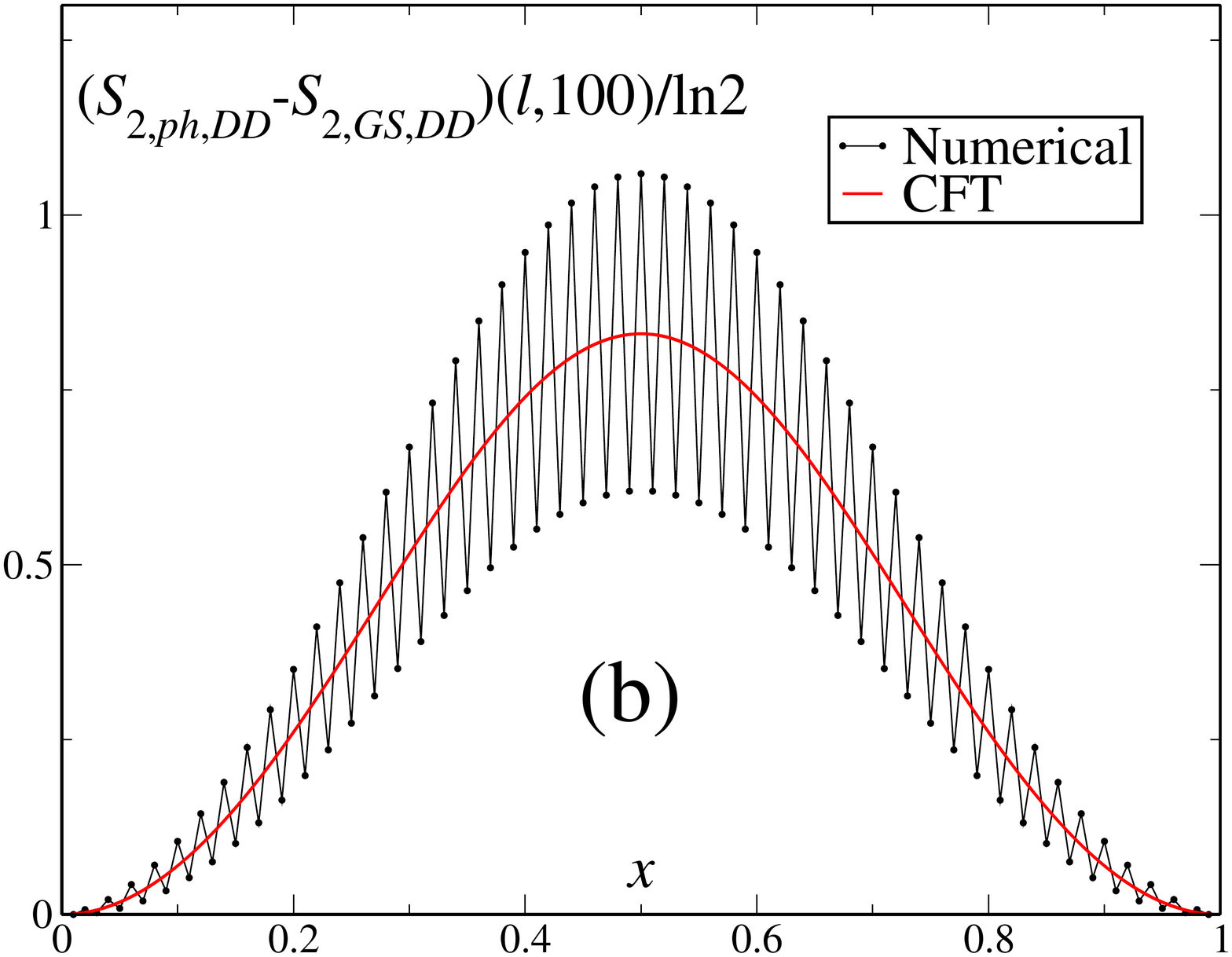}
  \caption{Numerical VNEE and $n=2$ REE of the first, starting from the half filling, particle-hole state of Hamiltonian (\ref{bilstein}) in the $DD$ case (\ref{realizationDD}), and comparison with the CFT prediction.\label{DD-ph}}
 \end{minipage}
\end{figure}

\begin{figure}[t]
 \begin{minipage}{\textwidth}
  \includegraphics[width=0.5\textwidth]{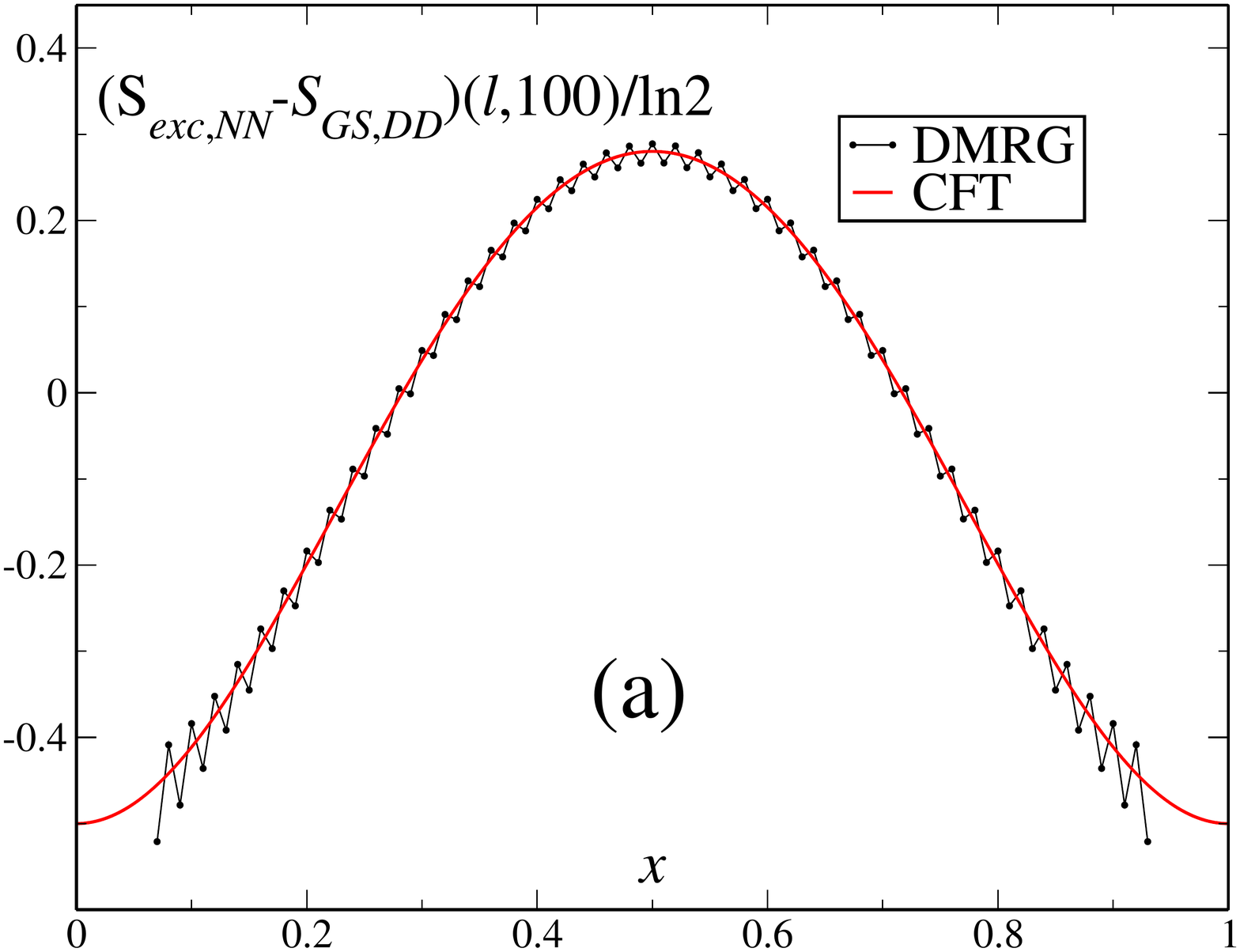}
  \includegraphics[width=0.5\textwidth]{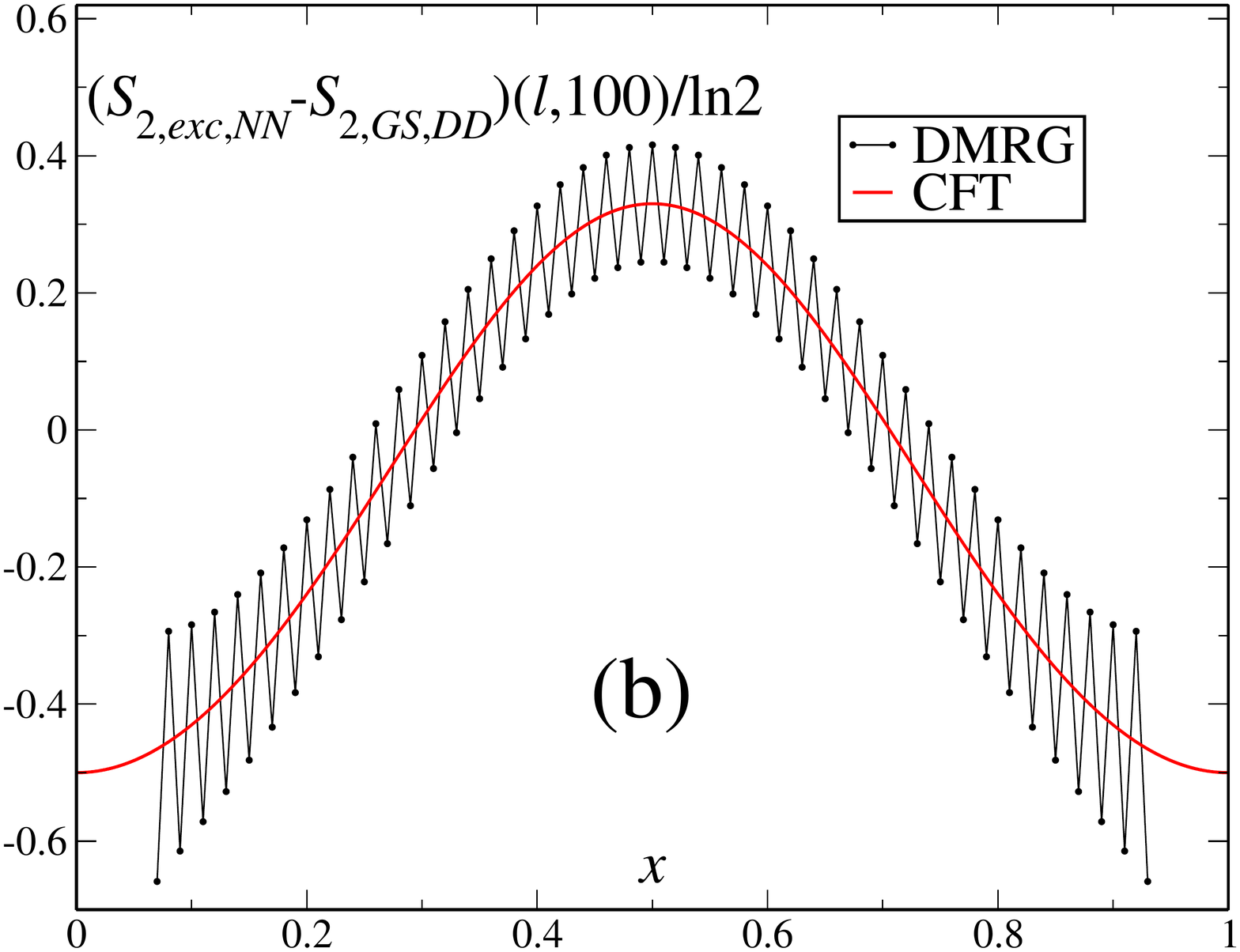}
  \caption{Numerical VNEE and $n=2$ REE of the first excited state of Hamiltonian (\ref{bilstein}) in the $NN$ case (\ref{realizationNNND}), and comparison with the CFT prediction.\label{NN-exc}}
 \end{minipage}
\end{figure}

The same corrections (up to the boundary entropy) shall be present even in the $NN$ case, for the first excited state, as it can be seen from the second of equations (\ref{Z_DD,NN}): they are given, again, by $F^{(n)}_{i\partial\phi}$. We performed simulations with multi-target DMRG \cite{DegliEspostiBoschiOrtolani2004}, using the usual number of sweeps and states per block, in order to achieve a maximum truncation error of $10^{-9}$: the results are displayed in figure \ref{NN-exc}: even in this case, the agreement is remarkable, up to the usual boundary entropy $-\frac{1}{2}\ln 2$ and oscillating contributions.

We then consider $ND$-boundary-conditions case: the partition function is now (\ref{Z_DN}), and it cannot be written as a sum of characters $K_\lambda(\tau)$. This suggests us to proceed in a quite different way. It can be seen by expanding (\ref{Z_DN}) in powers of $q$ around $q=0$ that
\begin{equation}\label{Z_ND_Ising}
 Z_{ND}(q)=\chi_{1/16}(q)[\chi_0(q)+\chi_{1/2}(q)]
\end{equation}
where the $\chi_h(q)$'s are the characters of the $c=1/2$ minimal CFT \cite{DiFrancescoMathieuSenechal1997}:
\begin{equation}
 \chi_{(r,s)}(q)=K_{r,s}^{(4,3)}(q)-K_{r,-s}^{(4,3)}(q)
\end{equation}
where $(r,s)=(1,1),\,(2,1),\,(1,2)$ for the identity, $\sigma$ and $\chi$ fields respectively, and
\begin{equation}
 K_{r,s}^{(4,3)}(q)\equiv\frac{1}{\eta(q)}\sum_{n\in\mathbb{Z}}q^{\frac{(24n+4r-3s)^2}{48}}
\end{equation}
Equation (\ref{Z_ND_Ising}) simply tells us that the states of the theory are the tensor product of two {\it chiral} states, one belonging to the Verma module of the $\sigma$ operator and one to the sum of the Verma modules of the identity and of the $\epsilon$. Therefore, the ground state in the $ND$ case should receive a correction of the form
\begin{equation}
 F^{(n)}_{\sigma\otimes\mathbb{I},p}(x)=F^{(n)}_{\sigma,p}(x)F^{(n)}_{\mathbb{I}}(x)=F^{(n)}_{\sigma,p}(x)
\end{equation}
This correction has already been computed in the previous section for $n=2,\,3$.

\begin{figure}[t]
 \begin{minipage}{\textwidth}
  \includegraphics[width=0.5\textwidth]{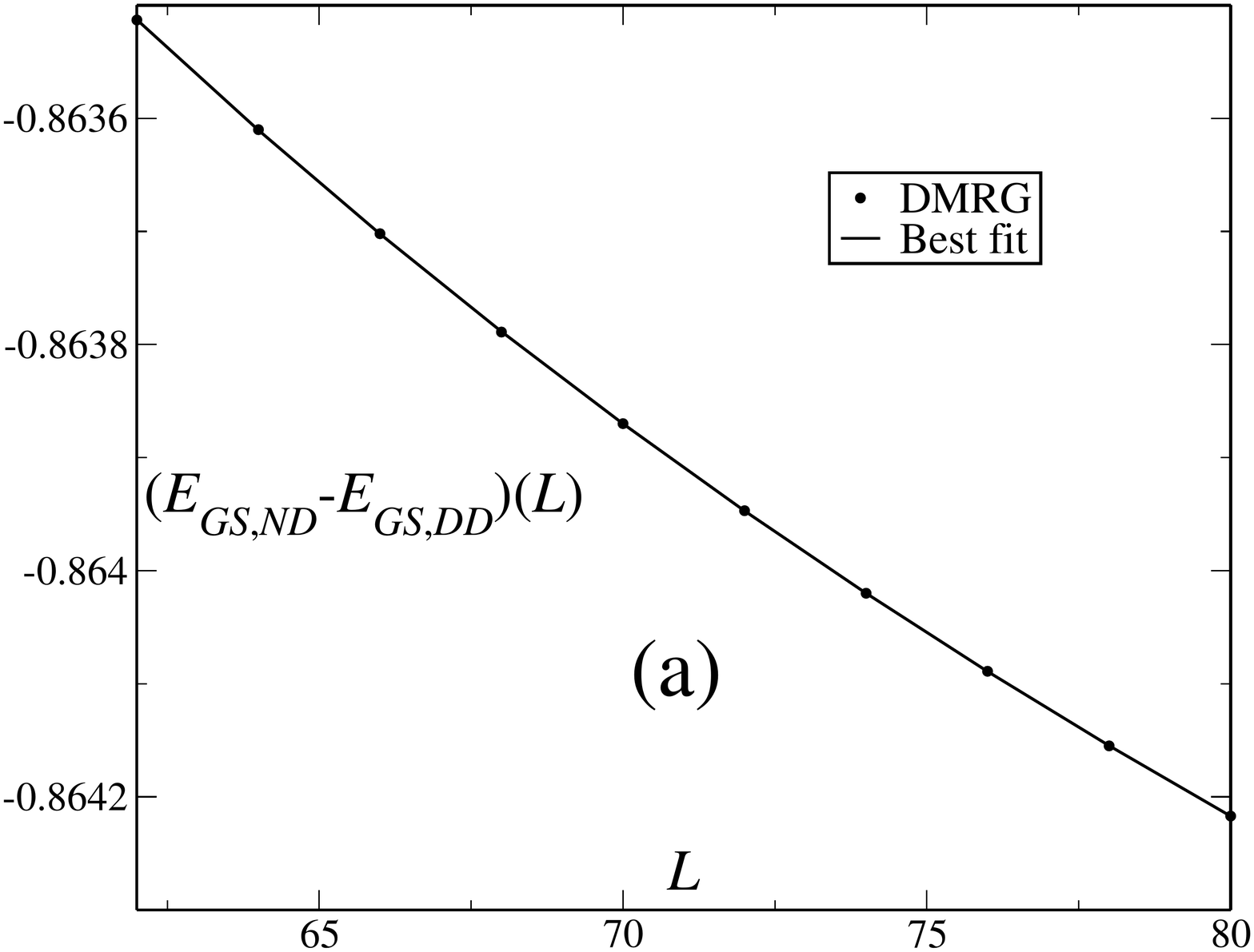}
  \includegraphics[width=0.5\textwidth]{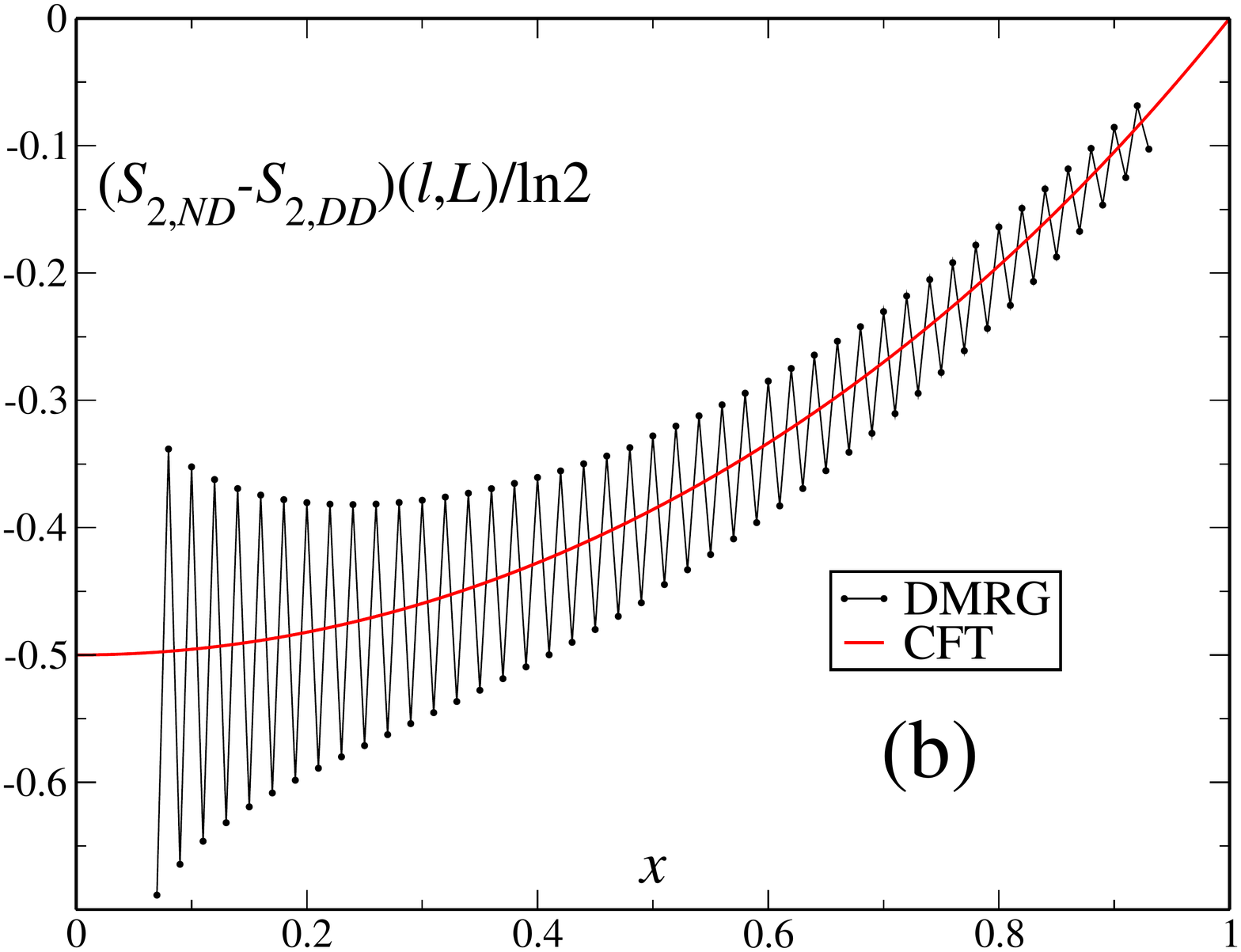}
 \end{minipage}
 \begin{minipage}{\textwidth}
  \centering
  \includegraphics[width=0.5\textwidth]{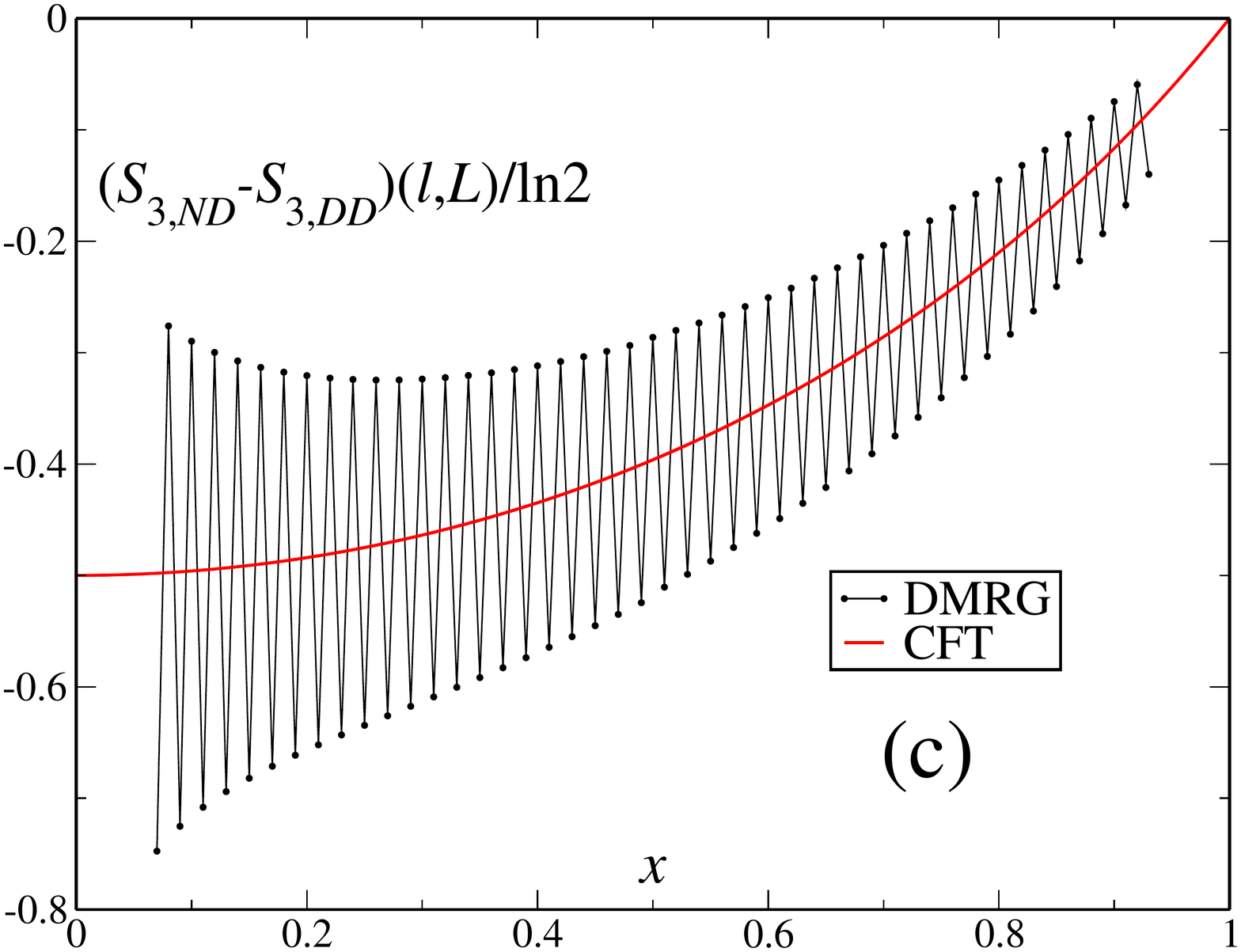}
 \end{minipage}
 \caption{Features of the ground state of Hamiltonian (\ref{bilstein}) in the $ND$ case (\ref{realizationNNND}). (a): scaling of the energy as a function of $L$; (b), (c): the $n=2,\,3$ REE's correction (see text).\label{DN-DD}}
\end{figure}

First of all, we verified with DMRG that $h=1/16$ is the conformal weight of the ground state; our strategy is to use the relations (\ref{E_h}) and (\ref{E_0}), taking as reference the ground state in the $DD$ case. In order to do it, we need to know the sound velocity $u$ of the system. Since it is a bulk property, we can get it from the simple $DD$ case, that we exactly solved in section \ref{FF}. The ground state energy is easily computed to be
\begin{equation}\label{E_GS_DD}
 \begin{split}
  E_0 &=-\sum_{m=1}^{L/2}\cos\frac{\pi m}{L+1}=\frac{1}{2}\left(1-\csc\frac{\pi}{2(L+1)}\right)=\\
  &=-\frac{L}{\pi}+\left(\frac{1}{2}-\frac{1}{\pi}\right)-\frac{\pi}{24L}+O(L^{-2})
 \end{split}
\end{equation}
that, compared to equation (\ref{E_0}), gives $cu=1$. Since we know that $c=1$, we conclude that $u=1$ too. To get the conformal dimension of the $ND$ ground state, we simulated with DMRG the system at different even sizes $L\in[62,80]$, with the usual number of sweeps and states per block, in order to achieve a maximum truncation error of the order of $10^{-10}$. In figure \ref{DN-DD}(a) we show the scaling of such numerical energy differences. By fitting the difference of the ground state energies with $ND$ and $DD$ boundary conditions using formula (\ref{E_h}), we get for $h$ a value of $0.061780$, that is very close to $1/16$, as expected. We then consider the $n=2,\,3$ REE's: in figure \ref{DN-DD}(b), (c) we plot the corrections, obtained via DMRG simulations (the truncation is of the order of $10^{-10}$ or less), and we compare it with the conformal predictions $F_{\sigma,0}^{(2)}$ (equation (\ref{F_sigma^2})) and $F_{\sigma,0}^{(3)}$ (equation (\ref{F_sigma^3})). Even in this case, the agreement between numerics and analytical predictions is excellent, up to the usual boundary-entropy constant $-\frac{1}{2}\ln 2$, that we had to add by hand.

\begin{figure}[t]
 \begin{minipage}{\textwidth}
  \includegraphics[width=0.5\textwidth]{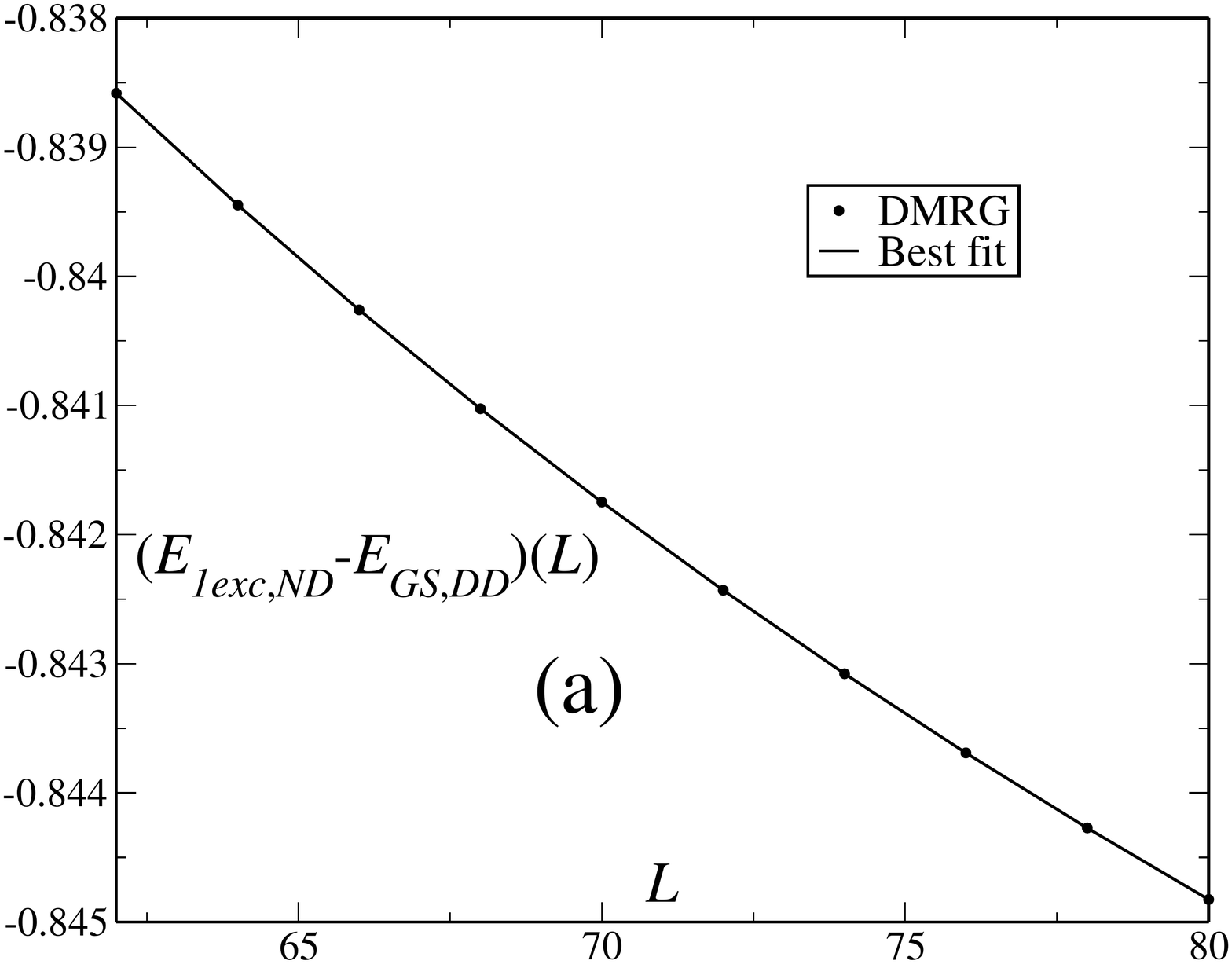}
  \includegraphics[width=0.5\textwidth]{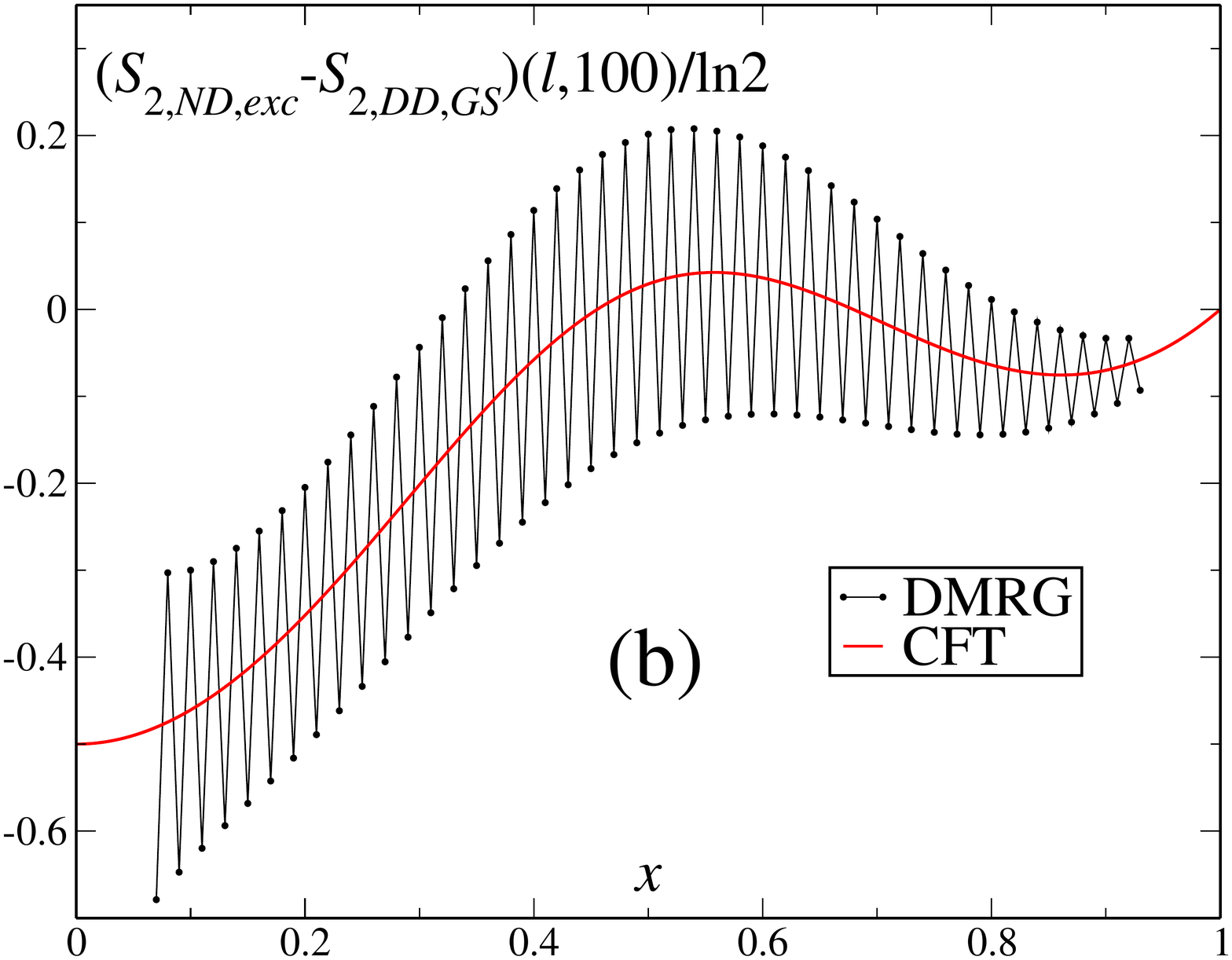}
 \end{minipage}
 \begin{minipage}{\textwidth}
  \centering
  \includegraphics[width=0.5\textwidth]{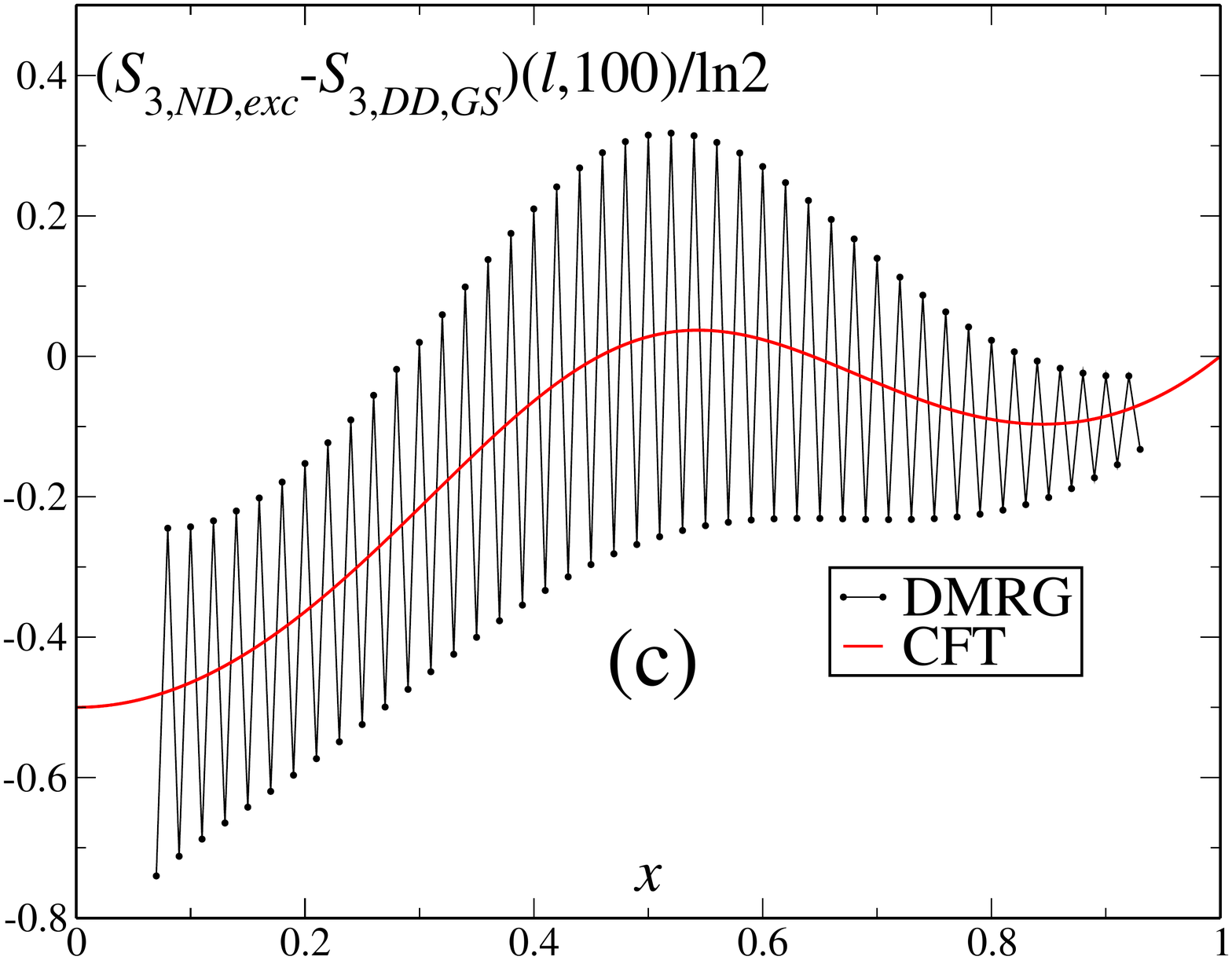}
 \end{minipage}
 \caption{Features of the first excited state of Hamiltonian (\ref{bilstein}) in the $ND$ case (\ref{realizationNNND}). (a): scaling of the energy as a function of $L$; (b), (c): the $n=2,\,3$ REE's correction (see text).\label{DN_exc-DD}}
\end{figure}

In this picture, that is related to the conformal blocks of the $c=1/2$ minimal CFT, one can even consider the first excited state, that should be associated with the conformal weight $h=9/16$, since it is created by the tensor product of primary fields $\chi\otimes\sigma$. First of all, we numerically verified that $h=9/16$ is indeed the conformal weight of the excited state, by means of a multi-target-DMRG simulation \cite{DegliEspostiBoschiOrtolani2004} and of the method used for to get the ground-state conformal weight, based on equations (\ref{E_h}) and (\ref{E_0}): we show the result in figure \ref{DN_exc-DD}(a). From this procedure we get the value $0.547667$, in reasonable agreement with the prediction $9/16$. By means of the same simulations, we computed the $n=2,\,3$ REE's. The correlation functions in equation (\ref{chiralF}), since the exciting operator is a tensor product, is simply the product of correlation functions of $\chi$ and $\sigma$ operators:
\begin{equation}
 F^{(n)}_{\sigma\otimes\chi,p}(x)=F^{(n)}_{\sigma,p}(x)F^{(n)}_{\chi}(x)
\end{equation}
For $n=2$, these quantities have already been computed in the previous section. We have
\begin{equation}
 F^{(2)}_{\sigma\otimes\chi,0}(x)=\cos\frac{\pi x}{4}\cdot\frac{1+\cos(2\pi x)}{8}
\end{equation}
In figure \ref{DN_exc-DD}(b) we show the correction we compute with DMRG, and we see that the CFT prediction fits very well the numerical data, up to the usual bothering oscillations and the constant boundary entropy $-\frac{1}{2}\ln 2$. The same happens for the $n=3$ REE (see figure \ref{DN_exc-DD}(c)), for which we have, from equations (\ref{Fs}) and (\ref{ECL})
\begin{equation}
 F_\chi^{(3)}=\frac{7+2\cos(2\pi x)}{9}
\end{equation}

Therefore, even in the $c=1$, highly non-trivial case, we were able to predict the corrections to the REE's in all the relevant cases our formalism allows to treat, with the exception of the constant boundary-entropies corrections, that, however, are under control.


\chapter{Conclusions and Outlook}\label{conc}

The original work of this thesis mainly regards R\'enyi entanglement entropies, and their relation with conformal field theory. In chapter \ref{Crossovers}, we developed a new method for the detection of the Luttinger parameter of a Luttinger-liquid system from DMRG simulations, while in chapter \ref{CFT+OBC} we analytically computed the corrections to the Calabrese-Cardy conformal behavior of R\'enyi entropies for open systems with general conformal boundary conditions.

The method we developed in chapter \ref{Crossovers} can, in principle, be applied to any lattice system which effective description is known to be a Luttinger liquid. However, as for all numerical methods, its results need comparisons with the ones obtained by other independent methods. Therefore, the research of new methods, possibly computationally convenient, is a necessary practical challenge. In recent times, several methods working on systems of {\it finite size} (i.e., without the need of extrapolating the thermodynamic limit) have been developed: we mention in particular the one introduced by Xavier and Alcaraz in \cite{XavierAlcaraz2011b}, based on entanglement entropies, and the one by You, Li and Gu \cite{YouLiGu2007}, based on the concept of {\it fidelity susceptibility}, for the detection of phase transitions. Moreover, we mentioned in the introduction that when a 1D system is critical but not conformal, something strange happens: in particular, the R\'enyi entanglement entropies seem to display an {\it essential singularity} around the considered critical point, as observed for the spin-1/2 XY \cite{FIJK2007} and XYZ chains \cite{EEFR2011}. Therefore, if one conjectures it to be a general property, can use it to determine the conformal invariance or not of a critical point. This could be the case in some non-trivial model, as the spin-1 $\lambda-D$ chain \cite{BotetJullienKolb1983}, where quantum critical points whose universality class is doubtful are present. This could even be a nice model to test the essential critical behavior of R\'enyi entanglement entropies.

From the analytical point of view, we have already stressed in the introduction the necessity to compute the explicit form of the corrections in the considered model. Regarding this, the most evident deficiency in literature is the absence of a clear analytical derivation of formula (\ref{CCEN}) for Luttinger liquids, even if partial proofs have been given by Cardy and Calabrese \cite{CardyCalabrese2010} and Swingle, McMinis and Tubman \cite{SwingleMcMinisTubman2012}. Another evident lack is a formula for the correction to the Calabrese-Cardy behavior for a {\it generic} excited state, not just for the ones created by primary operators: the main difficulty seems here to get a transformation rule, analogous to (\ref{primary}), for the considered operator. However, an immediate test can be the state associated to the chiral stress-energy operator, whose transformation rule (\ref{T-transformation}) is known.


\appendix

\chapter{Correlation Functions and Reduced Density Matrices}\label{peschel_method}

In this appendix we illustrate in detail a beautiful calculation allowing the computation of entanglement entropies in an exact and numerically efficient way. The method was developed parallely by Peschel \cite{Peschel2003} and Vidal and collaborators \cite{VLRK2003}.


\section{Reduced Density Matrices}\label{RDM}

Let us consider first a quadratic fermionic Hamiltonian of the form
\begin{equation}\label{H}
 H=\sum_{j,k=1}^Lc_j^\dagger H_{jk}c_k
\end{equation}
with $H$ hermitian and $c_j$ a fermionic annihilation operator. As it is well known, for free theories the Wick theorem \cite{DiFrancescoMathieuSenechal1997} holds: every correlation function must be expressible as a linear combination of products of two points correlators, i.e., of
\begin{equation}\label{wick_corr}
 C_{jk}\equiv\langle c_j^\dagger c_k\rangle,\;F_{jk}\equiv\langle c_jc_k\rangle
\end{equation}
If the Hamiltonian looks like (\ref{H}), the anomalous correlation $F$ is of course null, and therefore the physics of the system shall be given just by knowing $C$. In particular, we are interested in connecting to $C$ the reduced density matrix \cite{NielsenChuang2000} of any subsystem $A$ of the whole lattice.

If we restrict to $A$, we can compute its observables by using the density matrix of the system or the reduced density matrix of the subsystem, obtaining the same result \cite{NielsenChuang2000}; in particular, the Wick theorem for correlation functions must hold for both. Therefore, we can assume the reduced density matrix of $A$ to be of the form
\begin{equation}\label{rho_A}
 \rho_A\equiv\frac{1}{Z_A}e^{-h_A}
\end{equation}
being $h_A$ of the same form of $H$, i.e.,
\begin{equation}
 h_A\equiv\sum_{j=1}^lc_j^\dagger h_{jk}c_k
\end{equation}
Here $Z_A$ is the reduced partition function of $A$, $l$ is its size and of course $h_{jk}\neq H_{jk}$ in general.

The diagonalization procedure of $h_A$ is standard. It is done by introducing the fermionic operators $d$ defined as
\begin{equation}
 c_j\equiv\sum_{m=1}^lV_{jm}d_m
\end{equation}
being $V$ a unitary matrix. By means of these new operators, $h_A$ becomes
\begin{equation}
 h_A=\sum_{m,n=1}^ld_m^\dagger\left(\sum_{j,k=1}^lV_{jm}^*h_{jk}V_{kn}\right)d_n
\end{equation}
$V$ can be chosen in order to diagonalize $h$, i.e., the round bracket can be put to $\epsilon_m\delta_{mn}$. If we do it, we end up with
\begin{equation}
 h_A=\sum_{m=1}^l\epsilon_md_m^\dagger d_m
\end{equation}

From this form of $h_A$, it is immediate to see that the reduced density matrix $\rho_A$ becomes
\begin{equation}\label{kronecker}
 \rho_A=\bigotimes_{m=1}^l\frac{e^{-\epsilon_md_m^\dagger d_m}}{Z_m}
\end{equation}
Each single term in the Kronecker product can be interpreted as a single-particle reduced density matrix, and therefore $Z_m$ has to be chosen to normalize it. It is therefore
\begin{equation}\label{reduced partition function}
 Z_m=1+e^{-\epsilon_m}
\end{equation}
Our task is now to link $\epsilon_m$ directly to the correlation matrix $C$. We compute it according to its definition: if $j,k\in A$,
\begin{equation}
 C_{jk}=\mbox{Tr}_A\left[\rho_Ac_j^\dagger c_k\right]=\frac{1}{Z_A}\sum_{n,p=1}^lV_{jn}^*V_{kp}\mbox{Tr}_A\left[e^{-\sum_{m=1}^l\epsilon_md_m^\dagger d_m}d_n^\dagger d_p\right]
\end{equation}
In the second trace, all the terms different from the $n$-th and the $p$-th just cancel the respective factors in $Z_A$; moreover, the trace itself is zero if $n\neq p$. We are therefore left with the result
\begin{equation}
 \frac{1}{Z_A}\mbox{Tr}_A\left[e^{-\sum_{m=1}^l\epsilon_md_m^\dagger d_m}d_n^\dagger d_p\right]=\frac{\delta_{np}}{1+e^{\epsilon_n}}
\end{equation}
so that
\begin{equation}
 C_{jk}=\sum_{n=1}^lV_{jn}^*\frac{1}{1+e^{\epsilon_n}}V_{kn}
\end{equation}
Being $V$ unitary, the eigenvalues of $C$ are $\zeta_m=\left(1+e^{\epsilon_m}\right)^{-1}$, and inverting the relation we have
\begin{equation}\label{epsilon-zeta}
 \epsilon_m=\ln\frac{1-\zeta_m}{\zeta_m}
\end{equation}
Therefore, diagonalizing $C$, one can immediately get $\epsilon_m$, and so the reduced density matrix $\rho_A$.

Holding the Wick theorem for a generic free theory, one can consider the more general situation
\begin{equation}\label{HA}
 H=\sum_{j,k=1}^L\left\{c_j^\dagger H^H_{jk}c_k+\frac{1}{2}\left(c_j^\dagger H^A_{jk}c_h^\dagger+\mbox{h.c.}\right)\right\}
\end{equation}
being $H^H$ a hermitian and $H^A$ an anti-hermitian matrix. Now, the correlator $F$ is in general non-zero, and therefore we should be able to extract the reduced density matrix of any subsystem from the combined knowledge of $C$ and $F$. With exactly the same arguments as before, one can say that the reduced density matrix of the subsystem $A$ takes the form (\ref{rho_A}), this time with
\begin{equation}
 h_A=\sum_{j,k=1}^l\left\{c_j^\dagger h^H_{jk}c_k+\frac{1}{2}\left(c_j^\dagger h^A_{jk}c_h^\dagger+\mbox{h.c.}\right)\right\}
\end{equation}
with, again, $h^H$ hermitian and $h^A$ anti-hermitian. If one finds some operator $d_m$ diagonalizing $h_A$, of course the factorization (\ref{kronecker}) holds even this time, with the same normalization (\ref{reduced partition function}).

The solution to this problem is known. It has been seen by Lieb, Schultz and Mattis \cite{LiebSchultzMattis1961} that the right fermionic operators are now of the more complicated form, known as \emph{Bogolyubov transformation},
\begin{equation}\label{bogolyubov}
 d_m\equiv\sum_{j=1}^l\left(\frac{\phi_{mj}+\psi_{mj}}{2}c_j+\frac{\phi_{mj}-\psi_{mj}}{2}c_j^\dagger\right)
\end{equation}
where $\phi_m$ and $\psi_m$ satisfy the left eigenvalues equation
\begin{equation}
 \begin{cases}
  \phi_m(A-B)(A+B)=\epsilon_m^2\phi_m\\
  \psi_m(A+B)(A-B)=\epsilon_m^2\psi_m
 \end{cases}
\end{equation}
and can be chosen to be real and orthonormal, i.e.,
\begin{equation}\label{orthonormality}
 \sum_{j=1}^l\phi_{mj}\phi_{nj}=\sum_{j=1}^l\psi_{mj}\psi_{nj}=\delta_{mn}
\end{equation}

To compute now $C$ and $F$, we have to invert equation (\ref{bogolyubov}). It can be easily seen that the procedure gives
\begin{equation}
 c_j=\sum_{m=1}^l\left(\frac{\phi_{mj}+\psi_{mj}}{2}d_m+\frac{\phi_{mj}-\psi_{mj}}{2}d_m^\dagger\right)
\end{equation}
Putting the operators written this way into the first of the equations (\ref{wick_corr}), one has to compute four traces. The two anomalous traces, i.e., $\mbox{Tr}_A\left[\rho_Ad_m^\dagger d_n^\dagger\right]$ and $\mbox{Tr}_A\left[\rho_Ad_md_n\right]$ are of course zero, as it is immediately seen using the basis of the Hilbert space that diagonalizes the operators $d_m^\dagger d_m$. One is therefore left with the two ordinary traces, that give
\begin{equation}
 \mbox{Tr}_A\left[\rho_Ad_m^\dagger d_n\right]=\frac{\delta_{mn}}{1+e^{\epsilon_m}},\;\;\;\mbox{Tr}_A\left[\rho_Ad_md_n^\dagger\right]=\frac{\delta_{mn}}{1+e^{-\epsilon_m}}
\end{equation}
and therefore $C$ is seen to be, using relations (\ref{orthonormality}),
\begin{equation}
 C_{jk}=\frac{\delta_{jk}}{2}+\frac{1}{4}\sum_{m=1}^l\left(\phi_{mj}\psi_{mk}+\psi_{mj}\phi_{mk}\right)\tanh\frac{\epsilon_m}{2}
\end{equation}
In exactly the same way, $F$ is seen to be
\begin{equation}
 F_{jk}=\frac{1}{4}\sum_{m=1}^l\left(\phi_{mj}\psi_{mk}-\psi_{mj}\phi_{mk}\right)\tanh\frac{\epsilon_m}{2}
\end{equation}
Therefore, netiher $C$ nor $F$ is directly diagonalized. What is diagonalized is the combination
\begin{equation}
 \left[\left(C-\mathbb{I}/2-F\right)\left(C-\mathbb{I}/2+F\right)\right]_{jk}=\frac{1}{4}\sum_{m=1}^l\psi_{mj}\tanh^2\frac{\epsilon_m}{2}\psi_{mk}
\end{equation}
and therefore the eigenvalues of $\left(C-\mathbb{I}/2-F\right)\left(C-\mathbb{I}/2+F\right)$ are $\zeta_m\equiv\frac{1}{4}\tanh^2\frac{\epsilon_m}{2}$. The inversion of this relation is trivial, and gives
\begin{equation}
 \epsilon_m=2\tanh^{-1}\left(2\sqrt{\zeta_m}\right)
\end{equation}


\section{Entanglement Entropies}

The computation of entanglement entropies from $\epsilon_m$ is quite simple. To begin, let us consider the VNEE of a subsystem $A$. It is clear from its definition (\ref{VNE}) that it can be written in the form
\begin{equation}
 S(A)=-\sum_\lambda\lambda\ln\lambda
\end{equation}
where the sum is over all the eigenvalues of $\rho_A$. Since in (\ref{kronecker}) each particle is not correlated to the others, the VNEE can be seen as the sum of the VNEE's of the single particles. Therefore, it looks
\begin{equation}
 S(A)=\sum_{m=1}^lH_2\left(\frac{1}{1+e^{\epsilon_m}}\right)
\end{equation}
being $H_2(x)\equiv-x\ln x-(1-x)\ln(1-x)$ and since $(1+e^{\epsilon_m})^{-1}$ and $1-(1+e^{\epsilon_m})^{-1}$ are the two eigenvalues of $\rho_m$.

For the REE's the situation is similar. In terms of the eigenvalues of $\rho_A$, they look
\begin{equation}
 S_n(A)=\frac{1}{1-n}\ln\sum_\lambda\lambda^n
\end{equation}
and therefore they are expressible as
\begin{equation}
 S_n(A)=\frac{1}{1-n}\sum_{m=1}^l\ln\left[\left(\frac{1}{1+e^{\epsilon_m}}\right)^n+\left(1-\frac{1}{1+e^{\epsilon_m}}\right)^n\right]
\end{equation}

In the first case we have considered, i.e., when the Hamiltonian takes the form (\ref{H}), the situation is particularly simple. In fact, in this case we have, because of equation (\ref{epsilon-zeta}),
\begin{equation}
 \begin{cases}
  \frac{1}{1+e^{\epsilon_m}}=\zeta_m\\
  1-\frac{1}{1+e^{\epsilon_m}}=1-\zeta_m
 \end{cases}
\end{equation}
so that the VNEE looks
\begin{equation}
 S(A)=\sum_{m=1}^lH_2\left(\zeta_m\right)
\end{equation}
while the REE's look
\begin{equation}
 S_n(A)=\frac{1}{1-n}\sum_{m=1}^l\ln\left[\zeta_m^n+\left(1-\zeta_m\right)^n\right]
\end{equation}


\section{Two MATLAB Codes for the Computation of Entanglement Entropies}

In this section we provide the two numerical codes we wrote to compute entanglement entropies for free fermionic systems, with the method described in this appendix, using the software MATLAB\footnote{http://www.mathworks.it/products/matlab/.}.

The first code we report is suited to the spin-1/2 XX chain with OBC, but it is easily generalizable to any quadratic Hamiltonian of the form (\ref{H}). This chain can be seen (see section \ref{XY}) to be equivalent to the spinless fermionic Hamiltonian
\begin{equation}
 H=-\frac{1}{2}\sum_{j=1}^L\left(c_{j+1}^\dagger c_j+c_j^\dagger c_{j+1}\right)
\end{equation}
with OBC. Therefore the matrix $H_{jk}$ is of the form $H_{jk}=-\frac{1}{2}\left(\delta_{j,k+1}+\delta_{j+1,k}\right)$. We just remark that the correlation function is computed using formula (\ref{correlation function}). The rest of the code is just an implementation of what explained in the first part of section \ref{RDM}.

\begin{lstlisting}
% Constants

L=500;

% Hamiltonian Matrix

Ham=zeros(L);  % Inizialization
for i=1:L
    for j=1:L
        if ((i==j-1)||(i==j+1))
            Ham(i,j)=-1/2;
        end
    end
end

% Diagonalization of the Hamiltonian

[V,D]=eig(Ham); % V: eigenvectors (stored as columns); D: eigenvalues (D is a diagonal square matrix having the eigenvalues as diagonal entries)

% Correlation functions

C=zeros(L);
for j=1:L
    for k=1:L
        for m=1:L/2 % L/2 is there because of the half-filling
            C(j,k)=C(j,k)+V(j,m)*V(k,m);
        end
    end
end

% Entanglement entropies

VNE=zeros(L-1,1);
RE2=zeros(L-1,1);
RE3=zeros(L-1,1);
RE4=zeros(L-1,1);
RE5=zeros(L-1,1);
RE6=zeros(L-1,1);
RE7=zeros(L-1,1);
RE8=zeros(L-1,1);
RE9=zeros(L-1,1);
RE10=zeros(L-1,1);
for l=1:L-1
    TC=zeros(l);
    for j=1:l
        for k=1:l
            TC(j,k)=C(j,k);
        end
    end
    zeta=eig(TC);
    for m=1:l
        VNE(l,1)=VNE(l,1)+real(1/(1-1.0000001)*log(zeta(m,1)^1.0000001+(1-zeta(m,1))^1.0000001)); % VNE as RE limit
        RE2(l,1)=RE2(l,1)+real(1/(1-2)*log(zeta(m,1)^2+(1-zeta(m,1))^2));
        RE3(l,1)=RE3(l,1)+real(1/(1-3)*log(zeta(m,1)^3+(1-zeta(m,1))^3));
        RE4(l,1)=RE4(l,1)+real(1/(1-4)*log(zeta(m,1)^4+(1-zeta(m,1))^4));
        RE5(l,1)=RE5(l,1)+real(1/(1-5)*log(zeta(m,1)^5+(1-zeta(m,1))^5));
        RE6(l,1)=RE6(l,1)+real(1/(1-6)*log(zeta(m,1)^6+(1-zeta(m,1))^6));
        RE7(l,1)=RE7(l,1)+real(1/(1-7)*log(zeta(m,1)^7+(1-zeta(m,1))^7));
        RE8(l,1)=RE8(l,1)+real(1/(1-8)*log(zeta(m,1)^8+(1-zeta(m,1))^8));
        RE9(l,1)=RE9(l,1)+real(1/(1-9)*log(zeta(m,1)^9+(1-zeta(m,1))^9));
        RE10(l,1)=RE10(l,1)+real(1/(1-10)*log(zeta(m,1)^10+(1-zeta(m,1))^10));
    end
end
\end{lstlisting}

The second code we report is the one we used to compute the entanglement entropies for the Hamiltonian (\ref{HA}), in the case of the spin-1/2 XY chain (\ref{XY_ham}) with OBC (see section \ref{XY}). For the computation of correlation functions, we follow the recipe of Nielsen \cite{Nielsen2005}. Let us consider the Hamiltonian
\begin{equation}
 H=\sum_{j,k=1}^L\left(c_j^\dagger\alpha_{jk}c_k-c_j\alpha^*_{jk}c_k^\dagger+c_j\beta_{jk}c_k-c_j^\dagger\beta_{jk}c_k^\dagger\right)
\end{equation}
with $\alpha$ hermitian and $\beta$ anti-hermitian. In matrix terms, it is re-written as
\begin{equation}
 H=\left(\begin{array}{cc}
  c^\dagger, & c
 \end{array}\right)M\left(\begin{array}{c}
  c \\
  c^\dagger
 \end{array}\right)
\end{equation}
where
\begin{equation}
 M=\left(\begin{array}{cc}
  \alpha & -\beta \\
  \beta & -\alpha^*
 \end{array}\right)
\end{equation}
We write now the Bogolyubov transformation (\ref{bogolyubov}) in the form
\begin{equation}
 \left(\begin{array}{c}
  d \\
  d^\dagger
 \end{array}\right)=T\left(\begin{array}{c}
  c \\
  c^\dagger
 \end{array}\right)
\end{equation}
where the matrix
\begin{equation}
 T=\left(\begin{array}{cc}
  \gamma & \mu \\
  \mu^* & \gamma^*
 \end{array}\right)
\end{equation}
has to be unitary in order to make the $d$'s fermionic. In terms of the new operators, the Hamiltonian becomes
\begin{equation}
 H=\left(\begin{array}{cc}
  d^\dagger, & d
 \end{array}\right)TMT^\dagger\left(\begin{array}{c}
  d \\
  d^\dagger
 \end{array}\right)
\end{equation}
and $T$ is chosen in order to diagonalize $M$: it contains as columns its eigenvectors. Being the ground state the state $\left|0\right>$ that is annihilated by all the Bogolyubov annihilators, the correlation functions $C$ and $F$ are easily seen to be, for the ground state,
\begin{equation}
 C_{jk}^{\left|0\right>}=\sum_{m=1}^L\left(\mu^\dagger\right)_{jm}\left(\mu^T\right)_{km},\;F_{jk}^{\left|0\right>}=\sum_{m=1}^L\left(\gamma^\dagger\right)_{jm}\left(\mu^\dagger\right)_{km}
\end{equation}
Moreover, in this thesis we performed computations even for the first excited state $\left|1\right>$, i.e., for the state with one Bogolyubov fermion in the lowest energy level:
\begin{equation}
 d_m\left|1\right>=0,\ m=2,\cdots,L,\;\;\;d_1\left|1\right>=\left|0\right>
\end{equation}
whose two-points correlation functions are given by
\begin{equation}
 \begin{split}
  C_{jk}^{\left|1\right>} &=\left(\gamma^T\right)_{j1}\left(\gamma^\dagger\right)_{k1}+\sum_{m=2}^L\left(\mu^\dagger\right)_{jm}\left(\mu^T\right)_{km}\\
  F_{jk}^{\left|1\right>} &=\left(\mu^T\right)_{j1}\left(\gamma^\dagger\right)_{k1}+\sum_{m=2}^L\left(\gamma^\dagger\right)_{jm}\left(\mu^T\right)_{km}
 \end{split}
\end{equation}
This is the way the following code computes the correlation functions. The code is tailored on the spin-1/2 critical Ising chain, but is easily extendable to any quadratic Hamiltonian.
\begin{lstlisting}
%Constants

L=500;
J=-1/2;
h=1;
gamma=1;
h1=0;
hL=0;

%Hamiltonian Matrices

alpha=zeros(L);
for j=1:L
    for k=1:L
        if ((j==k-1)||(j==k+1))
            alpha(j,k)=J/2;
        end
        if (j==k)
            alpha(j,k)=-J*h;
        end
    end
end

beta=zeros(L);
for j=1:L
    for k=1:L
        if (j==k-1)
            beta(j,k)=-J*gamma/2;
        end
        if (j==k+1)
            beta(j,k)=J*gamma/2;
        end
    end
end

M=cat(1,cat(2,alpha,-beta),cat(2,beta,-conj(alpha)));

% Diagonalization of M

[V,D]=eig(M); % V=T^dagger, in Nielsen's notation
gammaT=zeros(L);
muT=zeros(L);
for j=1:L
    for k=1:L
        gammaT(j,k)=V(L+j,L+k);
        muT(j,k)=V(j,L+k);
    end
end

%Correlation functions: ground state

C=zeros(L);
for j=1:L
    for k=1:L
        for l=1:L
            C(j,k)=C(j,k)+conj(muT(j,l))*muT(k,l);
        end
    end
end

F=zeros(L);
for j=1:L
    for k=1:L
        for l=1:L
            F(j,k)=F(j,k)+conj(gammaT(j,l))*muT(k,l);
        end
    end
end

%Correlation functions: first excited state: uncomment to use

%C=zeros(L);
%for j=1:L
%    for k=1:L
%        for l=2:L
%            C(j,k)=C(j,k)+conj(muT(j,l))*muT(k,l);
%        end
%        C(j,k)=C(j,k)+gammaT(j,1)*conj(gammaT(k,1));
%    end
%end

%F=zeros(L);
%for j=1:L
%    for k=1:L
%        for l=2:L
%            F(j,k)=F(j,k)+conj(gammaT(j,l))*muT(k,l);
%        end
%        F(j,k)=F(j,k)+muT(j,1)*conj(gammaT(k,1));
%    end
%end

%Entanglement entropies

VNE=zeros(L-1,1);
RE2=zeros(L-1,1);
RE3=zeros(L-1,1);
RE4=zeros(L-1,1);
RE5=zeros(L-1,1);
RE6=zeros(L-1,1);
RE7=zeros(L-1,1);
RE8=zeros(L-1,1);
RE9=zeros(L-1,1);
RE10=zeros(L-1,1);
for l=1:L-1
    K1=-eye(l);
    K2=-eye(l);
    for j=1:l
        for k=1:l
            K1(j,k)=K1(j,k)+2*C(j,k)-2*F(j,k);
            K2(j,k)=K2(j,k)+2*C(j,k)+2*F(j,k);
        end
    end
    zeta=eig(K1*K2);
    epsilon=zeros(l,1);
    for j=1:l
        epsilon(j,1)=2*atanh(sqrt(zeta(j)));
    end
    rhoeig=zeros(l,1);
    for j=1:l
        rhoeig(j,1)=1/(1+exp(epsilon(j,1)));
    end
    for j=1:l
        VNE(l,1)=VLE(l,1)+real(1/(1-1.0000001)*log(rhoeig(j,1)^1.0000001+(1-rhoeig(j,1))^1.0000001));
        RE2(l,1)=RE2(l,1)+real(1/(1-2)*log(rhoeig(j,1)^2+(1-rhoeig(j,1))^2));
        RE3(l,1)=RE3(l,1)+real(1/(1-3)*log(rhoeig(j,1)^3+(1-rhoeig(j,1))^3));
        RE4(l,1)=RE4(l,1)+real(1/(1-4)*log(rhoeig(j,1)^4+(1-rhoeig(j,1))^4));
        RE5(l,1)=RE5(l,1)+real(1/(1-5)*log(rhoeig(j,1)^5+(1-rhoeig(j,1))^5));
        RE6(l,1)=RE6(l,1)+real(1/(1-6)*log(rhoeig(j,1)^6+(1-rhoeig(j,1))^6));
        RE7(l,1)=RE7(l,1)+real(1/(1-7)*log(rhoeig(j,1)^7+(1-rhoeig(j,1))^7));
        RE8(l,1)=RE8(l,1)+real(1/(1-8)*log(rhoeig(j,1)^8+(1-rhoeig(j,1))^8));
        RE9(l,1)=RE9(l,1)+real(1/(1-9)*log(rhoeig(j,1)^9+(1-rhoeig(j,1))^9));
        RE10(l,1)=RE10(l,1)+real(1/(1-10)*log(rhoeig(j,1)^10+(1-rhoeig(j,1))^10));
    end
end
\end{lstlisting}


\chapter{Numerical Methods for Luttinger-Parameter Estimations}\label{num_methods}

In this appendix, we describe the two numerical independent methods that allowed us to make comparisons to the predictions of the method we developed. The first, called {\it level spectroscopy} \cite{Giamarchi2003}, is very popular and reliable, while the second, the {\it bipartite-fluctuations} method, has been proposed very recently and a check of its working is by itself interesting.


\section{Level Spectroscopy}\label{level spectroscopy}

An important result of Luttinger liquids physics links the system compressibility, defined by \cite{Giamarchi2003}
\begin{equation}
 \kappa\equiv\frac{\partial n}{\partial\mu}
\end{equation}
i.e., the density change with respect to a chemical potential change, and the bosonization parameters. Explicitly, one has \cite{Giamarchi2003}:
\begin{equation}\label{K=upk}
 K=u\pi\kappa
\end{equation}
and therefore, knowing $u$ and $\kappa$, being $u$ the sound velocity of the system, one can get $K$. Luckily, one can relate $\kappa$ (at size $L$) to the energy spectrum of the system by the so called {\it level spectroscopy} formula \cite{Giamarchi2003}:
\begin{equation}\label{kappa}
 \kappa_L=L\left[E_L(N-1)+E_L(N+1)-2E_L(N)\right]
\end{equation}
being $E_L(N)$ the energy of the ground state of a system of size $L$, at filling $N/L$. Of course one wants to estimate $\kappa$ in the $L\rightarrow\infty$ limit, and has therefore to take a finite-size scaling of his data to extrapolate the true $\kappa$. A step is still missing, i.e., a way to get $u$. The question is answered by the well-known CFT relation \cite{BloteCardyNightingale1986, Affleck1986}
\begin{equation}\label{epsilon_CFT}
 e_{GS}(L)=e_0+\frac{f_\infty}{L}+\frac{uc\pi}{6\delta L^2}+o\left(L^{-2}\right)
\end{equation}
where $\delta=1/4$ for PBC/OBC, being $e_{GS}$ the energy density for the ground state, $e_0$ its thermodynamic-limit value, $f_\infty$ the surface free energy, vanishing for PBC, and $c$ the central charge of the CFT. Therefore, by this simple procedure, an estimate of $K$ can be given.


\section{The Bipartite-Fluctuations Method}\label{bipartite fluctuations}

The method we are going to describe was first theoretically founded by Song and collaborators in \cite{SongRachelLeHur2010,SRFKLLH2012}, and then applied by Dalmonte, Ercolessi and myself in \cite{DalmonteErcolessiTaddia2011,DalmonteErcolessiTaddia2012}, and, for different purposes, by Rachel and collaborators in \cite{RLSLH2012}.

Let us consider a 1D system, described by some quantum Hamiltonian having a U(1) symmetry, implying, e.g., particle-number conservation for fermionic or bosonic system, or total-magnetization for spin systems. These quantities are constants of motion, but their fluctuations in a subsystem $A$ are not. We introduce therefore the quantity
\begin{equation}\label{bipfluc}
 F_A\equiv\left<\left(\hat{N}_A-\left<\hat{N}_A\right>\right)^2\right>
\end{equation}
where $\hat{N}_A$ is the total-particle-number operator relative to subsystem $A$; for spin systems, it is replaced by the total magnetization relative to $A$. This quantity shares some interesting feature with the REE's. In particular, it is easily shown that $F_A=F_B$, where $B$ is the complementary of $A$: therefore, some kind of area law should hold in gapped systems, possibly logarithmically violated in critical situations.

Such logarithmical violations actually arise, and the main difference with VNEE resides in its prefactor. It can be shown that, for a generic CFT and a finite system, they take the form
\begin{equation}\label{LeHur formula}
 \pi^2F(l,L)\equiv\pi^2F_A=g\ln\left(\frac{L}{\pi}\sin\frac{\pi l}{L}\right)+f
\end{equation}
being $l$ and $L$ the sizes of $A$ and of the total system respectively, $f$ a constant non-universal contribution and $g$ given by
\begin{equation}
 g=\pi u\kappa
\end{equation}
being $u$ the sound velocity and $\kappa$ the compressibility (see section \ref{Bosonization}). It is clear from relation (\ref{K=upk}) that, for Luttinger liquids, one has $g=K$, and therefore formula (\ref{LeHur formula}) provides a practical tool to detect $K$. This is what we used.

A comment is in order. Our technique for the detection of the Luttinger parameter, basing itself on REE's, is very powerful if the data is taken with DMRG, i.e., mostly in 1D. On the other hand, the bipartite-fluctuations method, taking into account fluctuations of observables, is in principle less accurate than ours when the data are obtained by DMRG, but more powerful in higher dimension, where computations are usually performed with quantum Monte Carlo algorithms (see, e.g., \cite{RoscildeDegliEspostiBoschiDalmonte2012}), by which REE's are very difficult to get. Moreover, the present quantity has been conjectured to be easily experimentally accessible, differently from REE's \cite{RLSLH2012}.


\chapter{Chiral Conformal Blocks of the $c=1/2$ Minimal CFT}\label{CB}

In this appendix, we give the formulas, first derived by Ardonne and Sierra \cite{ArdonneSierra2010}, that we used in the $c=1/2$-minimal-CFT computations of chapter \ref{CFT+OBC}.

Let us start by the chiral correlator of $2n$ $\sigma$ fields, of conformal dimension 1/16, on the complex plane: in general, because of the $\sigma\times\sigma=1+\chi$ fusion rule, there will be $2^{n-1}$ different conformal blocks, indexed by the non-negative integer $p\in\{0,\cdots,2^{n-1}-1\}$. We write them as
\begin{equation}\label{S}
 S_{p}^{2n}(v_1,\cdots,v_{2n})\equiv\left<\sigma(v_1)\cdots\sigma(v_{2n})\right>_{\mathbb{C},p}
\end{equation}
We illustrate now the recipe for their computation. We divide the labels of the $2n$ coordinates $v_j$ into the $n$ {\it reference pairs} $(1,2)$, $(3,4)$, ..., $(2n-1,2n)$; we then define the $2^{n-1}$ couples of {\it macrogroups} of integers $\vec{l}\equiv(l_1,\cdots,l_n)$ and $\vec{l}'\equiv(l_1',\cdots,l_n')$ such that each of them does not contain coordinates in the same reference pair. E.g., for $n=3$, we will have the 4 couples of macrogroups (135)(246), (145)(236), (136)(245) and (146)(235). We label such macrogroup $\vec{l}$ by the integer $q\in\{0,\cdots,2^{n-1}-1\}$, with the exact recursive recipe
\begin{equation}
 \begin{split}
  l_1 &=1\\
  l_{k+1} &=l_k+2, q_k=0\\
 l_{k+1} &=l_k+1,\ q_k=1,\ l_k\ \mbox{even}\\
 l_{k+1} &=l_k+3,\ q_k=1,\ l_k\ \mbox{odd}
 \end{split}
\end{equation}
where $q_k,\ k\in\{1,\cdots,n-1\}$, is the $k$-th {\it binary digit} of $q=(q_1\cdots q_{n-1})_2$; $\vec{l}'$ is of course the complementary of $\vec{l}$. We then associate to each $\vec{l}$ the quantity
\begin{equation}
 v_{\vec{l}}\equiv\prod_{j<k}v_{l_jl_k},\ v_{l_jl_k}\equiv v_{l_j}-v_{l_k}
\end{equation}
and the same for $\vec{l}'$ (by convention, $v_{\vec{l}}=v_{\vec{l}'}=1$ for $n=1$). We are now able to give a general explicit form for (\ref{S}):
\begin{equation}\label{S_p^2n}
 S_{p}^{2n}(v_1,\cdots,v_{2n})=2^{-(n-1)/2}\prod_{a<b}v_{ab}^{-1/8}\sqrt{\sum_{q=0}^{2^{n-1}-1}\epsilon_{pq}\sqrt{v_{\vec{l}_q}v_{\vec{l}'_q}}}
\end{equation}
where
\begin{equation}
 \epsilon_{pq}\equiv(-1)^{\sum_{k=1}^{n-1}p_kq_k}
\end{equation}
being $p=(p_1,\cdots,p_{n-1})_2$ the binary representation of $p$. For $n=1$, there is just one conformal block, at $p=0$, and we recover, as it must be, the standard two-point correlator of primary operators \cite{DiFrancescoMathieuSenechal1997}:
\begin{equation}\label{2p}
 S^2_0(v_1,v_2)=v_{12}^{-1/8}
\end{equation}

We consider then the conformal block of $2n$ chiral $\chi$ operators, whose conformal dimension is 1/2, on the complex plane. We write it as
\begin{equation}
 C^{2n}(v_1,\cdots,v_{2n})\equiv\left<\chi(v_1)\cdots\chi(v_{2n})\right>_{\mathbb{C}}
\end{equation}
The situation is here much simpler. It is known \cite{DiFrancescoMathieuSenechal1997} that this conformal block takes, in general, the form
\begin{equation}
 C^{2n}(v_1,\cdots,v_{2n})=\mbox{Pf}\left[\frac{1}{v_i-v_j}\right]
\end{equation}
where the symbol $\mbox{Pf}$ denotes the {\it Pfaffian} of the $2n\times 2n$ antisymmetric matrix $\frac{1}{z_i-z_j}$ (its diagonal elements are taken to be zero). The Pfaffian can be easily computed by the relation
\begin{equation}
 \mbox{Pf}[A]=\sqrt{\det A}
\end{equation}
where $A$ is a general antisymmetric matrix.


\chapter{Spin Chains and Fixed Boundary Conditions}\label{Fixed_DMRG}

In this appendix, we explain how to write a lattice spin-1/2 Hamiltonian in order to fix automatically the boundary magnetization to $\pm 1/2$. The philosophy is very simple: if the magnetization on a site of a lattice is fixed to be, e.g., $+1/2$ in the $z$ direction, the Hilbert space of that site is reduced, and by means of the projector on the relevant part of the Hilbert space we can simplify the Hamiltonian of the full system.

Let us consider the $z$ component of a spin-1/2 $\vec{\sigma}$: its eigenstates are of course given by $(1,0)^T$, associated to the eigenvalue $+1$, and $(0,1)^T$, associated to the eigenvalue $-1$. The projectors on the eigenspaces relative, respectively, to $+1$ and $-1$, are
\begin{equation}
 \mathbb{P}^z_1=\left(\begin{array}{cc}
  1 & 0 \\
  0 & 0
 \end{array}\right),\ \mathbb{P}^z_{-1}=\left(\begin{array}{cc}
  0 & 0 \\
  0 & 1
 \end{array}\right)
\end{equation}
Let us consider now the exchange operator between sites $j$ and $j+1$, i.e., $\sigma_j^\alpha\sigma_{j+1}^\alpha$, with $\alpha=x,y,z$. According to the general rules of quantum mechanics \cite{CohenTannoudjiDiuLaloe1977}, the projector on the $\pm 1$ eigenspace of site $j$ can act, instead than on states, on the operators directly, modifying them to
\begin{equation}
 (\mathbb{P}^z_{\pm 1,j}\otimes\mathbb{I}_{j+1})^{-1}(\sigma^\alpha_j\otimes\sigma^\alpha_{j+1})(\mathbb{P}^z_{\pm 1,j}\otimes\mathbb{I}_{j+1})
\end{equation}
(we neglect all the trivial identity factors acting on the Hilbert spaces of the remaining sites). Expliciting the form of the projectors, we have
\begin{equation}
 \begin{split}
  (\mathbb{P}^z_{\pm 1,j}\otimes\mathbb{I}_{j+1})^{-1}(\sigma^z_j\otimes\sigma^z_{j+1})(\mathbb{P}^z_{\pm 1,j}\otimes\mathbb{I}_{j+1}) &=\pm\mathbb{P}^z_{\pm 1,j}\otimes\sigma_{j+1}^z\\
  (\mathbb{P}^z_{\pm 1,j}\otimes\mathbb{I}_{j+1})^{-1}(\sigma^x_j\otimes\sigma^x_{j+1})(\mathbb{P}^z_{\pm 1,j}\otimes\mathbb{I}_{j+1}) &=0\\
  (\mathbb{P}^z_{\pm 1,j}\otimes\mathbb{I}_{j+1})^{-1}(\sigma^y_j\otimes\sigma^y_{j+1})(\mathbb{P}^z_{\pm 1,j}\otimes\mathbb{I}_{j+1}) &=0  
 \end{split}
\end{equation}
with a similar expression when the projection is done on site $j+1$. Therefore, e.g., Hamiltonian (\ref{XY_ham}) with OBC becomes, fixing the first and the last spin-$x$ eigenvalue to $+1/2$ (in this case, the first and the last $z$ and $y$ exchange terms will be null, together with the first and last magnetic terms),
\begin{equation}
 \begin{split}
  H^{++}_{\gamma h} &=-\frac{1}{2}\frac{1+\gamma}{2}\mathbb{P}^x_{1,1}\sigma_2^x-\frac{1}{2}\sum_{j=2}^{L-2}\left[\frac{1+\gamma}{2}\sigma_j^x\sigma_{j+1}^x+\frac{1-\gamma}{2}\sigma_j^y\sigma_{j+1}^y+h\sigma_j^z\right]+\\
  &-\frac{1}{2}\frac{1+\gamma}{2}\sigma_{L-1}^x\mathbb{P}^x_{1,L}
 \end{split}
\end{equation}
while, if we fix the first spin-$x$ eigenvalue to $+1$ and the last to $-1$, the Hamiltonian is
\begin{equation}
 \begin{split}
  H^{+-}_{\gamma h} &=-\frac{1}{2}\frac{1+\gamma}{2}\mathbb{P}^x_{1,1}\sigma_2^x-\frac{1}{2}\sum_{j=2}^{L-2}\left[\frac{1+\gamma}{2}\sigma_j^x\sigma_{j+1}^x+\frac{1-\gamma}{2}\sigma_j^y\sigma_{j+1}^y+h\sigma_j^z\right]+\\
  &+\frac{1}{2}\frac{1+\gamma}{2}\sigma_{L-1}^x\mathbb{P}^x_{-1,L}
 \end{split}
\end{equation}

Therefore, the previous Hamiltonians are exactly simulable with DMRG by means of a lattice of $L-2$ site, where the first is indexed by $j=2$, and the last by $j=L-1$; sites 1 and $L$, appearing in the Hamiltonian just by means of projectors, are automatically implemented in this picture, and the first and the last term are simply writable as $\sigma_2^x$, $\sigma_{L-1}^x$, with the correct coefficients. We stress that, however, the presence of this single-spin terms breaks the quadraticity of the Hamiltonian, and therefore the applicability of the method of appendix \ref{peschel_method} for the computation of REE's, and one has therefore to rely on more complicated methods, as DMRG.


\backmatter


\begin{thebibliography}{99}
 \bibitem{Affleck1986} I. Affleck, Phys. Rev. Lett. {\bf 56}, 746 (1986).
 \bibitem{Affleck1989} I. Affleck, J. Phys.: Cond. Matt. {\bf 1}, 3047 (1989).
 \bibitem{AffleckBonner1990} I. Affleck, J.C. Bonner, Phys. Rev. B {\bf 42}, 954 (1990).
 \bibitem{AffleckLaflorencieSorensen2009} I. Affleck, N. Laflorencie, E.S. S{\o}rensen, J. Phys. A: Math. Theor. {\bf 42}, 504009 (2009).
 \bibitem{AffleckLudwig1991} I. Affleck, A.W.W. Ludwig, Phys. Rev. Lett. {\bf 67}, 161 (1991).
 \bibitem{AlcarazBerganzaSierra2011} F.C. Alcaraz, M.I. Berganza, G. Sierra, Phys. Rev. Lett. {\bf 106}, 201601 (2011).
 \bibitem{AlcarazMoreo1992} F.C. Alcaraz, A. Moreo, Phys. Rev. B {\bf 46}, 2896 (1992).
 \bibitem{AFOV2008} L. Amico, R. Fazio, A. Osterloh, V. Vedral, Rev. Mod. Phys. {\bf 80}, 517 (2008).
 \bibitem{Annett2004} J.F. Annett, {\it Superconductivity, Superfluids and Condensates}, Oxford University Press (2004).
 \bibitem{ArdonneSierra2010} E. Ardonne, G. Sierra, J. Phys. A: Math. Theor. {\bf 43} 505402 (2010).
 \bibitem{AshcroftMermin1976} N.W. Ashcroft, N.D. Mermin, {\it Solid State Physics}, Holt, Rinehart and Winston (1976).
 \bibitem{Auerbach1994} A. Auerbach, \emph{Interacting Electrons and Quantum Magnetism}, Springer-Verlag (1994).
 \bibitem{BCDDEBEO2010} L. Barbiero, M. Casadei, M. Dalmonte, C. Degli Esposti Boschi, E. Ercolessi, F. Ortolani, Phys. Rev. B {\bf 81}, 224512 (2010).
 \bibitem{BDRS1993} T. Barnes, E. Dagotto, J. Riera, E.S. Swanson, Phys. Rev. B {\bf 47}, 3196 (1993).
 \bibitem{BelavinPolyakovZamolodchikov1984} A.A. Belavin, A.M. Polyakov, A.B. Zamolodchikov, Nuclear Physics {\bf B241}, 333 (1984).
 \bibitem{BerganzaAlcarazSierra2012} M.I. Berganza, F.C. Alcaraz, G. Sierra, J. Stat. Mech. (2012), P01016.
 \bibitem{Bilstein2000} U. Bilstein, J. Phys. A: Math. Gen. {\bf 33}, 4437 (2000).
 \bibitem{BilsteinWehefritz1999} U. Bilstein, B. Wehefritz, J. Phys. A: Math. Gen. {\bf 32}, 191 (1999).
 \bibitem{BlochDalibardZwerger2008} I. Bloch, J. Dalibard, W. Zwerger, Rev. Mod. Phys. {\bf 80}, 885 (2008).
 \bibitem{BloteCardyNightingale1986} H.W.J. Bl\"ote, J.L. Cardy, M.P. Nightingale, Phys. Rev. Lett. {\bf 56}, 742 (1986).
 \bibitem{BotetJullienKolb1983} R. Botet, R. Jullien, M. Kolb, Phys. Rev. B {\bf 28}, 3914 (1983).
 \bibitem{CCEN2010} P. Calabrese, M. Campostrini, F. Essler, B. Nienhuis, Phys. Rev. Lett. {\bf 104}, 095701 (2010).
 \bibitem{CalabreseCardy2004} P. Calabrese, J. Cardy, J. Stat. Mech. (2004), P06002.
 \bibitem{CalabreseCardy2009} P. Calabrese, J. Cardy, J. Phys. A {\bf 42}, 504005 (2009).
 \bibitem{CalabreseCardyDoyon2009} P. Calabrese, J. Cardy, B. Doyon, J. Phys. A: Math. Theor. {\bf 42}, 500301 (2009).
 \bibitem{CalabreseEssler2010} P. Calabrese, F.H.L. Essler, J. Stat. Mech. (2010), P08029.
 \bibitem{CalabreseLefevre2008} P. Calabrese, A. Lefevre, Phys. Rev. A {\bf 78}, 032329 (2008).
 \bibitem{Cardy1986} J.L. Cardy, Nucl. Phys. {\bf B275}, 200 (1986).
 \bibitem{Cardy1989} J.L. Cardy, Nucl. Phys. {\bf B324}, 581 (1989).
 \bibitem{CardyCalabrese2010} J. Cardy, P. Calabrese, J. Stat. Mech. (2010), P04023.
 \bibitem{CohenTannoudjiDiuLaloe1977} C. Cohen-Tannoudji, B. Diu, F. Laloe, {\it Quantum Mechanics}, Wiley (1977).
 \bibitem{CreffieldHauslerMacDonald2001} C.E. Creffield, W. H\"ausler, A.H. MacDonald, Europhys. Lett. {\bf 53}, 221 (2001).
 \bibitem{Csontos2010} D. Csontos, Nature {\bf 464}, 175 (2010).
 \bibitem{DRCMNAO1994} E. Dagotto, J. Riera, Y.C. Chen, A. Moreo, A. Nazarenko, F. Alcaraz, F. Ortolani, Phys. Rev. B {\bf 49}, 3548 (1994).
 \bibitem{DDDBO2011} M. Dalmonte, M. Di Dio, L. Barbiero, F. Ortolani, Phys. Rev. B {\bf 83}, 155110 (2011).
 \bibitem{DEMOV2012} M. Dalmonte, E. Ercolessi, M. Mattioli, F. Ortolani, D. Vodola, Eur. Phys. J. Special Topics {\bf 217}, 13-27 (2013).
 \bibitem{DalmonteErcolessiTaddia2011} M. Dalmonte, E. Ercolessi, L. Taddia, Phys. Rev. B {\bf 84}, 085110 (2011).
 \bibitem{DalmonteErcolessiTaddia2012} M. Dalmonte, E. Ercolessi, L. Taddia, Phys. Rev. B {\bf 85}, 165112 (2012).
 \bibitem{DalmontePupilloZoller2010} M. Dalmonte, G. Pupillo, P. Zoller, Phys. Rev. Lett. {\bf 105}, 140401 (2010).
 \bibitem{DalmonteZollerPupillo2011} M. Dalmonte, P. Zoller, G. Pupillo, Phys. Rev. Lett. {\bf 107}, 163202 (2011).
 \bibitem{DegliEspostiBoschiOrtolani2004} C. Degli Esposti Boschi, F. Ortolani, Eur. J. Phys. B {\bf 41}, 503 (2004).
 \bibitem{DiFrancescoMathieuSenechal1997} P. Di Francesco, P. Mathieu, D. S\'en\'echal, \emph{Conformal Field Theory}, Springer (1997).
 \bibitem{DijkgraafVerlindeVerlinde1988} R. Dijkgraaf, E. Verlinde, H. Verlinde, Comm. Math. Phys. {\it 115}, 649 (1988).
 \bibitem{EinsteinPodolskyRosen1935} A. Einstein, B. Podolsky, N. Rosen, Phys. Rev. {\bf 47}, 777 (1935).
 \bibitem{EisertCramerPlenio2010} J. Eisert, M. Cramer, M.B. Plenio, Rev. Mod. Phys. {\bf 82}, 277 (2010).
 \bibitem{EloyXavier2012} D. Eloy, J.C. Xavier, Phys. Rev. B {\bf 86}, 064421 (2012).
 \bibitem{EEFR2011} E. Ercolessi, S. Evangelisti, F. Franchini, F. Ravanini, Phys. Rev. B {\bf 83}, 012402 (2011).
 \bibitem{ErikssonJohannesson2011} E. Eriksson, H. Johannesson, J. Stat. Mech. (2011), P02008.
 \bibitem{EFGKK2005} F.H.L. Essler, H. Frahm, F. G\"ohmann, A. Kl\"umper, V.E. Korepin, \emph{The One-Dimensional Hubbard Model}, Cambridge University Press (2005).
 \bibitem{EsslerLauchliCalabrese2012} F.H.L. Essler, A.M. L\"auchli, P. Calabrese,  Phys. Rev. Lett. {\bf 110}, 115701 (2013).
 \bibitem{FagottiCalabrese2011} M. Fagotti, P. Calabrese, J. Stat. Mech. (2011), P01017.
 \bibitem{FanoOrtolaniZiosi1998} G. Fano, F. Ortolani, L. Ziosi, J. Chem. Phys. {\bf 108}, 9246 (1998).
 \bibitem{Franchini2011} F. Franchini, {\it Notes on Bethe Ansatz Techniques}, http://people.sissa.it/\verb1~1ffranchi/BAClass.html/ (2011).
 \bibitem{FIJK2007} F. Franchini, A.R. Its, B.-Q. Jin,  V.E. Korepin, J. Phys. A: Math. Theor. 40 8467 (2007).
 \bibitem{Giamarchi2003} T. Giamarchi, \emph{Quantum Physics in One Dimension}, Oxford University Press (2003).
 \bibitem{GiamarchiShastry1995} T. Giamarchi, B.S. Shastry, Phys. Rev. B {\bf 51}, 10915 (1995).
 \bibitem{GMDADSI2012} S.M. Giampaolo, S. Montangero, F. Dell'Anno, S. De Siena, F. Illuminati, arXiv:1208.0735.
 \bibitem{GogolinNersesyanTsvelik1998} A.O. Gogolin, A.A. Nersesyan, A.M. Tsvelik, {\it Bosonization and Strongly Correlated Systems}, Cambridge University Press (1998).
 \bibitem{Haldane1981} F.D.M. Haldane, Phys. Rev. Lett. {\bf 47}, 1840 (1981).
 \bibitem{Haldane1982} F.D.M. Haldane, Bull. Am. Phys. Soc. {\bf 27}, 181 (1982).
 \bibitem{Haldane1983a} F.D.M. Haldane, Phys. Lett. A {\bf 93}, 464 (1983).
 \bibitem{Haldane1983b} F.D.M. Haldane, Phys. Rev. Lett. {\bf 50} 1153 (1983).
 \bibitem{HWHM1996} K. Hallberg, X.Q.G. Wang, P. Horsch, A. Moreo, Phys. Rev. Lett. {\bf 76}, 4955 (1996).
 \bibitem{HHMDRGDPN2010} E. Haller, R. Hart, M.J. Mark, J.G. Danzl, L. Reichs\"ollner, M. Gustavsson, M. Dalmonte, G. Pupillo, H.-C. N\"agerl, Nature {\bf 466}, 597 (2010).
 \bibitem{Hohenberg1967} P.C. Hohenberg, Phys. Rev. {\bf 158}, 383 (1967).
 \bibitem{HolzheyLarsenWilczek1994} C. Holzhey, F. Larsen, F. Wilczek, Nucl. Phys. B {\bf 424}, 443 (1994).
 \bibitem{Hubbard1963} J. Hubbard, Proc. Roy. Soc. Lond. {\bf 276 (1365)}, 238 (1963).
 \bibitem{IKET1995} S. Itoh, K. Kakurai, Y. Endoh, H. Tanaka, Physica B: Cond. Matt. {\bf 213}, 161 (1995).
 \bibitem{ItzyksonDrouffe1989} C. Itzykson, J.-M. Drouffe, {\it Statistical Field Theory}, Cambridge University Press (1989).
 \bibitem{JinKorepin2004} B.-Q. Jin, V.E. Korepin, Jour. Stat. Phys. {\bf 116}, 79 (2004).
 \bibitem{KCKIEFLDM2010} K. Kim, M.-S. Chang, S. Korenblit, R. Islam, E.E. Edwards, J.K. Freericks, G.-D. Lin, L.-M. Duan, C. Monroe, Nature {\bf 465}, 590 (2010).
 \bibitem{KinoshitaWengerWeiss2004} T. Kinoshita, T. Wenger, D.S. Weiss, Science {\bf 305}, 1125 (2004).
 \bibitem{LSCA2006} N. Laflorencie, E.S. S{\o}rensen, M.-S. Chang, I. Affleck, Phys. Rev. Lett. {\bf 96}, 100603 (2006).
 \bibitem{LiHaldane2008} H. Li, F.D.M. Haldane, Phys. Rev. Lett. {\bf 101}, 010504 (2008). 
 \bibitem{LiebSchultzMattis1961} E. Lieb, T. Schultz, D. Mattis, Ann. Phys. {\bf 16}, 407 (1961).
 \bibitem{LMSLP2009} T. Lahaye, C. Menotti, L. Santos, M. Lewenstein, T. Pfau, Rep. Prog. Phys. {\bf 72}, 126401 (2009).
 \bibitem{MajumdarGhosh1969} C.K. Majumdar, D.K. Ghosh, J. Math. Phys. {\bf 10}, 1388 (1969).
 \bibitem{MerminWagner1966} N.D. Mermin, H. Wagner, Phys. Rev. Lett. {\bf 17}, 1133 (1966).
 \bibitem{Montorsi1992} A. Montorsi (ed.), \emph{The Hubbard Model}, World Scientific (1992).
 \bibitem{MorandiNapoliErcolessi2001} G. Morandi, F. Napoli, E. Ercolessi, {\it Statistical Mechanics: An Intermediate Course}, World Scientific (2001).
 \bibitem{Mussardo2007} G. Mussardo, {\it Il Modello di Ising}, Bollati-Boringhieri (2007).
 \bibitem{MutkaPayenMolini1993}H. Mutka, C. Payen, P. Molini, Europhys. Lett. {\bf 21}, 623 (1993).
 \bibitem{NegeleOrland1998} J.W. Negele, H. Orland, {\it Quantum Many-Particle Systems}, Westview Press (1998).
 \bibitem{Nielsen2005} M.A. Nielsen, {\it The Fermionic canonical commutation relations and the Jordan-Wigner transform}, http://michaelnielsen.org/blog/complete-notes-on-fermions-and-the-jordan-wigner-transform/ (2005).
 \bibitem{NielsenChuang2000} M.A. Nielsen, I.L. Chuang, \emph{Quantum Computation and Quantum Information}, Cambridge University Press (2000).
 \bibitem{Nishimoto2011} S. Nishimoto, Phys. Rev. B {\bf 84}, 195108 (2011).
 \bibitem{Peschel2003} I. Peschel, J. Phys. A: Math. Gen. {\bf 36} (2003), L205.
 \bibitem{PitaevskiiStringari2003} L. Pitaevskii, S. Stringari, {\it Bose-Einstein Condensation}, Oxford University Press (2003).
 \bibitem{PrangeGirvin1987} R.E. Prange, S.M. Girvin (ed.), {\it The Quantum Hall Effect}, Springer-Verlag (1987).
 \bibitem{RLSLH2012} S. Rachel, N. Laflorencie, H.F. Song, K. Le Hur, Phys. Rev. Lett. {\bf 108}, 116401 (2012).
 \bibitem{Renyi2007} A. R\'enyi, \emph{Probability Theory}, Dover (2007).
 \bibitem{RoscildeDegliEspostiBoschiDalmonte2012} T. Roscilde, C. Degli Esposti Boschi, M. Dalmonte, Europhys. Lett. {\bf 97}, 23002 (2012).
 \bibitem{Rothe2012} H. Rothe, {\it Lattice Gauge Theories: An Introduction} (4th ed.), World Scientific (2012).
 \bibitem{Sachdev2011} S. Sachdev, {\it Quantum Phase Transitions}, Cambridge University Press (2011).
 \bibitem{Saleur1998} H. Saleur in A. Comtet, T. Jolicoeur, S. Ouvry, F. David (ed.), {\it Topological aspects of low dimensional systems}, Springer (1998); arXiv:cond-mat/9812110.
 \bibitem{SaleurBauer1989} H. Saleur, M. Bauer, Nucl. Phys. {\bf B320}, 591 (1989).
 \bibitem{Schollwock2005} U. Schollw\"ock, Rev. Mod. Phys. {\bf 77}, 259 (2005).
 \bibitem{SchollwockWhite2006} U. Schollw\"ock, S.R. White in G.G. Batrouni and D. Poilblanc (eds.), {\it Effective models for low-dimensional strongly correlated systems}, AIP Conference Proceedings (2006); arXiv:cond-mat/0606018.
 \bibitem{Schrieffer2007} J.R. Schrieffer, {\it Handbook of High-Temperature Superconductivity}, Springer (2007).
 \bibitem{Schulz1986} H.J. Schulz, Phys. Rev. B {\bf 34}, 6372 (1986).
 \bibitem{SRFKLLH2012} H.F. Song, S. Rachel, C. Flindt, I. Klich, N. Laflorencie, K. Le Hur, Phys. Rev. B {\bf 85}, 035409 (2012).
 \bibitem{SongRachelLeHur2010} H.F. Song, S. Rachel, K. Le Hur, Phys. Rev. B {\bf 82}, 012405 (2010).
 \bibitem{SwingleMcMinisTubman2012} B. Swingle, J. McMinis, N.M. Tubman, Phys. Rev. B {\bf 87}, 235112 (2013).
 \bibitem{TXAS2013} L. Taddia, J.C. Xavier, F.C. Alcaraz, G. Sierra, Phys. Rev. B {\bf 88}, 075112 (2013).
 \bibitem{Takahashi1999} M. Takahashi, \emph{Thermodynamics of One-Dimensional Solvable Models}, Cambridge University Press (1999).
 \bibitem{VerstraetePorrasCirac2004} F. Verstraete, D. Porras, J.I. Cirac, Phys. Rev. Lett. {\bf 93}, 227204 (2004).
 \bibitem{VLRK2003} G. Vidal, J.I. Latorre, E. Rico, A. Kitaev, Phys. Rev. Lett. {\bf 90}, 227902 (2003).
 \bibitem{Vojta2003} M. Vojta, Rep. Prog. Phys. {\bf 66}, 2069 (2003).
 \bibitem{White1992} S.R. White, Phys. Rev. Lett. {\bf 69}, 2863 (1992).
 \bibitem{White1993} S.R. White, Phys. Rev. B {\bf 48}, 10345 (1993).
 \bibitem{WhiteHuse1993} S.R. White, D. Huse, Phys. Rev. B {\bf 48}, 3844 (1993).
 \bibitem{Wilson1975} K.G. Wilson, Rev. Mod. Phys. {\bf 47}, 773 (1975).
 \bibitem{Wootters1998} W.K. Wootters, Phys. Rev. Lett. {\bf 80}, 2245 (1998).
 \bibitem{Xavier2010} J.C. Xavier, Phys. Rev. B {\bf 81}, 224404 (2010).
 \bibitem{XavierAlcaraz2011a} J.C. Xavier, F.C. Alcaraz, Phys. Rev. B {\bf 83}, 214425 (2011).
 \bibitem{XavierAlcaraz2011b} J.C. Xavier, F.C. Alcaraz, Phys. Rev. B {\bf 84}, 094410 (2011).
 \bibitem{XavierAlcaraz2012} J.C. Xavier, F.C. Alcaraz, Phys. Rev. B {\bf 85}, 024418 (2012).
 \bibitem{YouLiGu2007} W.-L. You, Y.-W. Li, S.-J. Gu, Phys. Rev. E {\bf 76}, 022101 (2007).
 \bibitem{ZanardiPaunkovic2006} P. Zanardi, N. Paunkovi\'c, Phys. Rev. E {\bf 74}, 031123 (2006) 
 \bibitem{ZBFS2006} H.-Q. Zhou, T. Barthel, J.O. Fj\ae restad, U. Schollw\"ock, Phys. Rev. A {\bf 74}, 050305 (2006).
\end{thebibliography}
\end{document}